\documentclass[11pt,oneside]{book}
\usepackage{microtype}
\usepackage[T1]{fontenc}
\usepackage[utf8]{inputenc}

\usepackage{autobreak}

\usepackage{epigraph}
\usepackage{datetime}
\newdateformat{monthyeardate}{\monthname[\THEMONTH], \THEYEAR}

\usepackage[usenames,dvipsnames]{xcolor}

\usepackage{graphicx}
\graphicspath{ {Figures/} }

\usepackage{enumitem}

\usepackage[colorinlistoftodos]{todonotes}
\setuptodonotes{fancyline, color=blue!20, figcolor=white}

\usepackage{lettrine}

\usepackage{mathtools}
\usepackage{bm}
\usepackage{amsmath,amsfonts,amssymb}
\allowdisplaybreaks

\usepackage{amsthm}
\newtheorem{theorem}{Theorem}[chapter]
\newtheorem{corollary}{Corollary}[chapter]
\newtheorem{conjecture}{Conjecture}[chapter]

\theoremstyle{definition}

\newtheorem{example}{Example}[chapter]
\AtEndEnvironment{example}{\null\hfill\qedsymbol}

\renewcommand{\qedsymbol}{$\blacksquare$}

\usepackage[a4paper]{geometry}

\usepackage{caption}
\usepackage{subcaption}

\usepackage{mfirstuc}

\usepackage{emptypage}
\usepackage{fancyhdr}
\usepackage{tocloft}

\usepackage{titlesec}

\newcommand{\overbar}[1]{{\mkern 1.5mu\overline{\mkern-1.5mu#1\mkern-1.5mu}\mkern 1.5mu}}

\titleformat{\chapter}[display]
{\normalfont\large}{
	\textsc{\chaptertitlename\ \thechapter}\centering
}{0pt}
{\Huge\bfseries\centering}

\titleformat{\section}[display]
{\normalfont}{\textsc{Section \thesection}}{0pt}{\Large\bfseries}

\titleformat{\subsection}
{\normalfont}{\thesubsection}{1em}{\large\bfseries}

\usepackage[
backend=bibtex,
style=numeric-comp,
sorting=none,
backref=true,
maxbibnames=99
]{biblatex}
\DefineBibliographyStrings{english}{%
	backrefpage = {page},
	backrefpages = {pages},
}

\usepackage{setspace}
\usepackage{tcolorbox}
\usepackage{array}
\usepackage{colortbl}
\usepackage{booktabs}

\usepackage{hyperref}
\hypersetup{
	colorlinks=true,
	urlcolor=blue,
	linkcolor=blue,
	citecolor=red,
	pdfdisplaydoctitle=true,
	pdfstartview={XYZ null null 1.00},
}

\usepackage[linesnumbered,algochapter,noline,ruled]{algorithm2e}
\SetArgSty{}
\DontPrintSemicolon

\usepackage[capitalise]{cleveref}
\crefname{conjecture}{Conjecture}{Conjectures}

\usepackage{ifthen}

\usepackage{xparse}

\newenvironment{axioms}[2][]{\begin{enumerate}[label=(#2\arabic*#1)]}
	{\end{enumerate}}
\crefalias{axiom}{enumi}
\crefname{axiom}{axiom}{axioms}
\Crefname{axiom}{Axiom}{Axioms}

\crefalias{step}{enumi}
\crefname{step}{step}{steps}
\Crefname{step}{Step}{Steps}

\DeclareMathAlphabet{\mathbbb}{U}{bbold}{m}{n}

\usepackage[htt]{hyphenat}

\usepackage{tikz-cd}

\usepackage{stackrel}

\newcommand{\lrangle}[1]{\ensuremath{\langle{#1}\rangle}}
\newcommand{\ket}[1]{\ensuremath{|{#1}\rangle}}
\DeclareMathOperator{\Tr}{Tr}
\DeclareMathOperator{\sign}{sign}

\DeclareMathOperator{\NC}{NC}

\DeclareMathOperator{\integer}{\mathbb{Z}}
\DeclareMathOperator{\real}{\mathbb{R}}
\DeclareMathOperator{\complex}{\mathbb{C}}
\DeclareMathOperator{\quarternion}{\mathbb{H}}
\DeclareMathOperator{\rational}{\mathbb{Q}}
\DeclareMathOperator{\projective}{\mathbb{P}}
\DeclareMathOperator{\CP}{\mathbb{CP}}
\DeclareMathOperator{\RP}{\mathbb{RP}}

\DeclareMathOperator{\symbolGL}{GL}
\newcommand{\GL}[2]{\symbolGL_{#1}(#2)}
\DeclareMathOperator{\symbolSL}{SL}
\newcommand{\SL}[2]{\symbolSL_{#1}(#2)}
\DeclareMathOperator{\symbolSU}{SU}
\newcommand{\SU}[1]{\symbolSU(#1)}
\DeclareMathOperator{\symbolU}{U}
\newcommand{\U}[1]{\symbolU(#1)}
\DeclareMathOperator{\symbolSpin}{Spin}
\newcommand{\Spin}[2]{\symbolSpin(#1,#2)}
\DeclareMathOperator{\symbolSO}{SO}
\newcommand{\SO}[1]{\symbolSO(#1)}

\DeclareMathOperator{\symbolMat}{Mat}
\newcommand{\Mat}[3]{\symbolMat_{#1,#2}(#3)}
\newcommand{\tpMat}[2]{\symbolMat_{#1,#2}^{\scriptscriptstyle>0}}
\newcommand{\tpTwistedMat}[2]{\symbolMat_{#1,#2}^{{\scriptscriptstyle>0},\tau}}

\DeclareMathOperator{\zero}{\mathcal{Z}}
\newcommand{\Tree}{\mathcal{T}}
\newcommand{\Forest}{\mathcal{F}}
\newcommand{\Graph}{\mathcal{G}}
\DeclareMathOperator{\tree}{tree}
\DeclareMathOperator{\forest}{forest}
\DeclareMathOperator{\Vertices}{\mathcal{V}}
\DeclareMathOperator{\Edges}{\mathcal{E}}
\DeclareMathOperator{\Trees}{Trees}

\newcommand{\tp}[1]{{#1}_{>0}}

\newcommand{\tpEnumerate}[2]{{#1}_{#2,>0}}
\newcommand{\tnn}[1]{{#1}_{\ge0}}

\newcommand{\tnnEnumerate}[2]{{#1}_{#2,\ge0}}

\DeclareMathOperator{\symbolG}{Gr}
\newcommand{\G}[2]{{\symbolG}_{#1,#2}}
\newcommand{\GReal}[2]{\G{#1}{#2}(\real)}
\newcommand{\tpG}[2]{{\symbolG}_{#1,#2}^{>0}}
\newcommand{\tnnG}[2]{{\symbolG}_{#1,#2}^{\ge0}}
\newcommand{\tnzG}[2]{{\symbolG}_{#1,#2}^{\ne0}}

\DeclareMathOperator{\symbolPos}{\Pi}
\newcommand{\PosCirc}[1]{\mathring{\symbolPos}_{#1}}
\newcommand{\Pos}[1]{\symbolPos_{#1}}
\newcommand{\tpPos}[1]{\tpEnumerate{\symbolPos}{#1}}
\newcommand{\tnnPos}[1]{\tnnEnumerate{\symbolPos}{#1}}

\DeclareMathOperator{\symbolOG}{OGr}
\newcommand{\OG}[2]{{\symbolOG}_{#1,#2}}
\newcommand{\OGReal}[2]{\OG{#1}{#2}(\real)}
\newcommand{\tpOG}[2]{{\symbolOG}_{#1,#2}^{>0}}
\newcommand{\tnnOG}[2]{{\symbolOG}_{#1,#2}^{\ge0}}
\newcommand{\tnzOG}[2]{{\symbolOG}_{#1,#2}^{\ne0}}

\DeclareMathOperator{\symbolOrth}{O}
\newcommand{\OrthCirc}[1]{\mathring{\symbolOrth}_{#1}}
\newcommand{\Orth}[1]{\symbolOrth_{#1}}
\newcommand{\tpOrth}[1]{\tpEnumerate{\symbolOrth}{#1}}
\newcommand{\tnnOrth}[1]{\tnnEnumerate{\symbolOrth}{#1}}

\DeclareMathOperator{\symbolOnShell}{\mathbb{O}}
\DeclareMathOperator{\symbolMandelstam}{\mathbb{K}}
\DeclareMathOperator{\symbolLittleGroup}{\mathbb{L}}
\DeclareMathOperator{\symbolGram}{\mathbb{G}}
\newcommand{\OnShellComplex}[2]{\symbolOnShell^{(#1)}_{#2}(\complex)}
\newcommand{\OnShell}[2]{\symbolOnShell^{(#1)}_{#2}}
\newcommand{\LittleGroup}[2]{\symbolLittleGroup^{(#1)}_{#2}}
\newcommand{\Gram}[2]{\symbolGram^{(#1)}_{#2}}
\newcommand{\Mandelstam}[1]{\symbolMandelstam_{#1}}

\newcommand{\mapOnShell}[2]{\mathbbb{o}^{(#1)}_{#2}}
\newcommand{\mapGram}[2]{\mathbbb{g}^{(#1)}_{#2}}
\newcommand{\idealGram}[2]{\mathfrak{g}^{(#1)}_{#2}}
\newcommand{\mapMandelstam}[2]{\mathbbb{k}^{(#1)}_{#2}}
\newcommand{\idealRestrict}[2]{\mathfrak{r}^{(#1)}_{#2}}
\newcommand{\idealRestrictPrime}[2]{\tilde{\mathfrak{r}}^{(#1)}_{#2}}

\newcommand{\setMandelstam}[1]{\mathcal{X}_{#1}}
\newcommand{\setRestrict}[2]{\mathcal{X}^{(#1)}_{#2}}

\NewDocumentCommand{\formOnShell}{O{} O{} m}{\Omega\ifthenelse{\equal{#1}{}}{}{^{(#1)}}_{#3\ifthenelse{\equal{#2}{}}{}{,#2}}}
\NewDocumentCommand{\formLittleGroup}{O{} O{} m}{\omega\ifthenelse{\equal{#1}{}}{}{^{(#1)}}_{#3\ifthenelse{\equal{#2}{}}{}{,#2}}}
\NewDocumentCommand{\formGram}{O{} O{} m}{\nu\ifthenelse{\equal{#1}{}}{}{^{(#1)}}_{#3\ifthenelse{\equal{#2}{}}{}{,#2}}}

\DeclareMathOperator{\symbolConf}{Conf}
\newcommand{\Conf}[2]{\symbolConf(#1,#2)}
\newcommand{\tpConf}[2]{\tp{\Conf{#1}{#2}}}

\DeclareMathOperator{\symbolAbhy}{\mathcal{A}}
\DeclareMathOperator{\symbolModuli}{\mathfrak{M}}
\DeclareMathOperator{\symbolMom}{\mathcal{M}}
\DeclareMathOperator{\symbolMomMap}{\Phi_{\symbolMom}}
\DeclareMathOperator{\symbolMomTwistorMap}{\Psi_{\symbolMom}}
\DeclareMathOperator{\symbolMomHatMap}{\Xi_{\symbolMom}}
\DeclareMathOperator{\symbolOMom}{\mathcal{O}}
\DeclareMathOperator{\symbolOMomMap}{\Phi_{\symbolOMom}}
\DeclareMathOperator{\symbolOMomTwistorMap}{\Psi_{\symbolOMom}}
\DeclareMathOperator{\symbolOMomHatMap}{\Xi_{\symbolOMom}}
\DeclareMathOperator{\symbolV}{\mathcal{V}}
\DeclareMathOperator{\symbolW}{\mathcal{W}}
\DeclareMathOperator{\symbolAmp}{\mathcal{W}}

\newcommand{\Amp}[2]{\symbolAmp_{#2,#1}}

\newcommand{\Abhy}[1]{\symbolAbhy_{#1}}
\newcommand{\tpAbhy}[1]{\symbolAbhy_{#1}^\circ}

\newcommand{\Moduli}[2]{\symbolModuli_{#1,#2}}
\newcommand{\ModuliReal}[2]{\symbolModuli_{#1,#2}(\real)}
\newcommand{\tpModuli}[2]{\symbolModuli_{#1,#2}^+}
\newcommand{\ModuliCompact}[2]{\overline{\symbolModuli}_{#1,#2}(\real)}
\newcommand{\tnnModuli}[2]{\symbolModuli_{#1,#2}^{\prime}(\real)}

\newcommand{\MomMap}[1]{\symbolMomMap(#1)}
\newcommand{\Mom}[2]{\MomMap{\G{#1}{#2}}}
\newcommand{\MomReal}[2]{\MomMap{\GReal{#1}{#2}}}
\newcommand{\tpMom}[2]{\symbolMom_{#2,#1}^\circ}
\newcommand{\tnnMom}[2]{\symbolMom_{#2,#1}}

\newcommand{\MomHatMap}[1]{\symbolMomHatMap(#1)}

\newcommand{\tpMomHat}[2]{\widehat{\symbolMom}{}_{#2,#1}^\circ}
\newcommand{\tnnMomHat}[2]{\widehat{\symbolMom}_{#2,#1}}

\newcommand{\MomV}[2]{\symbolV_{#2,#1}}
\newcommand{\MomVReal}[2]{\symbolV_{#2,#1}(\real)}
\newcommand{\MomW}[2]{\symbolW_{#2,#1}}

\newcommand{\MomHatV}[2]{\widehat{\symbolV}_{#2,#1}}
\newcommand{\MomHatVReal}[2]{\widehat{\symbolV}_{#2,#1}(\real)}
\newcommand{\MomHatW}[2]{\widehat{\symbolW}_{#2,#1}}

\newcommand{\OMomMap}[1]{\symbolMomMap(#1)}
\newcommand{\tpOMom}[1]{\symbolOMom_{#1}^\circ}
\newcommand{\tnnOMom}[1]{\symbolOMom_{#1}}

\newcommand{\OMomHatMap}[1]{\symbolMomHatMap(#1)}
\newcommand{\tpOMomHat}[1]{\widehat{\symbolOMom}{}_{#1}^\circ}
\newcommand{\tnnOMomHat}[1]{\widehat{\symbolOMom}_{#1}}

\newcommand{\OMomV}[1]{\symbolV_{#1}}
\newcommand{\OMomVReal}[1]{\symbolV_{#1}(\real)}
\newcommand{\OMomW}[1]{\symbolW_{#1}}

\newcommand{\OMomHatV}[1]{\widehat{\symbolV}_{#1}}
\newcommand{\OMomHatVReal}[1]{\widehat{\symbolV}_{#1}(\real)}
\newcommand{\OMomHatW}[1]{\widehat{\symbolW}_{#1}}

\renewcommand{\vec}[1]{\bm{#1}}

\DeclareMathOperator{\var}{var}
\DeclareMathOperator{\rank}{rank}
\newcommand{\precdot}{\prec\mathrel{\mkern-5mu}\mathrel{\cdot}}
\DeclareMathOperator{\Res}{Res}
\DeclareMathOperator{\vol}{vol}

\newcommand{\Mathematica}{\textsc{Mathematica}}
\newcommand{\positroids}{\texttt{positroids}}
\newcommand{\orthitroids}{\texttt{orthitroids}}
\newcommand{\amplituhedronBoundaries}{\texttt{amplituhedronBoundaries}}

\newcommand{\checktoopen}{
	\if@openright\cleardoublepage\else\clearpage\fi
	\ifdef{\phantomsection}{\phantomsection}{}
}

\makeatletter

\def\displaytitle#1{\gdef\@displaytitle{#1}}
\def\subtitle#1{\gdef\@subtitle{#1}}
\DeclareDocumentCommand{\author}{O{} m}{\gdef\@authorURL{#1}\gdef\@author{#2}}
\DeclareDocumentCommand{\supervisor}{O{} m}{\gdef\@supervisorURL{#1}\gdef\@supervisor{#2}}
\DeclareDocumentCommand{\university}{O{} m}{\gdef\@universityURL{#1}\gdef\@university{#2}}
\DeclareDocumentCommand{\school}{O{} m}{\gdef\@schoolURL{#1}\gdef\@school{#2}}
\DeclareDocumentCommand{\group}{O{} m}{\gdef\@groupURL{#1}\gdef\@group{#2}}
\def\degree#1{\gdef\@degree{#1}}
\def\subject#1{\gdef\@subject{#1}}

\newcommand{\abstractname}{Abstract}
\newenvironment{abstract}{
	\checktoopen
	\addcontentsline{toc}{chapter}{\abstractname}	
	\thispagestyle{plain}
	\markboth{\abstractname}{\abstractname}	
	\null
	\vfill
	\begin{center}
		{\normalsize{\href{\@universityURL}{\@university}}\par}
		\bigskip
		{\huge\itshape\abstractname\par}
		\bigskip
		{\normalsize\@degree{} in \@subject{}\par}
		\bigskip
		{\normalsize\bfseries\@title\par}
		\bigskip
		{\normalsize\href{\@authorURL}{\@author}\par}
		\bigskip
	\end{center}
}{
	\vfill
	\null
}

\newcommand{\declarationname}{Author's Declaration}
\newenvironment{declaration}{
	\checktoopen
	\addcontentsline{toc}{chapter}{\declarationname}
	\thispagestyle{plain}
	\markboth{\declarationname}{\declarationname}	
	\begin{center}{\huge\itshape\declarationname\par}\end{center}
	\bigskip
	I hereby declare that I am the sole author of this dissertation. The work presented herein is based on research conducted within the \@school{} at the \@university, under the supervision of \@supervisor{}.
}{}

\newcommand{\acknowledgementsname}{Acknowledgements}
\newenvironment{acknowledgements}{
	\checktoopen
	\addcontentsline{toc}{chapter}{\acknowledgementsname}
	\thispagestyle{plain}
	\markboth{\acknowledgementsname}{\acknowledgementsname}	
	\begin{center}{\huge\itshape\acknowledgementsname\par}\end{center}
	\bigskip
}{}

\newcommand{\dedicationname}{Dedication}
\newenvironment{dedication}{
	\checktoopen
	\addcontentsline{toc}{chapter}{\dedicationname}
	\thispagestyle{plain}
	\markboth{\dedicationname}{\dedicationname}	
	\null\vfill
	\begin{center}\Large\slshape
}{
	\end{center}
	\vfill\null
}

\makeatother

\displaytitle{Positive Geometries\\for Scattering Amplitudes\\in Momentum Space}
\title{Positive Geometries for Scattering Amplitudes in Momentum Space}
\subtitle{How Positivity Breathes Combinatorial Life}
\author[https://go.herts.ac.uk/robert_moerman-redding]{Robert William Moerman}
\supervisor[https://go.herts.ac.uk/tomasz_lukowski]{Tomasz {\L}ukowski}
\university[https://www.herts.ac.uk/]{University of Hertfordshire} 
\school[https://www.herts.ac.uk/study/schools-of-study/physics-engineering-and-computer-science]{School of Physics, Engineering and Computer Science}
\group[https://www.herts.ac.uk/research/groups-and-units/mathematics-and-mathematical-physics]{Mathematics and Theoretical Physics}
\degree{Doctor of Philosophy}
\subject{Physics}

\makeatletter
\AtBeginDocument{
	\hypersetup{pdftitle=\@title} 
	\hypersetup{pdfauthor=\@author} 
}
\makeatother

\DeclareBibliographyCategory{included}
\nocite{Lukowski:2020bya,Ferro:2020lgp,Damgaard:2020eox,Moerman:2021cjg,Lukowski:2021fkf,Lukowski:2022fwz}
\addtocategory{included}{Lukowski:2020bya}
\addtocategory{included}{Ferro:2020lgp}
\addtocategory{included}{Damgaard:2020eox}
\addtocategory{included}{Moerman:2021cjg}
\addtocategory{included}{Lukowski:2021fkf}
\addtocategory{included}{Lukowski:2022fwz}

\DeclareBibliographyCategory{excluded}
\nocite{Damgaard:2021qbi,Ferro:2021dub}
\addtocategory{excluded}{Damgaard:2021qbi}
\addtocategory{excluded}{Ferro:2021dub}

\newcommand{\nameuse}[1]{%
	\def\do##1{\settoggle{blx@use##1}{#1}}%
	\dolistcsloop{blx@datamodel@names}}

\newcommand{\nameusesave}{%
	\def\do##1{%
		\providetoggle{blx@save@use##1}%
		\iftoggle{blx@use##1}{\toggletrue{blx@save@use##1}}{\togglefalse{blx@save@use##1}}%
	}%
	\dolistcsloop{blx@datamodel@names}}

\newcommand{\nameuserestore}{%
	\def\do##1{%
		\iftoggle{blx@save@use##1}{\toggletrue{blx@use##1}}{\togglefalse{blx@use##1}}%
	}%
	\dolistcsloop{blx@datamodel@names}}

\begin{document}
	
	\frontmatter
	
\makeatletter
\begin{titlepage}
	\begin{center}
		\null
		\vfill
		
		{\Huge\bfseries\@displaytitle\par}
		
		\vspace{1.5cm}
		
		{\LARGE\itshape\@subtitle\par}
		
		\vspace{1.5cm}
		
		{\Large\href{\@authorURL}{\@author}\par}
		
		\vspace{1.5cm}
		
		{\large\itshape A dissertation submitted to the\\\@university{}\\ in partial fulfilment of the requirements for the degree of\\\@degree{}\par}
		
		\vspace{1.5cm}
		
		{\large{\href{\@groupURL}{\@group}}\par}
		{\large{\href{\@schoolURL}{\@school}}\par}
		{\large{\href{\@universityURL}{\@university}}\par}
		
		\vspace{1.5cm}
				
		{\large January, 2023\par}
		
		\vfill
		\null
	\end{center}
\end{titlepage}
\makeatother	
	
	\begin{abstract}
Positive geometries provide a purely geometric point of departure for studying scattering amplitudes in quantum field theory. A positive geometry is a specific semi-algebraic set equipped with a unique rational top form---the canonical form. There are known examples where the canonical form of some positive geometry, defined in some kinematic space, encodes a scattering amplitude in some theory. Remarkably, the boundaries of the positive geometry are in bijection with the physical singularities of the scattering amplitude. The Amplituhedron, discovered by Arkani-Hamed and Trnka, is a prototypical positive geometry. It lives in momentum twistor space and describes tree-level (and the integrands of planar loop-level) scattering amplitudes in maximally supersymmetric Yang-Mills theory. 

In this dissertation, we study three positive geometries defined in on-shell momentum space: the Arkani-Hamed--Bai--He--Yan (ABHY) associahedron, the Momentum Amplituhedron, and the orthogonal Momentum Amplituhedron. Each describes tree-level scattering amplitudes for different theories in different spacetime dimensions. The three positive geometries share a series of interrelations in terms of their boundary posets and canonical forms. We review these relationships in detail, highlighting the author’s contributions. We study their boundary posets, classifying all boundaries and hence all physical singularities at the tree level. We develop new combinatorial results to derive rank-generating functions which enumerate boundaries according to their dimension. These generating functions allow us to prove that the Euler characteristics of the three positive geometries are one. In addition, we discuss methods for manipulating canonical forms using ideas from computational algebraic geometry.
\end{abstract}
	
	
	\begin{declaration}
This dissertation summaries results from the following articles co-authored by me:
\nameusesave
\nameuse{false}
\printbibliography[heading=none,category=included]
\newpage
During the completion of my PhD degree, I also produced the following works which are not included in this dissertation:
\printbibliography[heading=none,category=excluded]
\nameuserestore
\end{declaration}
	
	\begin{acknowledgements}
\epigraph{\itshape``If you want to go fast, go alone. If you want to go far, go together.''}{An African Proverb}

Albert Einstein once said, ``If we knew what we were doing, it would not be called research, would it?'' This quote expresses the idea that research involves a certain level of uncertainty and exploration. The research process is not simply about finding pre-existing answers but a journey of discovery and learning, where the outcome is often unknown. Seldom is such a journey undertaken alone. Indeed, the research presented in this dissertation is in no small part thanks to the help and support of many individuals, without whom I would not have gone as far as I have. 

Firstly, I am indebted to my supervisor, Tomasz {\L}ukowski. You are an exemplary supervisor. Thank you for developing me as a researcher, teaching me to ask good research questions and equipping me with the tools to analyse and answer them. I am grateful for our many discussions on topics relating to research and life. Thank you for your patience and for encouraging me to pursue various opportunities. I hope to continue collaborating with you in the future. I also wish to thank James Collett and Charles Young, members of my supervisory team, for their continual encouragement and always showing interest in my research. 

It has been an honour to be associated with the Mathematics and Theoretical Physics group at the University of Hertfordshire. I am grateful to its staff for stimulating conversations, often at lunch-time, including Luigi Alfonsi, Leron Borsten, Livia Ferro, Tomasz {\L}ukowski, Yann Peresse, Vidas Regelskis, Charles Strickland-Constable, and Charles Young. I am also thankful to the following PhD students for their camaraderie, for sharing their knowledge, and for asking helpful questions during research seminars: Martin Christensen, Tommaso Franzini, Simon Jonsson, and Jonah Stalknecht. 

During the completion of my PhD degree, I have had the privilege of collaborating with multiple researchers on eight research projects, all of which lead to publications. I want to thank my supervisor for the opportunity to work with him on many of these projects. In addition, I am enormously thankful to Livia Ferro and Lauren Williams, two top-tier researchers, for the opportunity to work with and learn from them. I am grateful to Lauren Williams for inviting me to Bristol when she was visiting there. Furthermore, I thank David Damgaard and Jonah Stalknecht, fellow PhD students, for innumerable discussions. Their close collaborations have helped me hone my understanding of various topics.

Part of my PhD degree coincided with the coronavirus pandemic when attending conferences and workshops in person was impossible. During this time, I benefited immensely from the fortnightly online meetings known as the Geometry and Scattering Amplitudes Journal Club. I thank David Damgaard and (later) Henrik Munch for co-organising these meetings with me and Jonah Stalknecht for taking over from me. Moreover, thank you to everyone who attended these meetings and contributed to the energising discussions. 

I am grateful to Erik Panzer for inviting me to Oxford to present my research at the Mathematical Insititute. I very much appreciated the hospitality of the Mathematical Institute and St Peter's College. Furthermore, I am thankful to Hadleigh Frost, Lionel Mason, Erik Panzer, and Atul Sharma for many interesting discussions.

The past three and a half years would not have been possible without the generous support of family, friends, and the community at St Paul's Church in St Albans. There are too many people to thank by name. I am particularly thankful to my parents and sister for their continued encouragement and unwavering love. Moreover, I owe an infinite debt of gratitude to my wife and best friend, Tori. Thank you for always being there for me on good and not-so-good research days. As a Christian, I thank God for this opportunity and for all the above-referenced people.

Finally, I am thankful to Tomasz {\L}ukowski for reviewing my dissertation, and to Leron Borsten and Arthur Lipstein, my PhD examiners, for their questions, comments and corrections.
\end{acknowledgements}
	
	\begin{dedication}
To my wife and best friend, Tori
\end{dedication}
	
	\tableofcontents
	
	\cleardoublepage
	\phantomsection
	\addcontentsline{toc}{chapter}{\listfigurename}
	\listoffigures
	
	
	
	\mainmatter
	
	\chapter{Introduction}
\label{chp:intro}

\epigraph{\itshape``Evolution goes beyond what went before, but because it must embrace what went before, then its very nature is to transcend and include and thus it has an inherent directionality, a secret impulse, toward increasing depth, increasing intrinsic value, increasing consciousness. In order for evolution to move at all, it must move in those directions --- there's no place else for it to go!''}{\citeauthor{wilber2017brief}, \citetitle{wilber2017brief} \cite{wilber2017brief}}

\lettrine{I}{t is difficult to overstate} the success of the \emph{Standard Model (SM)} in its applications to fundamental particle physics. It describes interactions between normal matter’s most primitive pieces, providing a unifying framework for three of nature’s four fundamental forces. Moreover, it predicts a wide variety of phenomena with unparalleled precision. Most notably, it predicted the existence of the Higgs boson \cite{Englert:1964et,Guralnik:1964eu,Higgs:1964pj} 48 years before its discovery in 2012 \cite{ATLAS:2012yve,CMS:2012qbp}. Thus the SM has earned the epithet of ``the most successful scientific theory ever'' \cite{buder2021}.

The SM is an example of a \emph{quantum field theory (QFT)}. Its origins\footnote{For a historical account of the development of QFT, see \cite{weinberg1977search}, the first chapter of \cite{Weinberg:1995mt}, \cite{sep-quantum-field-theory}, and references therein.} date back to \citeauthor{Dirac:1927dy}’s \citeyear{Dirac:1927dy} publication \citetitle{Dirac:1927dy} \cite{Dirac:1927dy}. The theory of quantum fields combines the probabilistic perspective of quantum mechanics with \citeauthor{einstein1905elektrodynamik}’s special theory of relativity \cite{einstein1905elektrodynamik}. The foundations of QFT include the following triumvirate:
\begin{itemize}
	\item \emph{Causality:} Each spacetime point has an associated light cone which designates the region of causal influence. In particular, causally connected events are timelike-separated (i.e. one event is in the past/future light cone of the other). 
	\item \emph{Locality:} Two spacetime points interact via a carrier (i.e.\ a particle or wave) that mediates the interaction.
	\item \emph{Unitarity:} The probabilities of outcomes associated with a quantum event sum to unity.
\end{itemize}
To define a QFT, one specifies a \emph{Lagrangian density}\footnote{While this definition is (believed to be) too restrictive in general, the study of the S-matrix restricts us to QFTs which admit a Lagrangian density.}. The latter depends only on \emph{local operators} as per the locality principle. Local operators refer to finite combinations of products of quantum fields and their derivatives evaluated at the same spacetime point. The principle of causality requires that commutators between spacelike-separated local operators vanish. The Lagrangian density determines the QFT's dynamics via a path integral which averages over all quantum field configurations. It is then (relatively) straightforward to obtain observables from the Lagrangian density as covered in standard textbooks; see \cite{Peskin:1995ev,Srednicki:2007qs,Weinberg:1995mt} for details. 

There are two types of observables in high-energy particle physics: \emph{cross-sections} and \emph{decay rates}. The former is especially relevant for accelerator-based experiments, such as those conducted at the Large Hadron Collider. Both observables depend on (the squared modulus of) scattering amplitudes which constitute the central quantities of interest in this dissertation. Roughly speaking, a \emph{scattering amplitude} (or simply \emph{amplitude}) gives the probability for one state of particles --- with prescribed momenta and other quantum numbers --- to evolve into another via some scattering process. \citeauthor{Wheeler:1937zz} \cite{Wheeler:1937zz} and \citeauthor{Heisenberg:1943zz} \cite{Heisenberg:1943zz}, independent of each other, introduced the \emph{scattering matrix} (or simply \emph{S-matrix}), a unitary matrix collecting amplitudes associated with pairs of initial and final states. In perturbative calculations, one expands the S-matrix in powers of the coupling constant which appears in the Lagrangian density, producing an asymptotic series. One designates the lowest-order terms in the expansion as ``tree-level'', while ``loop-level'' describes higher-order contributions; these adjectives refer to different Feynman diagram topologies introduced below. In this dissertation, I will focus exclusively on tree-level amplitudes.

After the complexification of momenta, amplitudes become complex functions with a complicated singularity structure. But the principles of locality and unitarity provide stringent constraints. Locality dictates the location of singularities, while unitarity determines their behaviour. Let us consider tree-level amplitudes, for example. Locality produces simple poles corresponding to on-shell momenta. Unitarity requires the associated residues to factorise as a pair of amplitudes involving fewer particles. 

In \citeyear{Feynman:1949zx}, \citeauthor{Feynman:1949zx} introduced a pictorial bookkeeping procedure for calculating amplitudes, expressing them as a sum of \emph{Feynman diagrams} \cite{Feynman:1949zx}. His approach quickly became what \citeauthor{kuhn1970structure} would describe as the \emph{dominant paradigm} \cite{kuhn1970structure}. \citeauthor{kaiser2005physics} articulates \cite{kaiser2005physics}, ``Since the middle of the 20th century, theoretical physicists have increasingly turned to this tool to help them undertake critical calculations. Feynman diagrams have revolutionised nearly every aspect of theoretical physics.''  

While ubiquitous, Feynman diagrams have drawbacks, often obfuscating the amplitude's simplicity and structure. Feynman diagrams introduce \emph{virtual particles} corresponding to off-shell degrees of freedom to maintain locality and unitarity. Unfortunately, this causes computational complexity to compound: the number of Feynman diagrams for an amplitude rapidly grows with external particle number. This growth in the number of summands is an artefact frequently not reflected in the final result. \citeauthor{Parke:1986gb} famously demonstrated this point in \citeyear{Parke:1986gb} by proposing a concise all-multiplicity formula for \emph{maximally helicity violating (MHV)} gluon amplitudes \cite{Parke:1985ax,Parke:1986gb}. MHV gluon amplitudes are those involving all-but-two positive helicity gluons. Their work simplified multiple pages\footnote{The number of Feynman diagrams needed to calculate the $n$-particle MHV gluon amplitude equals the number of dissections of a convex $n$-gon into triangles and quadrilaterals by non-intersecting diagonals \cite{Elvang:2013cua,oeis:A001002}.} of Feynman calculus into one line. The inability of the then-dominant paradigm to account for these simplifications inspired the development of modern on-shell methods; see \cite{Elvang:2013cua,Henn:2014yza} for reviews.

The  \emph{Britto--Cachazo--Feng--Witten (BCFW) recursion relations} for tree-level amplitudes \cite{Britto:2004ap,Britto:2005fq} (and their loop-level generalisations \cite{Arkani-Hamed:2010zjl,Boels:2010nw}) are a principal outcome of this modern approach. They express amplitudes as a sum of \emph{on-shell diagrams}. Unlike their Feynman counterparts, 
they compute amplitudes directly in terms of on-shell processes. On-shell diagrams identify ``cells'' in the non-negative Grassmannian\footnote{Often ambiguously referred to as the ``positive Grassmannian'' by physicists.}. The \emph{non-negative Grassmannian} generalises the projective simplex to the Grassmannian. This object has been the subject of intense mathematical inquiry for the past 20 years \cite{Postnikov:2006kva,Arkani-Hamed:2012zlh,Postnikov:2018jfq}. On-shell diagrams respect unitarity but not locality since they introduce \emph{spurious poles} not present in the final answer. Remarkably, spurious poles cancel in particular combinations, i.e.\ those appearing in solutions to the BCFW recursion relations. But why these combinations? While the locality principle ensures the cancellation of spurious poles, how does this emerge from the non-negative Grassmannian? 

The framework of positive geometries \cite{Arkani-Hamed:2017tmz} presents a simple yet surprising solution to this problem. A \emph{positive geometry} is a non-empty closed semi-algebraic set. It admits a unique non-zero top-dimensional rational form called the \emph{canonical form}, satisfying some recursive axioms. For some QFTs, there is a family of positive geometries, defined via positivity conditions in some on-shell kinematic space, whose canonical forms encode amplitudes. In this setting, \emph{tilings}, which are akin to ``triangulations'', of the positive geometry coincide with equivalent representations of the same amplitude; each tile in a tiling corresponds to an on-shell diagram. This description restores manifest locality with spurious poles corresponding to shared internal boundaries of some tiling. In addition, the faces of the positive geometry encode the amplitude’s entire singularity structure. In other words, locality and unitarity emerge as consequences of \emph{positivity} in the kinematic space without reference to quantum fields, strings, or any notion of spacetime evolution. 

The prototypical example of such a family is the \emph{Amplituhedron}\footnote{We adopt non-standard notation for the Amplituhedron to avoid confusion with the ABHY associahedron. The notation $\Amp{k}{n}$ emphasises that the Amplituhedron encodes the expectation value of the dual null polygonal Wilson loop usually denoted by $W_{n,k}$.} $\Amp{k}{n}$, discovered by \citeauthor{Arkani-Hamed:2013jha} in \citeyear{Arkani-Hamed:2013jha}. It encodes tree-level amplitudes and (the integrands of) loop-level planar amplitudes in \emph{$\mathcal{N}=4$ supersymmetric Yang-Mills (SYM) theory} \cite{Arkani-Hamed:2013jha}. The Amplituhedron generalises \citeauthor{Hodges:2009hk}’s idea \cite{Hodges:2009hk}, interpreting the amplitude as the volume of (the dual of) some canonical region in some space. The Amplituhedron was initially defined as the image of the non-negative Grassmannian via a linear map  \cite{Arkani-Hamed:2013jha}, providing a direct connection to on-shell diagrams. Later, \citeauthor{Arkani-Hamed:2017vfh} gave an equivalent definition directly in the space of so-called momentum twistor variables demarcated by positivity conditions \cite{Arkani-Hamed:2017vfh}. 

Around the time of the Amplituhedron's discovery, \citeauthor{Cachazo:2013iaa} presented an altogether contrasting characterisation of amplitudes: the \emph{Cachazo--He--Yuan (CHY)} formalism \cite{Cachazo:2013iaa,Cachazo:2013gna,Cachazo:2013hca,Cachazo:2013iea}. It originates from \citeauthor{Witten:2003nn}’s twistor-string theory, which produced new formulae for tree-level amplitudes in $\mathcal{N}=4$ SYM \cite{Witten:2003nn,Roiban:2004vt,Roiban:2004yf} and $\mathcal{N}=8$ supersymmetric gravity \cite{Cachazo:2012da,Cachazo:2012kg,Cachazo:2012pz}. The CHY formalism applies to massless scattering in a broad collection of QFTs \cite{Cachazo:2013iaa,Cachazo:2013gna,Cachazo:2013hca,Cachazo:2013iea,Cachazo:2014nsa,Cachazo:2014xea} in arbitrary spacetime dimensions \cite{Cachazo:2013hca}. It expresses amplitudes as integrals over a space of worldsheet moduli. These integrals fully localise on the support of the scattering equations. The \emph{scattering equations} are a finite system of coupled equations identifying points in the relevant worldsheet moduli space with on-shell kinematical invariants in a many-to-one manner. There also exist CHY formulae for amplitude-like objects, e.g. the Cachazo--Early--Guevara--Mizera (CEGM) amplitudes \cite{Cachazo:2019ngv}, that do not have a known QFT description.

\begin{figure}
	\centering
	\includegraphics[scale=0.6]{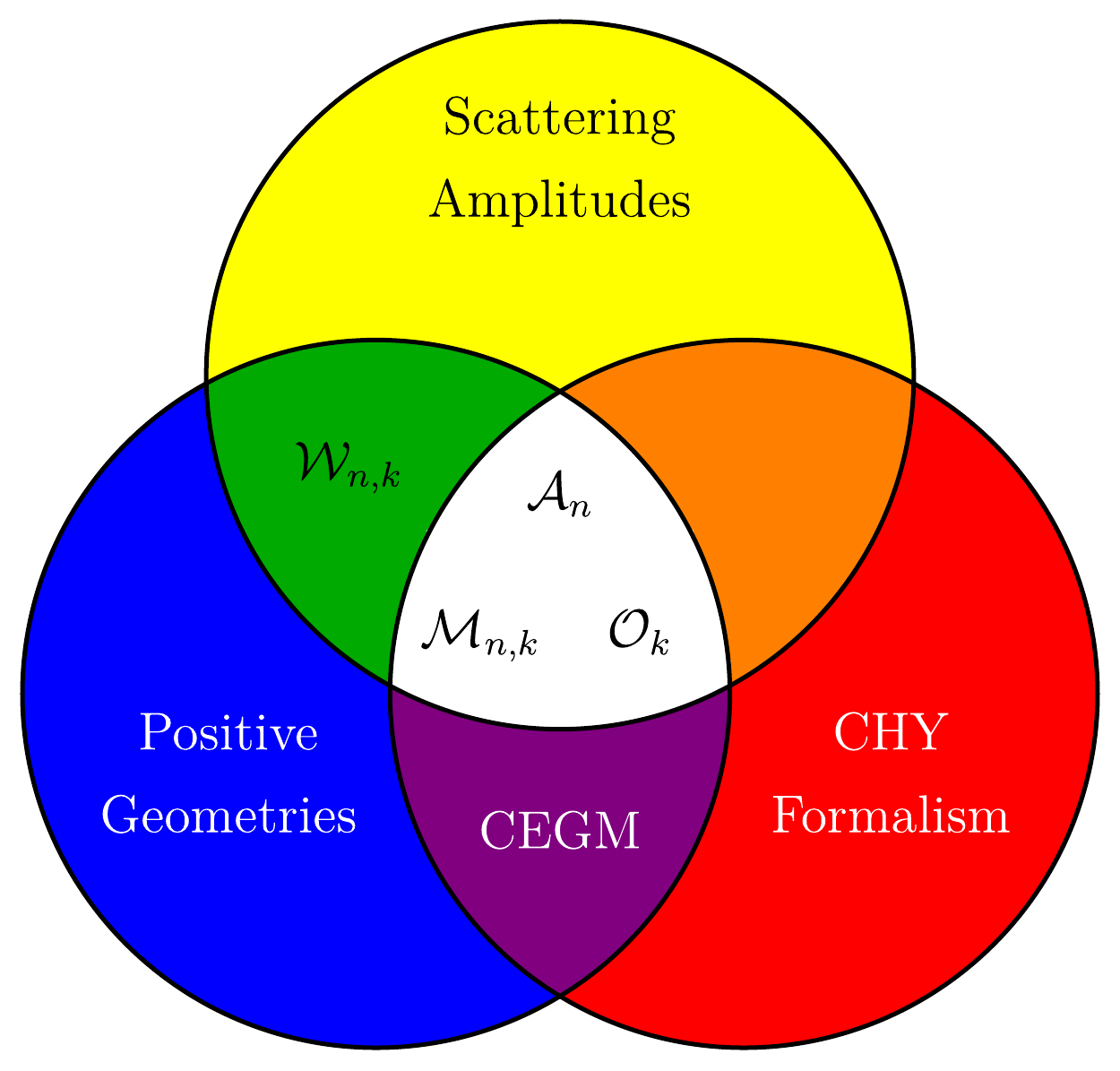}
	\caption{Venn diagram showing scattering amplitudes, positive geometries and the CHY formalism.}
	\label{fig:intro-venn}	
\end{figure}

There are no CHY formulae for amplitudes in terms of momentum twistor variables. However, there are examples of tree-level amplitudes described by families of positive geometries, which also have a CHY representation. In this dissertation, we investigate three theories which admit both descriptions. In particular, we study the \emph{Arkani-Hamed--Bai--He--Yuan (ABHY) associahedron} $\Abhy{n}$ \cite{Arkani-Hamed:2017mur}, the \emph{Momentum Amplituhedron} $\tnnMom{k}{n}$ \cite{Damgaard:2019ztj}, and the \emph{orthogonal Momentum Amplituhedron} $\tnnOMom{k}$ \cite{Huang:2021jlh,He:2021llb}. All three positive geometries express amplitudes in on-shell momentum space. The ABHY associahedron encodes the tree-level S-matrix of \emph{bi-adjoint scalar (BAS) theory} in terms of Mandelstam invariants. The Momentum Amplituhedron, like the Amplituhedron, encodes the tree-level S-matrix of $\mathcal{N}=4$ SYM but directly in terms of four-dimensional spinor-helicity variables. Similarly, the orthognal Momentum Amplituhedron uses three-dimensional spinor-helicity variables to describe tree-level amplitudes in \emph{Aharony--Bergman--Jafferis--Maldacena (ABJM) theory}. In each case, the relevant family of positive geometries connects directly to its corresponding CHY representation via the scattering equations: the momentum-space-based positive geometry is the image of a positive geometry in the corresponding worldsheet moduli space via the scattering equations \cite{He:2021llb}. \cref{fig:intro-venn} displays the qualitative overlap between scattering amplitudes, positive geometries and the CHY formalism.

There are many examples of positive geometries relevant to high-energy physics which are beyond the scope of this dissertation. The construction of ABHY, which describes tree-level scattering between massless particles via cubic interactions, extends to higher polynomial interactions via Stokes polytopes and the Accordiahedron \cite{Banerjee:2018tun,Aneesh:2019ddi,Aneesh:2019cvt,Kojima:2020tox}, to multiple fields \cite{Jagadale:2021iab}, to massive particles \cite{Jagadale:2022rbl}, and even to loop-level via the Halohedron \cite{Salvatori:2018aha} and cluster polytopes \cite{Arkani-Hamed:2019vag}. The Correlahedron \cite{Eden:2017fow} conjecturally encodes stress-energy correlators in planar $\mathcal{N} = 4$ SYM. While it is not a positive geometry, it coincides with the square of the Amplituhedron in a certain limit \cite{Dian:2021idl}. Aspects of positive geometries apply more generally to conformal field theories, including the conformal bootstrap programme \cite{Arkani-Hamed:2018ign}. The EFT-hedron places positivity bounds on the low-energy effective field theory (EFT) expansion of amplitudes \cite{Arkani-Hamed:2020blm}. In cosmology, cosmological polytopes compute contributions to the cosmological wavefunction for a class of toy models  \cite{Arkani-Hamed:2017fdk,Benincasa:2022gtd}. For a more comprehensive review of these and other developments, see \cite{Ferro:2020ygk,Herrmann:2022nkh}.

In this dissertation, we focus primarily on how \emph{the boundaries of positive geometries encode information about physical singularities} \cite{Ferro:2020lgp,Lukowski:2021fkf}. Through this encoding, positive geometries provide a natural mechanism for organising and classifying this information, an intractable feat by traditional perturbative approaches. Moreover, it facilitates rigorous analysis, stimulating novel pure mathematics research. For example, to enumerate the physical singularities of tree-level amplitudes in $\mathcal{N}=4$ SYM, we derive new analogues of the \emph{Exponential formula}, a well-known theorem relevant for constructing generating functions \cite{stanley1999enumerative,bergeron1998combinatorial,ardila2015algebraic}. In particular, we derive analogues for series-reduced planar trees and forests \cite{Moerman:2021cjg} and apply these results to the Momentum Amplituhedron \cite{Moerman:2021cjg} and the orthogonal Momentum Amplituhedron \cite{Lukowski:2021fkf}. In each case, we show that the poset of physical singularities has Euler characteristic equal to one\footnote{This result, while mathematically interesting, has no known physical interpretation.}. In addition, we describe an intricate web of connections between the three above-mentioned positive geometries. This web encompasses relationships between boundaries \cite{Ferro:2020lgp,Moerman:2021cjg,Lukowski:2021fkf} and canonical forms \cite{Damgaard:2020eox,He:2021llb,Lukowski:2021fkf}. We also develop methods for manipulating canonical forms using ideas from computational algebraic geometry \cite{Lukowski:2022fwz}. These methods are useful not only for calculating canonical forms but also for verifying relationships between them.  

The framework of positive geometries marks the dawn of a ``scientific revolution'' that transcends and includes the principles underpinning QFT. While bearing no resemblance to its predecessor, the paradigm of positive geometries, using positivity alone, captures the richness of scattering amplitudes in momentum space. In the remaining pages of this dissertation, we explore some of the author’s contributions to this exciting new research endeavour.

\section{Outline}

There are two parts to this dissertation. \cref{chp:pos,chp:enum,chp:grass,chp:push} are technical, consisting primarily of definitions and results. This presentation highlights some original contributions and serves as a primer for the dissertation's latter half. Thereafter, \cref{chp:abhy,chp:mom,chp:omom,chp:meet} introduce the three positive geometries investigated in this dissertation, emphasising their boundary posets. In particular, we organise and classify boundaries, commenting on their physical and mathematical significance. Along the way, we build an intricate web of connections at the level of boundary posets and canonical forms. \cref{chp:meet} completes this web and speculates on additional strands for future work. The following paragraphs give further details on the content of each chapter.

\cref{chp:pos} reviews the framework of \emph{positive geometries} and particular maps between them called \emph{morphisms}. Morphisms play a central role in this thesis and the literature, with many physically motivated positive geometries defined via these maps. In this chapter, we modify the original definition of morphisms, given in \cite{Arkani-Hamed:2017tmz}, to facilitate the definition of \emph{morphism-induced boundaries}. In addition, we describe an algorithm for generating boundary posets via morphisms. This algorithm first appeared in \cite{Lukowski:2019kqi} and was later applied in \cite{Lukowski:2020bya,Ferro:2020lgp,Lukowski:2021fkf}. Our presentation extracts core features from \cite{Lukowski:2019kqi}, expressing them in general terms. Applications of the algorithm appear in \cref{chp:mom,chp:omom}.  

In \cref{chp:enum}, we present results from enumerative combinatorics. These results, developed in \cite{Moerman:2021cjg}, facilitate the construction of \emph{generating functions}, as demonstrated in \cite{Moerman:2021cjg,Lukowski:2021fkf}. They are useful for enumerating acyclic labels such as tree-level Feynman/on-shell diagrams. Moreover, they lead to the enumeration formulae presented in \cref{chp:abhy,chp:mom,chp:omom}. The tools developed in this chapter allow us to answer questions about the total number of physical singularities and the poset's Euler characteristic, for example.

We review the \emph{Grassmannian} and the \emph{orthogonal Grassmannian} in  \cref{chp:grass}. Their non-negative parts, defined in \cite{Postnikov:2006kva,Huang:2013owa,Huang:2014xza,Galashin:2020jvd}, are central to \cref{chp:mom,chp:omom} for the definitions of the \emph{Momentum Amplituhedron} and the \emph{orthogonal Momentum Amplituhedron}, respectively. We discuss the combinatorics of \emph{positroid} cells \cite{Postnikov:2006kva} and \emph{orthitroid} cells \cite{Kim:2014hva}. \emph{Grassmannian graphs} \cite{Postnikov:2018jfq} and \emph{orthogonal Grassmannian graphs} \cite{Lukowski:2021fkf} provide labels for these cells. These graphs coincide with on-shell diagrams and are relevant to the investigations of \cref{chp:mom,chp:omom}.

In \cref{chp:push}, we study the problem of evaluating \emph{push forwards}, an operation on rational forms, via rational maps without needing to find local inverses. This problem is relevant for determining the canonical forms of the Momentum Amplituhedron and the orthogonal Momentum Amplituhedron, as discussed in \cref{chp:mom,chp:omom}, respectively. We formulate this problem in the language of \emph{polynomials ideals} and \emph{algebraic sets}, and present three solutions employing tools from computational algebraic geometry \cite{Lukowski:2022fwz}. 

\cref{chp:abhy} concerns the \emph{Arkani-Hamed--Bai--He--Yuan (ABHY) associahedron} and its description of tree-level amplitudes in the \emph{bi-adjoint scalar (BAS) theory} \cite{Arkani-Hamed:2017mur}. We outline its connection to the \emph{worldsheet associahedron} via the \emph{scattering equations} \cite{Arkani-Hamed:2017mur}. Then, as a simple application of the results of \cref{chp:enum}, we derive a rank generating function which enumerates its boundaries. The ABHY associahedron features prominently in \cref{chp:meet}.

In \cref{chp:mom}, we focus on the Momentum Amplituhedron. Its canonical form encodes the tree-level amplitudes of \emph{$\mathcal{N}=4$ supersymmetric Yang-Mills theory (SYM)} in terms of four-dimensional spinor-helicity variables \cite{Damgaard:2019ztj}. The Momentum Amplituhedron is related to two positive geometries. It relates to the non-negative part of the four-dimensional worldsheet moduli space \cite{He:2021llb}. In this case, the four-dimensional scattering equations \cite{Geyer:2014fka} define a morphism of positive geometries. The Momentum Amplituhedorn is also the image of the non-negative Grassmannian via a linear map. The second definition allows us to study the Momentum Amplituhedron’s boundary stratification \cite{Ferro:2020lgp}. Remarkably, the boundaries admit a simple classification: they are in bijection with \emph{contracted Grassmannian forests} \cite{Moerman:2021cjg}. In fact, the boundary poset forms an induced subposet of the positroid stratification. We give a rank generating function for Momentum Amplituhedron boundaries and use it to show that the Momentum Amplituhedron's Euler characteristic is one \cite{Moerman:2021cjg}.

We investigate the orthogonal Momentum Amplituhedron, the positive geometry of \emph{Aharony--Bergman--Jafferis--Maldacena (ABJM) theory} \cite{Huang:2021jlh,He:2021llb}, in  \cref{chp:omom}. The relevant on-shell momentum space consists of three-dimension spinor-helicity variables. It relates to the worldsheet associahedron via the three-dimensional scattering equations \cite{He:2021llb}. Here \emph{orthogonal Grassmannian forests} label the orthogonal Momentum Amplituhedron’s boundaries \cite{Lukowski:2021fkf}. This observation is analogous to the findings of \cref{chp:mom}. There are two ways to characterise its boundary stratification. It is an induced subposet of the orthitroid stratification. At the same time, it is an induced subposet of the Momentum Amplituhedron’s boundary stratification. In this sense, the Momentum Amplituhedron contains the orthogonal Momentum Amplituhedron. We present a rank generating function for the boundaries using \cref{chp:enum}’s results and show that the Euler characteristic of the orthogonal Momentum Amplituhedron is one \cite{Lukowski:2021fkf}. We also define a diagrammatic map from the worldsheet associahedron’s boundaries to the boundaries of the orthogonal Momentum Amplituhedron \cite{Lukowski:2021fkf}. This partial-order preserving map provides an alternative heuristic for studying the physical singularities of ABJM.

In \cref{chp:meet}, we consider restricting the ABHY associahedron to specific spacetime dimensions. We show how to obtain these restrictions starting from the Momentum Amplituhedron \cite{Damgaard:2020eox} and the orthogonal Momentum Amplituheron.  \cref{chp:meet} introduces additional strands in our web of connections, reflecting the shared singularity structure of the various theories corresponding to vanishing planar Mandelstam variables. We also speculate on possible six-dimensional analogues of these results as the source of future work.

Finally, this dissertation closes with some conclusions and an outlook on open research problems in \cref{chp:concl}.	
	\chapter{Positive Geometries}
\label{chp:pos}

\lettrine{T}{he framework of positive geometries} forms the foundation for this dissertation. In this chapter, we define positive geometries and their canonical forms, summarising ideas from \cite{Arkani-Hamed:2017tmz} relevant for later chapters. For a recent mathematical survey of this topic, see \cite{Lam:2022yly}. Importantly, we define (categorical) morphisms between positive geometries. In applications, morphisms facilitate the definition of new positive geometries from old ones. We will encounter multiple examples of morphisms in \cref{chp:abhy,chp:mom,chp:omom,chp:meet}, so we define them formally here. Our definition modifies the original definition of \citeauthor{Arkani-Hamed:2017tmz} in \cite{Arkani-Hamed:2017tmz}. It allows for a general discussion about the image of boundaries via morphisms. For physically relevant examples, boundaries correspond to physical singularities of some S-matrix. In particular, we introduce the notion of morphism-induced boundaries. This notion is pertinent to \cref{chp:mom,chp:omom}, where we investigate the boundary stratifications of the Momentum Amplituhedron and the orthogonal Momentum Amplituhedron, respectively. 

\section{Positive Geometries and Canonical Forms}
\label{sec:pos-pos}

To understand the definition of positive geometries, we first need to define the residue of a rational form. To this end, we introduce the following objects. Suppose $X$ is a $d$-dimensional complex variety\footnote{By ``variety'', we mean an irreducible algebraic set.}, $\omega$ is a rational $d$-form on $X$, and $H$ is a hypersurface\footnote{$H$ is the zero-set of an irreducible polynomial.} in $X$. Let $z$ be a holomorphic coordinate whose zero-set locally parametrises $H$. We say that $\omega$ has a \emph{simple pole} at $H$ if 
\begin{align}\label{eq:pos-simple-pole}
	\omega = \frac{dz}{z}\wedge\eta + \eta'
\end{align}
for some $(d-1)$-form $\eta$ and $d$-form $\eta'$, both holomorphic on $H$. In this case, we define the \emph{residue} of $\omega$ at $H$ as the following restriction:
\begin{align}\label{eq:pos-residue}
	\Res_H\omega\coloneqq \eta|_H\,.
\end{align}
This residue is a well-defined $(d-1)$-form on $H$, independent of $z,\eta,\eta'$. If no such simple pole exists, then the residue equates to zero.

For the remainder of this section, we assume $X$ is defined over $\real$, i.e.\ generated by \emph{real} homogeneous polynomials. We equip $X(\real)$, the \emph{real part} of $X$, with the standard topology. Let $\tnn{X}$ be a non-empty closed semialgebraic set in $X(\real)$. We will assume that $\tp{X}$, the \emph{interior} of $\tnn{X}$, is an oriented $d$-dimensional manifold whose \emph{closure} recovers $\tnn{X}$. Let $\partial\tnn{X}$ be the \emph{boundary} $\tnn{X}\setminus\tp{X}$. The symbol $\partial X$ denotes the \emph{Zariski closure} of $\partial\tnn{X}$ in $X$. Let $C_1,\ldots,C_{b_d}$ designate the \emph{irreducible components}\footnote{While we assume $X$ is generically smooth along each $C_i$, see \cite[Remark 3]{Lam:2022yly}.} of $\partial X$. We symbolize by $\tnnEnumerate{C}{i}$ the closure of the interior of $C_i\cap\partial X_{\ge0}$ in $C_i(\real)$. Collectively, the sets $\tnnEnumerate{C}{1},\ldots,\tnnEnumerate{C}{b_d}$ define the \emph{boundary components} or \emph{facets} of $\tnn{X}$.

The pair $(X,\tnn{X})$ is called a $d$-dimensional \emph{positive geometry} if there exists a unique non-zero rational $d$-form $\Omega(X,\tnn{X})$ on $X$ called the \emph{canonical form}, satisfying the following recursive axioms \cite{Arkani-Hamed:2017tmz}:
\begin{itemize}
	\item If $d=0$, then $X$ is a point, $\tnn{X}=X$, and $\Omega(X,\tnn{X})=\pm1$ where orientation determines the sign.
	\item If $d>0$, then
	\begin{axioms}{P}
		\item Every boundary component $(C,\tnn{C})$ of $(X,\tnn{X})$ is a positive geometry of dimension $(d-1)$.
		\item $\Omega(X,\tnn{X})$ only has simple poles along the irreducible components of $\partial X$. For each boundary component $(C,\tnn{C})$ of $(X,\tnn{X})$
		\begin{align}\label{eq:pos-residue-canonical-form}
			\Res_{C}\Omega(X,\tnn{X}) = \Omega(C,\tnn{C})\,.
		\end{align} 
	\end{axioms}
\end{itemize}

Allowing $\tnn{X}$ to be empty leads to the definition of \emph{pseudo-positive geometries}. The latter will, however, not feature in this dissertation. The variety $X$ is called the \emph{embedding space} and $\tnn{X}$ (resp., $\tp{X}$) is called the \emph{non-negative part} (resp., \emph{positive part}) of $X$. We will frequently omit the embedding space, referring to $\tnn{X}$ as the positive geometry and $\Omega(\tnn{X})$ as its canonical form. The \emph{leading residues} of $\tnn{X}$ --- the non-zero $d$-fold residues of $\Omega(\tnn{X})$ --- are $\pm1$. The orientation of each boundary component $\tnn{C}$ of $\tnn{X}$ is induced from $\tnn{X}$; details about orientation can be found in \cite{Arkani-Hamed:2017tmz}. Every positive geometry appearing in the recursive definition of $\tnn{X}$ --- $\tnn{X}$, its boundary components, their boundary components, etc. --- is called a \emph{boundary (stratum)} or \emph{face} of $\tnn{X}$. Occasionally we will refer to the boundary $\tnn{C}$ (resp., $\tp{C}$) as closed (resp., open). 

The \emph{boundary stratification} of $\tnn{X}$ is the set of all boundaries of $\tnn{X}$ denoted by $\mathcal{B}(\tnn{X})$. It is a \emph{partially ordered set (poset)} with the \emph{partial order} $\preceq_X$ defined via set inclusion: 
\begin{align}
	C\preceq_X D\iff C\subseteq D\,.
\end{align}
The \emph{proper boundary stratification} of $\tnn{X}$ is the set of positive codimension boundaries of $\tnn{X}$ denoted by $\mathcal{P}(\tnn{X})\coloneqq\mathcal{B}(\tnn{X})\setminus\{\tnn{X}\}$. For each boundary $\tnn{C}$ of $\tnn{X}$, let $\mathcal{B}^{-1}_{\tnn{X}}(\tnn{C})$ be the subset of $\mathcal{B}(\tnn{X})$ for which $\tnn{C}$ is a boundary. We call $\mathcal{B}^{-1}_{\tnn{X}}(\tnn{C})$ the \emph{inverse boundary stratification} of $\tnn{C}$ in $\tnn{X}$. The \emph{proper inverse boundary stratification} of $\tnn{C}$ in $\tnn{X}$ is given by $\mathcal{P}^{-1}_{\tnn{X}}(\tnn{C})\coloneqq\mathcal{B}^{-1}_{\tnn{X}}(\tnn{C})\setminus\{\tnn{C}\}$.

With a suitable definition for addition, positive geometries generalize to \emph{weighted positive geometries} \cite{Dian:2022tpf}. Weighted positive geometries allow leading residues to have arbitrary integer values. Their definition naturally extends to \emph{weighted pseudo-positive geometries}. While these categories are beyond the scope of this dissertation, we anticipate that they formalise the notion of \emph{oriented sums} discussed in \cite{Damgaard:2021qbi}.

\begin{figure}
	\centering
	\includegraphics[width=0.4\textwidth]{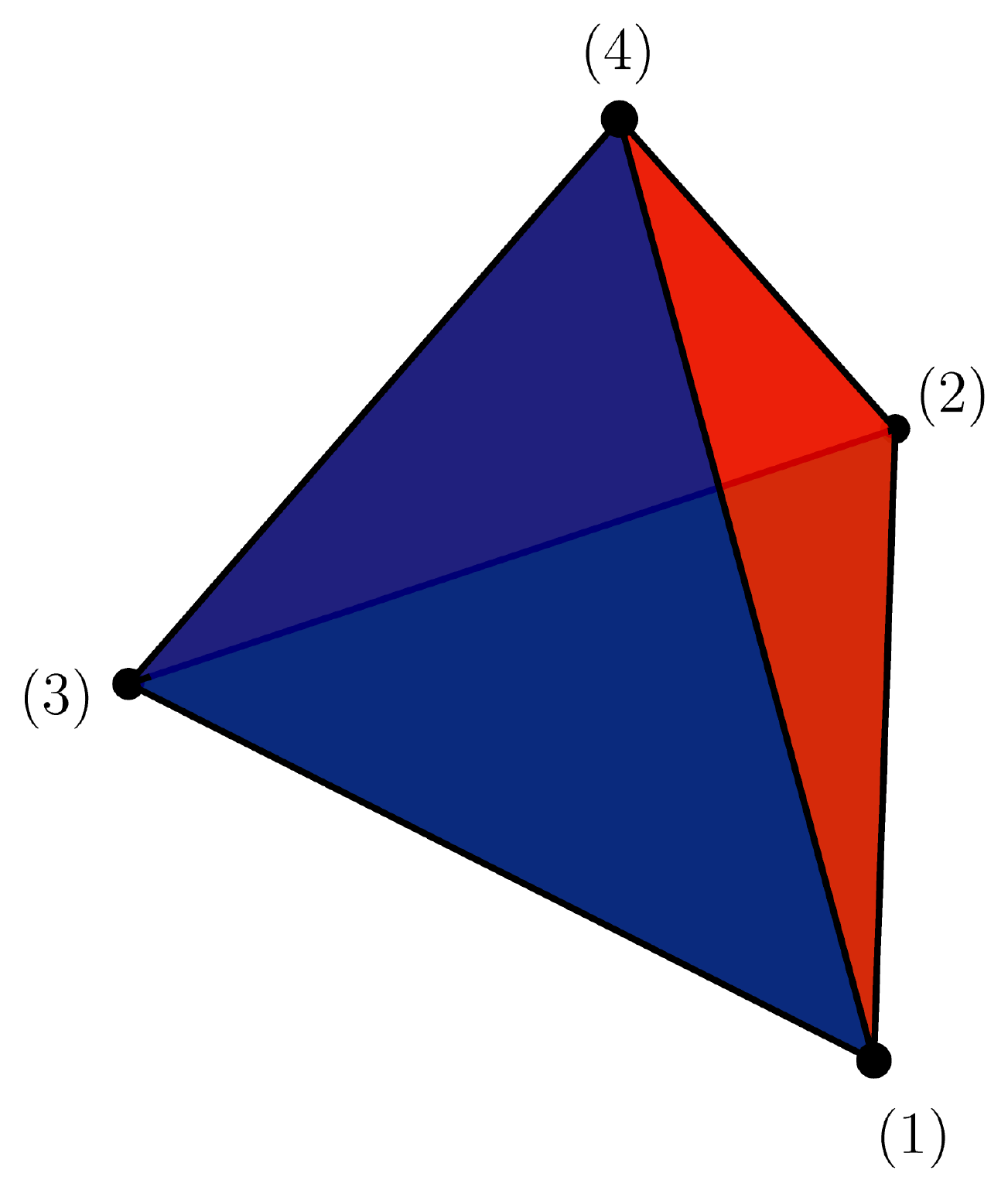}
	\caption{The standard $3$-simplex or tetrahedron with labelled vertices.}
	\label{fig:pos-Delta-3}
\end{figure}

\begin{example}[The Standard Simplex]
	A classic example of a positive geometry is the standard $n$-simplex $(\Delta_n,\Delta_n^{\ge0})$ (see \cref{fig:pos-Delta-3}). The embedding space is the affine hyperplane in $\real^{n+1}$ defined as
	\begin{align}
		\Delta_n\coloneqq \left\{(x_1,\ldots,x_{n+1})\in\real^{n+1}:\sum_{i=1}^{n+1}x_i=1
		\right\},
	\end{align}
	while the non-negative part is given by
	\begin{align}
		\Delta_n^{\ge0} \coloneqq \real_{\ge0}^{n+1}\cap\Delta_n\,.
	\end{align}
	The above notation, while unconventional\footnote{The standard $n$-simplex is usually denoted by $\Delta_n$ and not $\Delta_n^{\ge0}$.}, highlights the standard $n$-simplex as a semi-algebraic subset of some embedding space. The positive geometry inherits the standard orientation of $\real^{n+1}$. Its canonical form is given by
	\begin{align}
		\Omega(\Delta_n, \Delta_n^{\ge0})=d\log\frac{x_1}{x_2}\wedge d\log\frac{x_2}{x_3}\wedge\cdots\wedge d\log\frac{x_n}{x_{n+1}}\,.
	\end{align}
	
	Equivalently, the standard $n$-simplex is the convex hull of all vectors in the standard basis of $\real^{n+1}$: $\Delta_n^{\ge0}=\text{Conv}(e_1,\ldots,e_n)$ where $\text{Conv}(\cdot)$ denotes the convex hull and $e_i$ denotes the $i\textsuperscript{th}$ standard basis vector. Every boundary is the convex hull of some subset of standard basis vectors. For brevity, we will identify the boundary $\text{Conv}(e_{i_1},\ldots,e_{i_p})$ using the notation $(i_1\cdots i_p)$ where $\{i_1,\ldots,i_p\}\subseteq[n+1]$. Moreover, the partial order $\preceq_{\Delta}$ for $\mathcal{B}(\Delta_n^{\ge0})$ reads $(i_1\cdots i_p)\preceq_{\Delta}(j_1\cdots j_q)\iff\{i_1,\ldots,i_p\}\subseteq\{j_1,\ldots,j_q\}$.
\end{example}

\section{Morphisms}
\label{sec:pos-morphisms}

Having defined positive geometries, we now define particular maps between them called morphisms. To this end, we also introduce dissections and tilings. We will encounter multiple examples of morphisms in \cref{chp:abhy,chp:mom,chp:omom}. While our definition of morphisms differs from the one in \cite{Arkani-Hamed:2017tmz}, it allows us to present a general algorithm for finding boundaries in \cref{sec:pos-boundaries}. For the remainder of this chapter, let $(X,\tnn{X})$ and $(Y,\tnn{Y})$ be positive geometries of (possibly differing) dimensions. 

A \emph{rational map} $\Phi(X,\tnn{X})\dashrightarrow(Y,\tnn{Y})$ between positive geometries is a rational map $\Phi:X\dashrightarrow Y$ between complex varieties, well-defined on each open boundary of $\tnn{X}$, such that $\Phi(\tp{X})=\tp{Y}$. The following definitions encapsulate those given in \cite{Moerman:2021cjg}. For each boundary $\tnn{C}$ of $\tnn{X}$, let $\Phi(\tnn{C})$ denote the closure of $\Phi(\tp{C})$ in $\Phi(X)(\real)$. We refer to $\Phi(\tnn{C})$ as a \emph{$\Phi$-induced stratum} of $\tnn{Y}$ and $\dim(\Phi(\tnn{C}))$ as the \emph{$\Phi$-dimension} of $\tnn{C}$. 

One can compute $\dim(\Phi(\tnn{C}))$ using a chart/parametrisation as follows. Let $d_Y$ be the dimension of $\tnn{Y}$ and suppose $\tnn{C}$ is a $d_C$-dimensional boundary of $\tnn{X}$. Let $\phi:U\subset C(\real)\to\phi(U)\subset\real^{d_C}$ and $\psi:V\subset Y(\real)\to\psi(V)\subset\real^{d_Y}$ be affine charts such that $\tnn{C}\cap U\cap\Phi^{-1}(V)\neq\emptyset$. If $\psi\circ\Phi\circ\phi^{-1}$ is smooth on $\tnn{C}\cap U\cap\Phi^{-1}(V)$, then its differential is a linear map whose rank gives $\dim(\Phi(\tnn{C}))$. 

Next, we introduce dissections and tilings as analogues of  subdivisions and triangulations from polytopal geometry. The following definition is motivated by one found in \cite{Lukowski:2020dpn}. Let $I$ be a finite indexing set. We say a collection $\mathcal{C}=\{\tnnEnumerate{C}{i}\}_{i\in I}\subset\mathcal{B}(\tnn{X})$ is a \emph{$\Phi$-induced dissection} of $\tnn{Y}$ if the following conditions are met:
\begin{axioms}{D}
	\item\label[axiom]{axiom:pos-dissection-1} $\dim(\Phi(\tnnEnumerate{C}{i}))=\dim(\tnn{Y})$ for every $i\in I$.
	\item\label[axiom]{axiom:pos-dissection-2} The images of distinct open boundaries in $\mathcal{C}$ are disjoint: $\Phi(\tpEnumerate{C}{i})\cap\Phi(\tpEnumerate{C}{j})=\emptyset$ for $i\ne j\in I$.
	\item\label[axiom]{axiom:pos-dissection-3} The union of images of open boundaries in $\mathcal{C}$ is dense in $\tnn{Y}$: $\bigcup_{i\in I}\Phi(\tnnEnumerate{C}{i}) = \tnn{Y}$.
\end{axioms} 
Clearly, $\{\tnn{X}\}$ constitutes a trivial $\Phi$-induced dissection of $\tnn{Y}$. In addition, if $\Phi|_{\tpEnumerate{C}{i}}$ is injective for every $i\in I$, then $\mathcal{C}$ is called a \emph{$\Phi$-induced tiling} of $\tnn{Y}$. We will refer to the elements of a $\Phi$-induced tiling $\mathcal{C}$ as \emph{$\Phi$-induced tiles}. A tiling is said to be \emph{orientation-preserving} (resp., \emph{orientation-reversing}) if $\Phi_{C_{i,\ge0}}$ is orientation-preserving (resp., orientation-reversing) for every $i\in I$. In either case, we say that the tiling has \emph{uniform orientation}.

Finally, we define a \emph{morphism} $\Phi:(X,\tnn{X})\to(Y,\tnn{Y})$ between positive geometries as a rational map $\Phi(X,\tnn{X})\dashrightarrow(Y,\tnn{Y})$, satisfying the following assumptions:
\begin{axioms}{M}
	\item\label[axiom]{axiom:pos-morphism-1} It admits at least one $\Phi$-induced tiling of $\tnn{Y}$.
	\item\label[axiom]{axiom:pos-morphism-2} Every $\Phi$-induced tiling of $\tnn{Y}$ has uniform orientation.
\end{axioms}

When $\tnn{X}$ and $\tnn{Y}$ have the same dimension, our definition coincides with that of \cite{Arkani-Hamed:2017tmz}. To confirm this remark, consider the following argument. \Cref{axiom:pos-morphism-1} guarantees at least one $\Phi$-induced tiling of $\tnn{Y}$. Note that for each boundary $\tnn{C}$ of $\tnn{X}$, $\dim(\Phi(\tnn{C}))\le\dim(\tnn{C})$. Since $\tnn{X}$ and $\tnn{Y}$ have the same dimension (by assumption), $\tp{X}$ is the only open boundary whose $\Phi$-dimension equals $\dim(\tp{Y})$. \Cref{axiom:pos-dissection-1} implies that $\{\tnn{X}\}$ is the only $\Phi$-induced tiling of $\tnn{Y}$. Consequently, $\Phi|_{\tp{X}}:\tp{X}\to\tp{Y}$ is an orientation-preserving diffeomorphism according to \cref{axiom:pos-morphism-2}. The latter implication recovers the definition found in \cite{Arkani-Hamed:2017tmz}. 

\begin{figure}
\centering
\hfill
\begin{subfigure}{0.4\textwidth}
	\includegraphics[width=\textwidth]{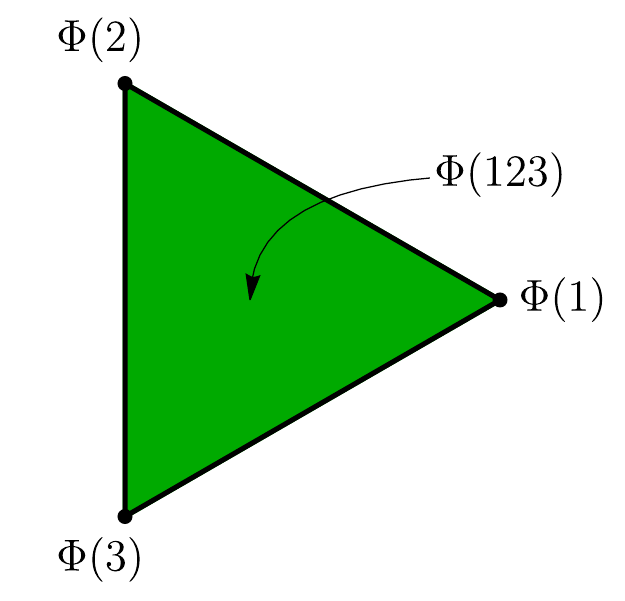}
	\caption{$\{(123)\}$.}
	\label{fig:pos-tiling-1}
\end{subfigure}
\hfill
\begin{subfigure}{0.4\textwidth}
	\includegraphics[width=\textwidth]{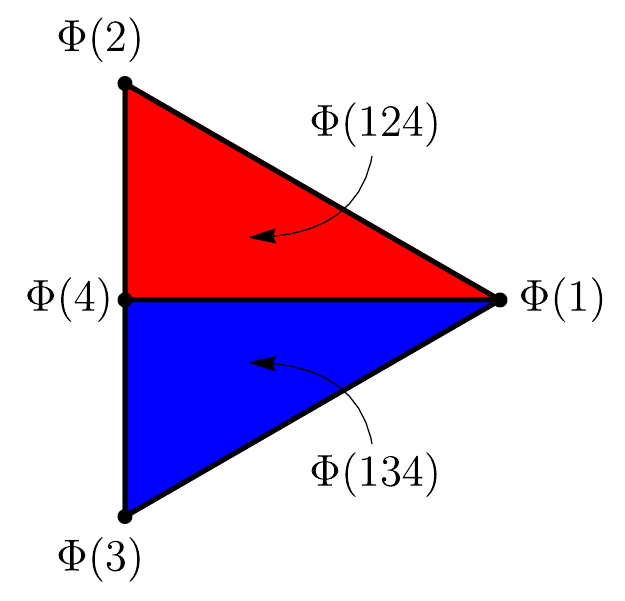}
	\caption{$\{(124),(134)\}$.}
	\label{fig:pos-tiling-2}
\end{subfigure}
\hfill
\null
\caption{Two $\Phi$-induced tilings of $\Delta_2^{\ge0}$ where $\Phi$ is given by \eqref{eq:pos-phi}. \cref{fig:pos-tiling-1} is orientation-reversing and \cref{fig:pos-tiling-2} is orientation-preserving.}
\label{fig:pos-tiling}
\end{figure}

\begin{example}[Morphisms]\label{ex:pos-morphism}
	 Consider the rational map $\Phi:(\Delta_3,\Delta_3^{\ge0})\dashrightarrow(\Delta_2,\Delta_2^{\ge0})$ defined as
	\begin{align}
		\label{eq:pos-phi}
		\Phi(x_1,x_2,x_3,x_4)\coloneqq(x_1,x_2+\frac{x_4}{2},x_3+\frac{x_4}{2})\,.
	\end{align}
	\cref{fig:pos-Delta-3} depicts the domain. The rational map induces two tilings of $\Delta_2^{\ge0}$, namely $\{(123)\}$ and $\{(124),(134)\}$ (see \cref{fig:pos-tiling}). The first tiling is orientation-reversing while the second is orientation-preserving. Since both tilings have uniform orientation, $\Phi$ defines a morphism of positive geometries.
\end{example}

To summarise, a morphism is a map between positive geometries that generates at least one tiling such that, on each tile, the map is an orientation-preserving (or orientation-reversing) diffeomorphism. In subsequent chapters, we will encounter examples of positive geometries defined via morphisms, including the Momentum Amplituhedron in \cref{chp:mom} and the orthogonal Momentum Amplituhedron in \cref{chp:omom}. These examples motivate the definitions presented in this section. In the next section, we consider the action of morphisms on canonical forms via the pushforward operation. This operation makes morphisms what they are in the category-theoretic sense: the pushforward of a canonical form via a morphism is again a canonical form.

\section{Pushforwards}
\label{sec:pos-push}

The pushforward relates the canonical forms of two positive geometries connected via a morphism. The operation consists of multiple pullbacks, so we will define pullbacks and pushforwards in this section beginning with the former. Let $M$ and $N$ be complex manifolds of the same dimension, let $\omega$ be a meromorphic form on $M$ (not necessarily top-dimensional), and let $y$ be a point in $N$. Suppose $\psi:N\to M$ is a differentiable map and let $x=\psi(y)$. The \emph{pullback} of $\omega$ via $\psi$, denoted by $\psi^\ast\omega$, is the $p$-form on $N$ defined by
\begin{align}\label{eq:pos-pullback}
	(\psi^\ast\omega)(y) = \omega(\psi(y))\,.
\end{align}
Said differently, the pullback operation simply evaluates $\omega$ on $\psi$. In the opposite direction, suppose $\phi:M\to N$ is a surjective meromorphic map of degree $d$. Then $y$ has precisely $d$ pre-images under $\phi$ labelled $\phi^{-1}(\{y\})=\{x^{(\alpha)}\}_{\alpha=1}^d$. For each $\alpha\in[d]$ there exists an open subset $U_\alpha$ containing $x^{(\alpha)}$ such that $\phi|_{U_\alpha}$ is invertible. Let $\xi^{(\alpha)}\coloneqq\phi|_{U_\alpha}^{-1}$ designate the local inverse of $\phi$ restricted to $U_\alpha$. The \emph{pushforward} of $\omega$ via $\phi$, denoted by $\phi_\ast\omega$, is the meromorphic $p$-form on $N$ defined by 
\begin{align}\label{eq:pos-pushforward}
	(\phi_\ast\omega)(y) = \sum_{\alpha=1}^{d} (\xi^{(\alpha)\ast}\omega)(y) = \sum_{\alpha=1}^{d} \omega(\xi^{(\alpha)}(y))\,.
\end{align}

Having modified the definition of morphisms found in \cite{Arkani-Hamed:2017tmz}, we need to clarify our definition of the pushforward in this context. Given a morphism $\Phi:(X,\tnn{X})\to(Y,\tnn{Y})$, let $\mathcal{C}=\{\tnnEnumerate{C}{i}\}_{i\in I}$ be any tiling of $\tnn{Y}$. For each $i\in I$, $\Phi|_{\tpEnumerate{C}{i}}:\tpEnumerate{C}{i}\to\Phi(\tpEnumerate{C}{i})$ is an orientation-preserving diffeomorphism. We define the \emph{pushforward} of $\Omega(\tnn{X})$ via $\Phi$ as
\begin{align}\label{eq:pos-pushforward-morphism}
	\Phi_\ast\Omega(\tnn{X})\coloneqq\sign(\mathcal{C})\sum_{i\in I}\Phi_\ast\Omega(\tnnEnumerate{C}{i})\,,
\end{align}
where $\sign(\mathcal{C})=1$ (resp., $\sign(\mathcal{C})=-1$) if $\mathcal{C}$ is orientation-preserving (resp., orientation-reversing). Remarkably, $\Phi_\ast\Omega(\tnn{X})=\Omega(\tnn{Y})$ \cite{Arkani-Hamed:2017tmz}: one can calculate the canonical form of any morphism image via a pushforward. We will explore the pushforward operation from a computational point of view in \cref{chp:push}.


\begin{example}[Pushforwards]
	Let us compute $\Phi_\ast\Omega(\Delta_3^{\ge0})$ using the one-element tiling $\{(123)\}$. The boundary $(123)$ is an algebraic subset of $\Delta_3$ for which $x_4=0$. Its canonical form is given by
	\begin{align}
		\Omega(123)=\text{Res}|_{x_4=0}\Omega(\Delta_3^{\ge0}) = -d\log\frac{x_1}{x_2}\wedge d\log\frac{x_2}{x_3}\,.
	\end{align}
	The restriction of $\Phi$ to $(123)$ is simply
	\begin{align}
		 \Phi|_{(123)}(x_1,x_2,x_3,0) = (x_1,x_2,x_3)\,.
	\end{align}
	Hence,
	\begin{align}
		\Phi_\ast\Omega(\Delta_3^{\ge0}) = -\Phi|_{(123)\ast}\Omega(123) = \Omega(\Delta_2^{\ge0})\,,
	\end{align}
	where the minus sign after the first equality appears because $\{(123)\}$ is an orientation-reversing tiling. Alternatively, we can use the two-element tiling $\{(124),(134)\}$. The canonical forms for each tile are given by
	\begin{subequations}
		\begin{align}
			\Omega(124) &= \text{Res}|_{x_3=0}\Omega(\Delta_3^{\ge0}) =  d\log\frac{x_1}{x_2}\wedge d\log\frac{x_2}{x_4}\,,\\ 
			\Omega(134) &= \text{Res}|_{x_2=0}\Omega(\Delta_3^{\ge0}) =  -d\log\frac{x_1}{x_3}\wedge d\log\frac{x_3}{x_4}\,.
		\end{align}
	\end{subequations}
	The restriction of $\Phi$ to each tile is given by
	\begin{subequations}
		\begin{align}
			\Phi|_{(124)}(x_1,x_2,0,x_4) &= (x_1,x_2+\frac{x_4}{2},\frac{x_4}{2})\,,\\
			\Phi|_{(134)}(x_1,0,x_3,x_4) &= (x_1,\frac{x_4}{2},x_3+\frac{x_4}{2})\,.
		\end{align}
	\end{subequations}
	Consequently,
	\begin{align}
		\begin{split}
		\Phi_\ast\Omega(\Delta_3^{\ge0}) 
		&= \Phi|_{(124)\ast}\Omega(124) + \Phi|_{(134)\ast}\Omega(134) \\
		&= d\log\frac{x_1}{x_2-x_3}\wedge d\log\frac{x_2-x_3}{2x_3} - d\log\frac{x_1}{x_3-x_2}\wedge d\log\frac{x_3-x_2}{x_2} \\
		&= d\log\frac{x_1}{x_2-x_3}\wedge d\log\frac{x_2}{x_3} = d\log\frac{x_1}{x_2}\wedge d\log\frac{x_2}{x_3} = \Omega(\Delta_2^{\ge0})\,,
		\end{split}
	\end{align}
	as expected.
\end{example}

\section{Boundaries}
\label{sec:pos-boundaries}

In this section, we study the boundaries of positive geometries, focussing on how morphisms repackage this information. We present a simple combinatorial algorithm that determines the boundary stratification of any morphism image: given two positive geometries, where one is the image of another via a morphism, and the latter’s boundaries are known, one can recursively deduce the faces of the former. This algorithm first appeared in \cite{Lukowski:2019kqi}. The Momentum Amplituhedron and the orthogonal Momentum Amplituhedron were studied using this method in \cite{Ferro:2020lgp, Lukowski:2021fkf}. We review these applications in \cref{chp:mom,chp:omom}.

The following definition is inspired by a similar one given in \cite{Moerman:2021cjg}. For each proper boundary $\tnn{C}$ of $\tnn{X}$, we call $\Phi(\tnn{C})$ a \emph{proper $\Phi$-induced boundary (stratum)} or \emph{face} of $\tnn{Y}$ if it adheres to the following requirements:
\begin{axioms}{B}
	\item\label[axiom]{axiom:pos-boundary-1} The image of $\tp{C}$ is not in the interior of $\tnn{Y}$: $\Phi(\tp{C})\cap\tp{Y}=\emptyset$.
	\item\label[axiom]{axiom:pos-boundary-2} The $\Phi$-dimension of every inverse boundary of $\tnn{C}$ is larger than that of $\tnn{C}$: $\dim(\Phi(\tp{C}))<\dim(\Phi(\tp{D}))$ for every $\tnn{D}\in\mathcal{P}^{-1}_{\tnn{X}}(\tnn{C})$. 
\end{axioms}
Let $\mathcal{P}[\Phi](\tnn{X})$ designate the subset of proper boundaries of $\tnn{X}$ corresponding to proper $\Phi$-induced boundaries of $\tnn{Y}$ and declare $\mathcal{B}[\Phi](\tnn{X})\coloneqq\mathcal{P}[\Phi](\tnn{X})\cup\{\tnn{X}\}$. We will refer to $\Phi\circ\mathcal{B}[\Phi](\tnn{X})$ (resp., $\Phi\circ\mathcal{P}[\Phi](\tnn{X})$) as the \emph{$\Phi$-induced boundary stratification} (resp., \emph{proper $\Phi$-induced boundary stratification}) of $\tnn{Y}$. The following conjecture is central to the analysis of \cref{chp:mom,chp:omom}, and implicitly assumed in \cite{Lukowski:2019kqi,Ferro:2020lgp,Lukowski:2021fkf}.

\begin{conjecture}\label{conj:pos-boundaries}
The $\Phi$-induced boundary stratification of $\tnn{Y}$ equals the boundary stratification of $\tnn{Y}$:
\begin{align}
\Phi\circ\mathcal{B}[\Phi](\tnn{X}) = \mathcal{B}(\tnn{Y})\,.
\end{align}
\end{conjecture}

The definition of $\Phi$-induced boundaries is challenging to use in practice. In particular, \cref{axiom:pos-boundary-1} is topological and non-trivially demonstrated, whereas the \cref{axiom:pos-boundary-2} is a purely combinatorial condition and hence straightforward to verify. Fortunately, the algorithm of \cite{Lukowski:2019kqi} provides a practical solution which sidesteps the difficulties associated with \cref{axiom:pos-boundary-1}. The algorithm has two parts. The first step determines all $\Phi$-induced facets starting from any $\Phi$-induced tiling. The second step builds on the previous step to recursively find all higher codimension $\Phi$-induced boundaries. 

To clarify the algorithm's details, consider the following definitions. Let
\begin{align}
	\tilde{\mathcal{B}}[\Phi,\delta](\tnn{X})\coloneqq\left\{\tnn{C}\in\mathcal{B}(\tnn{X}):\dim(\Phi(\tnn{C}))=\dim(\tnn{Y})-\delta\right\},
\end{align}
be the subset of boundaries of $\tnn{X}$ of $\Phi$-codimension-$\delta$. For each boundary $\tnn{C}$ of $\tnn{X}$, let
\begin{align}
	\tilde{\mathcal{B}}_{\tnn{X}}^{-1}[\Phi,\delta](\tnn{C})\coloneqq\left\{\tnn{D}\in\mathcal{B}_{\tnn{X}}^{-1}(\tnn{C}):\dim(\Phi(\tnn{D}))=\dim(\Phi(\tnn{C}))+\delta\right\},
\end{align}
be the subset of inverse boundaries of $\tnn{C}$ in $\tnn{X}$ with $\Phi$-dimension different from $\dim(\Phi(\tnn{C}))$ by $\delta$. We say that $\tnn{C}$ is \emph{$\Phi$-maximal} if  $\tilde{\mathcal{B}}_{\tnn{X}}^{-1}[\Phi,0](\tnn{C})=\{\tnn{C}\}$. We hypothesise that $\tilde{\mathcal{B}}_{\tnn{X}}^{-1}[\Phi,0](\tnn{C})$ always contains a unique $\Phi$-maximal element called the \emph{$\Phi$-maximal cover} of $\tnn{C}$. Finally, let $\mathcal{B}[\Phi,\delta](\tnn{X})$ be the subset of $\mathcal{B}[\Phi](\tnn{X})$ of $\Phi$-codimension-$\delta$, i.e.\ $\Phi\circ\mathcal{B}[\Phi,\delta](\tnn{X})$ are codimensional-$\delta$ $\Phi$-induced boundaries of $\tnn{Y}$. On the basis of \cref{conj:pos-boundaries}, we will omit the adjective ``$\Phi$-induced'' in the following discussion. 

Having established the above definitions, we can now discuss the algorithm for identifying the facets of $\tnn{Y}$. To this end, let $\mathcal{C}$ be any tiling of $\tnn{Y}$. All $\Phi$-codimension-$1$ boundaries of tiles in $\mathcal{C}$ are possible candidates for facets. If a candidate $\tnn{C}$ is a $\Phi$-codimension-$1$ boundary of two or more tiles, its image must lie in $\tp{Y}$. In this instance, $\Phi(\tnn{C})$ is called a \emph{spurious} facet of $\tnn{Y}$, in analogy with the definition of spurious facets in polytopal subdivisions \cite{Arkani-Hamed:2017tmz}. These candidates are excluded. Thereafter, we identify the $\Phi$-maximal covers of the remaining candidates with $\mathcal{B}[\Phi,1](\tnn{X})$. The argument for this identification is as follows. If a candidate $\tnn{C}$ is not $\Phi$-maximal, $\Phi(\tnn{C})$ is necessarily a subset of $\Phi(\tnn{D})$, the image of its $\Phi$-maximal cover $\tnn{D}$, and hence an a element of some dissection of $\Phi(\tnn{D})$.

The method for determining higher codimension boundaries of $\tnn{Y}$ is even simpler. Suppose we know $\mathcal{B}[\Phi,\delta](\tnn{X})$ for some $\delta>0$. To determine $\mathcal{B}[\Phi,\delta+1](\tnn{X})$, we consider all elements in $\tilde{\mathcal{B}}[\Phi,\delta+1](\tnn{X})$ as possible candidates. If a candidate $\tnn{C}$ is not $\Phi$-maximal, $\Phi(\tnn{C})$ must be a subset of the image of its maximal cover. Consequently, we exclude all candidates which are not $\Phi$-maximal. We further exclude any candidate which is a boundary of only one element in $\mathcal{B}[\Phi,\delta](\tnn{X})$; such candidates correspond to higher-codimension spurious boundaries of $\tnn{Y}$. The remaining candidates (conjecturally) give $\mathcal{B}[\Phi,\delta+1](\tnn{X})$. The recursive application of this method terminates after a finite number of iterations when $\mathcal{B}[\Phi,\delta+1](\tnn{X})=\emptyset$. 

This algorithm is implemented in the \Mathematica{} packages \amplituhedronBoundaries{} \cite{Lukowski:2020bya} and \orthitroids{} \cite{Lukowski:2021fkf} for the Momentum Amplituhedron and the orthogonal Momentum Amplituhedron, respectively.

The subset $\mathcal{B}[\Phi](\tnn{X})$ is an induced subposet \cite{stanley2011enumerative} of $\mathcal{B}(\tnn{X})$, inheriting the partial order $\preceq_X$ on $\mathcal{B}(\tnn{X})$. This naturally extends to a partial order on $\Phi\circ\mathcal{B}[\Phi](\tnn{X})$ as follows:
\begin{align}
	\Phi(\tnn{C})\preceq_X\Phi(\tnn{D})\iff\tnn{C}\preceq_X\tnn{D}\,,
\end{align}
for all boundaries $\tnn{C},\tnn{D}\in\mathcal{B}[\Phi](\tnn{X})$. With the above definition in mind, we say that the $\Phi$-induced boundary stratification of $\tnn{Y}$ is an \emph{induced subposet} of $\tnn{X}$. Furthermore, on the support of \cref{conj:pos-boundaries} we speculate that the partial order on $\mathcal{B}(\tnn{Y})$ coincides with the one it inherits from $\mathcal{B}(\tnn{X})$. We will, therefore, refer to the boundary stratification of $\tnn{Y}$ as an \emph{induced subposet} of $\tnn{X}$. In \cref{chp:mom,chp:omom}, we will see that the boundary stratifications of the Momentum Amplituhedron and the orthogonal Momentum Amplituhedron are induced subposets of the non-negative Grassmannian and non-negative orthogonal Grassmannian, respectively. Before finishing this chapter, let us apply the above algorithm to \cref{ex:pos-morphism}.

\begin{figure}[t]
	\centering
	\includegraphics[scale=0.7]{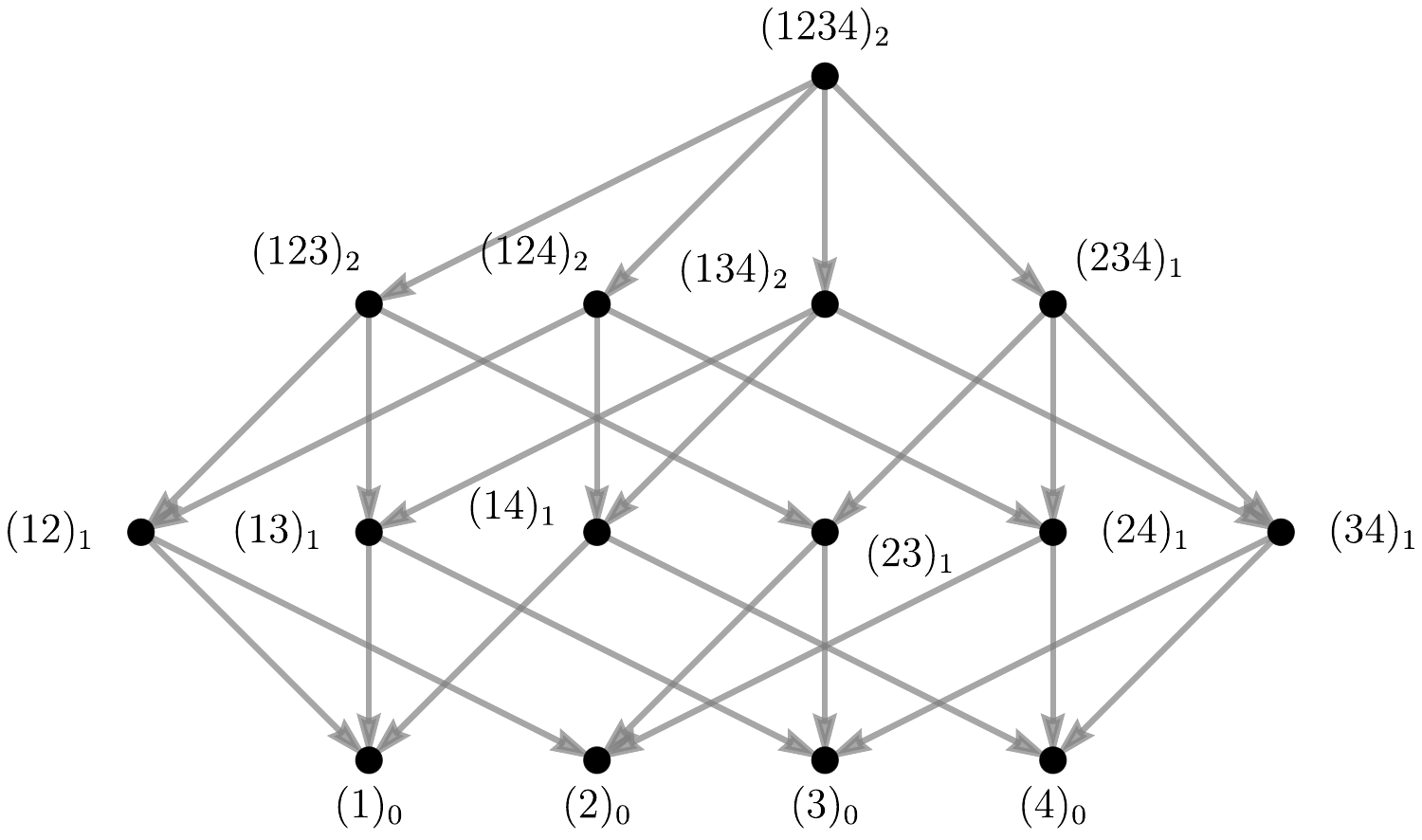}
	\caption{Hasse diagram for $\mathcal{B}(\Delta_3^{\ge0})$. The subscript on each label denotes the $\Phi$-dimension of the corresponding boundary where $\Phi$ is given by \eqref{eq:pos-phi}.}
	\label{fig:pos-poset-Delta-3}
\end{figure}

\begin{figure}[t]
	\centering
	\includegraphics[scale=0.7]{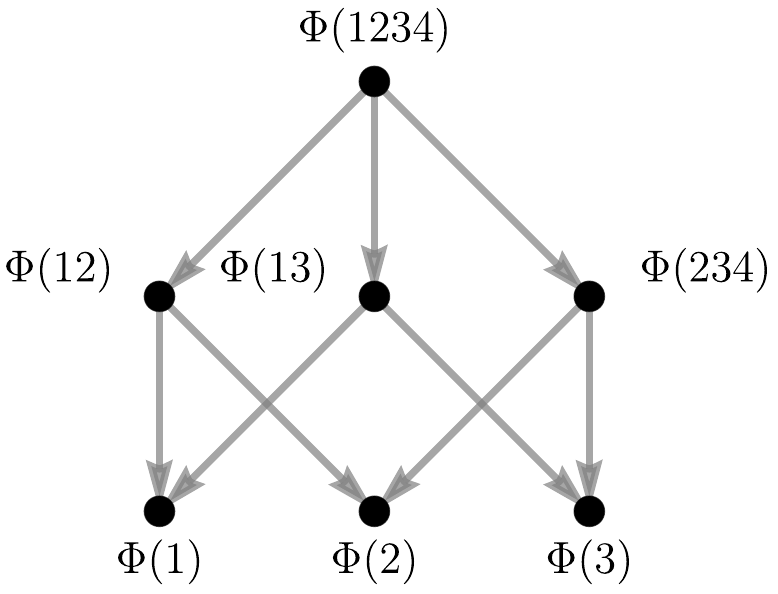}
	\caption{Hasse diagram for $\Phi\circ\mathcal{B}[\Phi](\Delta_2^{\ge0})$ where $\Phi$ is given by \eqref{eq:pos-phi}.}
	\label{fig:pos-poset-induced}
\end{figure}

\begin{example}[Boundaries]
	The above algorithm allows us to determine the $\Phi$-induced boundary stratification of $\Delta_2^{\ge0}$ knowing the boundary stratification of $\Delta_3^{\ge0}$. \cref{fig:pos-poset-Delta-3} displays the Hasse diagram for $\mathcal{B}(\Delta_3^{\ge0})$ together with the $\Phi$-dimension of each boundary. The $\Phi$-dimension of each boundary is computed as follows. Consider $(234)$, for example, where $x_1=0$ and $x_2+x_3+x_4=1$. Solving for $x_4$ in terms of $x_2$ and $x_3$ yields the following parametrisation for $\Phi(234)$:
	\begin{align}
		\Phi(234) = (0,\frac{1}{2}(1+x_2-x_3),\frac{1}{2}(1-x_2+x_3))\,. 
	\end{align}
	The corresponding Jacobian matrix is given by
	\begin{align}
		\frac{\partial\Phi(234)}{\partial(x_2,x_3)} = 
		\begin{pmatrix}
			0 & +\frac12 & -\frac12\\
			0 & -\frac12 & +\frac 12
		\end{pmatrix},
	\end{align}
	from which we obtain 
	\begin{align}
		\dim\Phi(234) = \text{rank}\frac{\partial\Phi(234)}{\partial(x_2,x_3)} = 1\,.
	\end{align}
	This result agrees with the fact that $\Phi$ maps $(234)$ to a line segment. 
	
	Once every boundary's $\Phi$-dimension is known, we need a $\Phi$-induced tiling of $\Delta_2^{\ge0}$ to determine $\mathcal{B}[\Phi,1](\Delta_3^{\ge0})$, the boundaries of $\Delta_3^{\ge0}$ whose images are $\Phi$-induced facets of $\Delta_2^{\ge0}$. Let us choose the two-element tiling $\{(124),(134)\}$. The $\Phi$-codimension-one boundaries of each tile are given by
	\begin{align}
		\tilde{\mathcal{B}}[\Phi,1](124)=\{(12),(14),(24)\}\,,\qquad
		\tilde{\mathcal{B}}[\Phi,1](134)=\{(13),(14),(34)\}\,.
	\end{align}
	Since both tiles share $(14)$ as a $\Phi$-codimension-one boundary, it is spurious and hence discarded. We then identity the $\Phi$-maximal covers of the remaining candidates with $\mathcal{B}[\Phi,1](\Delta_3^{\ge0})$. Both $(12)$ and $(13)$ are $\Phi$-maximal, whereas $(234)$ is the $\Phi$-maximal cover of $(24)$ and $(34)$. Therefore
	\begin{align}
		\mathcal{B}[\Phi,1](\Delta_3^{\ge0}) = \{(12),(13),(234)\}\,.
	\end{align}
	Next we determine $\mathcal{B}[\Phi,2](\Delta_3^{\ge0})$, the boundaries of $\Delta_3^{\ge0}$ whose images are the $\Phi$-induced vertices of $\Delta_2^{\ge0}$. To this end, consider
	\begin{align}
		\tilde{\mathcal{B}}[\Phi,2](1234)=\{(1),(2),(3),(4)\}\,.
	\end{align}
	We discard $(4)$ since it is the boundary of only one element in $\mathcal{B}[\Phi,1](\Delta_3^{\ge0})$, namely $(234)$. The $\Phi$-maximal covers of the remaining candidates give
	\begin{align}
		\mathcal{B}[\Phi,2](\Delta_3^{\ge0}) = \{(1),(2),(3)\}\,.
	\end{align}
	\cref{fig:pos-poset-induced} displays the Hasse diagram for $\Phi\circ\mathcal{B}[\Phi](\Delta_3^{\ge0})$. Moreover, one can easily verify that $\Phi\circ\mathcal{B}[\Phi](\Delta_3^{\ge0}) = \mathcal{B}(\Delta_2^{\ge0})$.
\end{example}

\section{Summary}
In this chapter, we introduced the category of positive geometries (and their canonical forms). We will encounter several examples of physically-relevant positive geometries, including the ABHY associahedron in \cref{chp:abhy}, the Momentum Amplituhedron in \cref{chp:mom} and the orthogonal Momentum Amplituhedron in \cref{chp:omom}. We then introduced morphisms between positive geometries, extending the original definition of \citeauthor{Arkani-Hamed:2017tmz} in \cite{Arkani-Hamed:2017tmz} to positive geometries of possibly differing dimensions. This generalisation is motivated by the original Grassmannian-based constructions of the Momentum Amplituhedron \cite{Damgaard:2019ztj} and the orthogonal Momentum Amplituheron \cite{Huang:2021jlh,He:2021llb}, which we review in \cref{chp:mom,chp:omom}, respectively. We further explored the utility of morphisms for calculating canonical forms and determining boundary stratifications. Given two positive geometries related via a morphism, the canonical form of the image is the pushforward of the domain’s canonical form. We will unpack the pushforward operation in more detail in \cref{chp:push}. 
We also formalised a crucial conjecture: the boundary stratification of a morphism’s image coincides with the boundary stratification induced by the morphism. 
In addition, we presented an algorithm for generating morphism-induced boundary stratifications. We will review two applications of this algorithm in \cref{chp:mom,chp:omom}. Concretely, we will see how this algorithm gives access to all physical singularities of tree-level amplitudes in the relevant QFT. While this chapter provides algorithmic tools for identifying morphism-induced boundaries, the next chapter develops methods for systematically enumerating them. 
	\chapter{Elements of Enumerative Combinatorics}
\label{chp:enum}

\lettrine{T}{he Exponential Formula} is a classical result in enumerative combinatorics; see e.g.\ \cite{stanley1999enumerative,bergeron1998combinatorial,ardila2015algebraic}. It allows one to construct the (exponential) generating function for any class of combinatorial objects built by choosing a partition and placing a structure on each block which depends only on the block’s size. A generating function is a useful book-keeping device for encoding an infinite sequence of numbers as coefficients of a formal power series. There are several variants of the Exponential formula. They include Speicher's result for non-crossing partitions \cite{speicher1994multiplicative} and analogues for series-reduced planar trees and forests. We focus on the latter two results in this chapter, following the presentation of \cite{Moerman:2021cjg}. These results are relevant to \cref{chp:abhy,chp:mom,chp:omom}, where we apply them to derive rank generating functions for the boundaries of various positive geometries in on-shell momentum space. We will also reuse some of this chapter’s definitions in \cref{chp:grass}.

\section{Generating Functions}
\label{sec:enum-gen}

Given a field $k$, let $k[[x]]$ denote the set of all (formal) power series in $x$ over $k$. The subset of all power series with zero constant term is designated by $x\,k[[x]]$. Under the operation of functional composition, $xk[[x]]$ forms a monoid where the identity element is the power series $x$. Let $f\in k[[x]]$ be a power series. The notation $[x^n]f(x)$ denotes the coefficient of the monomial $x^n$. A power series $g\in k[[x]]$ is called a \emph{compositional inverse} of $f$ if $(f\circ g)(x) = (g\circ f)(x) =x$, in which case we write $g(x)=f^{\lrangle{-1}}(x)$. 

One can iteratively determine the coefficients of a power series' compositional inverse\footnote{In \Mathematica{}, one can use the built-in function \texttt{InverseSeries} to perform series reversion.} using the \emph{Lagrange inversion formula} \cite[Theorem 5.4.2]{stanley1999enumerative}. In particular, let $k$ have a zero characteristic and let $f\in xk[[x]]$ be a power series such that $[x]f(x)$ is non-zero. Then for positive integers $k$ and $n$, we have that
\begin{align}\label{eq:enum-LIF}
	n [x^n] f^{\lrangle{-1}}(x)^{k}=k[x^{n-k}]\left(\frac{x}{f(x)}\right)^n.
\end{align}
We will repeatedly use this formula in subsequent sections.

\section{Variations of the Exponential Formula}
\label{eq:enum-var}

Many combinatorial objects are built by choosing a set of connected components and placing some structure on each piece that depends only on the piece's size. In this case, if one can enumerate all possible placements of said structure based on size, then the well-known \emph{Exponential Formula} (see e.g.\ \cite{stanley1999enumerative,bergeron1998combinatorial,ardila2015algebraic}) enumerates the resulting class of combinatorial objects. Speicher formulated an analogue of the Exponential Formula that applies to combinatorial objects built by choosing a non-crossing partition and placing a structure on each block \cite{speicher1994multiplicative} (see also \cite[Exercise 5.35b]{stanley1999enumerative}). There are also analogues of the Exponential Formula for series-reduced planar trees and forests \cite{Moerman:2021cjg}. In this section, we review these results. We will use them in later chapters to enumerate the boundaries of positive geometries (equivalently, the physical singularities of tree-level amplitudes). 

\subsection{Non-Crossing Partitions and Speicher's Result}
\label{sec:enum-ncp}

\begin{figure}
	\centering
	\includegraphics[scale=0.45]{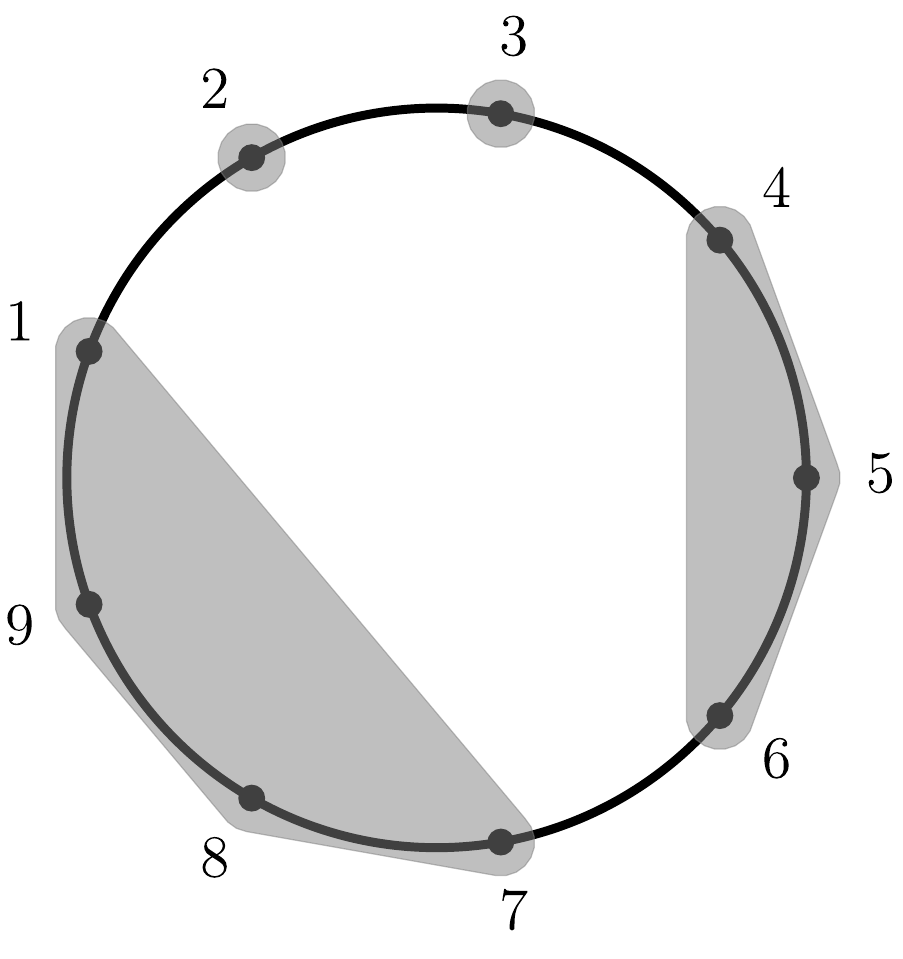}
	\caption{The non-crossing partition $\{\{1, 7, 8, 9\},\{2\}, \{3\}, \{4, 5, 6\}\}$ of $[9]$.}
	\label{fig:enum-ncp}
\end{figure}

A \emph{non-crossing partition} of $[n]$ is a partition $\pi$ of $[n]$ into blocks $B_1,\ldots,B_t$ satisfying the following requirement: if $a<b<c<d$ and $B$ and $B'$ are blocks such that $a,c\in B$ and $b,d\in B'$, then $B=B'$. Alternatively, consider $n$ points on a circle, labelled $\{1,\ldots,n\}$ in order. Then $\pi$ is a non-crossing partition of $[n]$ if the convex hull of the points labelled by each block in $\pi$ produce non-overlapping polygons. \cref{fig:enum-ncp} displays an example of a non-crossing partition.

Let $\NC_n$ denote the set of non-crossing partitions of $[n]$. Let $\mathcal{H}_{\NC}$ be a class of combinatorial objects built by choosing a non-crossing partition and independently placing some structure on each block. Let $f$ be a function on $\integer_{>0}$ where $f(d)$ enumerates all possible placements of said structure on a block of size $d$. Then the generating function $H_{\NC}(x)$ for $\mathcal{H}_{\NC}$ depends on the generating function $F(x)$ for $f$ as stated by \cref{thm:enum-speicher}.

\begin{theorem}[{Speicher's analogue of the Exponential Formula \cite{speicher1994multiplicative}}]
	\label{thm:enum-speicher}
	Let $k$ be a field. Given some function $f:\integer_{>0} \to k$, define a new function $h:\integer_{>0} \to k$ by 
	\begin{align}
		h(n)=\sum_{\pi = \{B_1,\dots,B_\ell\}\in \NC_n} f(\# B_1) f(\# B_2)\dots f(\# B_\ell)\,,
	\end{align} 
	where $\# B_i$ denotes the cardinality of block $B_i$. Define $F(x)=1+\sum_{n\geq 1}f(n)x^n$ and $H_{\NC}(x) = 1+\sum_{n \geq 1} h(n)x^n$. Then
	\begin{align}
		x H_{\NC}(x) = \left(\frac{x}{F(x)}\right)^{\lrangle{-1}}.
	\end{align}	
\end{theorem}

\subsection{Series-Reduced Planar Trees and Forests}
\label{sec:enum-graphs}

In perturbative calculations, one usually expresses amplitudes as a sum of Feynman diagrams or on-shell diagrams --- graphs with additional data assigned to edges and vertices. We will focus exclusively on tree-level amplitudes in this dissertation. Consequently, we introduce planar trees and forests. These objects will be relevant for \cref{chp:grass,chp:abhy,chp:mom,chp:omom}.

A \emph{planar graph} $\Gamma$ on $n$ leaves is a graph with \emph{vertices} $\Vertices(\Gamma)$ and \emph{edges} $\Edges(\Gamma)$, properly embedded in a disk (and considered up to homeomorphism). It has $n$ \emph{boundary vertices} (i.e.\ vertices of degree one) on the boundary of the disk labelled $\{1,2,\ldots,n\}$ in clockwise order. Let $\Vertices_\text{ext}(\Gamma)\coloneqq\{b_1,b_2,\ldots,b_n\}$ denote the boundary vertices of $\Gamma$. The set of \emph{internal vertices} of $\Gamma$ is designated by $\Vertices_\text{int}(\Gamma)\coloneqq\Vertices(\Gamma)\setminus\Vertices_\text{ext}(\Gamma)$. The internal vertices of $\Gamma$ must be path-connected to the boundary of the disk. The \emph{internal edges} of $\Gamma$ are those which are non-adjacent to boundary vertices. Let $\Edges_\text{int}(\Gamma)$ denote the set of internal edges of $\Gamma$. The \emph{internal subgraph} of $\Gamma$, denoted by $\Gamma_\text{int}$, is the graph with vertices $\Vertices(\Gamma_\text{int})=\Vertices_\text{int}(\Gamma)$ and edges $\Edges(\Gamma_\text{int})=\Edges_\text{int}(\Gamma)$. A planar graph is called \emph{finite} if it has a finite number of vertices and edges. It is called \emph{series-reduced} if it contains no internal vertices of degree $2$. 

\begin{figure}
	\centering
	\includegraphics[scale=0.45]{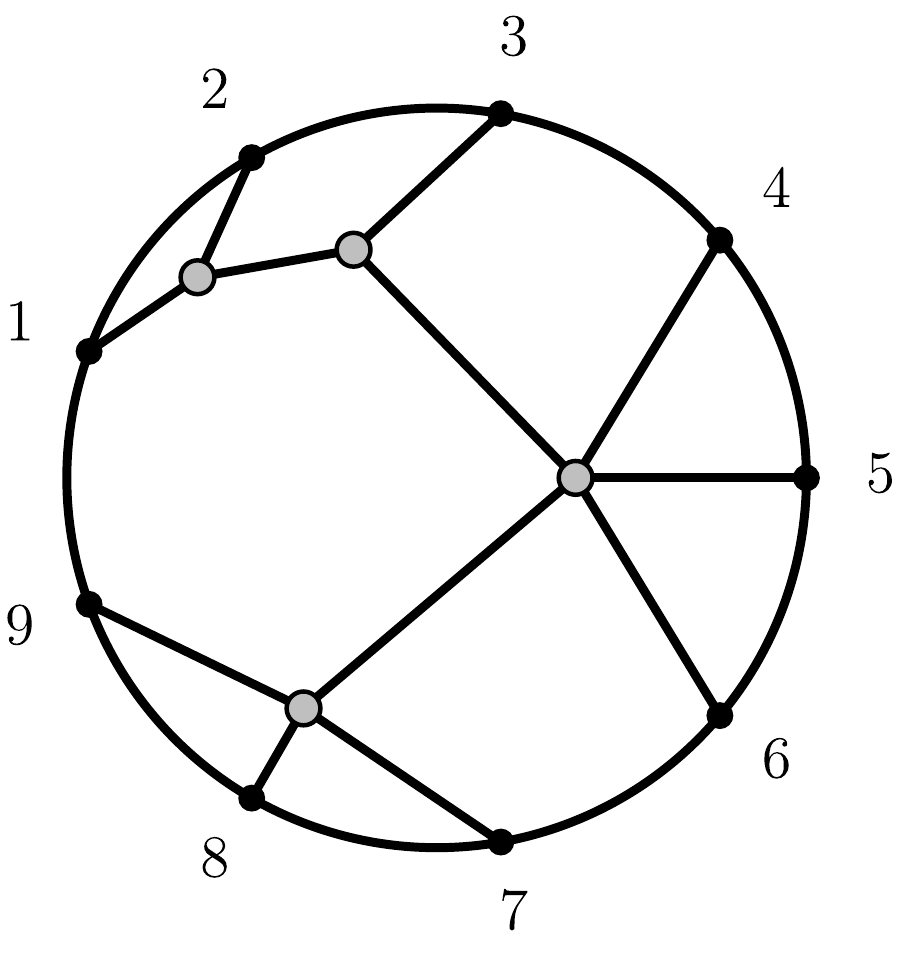}
	\caption{A series-reduced planar tree on $9$ leaves of type $(2, 1, 1, 0, 0, 0, 0)$.}
	\label{fig:enum-trees}
\end{figure}

A \emph{planar forest} is an acyclic planar graph while a \emph{planar tree} is a connected planar forest. The set of trees contained in a planar forest $F$ is denoted by $\Trees(F)$. Let $\Tree_n$ (resp., $\Forest_n$) denote the set of series-reduced planar trees (resp.\ forests) on $n$-leaves. Both $\Tree_n$ and $\Forest_n$ are finite due to the restriction on the degrees of internal vertices. A tree $T\in\Tree_n$ is said to be of \emph{type} $(r_3,\ldots,r_n)$ if it has $r_i$ internal vertices of degree $i$. We denote the subset of $\Tree_n$ of type $\vec{r}=(r_3,\ldots,r_n)$ by $\Tree_n(\vec{r})$. Its size, denoted by $t_n(\vec{r}) \coloneqq |\Tree_n(\vec{r})|$, is given by \cite{Moerman:2021cjg}
\begin{align}
	\label{eq:enum-tree-count}
	t_n(\vec{r}) = \frac{(n+|\vec{r}|-2)!}{(n-1)! r_3! \cdots r_n!}\,.
\end{align}
where $|\vec{r}|=\sum_{i=3}^nr_i = r_3 + \ldots + r_n$. An example of a series-reduced planar tree is shown in \cref{fig:enum-trees}.

With the above definitions established, we now consider two important analogues of the Exponential Formula which we will frequently use in the following chapters.

Let $\mathcal{H}_{\tree}$ be a class of combinatorial objects built by choosing a series-reduced planar tree in $\Tree_n$ and independently placing a structure on each internal vertex. Let $f$ be a function on $\integer_{\ge3}$ where $f(d)$ enumerates all possible placements of said structure on an internal vertex of degree $d$. Then the generating function $H_{\tree}(x)$ for $\mathcal{H}_{\tree}$ can be obtained from the generating function $F(x)$ for the function $f$ as per \cref{thm:enum-tree}.
\pagebreak
\begin{theorem}[{Series-reduced planar tree analogue of the Exponential Formula \cite{Moerman:2021cjg}}]
	\label{thm:enum-tree}
	Let $k$ be a field. Given a function $f:\integer_{\ge 3} \to k$, define a new function $h:\integer_{\ge 3} \to k$ by
	\begin{align}
		\label{eq:enum-tree-h}
		h(n)=\sum_{T \in \Tree_n} \prod_{v\in \Vertices_\text{int}(T)} f(\deg(v))\,.
	\end{align}
	Let $F(x)=\sum_{n\geq 3} f(n)x^n$ and 
	$H_{\tree}(x) = x^2+ \sum_{n \geq 3} h(n)x^n$.  Then
	\begin{align} 
		\label{eq:enum-tree-H}
		\frac{1}{x}H_{\tree}(x) = {\left(x-\frac{1}{x}F(x) \right)}^{\lrangle{-1}}.
	\end{align}
	The inclusion of $x^2$ in $H_{\tree}(x)$ reflects the unique tree in $\Tree_2$. Modifying $H_{\tree}(x)$ by 
	\begin{align}
		\label{eq:enum-tree-Hhat}
		\hat{H}_{\tree}(x) = h(1)x + H_{\tree}(x)\,,
	\end{align}
	we can account for the unique tree in $\Tree_1$ (which has a single internal vertex). In this case, $h(1)=f(1)$ enumerates all possible structures that can be placed on the single internal vertex.
\end{theorem}

The authors of \cite{Moerman:2021cjg} remark on the appearance of \cref{thm:enum-tree} in various references: \textit{``for example, it fits naturally into the theory of species and is closely related to \cite[page 168, Equation (18)]{bergeron1998combinatorial}; it can also be reformulated using Schr\"{o}der trees and viewed as a combinatorial interpretation of Lagrange Inversion, see \cite[Theorem 2.2.1]{ardila2015algebraic}.''}

\begin{example}[Feynman Diagrams for MHV Gluon Amplitudes]
	As a simple application of \cref{thm:enum-tree}, let us derive a generating function for the number of Feynman diagrams needed to calculate the $n$-particle MHV gluon amplitude. The relevant Feynman diagrams are series-reduced planar trees on $n$-leaves that have only trivalent and tetravalent internal vertices. To this end, define the function $f:\integer_{\ge3}\to\real$ by
	\begin{align}
		f(d)\coloneqq
		\begin{cases}
			1,& \text{if}~d=3~\text{or}~4\\
			0,& \text{otherwise}
		\end{cases},
	\end{align}
	i.e.\ $f$ only keeps track of trivalent and tetravalent internal vertices. Let $F(x)$ and $H_\text{tree}(x)$ be as they are in \cref{thm:enum-tree}. Then $F(x) = x^3+x^4$ and 	$H_\text{tree}(x)=x(x-x^2-x^3)^{\lrangle{-1}}$. The first few coefficients in the power series for $H_\text{tree}(x)$ in $x$ are given in \cref{tbl:enum-FD}. These coefficients correctly count the relevant Feynman diagrams \cite{Elvang:2013cua,oeis:A001002}.
\end{example}

\begin{table}
	\centering
	\begin{tabular}{c| cccccccccc}
		\toprule
		$n$ & $4$ & $5$ & $6$ & $7$ & $8$ & $9$ & $10$ & $11$ & $12$ & $13$\\
		\midrule
		$[x^n]H_\text{tree}(x)$ & $3$ & $10$ & $38$ & $154$ & $654$ & $2871$ & $12925$ & $59345$ & $276835$ & $1308320$ \\
		\bottomrule
	\end{tabular}
	\caption{The coefficient of $x^n$ in $H_\text{tree}(x)=x(x-x^2-x^3)^{\lrangle{-1}}$, denoted by $[x^n]H_\text{tree}(x)$, gives the number of Feynman diagrams contributing to the $n$-particle MHV gluon amplitude.}
	\label{tbl:enum-FD}
\end{table}

Reference \cite{Moerman:2021cjg} gives a proof for \cref{thm:enum-tree}, but one can also construct a proof along the lines of \cite[Exercise 5.35b]{stanley1999enumerative} which we do below. 

\begin{proof}[Alternative proof of \cref{thm:enum-tree}]
	First we must check that the coefficient of $x^1$ agrees on both sides of \eqref{eq:enum-tree-H}. On the left-hand side, $[x^1] \frac{1}{x} H(x) = [x^2]H(x) = 1$ by construction. On the right-hand side, using \eqref{eq:enum-LIF} we obtain
	\begin{align*}
		[x^1]\left(x-\frac{1}{x} F(x)\right)^{\lrangle{-1}} =
		[x^0] \left(\frac{x}{x-\frac{F(x)}{x}}\right) =
		[x^0] \left(\frac{1}{1-\frac{F(x)}{x^2}}\right) 
		= 1.
	\end{align*} 
	For $n\geq 3$, the coefficient of $x^{n-1}$ on the left-hand side of \eqref{eq:enum-tree-H} is $h(n)$. In  \eqref{eq:enum-tree-h}, the formula for $h(n)$, the sum over trees $T\in\Tree_n$ can be re-expressed as a sum over trees $T\in\Tree_n(\vec{r})$ multiplied by $t_n(\vec{r})$ for all tree types $\vec{r}=(r_3,\ldots,r_n)$, i.e.\ all possible partitions $\lrangle{1^{r_3},\ldots,(n-2)^{r_n}}\vdash(n-2)$. Consequently,
	\begin{align*}
		h(n) &= \sum_{\lrangle{1^{r_3},\dots ,(n-2)^{r_n}} \vdash n-2}f(3)^{r_3}\cdots f(n)^{r_n}\, t_n(\vec{r})\\
		&= \sum_{\lrangle{1^{r_3},\dots ,(n-2)^{r_n}} \vdash n-2}f(3)^{r_3}\cdots f(n)^{r_n} \frac{(n+|\vec{r}|-2)!}{(n-1)! r_3!\cdots r_n!}\\
		&= \sum_{\lrangle{1^{r_3},\dots ,(n-2)^{r_n}} \vdash n-2}
		\frac{(n+|\vec{r}|-2)!}{(n-1)!} \cdot \frac{1}{|\vec{r}|!} 
		\cdot f(3)^{r_3}\cdots f(n)^{r_n} \binom{|\vec{r}|}{r_3,\cdots, r_n}\\
		&=\frac{1}{n-1}\sum_{|\vec{r}|=1}^{n-2}\binom{n+|\vec{r}|-2}{|\vec{r}|}[x^{n-2}]\left(\frac{F(x)}{x^2}\right)^{|\vec{r}|}\\
		&=\frac{1}{n-1}[x^{n-2}]\sum_{|\vec{r}|=0}^{\infty}\binom{n+|\vec{r}|-2}{|\vec{r}|}\left(\frac{F(x)}{x^2}\right)^{|\vec{r}|}\\
		&=\frac{1}{n-1}[x^{n-2}]\left(\frac{1}{1-\frac{F(x)}{x^2}}\right)^{n-1}\\
		&=\frac{1}{n-1}[x^{n-2}]\left(\frac{x}{x-\frac{F(x)}{x}}\right)^{n-1}\\
		&=[x^{n-1}]\left(x-\frac{1}{x}F(x)\right)^{\lrangle{-1}},
	\end{align*}
	where in the last line we used \eqref{eq:enum-LIF}.
\end{proof}

Combining \cref{thm:enum-tree,thm:enum-speicher} produces \cref{thm:enum-forest}, a series-reduced planar forest analogue of the Exponential Formula \cite{Moerman:2021cjg}. It has the following interpretation. Suppose $\mathcal{H}_{\forest}$ is a class of combinatorial objects built by choosing a series-reduced planar forest in $\Forest_n$ and independently placing a structure on each internal vertex. Let $g$ be a function on $\integer_{\ge 3}$ where $g(d)$ enumerates all possible placements of said structure on an internal vertex of degree $d$. Then the generating function $H_{\forest}(x)$ for $\mathcal{H}_{\forest}$ can be obtained from the generating function $G(x)$ for the function $g$ as per \cref{thm:enum-forest}.

\begin{corollary}[{Series-reduced planar forest analogue of the exponential formula \cite{Moerman:2021cjg}}]
	\label{thm:enum-forest}
	Let $k$ be a field. Given a function $g:\integer_{>0} \to k$, define a new function $h:\integer_{>0} \to k$  by
	\begin{align} 
		\label{eq:enum-forest-h}
		h(n)=\sum_{F \in \Forest_n} \prod_{v\in \Vertices_\text{int}(F)} g(\deg(v))\,.
	\end{align}
	Note that $h(1)=g(1)$, and that $g(2)$ is irrelevant for series-reduced planar forests. Let $G(x)=\sum_{n\geq 3} g(n)x^n$ and $H_{\forest}(x) = 1+\sum_{n \geq 1} h(n)x^n$. Then	
	\begin{align} 
		\label{eq:enum-forest-H}
		x H_{\forest}(x) 
		= {\left( \frac{x}{1+\hat{H}_{\tree}(x)} \right)}^{\lrangle{-1}} 
		= {\left( \frac{x}{1+h(1)x+ x\left( x-\frac{G(x)}{x} \right)^{\lrangle{-1}} } \right)}^{\lrangle{-1}}, 
	\end{align}
	where $\hat{H}_{\tree}(x)$ is given by \eqref{eq:enum-tree-Hhat}.
\end{corollary}

\section{Summary}
The main results of this chapter are \cref{thm:enum-tree,thm:enum-forest}, which provide procedures for finding the rank generating functions for any class of combinatorial objects built from series-reduced trees and forests. In particular, they facilitate the counting of Feynman diagrams and on-shell diagrams. We will employ these results in \cref{chp:abhy,chp:mom,chp:omom} to calculate rank generating functions for boundaries of various positive geometries in on-shell momentum space. These rank generating functions effectively enumerate all physical singularities of tree-level amplitudes according to some hierarchy. Crucially, \cref{thm:enum-tree,thm:enum-forest} allow us to bootstrap our calculations: we need only categorise all possible structures on vertices as a function of vertex degree. In the following chapter, we introduce the combinatorial objects relevant to \cref{chp:mom,chp:omom}. 
	\chapter{Grassmannian Spaces}
\label{chp:grass}

\lettrine{T}{he Grassmannian} is a sophisticated mathematical object which generalises complex projective space. It features prominently in the on-shell description of amplitudes in various theories (see \cite{Arkani-Hamed:2012zlh} and references therein). This chapter reviews the Grassmannian and its orthogonal counterpart, the orthogonal Grassmannian, before introducing their non-negative parts. The non-negative (orthogonal) Grassmannian admits a cell decomposition labelled by (orthogonal) Grassmannian graphs. These non-negative parts are pertinent to the Momentum Amplituhedron and the orthogonal Momentum Amplituhedron, as we will explain in \cref{chp:mom,chp:omom}. Our discussion will focus on (orthogonal) Grassmannian forests --- acyclic (orthogonal) Grassmannian graphs --- which label the boundaries of the above-mentioned positive geometries. Using the results of \cref{chp:enum}, we will enumerate Grassmannian forests (resp., orthogonal Grassmannian forests) in  \cref{chp:mom} (resp., \cref{chp:omom}) as a proxy for the boundaries of the Momentum Amplituhdron (resp., orthogonal Momentum Amplituhedron). This chapter also includes the covering relations for (orthogonal) Grassmannian forests, which characterise the corresponding partial order. 

\section{The Grassmannian}
\label{sec:grass}

\subsection{Definition}
\label{sec:grass-def}

To define the Grassmannian, let $k$ and $n$ be positive integers such that $k<n$. The \emph{Grassmannian} $\G{k}{n}$ is the set of all $k$-dimensional linear subspaces of $\complex^n$. The \emph{complex projective $n$-space} $\projective^{n-1}=\G{1}{n}$ is a special case of the Grassmannian. Each element $V\in\G{k}{n}$ can be represented by a full rank $k\times n$ matrix (modulo invertible row operations) whose rows span $V$. Consequently,
\begin{align}\label{eq:grass-def}
	\G{k}{n}\coloneqq\left\{C\in\Mat{k}{n}{\complex}:\rank(C)=k\right\}/\GL{k}{\complex}\,,
\end{align}
where $\GL{k}{\complex}$ acts on $\Mat{k}{n}{\complex}$ by left multiplication. The Grassmannian's dimension is given by
\begin{align}\label{eq:grass-dimension}
	\dim(\G{k}{n}) = \dim(\Mat{k}{n}{\complex}) - \dim(\GL{k}{\complex}) = k(n-k)\,.
\end{align}
The notation $[C]=\GL{k}{\complex}\cdot C$ denotes an element in $\G{k}{n}$ represented by the matrix $C$. We label the rows of $C$ by  $C_1,\ldots,C_k\in\complex^n$ (i.e.\ using uppercase letters) and the columns of $C$ by $c_1,\ldots,c_n\in\complex^k$ (i.e.\ using lowercase letters). 

The \emph{Pl\"{u}cker map} embeds the Grassmannian $\G{k}{n}$ in $\projective(\wedge^k\complex)$ by mapping $V=[C]\in\G{k}{n}$ to $[C_1\wedge\cdots\wedge C_k]$, the projective equivalence class of $C_1\wedge\cdots\wedge C_k$. This embedding, known as the \emph{Pl\"{u}cker embedding}, depends on the equivalence class and not on any representative. It defines a natural choice of coordinates for $\G{k}{n}$ in $\projective(\wedge^k\complex)\cong\projective^{\binom{n}{k}-1}$. To see this, let $\{e_i\}_{i=1}^n$ be the standard basis for $\complex^n$ and for each ordered set $I\in\binom{[n]}{k}$ define $e_I\coloneqq\wedge_{i\in I}e_i$. Then
\begin{align}\label{eq:grass-wedge}
	C_1\wedge\cdots\wedge C_k = \sum_{I\in\binom{[n]}{k}}p_I(C)e_I\,,
\end{align}
where $p_I(C)$ is the maximal minor of $C$ located in the column set $I$. These maximal minors furnish projective coordinates for $V$ called \emph{Pl\"{u}cker coordinates}, independent of the choice of representative (up to simultaneous rescaling by a constant). Consequently, we write $p_I(V)$ to mean $p_I(C)$. In the physics literature, $p_I(C)$ is often written as $(i_1,\ldots,i_k)$ where $I=\{i_1,\ldots,i_k\}$. We will adopt the latter notation in later chapters.

The Pl\"{u}cker embedding of $\G{k}{n}$ is an algebraic set in $\projective(\wedge^k\complex)\cong\projective^{\binom{n}{k}-1}$ defined by a family of quadratic relations called the \emph{Pl\"{u}cker relations}. In terms of Pl\"{u}cker coordinates, these relations for $V\in\G{k}{n}$ are given by
\begin{align}\label{eq:grass-plucker-relations}
	\sum_{\ell=1}^{k+1}(-1)^\ell\,p_{I\cup\{j_l\}}(V)\,p_{J\setminus\{j_\ell\}}(V) = 0\,,
\end{align}
where $I\in\binom{[n]}{k-1}$ and $J=\{j_1,\ldots,j_{k+1}\}\in\binom{[n]}{k+1}$ are ordered sets.

The Grassmannian admits a decomposition into strata labelled by affine permutations (or equivalently decorated permutations, as discussed later). An \emph{affine permutation} $f$ on $[n]$ is a bijection $f:\integer\to\integer$ for which $f(i+n)=f(i)+n$ and $i\le f(i)\le i+n$ for all $i\in\integer$ \cite{knutson2013positroid}. We say that $f$ is \emph{$(k,n)$-bounded} if $\sum_{i=1}^n(f(i)-i)=kn$ \cite{knutson2013positroid}. We can associate a $(k,n)$-bounded affine permutation $f$ to each element $[C]\in\G{k}{n}$ as follows \cite{Postnikov:2006kva,Arkani-Hamed:2012zlh,knutson2013positroid}. For each pair $(i,j)\in\integer^2$ define $r(i,j)$ as the rank of the matrix with columns $c_i,c_{i+1}\ldots,c_j$ if $i\le j$ (the indices labelling the columns of $C$ are understood modulo $n$) and zero otherwise. For each $i\in\integer$, define 
\begin{align}\label{eq:grass-affine-permutation}
	f(i)\coloneqq\min\{j\ge i: r(i,j)=r(i+1,j)\}\,.
\end{align}
If $c_i$ is a zero vector, $f(i)=i$, and if $c_i$ is not in the span of the other columns of $C$, $f(i)=i+n$. The \emph{positroid stratification} of the Grassmannian $\G{k}{n}$ is the decomposition
\begin{align}\label{eq:grass-prositroid-stratification}
	\G{k}{n}=\bigsqcup_{f}\PosCirc{f}\,,
\end{align} 
where $\PosCirc{f}$ denotes the subset of $\G{k}{n}$ labelled by the affine permutation $f$. The \emph{positroid variety} $\Pos{f}$, the closure of $\PosCirc{f}$ in $\G{k}{n}$ \cite{Arkani-Hamed:2017tmz}, is a complex projective variety \cite{knutson2013positroid} whose dimension is given by \cite{Arkani-Hamed:2012zlh}
\begin{align}\label{eq:grass-positroid-dimension}
	\dim(\Pos{f})=\left(\sum_{i=1}^nr(i,f(i))\right)-k^2\,.
\end{align}

\subsection{The Non-negative Grassmannian}
\label{sec:grass-tnn}

The real Grassmannian $\GReal{k}{n}$ is the real part of $\G{k}{n}$. The \emph{positive Grassmannian} $\tpG{k}{n}$ (resp., \emph{non-negative Grassmannian} $\tnnG{k}{n}$) is the semi-algebraic set of elements of $\GReal{k}{n}$ whose Pl\"{u}cker coordinates are all positive (resp., non-negative) \cite{Postnikov:2006kva}. Given a $(k,n)$-bounded affine permutation $f$, the intersections
\begin{align}\label{eq:grass-positroid}
	\tpPos{f}\coloneqq\tnnG{k}{n}\cap\PosCirc{f}\,,\qquad
	\tnnPos{f}\coloneqq\tnnG{k}{n}\cap\Pos{f}\,,
\end{align}
are called open and closed \emph{positroid cells}, respectively \cite{Arkani-Hamed:2017tmz}. Moreover,  $(\Pos{f},\tnnPos{f})$ is a positive geometry \cite{Arkani-Hamed:2017tmz} whose canonical form was studied in \cite{Arkani-Hamed:2012zlh,knutson2013positroid}. The authors of \cite{Arkani-Hamed:2012zlh} construct a particular parametrization for $\Pos{f}$ using \emph{``BCFW bridges''}. This parametrization furnishes \emph{canonically positive coordinates} for $\tpPos{f}$: they construct a map $C_f:\real^d\to\Mat{k}{n}{\real}$, where $d=\dim(\Pos{f})$ and $[C_f(\real^d)]\subset\Pos{f}$, such that $[C_f(\alpha)]\in\tpPos{f}$ if and only if $\alpha\in\real_{>0}^d$. We refer the reader to \cite{Arkani-Hamed:2012zlh} for further details. This canonically positive parametrization is implemented in the \Mathematica{} package \positroids{} \cite{Bourjaily:2012gy}. Using canonically positive coordinates $\alpha=(\alpha_1,\ldots,\alpha_d)$, the canonical form $\Omega(\tnnPos{f})$ is given by \cite{Arkani-Hamed:2012zlh}
\begin{align}\label{eq:grass-positroid-Omega}
\Omega(\tnnPos{f})=\bigwedge_{i=1}^dd\log\alpha_i\,.
\end{align}

Postnikov proved that the positroid cells of $\tnnG{k}{n}$ are in bijection with various combinatorial objects including so-called move-equivalence classes of reduced plabic graphs $\Gamma$ of type $(k,n)$ \cite{Postnikov:2006kva}. The notion of plabic graphs was later generalized to Grassmannian graphs, first implicitly in \cite{Arkani-Hamed:2012zlh}, and then formally in \cite{Postnikov:2018jfq}. The latter reference establishes a bijection between positroid cells of $\tnnG{k}{n}$ and refinement-equivalence classes of reduced Grassmannian graphs $\Gamma$ of type $(k,n)$. Since they provide unambiguous labels for positroid cells, we denote by $\tpPos{\Gamma}$ (resp., $\tnnPos{\Gamma}$) the open positroid cell (resp., closed positroid cell) labelled by $\Gamma$. 

\subsection{Grassmannian Graphs, Trees and Forests}
\label{sec:grass-graphs}

In this subsection, we introduce Grassmannian graphs. Then we define Grassmannian trees and forests, which will feature prominently in \cref{chp:mom}. Most of the definitions and results presented here paraphrase those of \cite{Postnikov:2018jfq}, unless otherwise stated.

A \emph{Grassmannian graph} $\Gamma$ on $n$ leaves is a finite series-reduced planar graph, where each internal vertex $v$ is equipped with a non-negative integer $h(v)$ called the \emph{helicity} of $v$, satisfying $\min\{1,\deg(v)-1\}\le h(v)\le\max\{1,\deg(v)-1\}$. If $v$ is a \emph{boundary leaf}, an internal vertex of degree $1$ connected to a boundary vertex, then $h(v) \in\{0,1\}$, otherwise $1 \leq h(v) \leq \deg(v)-1$. We say that $v$ is of \emph{type $(h,d)$} if $h=h(v)$ and $d=\deg(v)$. We refer to $v$ as \emph{white} if $h(v)=1$, \emph{black} if $h(v)=\deg(v)-1$, and \emph{generic} otherwise. The \emph{helicity} of $\Gamma$ is an integer, given by 
\begin{align}\label{eq:grass-helicity-graph}
	h(\Gamma)\coloneqq\sum_{v\in\Vertices_\text{int}(\Gamma)}\left(h(v)-\frac{\deg(v)}{2}\right) +\frac{n}{2}\,.
\end{align}  
and bounded by $0\le h(\Gamma)\le n$. We say that a Grassmannian graph $\Gamma$ on $n$ leaves is of \emph{type $(k,n)$} if $k=h(\Gamma)$.

The above definition, taken from \cite{Moerman:2021cjg}, modifies Postnikov's original definition \cite{Postnikov:2018jfq} in two aspects. Firstly, it disallows degree two internal vertices. Secondly, it restricts the helicity of a vertex $v$ to exclude $0$ and $\deg(v)$ except when $v$ is a boundary leaf. As a consequence of these changes, Grassmannian forests (defined later) are automatically reduced (also defined later). 

To each Grassmannian graph $\Gamma$, one can assign a collection of one-way strands. A \emph{one-way strand} $\alpha$ in $\Gamma$ is a directed path that either starts and ends at boundary vertices, or forms a closed path in $\Gamma_\text{int}$. It is defined according to the following rules-of-the-road. For each internal vertex $v\in\mathcal{V}_\text{int}(\Gamma)$ of degree $d$, with adjacent edges labelled $e_1,\ldots,e_d$ in clockwise order, if $\alpha$ enters $v$ through the edge $e_i$, it exits $v$ through the edge $e_j$ where $j=i+h(v)~(\text{mod}~d)$. We say that a Grassmannian graph $\Gamma$ is \emph{reduced} if its one-way strands satisfy the following requirements:
\begin{enumerate}[label=(\arabic*)]
	\item There are no strands forming closed loops in $\Gamma_\text{int}$.
	\item All strands in $\Gamma$ are simple curves without self-intersections. The only exceptions are strands $b_i \to v \to b_i$ where $v\in\Vertices_\text{int}(\Gamma)$ is a boundary leaf adjacent to the boundary vertex $b_i$.
	\item Any pair of strands $\alpha\ne\beta$ cannot have a \emph{bad double crossing}, a pair of vertices $u\ne v$ such that both $\alpha$ and $\beta$ pass from $u$ to $v$.
\end{enumerate}

The following definition is taken from \cite{Postnikov:2006kva,Karp:2017ouj}. A \emph{decorated permutation} $\pi$ on $[n]$ is a bijection $\pi:[n]\to[n]$ with fixed points coloured either black or white. We denote a \emph{black fixed point (loop)} $i$ by $\pi(i)=\underline{i}$ and a \emph{white fixed point (coloop)} $i$ by $\pi(i)=\overline{i}$. An \emph{anti-excedance} of $\pi$ is an element $i\in[n]$ for which $i$ is a white fixed point or $\pi^{-1}(i)>i$. We say that $\pi$ is of \emph{type $(k,n)$} if it has $k$ anti-excedances. 

Given a reduced Grassmannian graph $\Gamma$ on $n$ leaves, one can assign a decorated permutation $\pi_\Gamma$ on $[n]$. For each $i\in[n]$, if the one-way strand that starts at the boundary vertex $b_i$ ends at a different boundary vertex $b_j$, then $\pi_\Gamma(i)=j$. Otherwise, $i$ is a fixed point of $\pi_\Gamma$, coloured according to the colour of the boundary leaf $v$ adjacent to $b_i$. Moreover, since the number of anti-excedances of $\pi_\Gamma$ equals the helicity of $\Gamma$, $\pi_\Gamma$ is of type $(k,n)$ where $k=h(\Gamma)$. A \emph{complete} reduced Grassmannian graph $\Gamma$ of type $(k, n)$, with $0<k<n$, is one whose decorated strand permutation is given by $\sigma_\Gamma(i) = i + k~(\text{mod}~n)$ for all $i\in[n]$.

Given two Grassmannian graphs $\Gamma$ and $\Gamma'$, we say that $\Gamma$ \emph{refines} $\Gamma'$ (and that $\Gamma'$ \emph{coarsens} $\Gamma$) if $\Gamma$ can be obtained from $\Gamma'$ by a repeatedly applying the following operation: replace an internal vertex of type $(h,d)$ by a complete reduced Grassmannian graph of type $(h,d)$. The \emph{refinement order} on Grassmannian graphs is the partial order $\preceq_\text{ref}$ where $\Gamma\preceq_\text{ref}\Gamma'$ if $\Gamma$ refines $\Gamma'$. We say that $\Gamma'$ covers $\Gamma$, denoted by $\Gamma\precdot_\text{ref}\Gamma'$, if $\Gamma'$ covers $\Gamma$ in the refinement order. Two Grassmannian graphs $\Gamma$ and $\Gamma'$ are \emph{refinement-equivalent} if they are in the same connected component of the refinement order $\preceq_\text{ref}$. Moreover, two reduced Grassmannian graphs $\Gamma$ and $\Gamma'$ are refinement-equivalent if and only if their decorated permutations coincide.

The remaining definitions and results come from \cite{Moerman:2021cjg}. An acyclic Grassmannian graph is called a \emph{Grassmannian forest} and a connected Grassmannian forest is called a \emph{Grassmannian tree}. A Grassmannian forest is automatically reduced since it contains no internal cycles.	Given a Grassmannian tree $T$ on $n$ leaves, we have that
\begin{align}\label{eq:grass-multiplicity-T}	
	n=\sum_{v\in\Vertices_\text{int}(T)}\deg(v)-2(|{\Vertices}_\text{int}(T)|-1)=2+\sum_{v\in\Vertices_\text{int}(T)}(\deg(v)-2)\,.
\end{align}
Substituting \eqref{eq:grass-multiplicity-T} into \eqref{eq:grass-helicity-graph}, we find that its helicity can be expressed as
\begin{align}\label{eq:grass-helicity-T}		
	h(T)=\sum_{v\in\Vertices_\text{int}(T)}h(v)-(|{\Vertices}_\text{int}(T)|-1)=1+\sum_{v\in\Vertices_\text{int}(T)}(h(v)-1)\,.
\end{align}

\begin{figure}
	\centering
	\null
	\hfill
	\begin{subfigure}{0.49\textwidth}
		\begin{align*}
			\vcenter{\hbox{\includegraphics[scale=0.3]{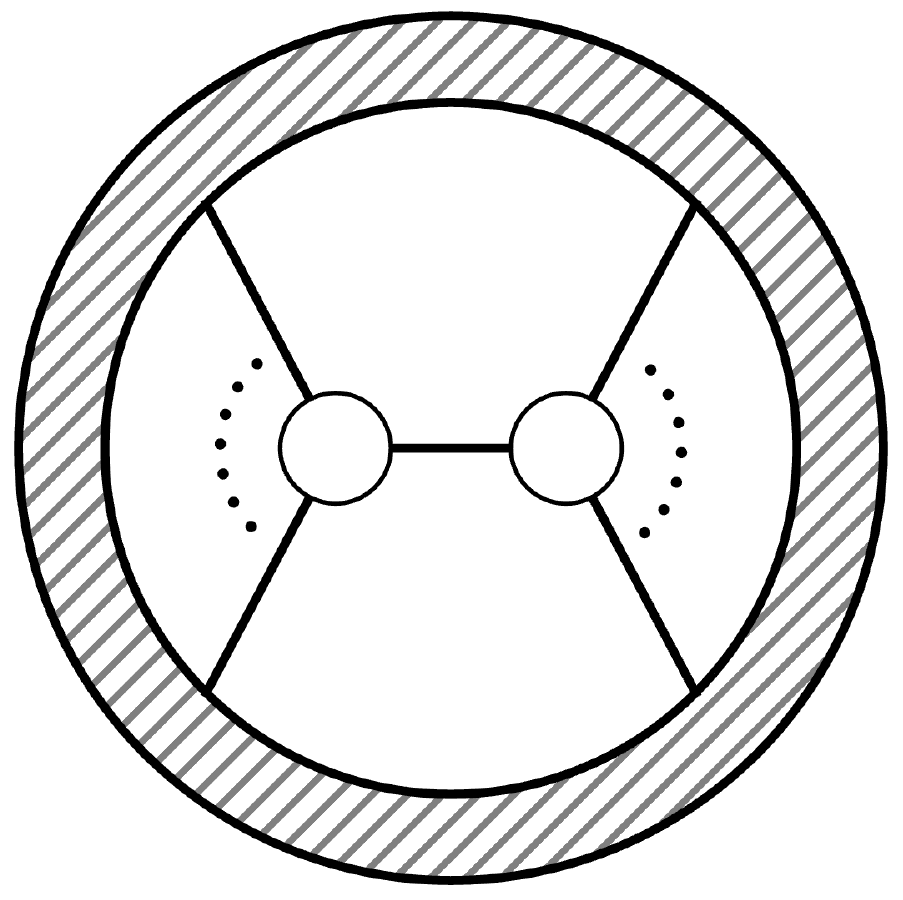}}}
			\mapsto	
			\vcenter{\hbox{\includegraphics[scale=0.3]{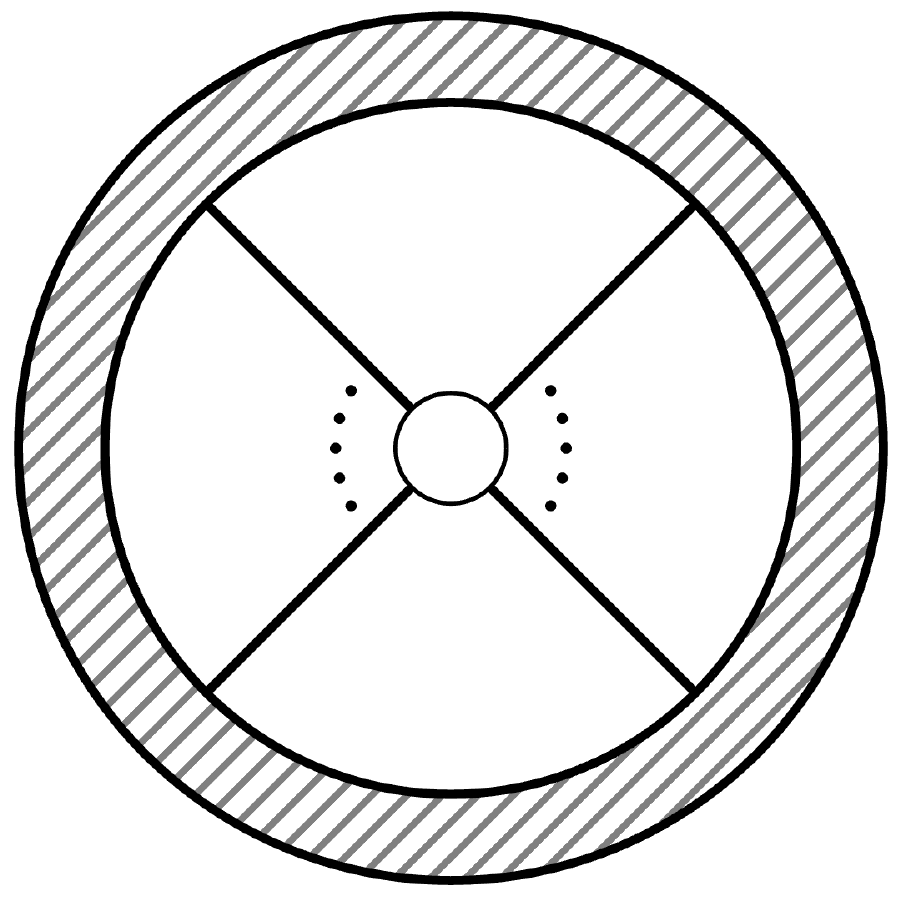}}}
		\end{align*}
		\caption{Contracting adjacent white vertices.}
	\end{subfigure}
	\hfill
	\begin{subfigure}{0.49\textwidth}
		\begin{align*}
			\vcenter{\hbox{\includegraphics[scale=0.3]{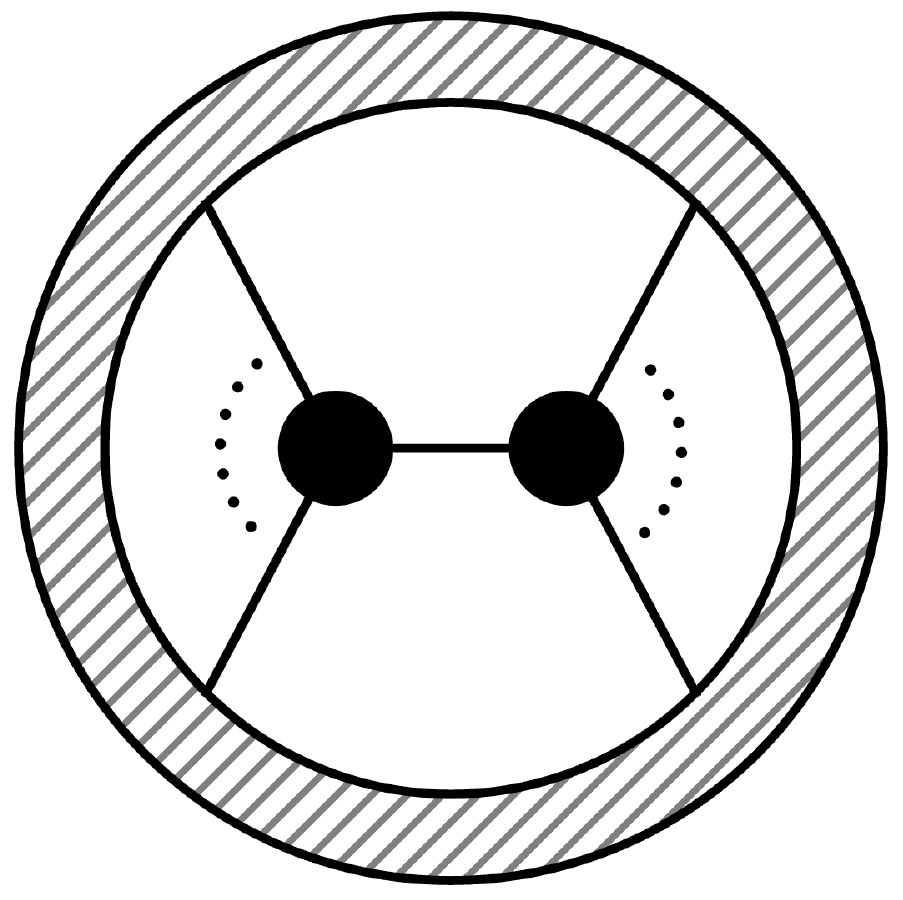}}}
			\mapsto
			\vcenter{\hbox{\includegraphics[scale=0.3]{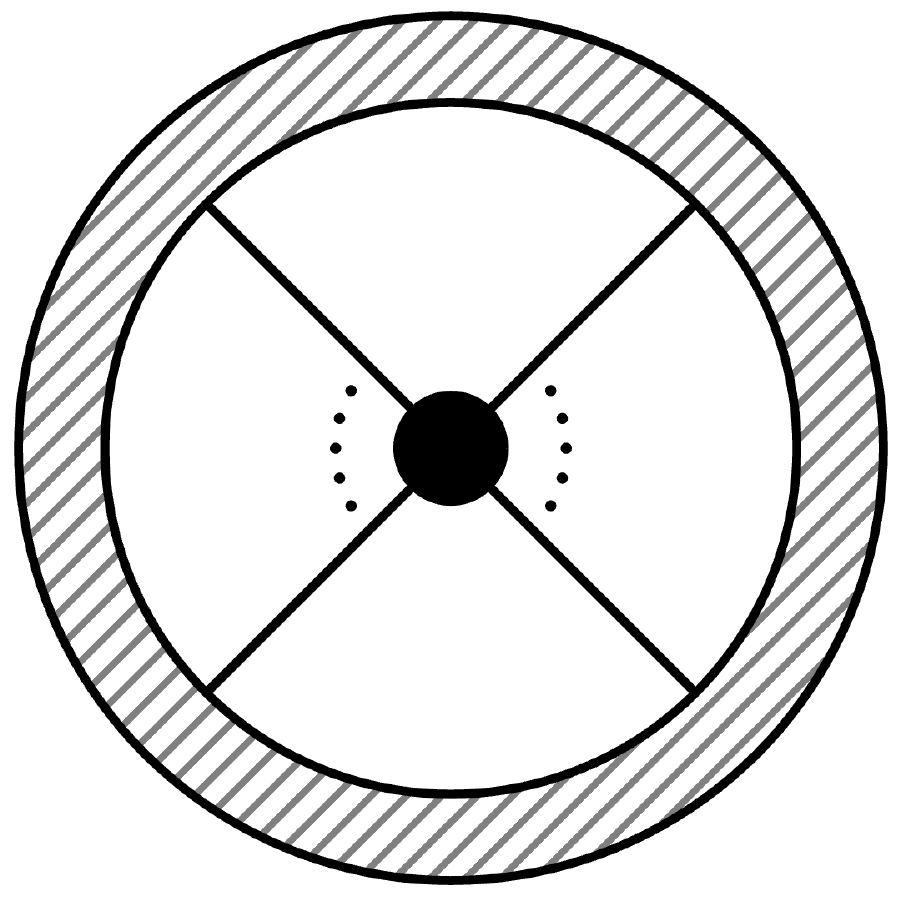}}}
		\end{align*}
		\caption{Contracting adjacent black vertices.}
	\end{subfigure}
	\hfill
	\null
	\caption{Vertex contraction moves. Shaded regions represent unaffected graph components.}
	\label{fig:grass-contraction}
\end{figure}

The refinement-equivalence class of a Grassmannian tree (resp., forest) only contains Grassmannian trees (resp., forests) since vertices connected via a path remain so after refinements/coarsenings and these operations never introduce cycles. Consequently, restricting the refinement order to Grassmannian trees (resp., forests) is well-defined. This restriction can be equivalently formulated as follows. Given two Grassmannian forests $F$ and $F'$, we say that $F'$ \emph{coarsens} $F$ (and that $F$ \emph{refines} $F'$) if $F'$ can be obtained from $F$ by applying a sequence of \emph{vertex contraction moves}:  contract two adjacent internal white vertices (or two adjacent internal black vertices) into a single white (or black) vertex, as demonstrated in \cref{fig:grass-contraction}. A Grassmannian forest is said to be \emph{contracted} or \emph{maximal} if it is maximal with respect to the refinement order. Each refinement-equivalence class of Grassmannian trees (resp., forests) contains a unique contracted Grassmannian tree (resp., forest) which provides a canonical choice of representative for that equivalence class. Note that a Grassmannian forest is contracted if and only if it has no adjacent white or black vertices.

\begin{figure}[t]
	\centering
	\begin{subfigure}{\textwidth}
		\begin{align*}
			\vcenter{\hbox{\includegraphics[scale=0.3]{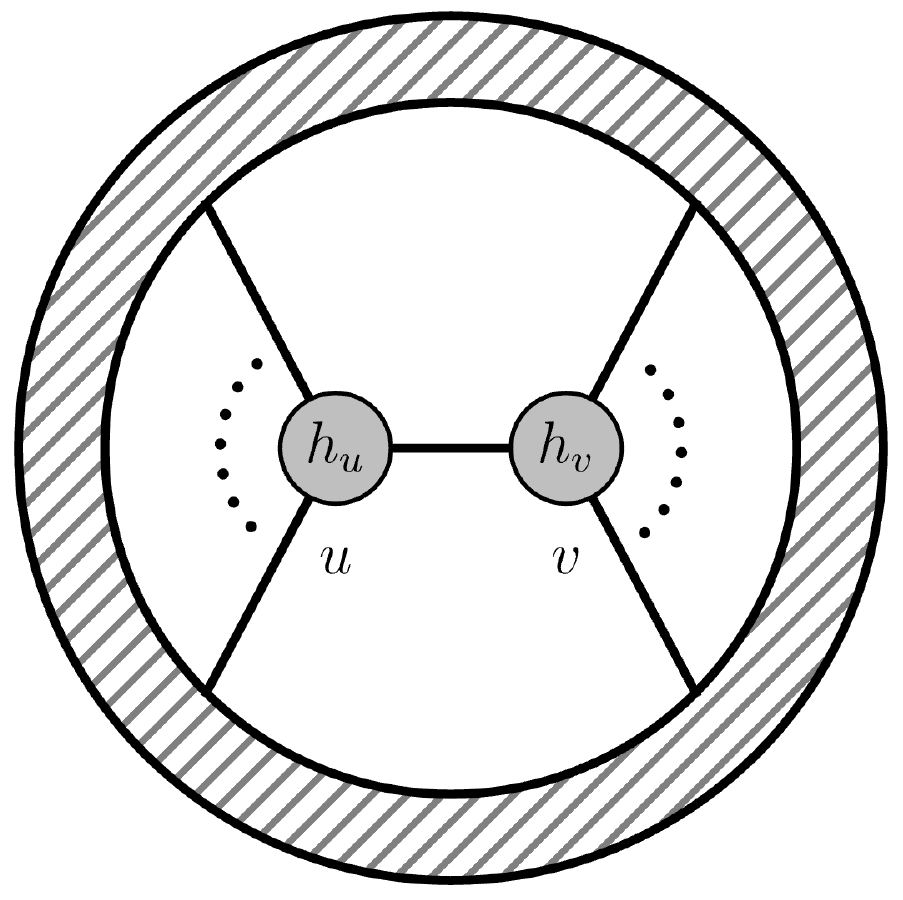}}}
			\precdot_{\symbolG}	
			\vcenter{\hbox{\includegraphics[scale=0.3]{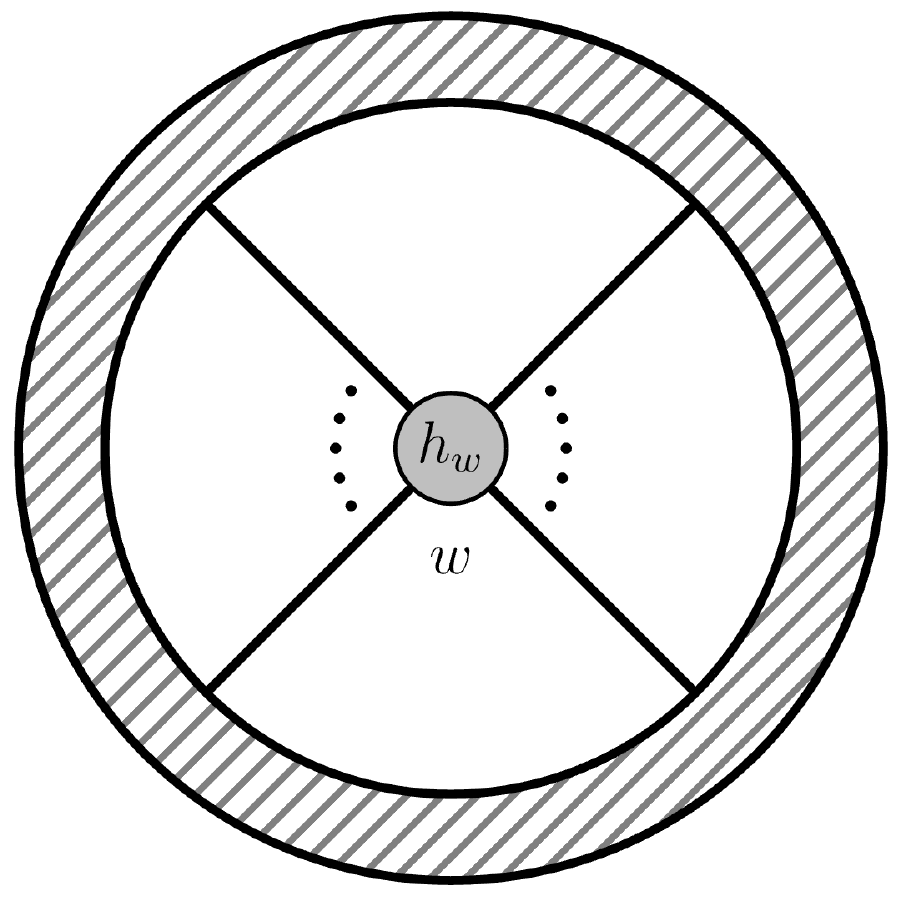}}}
		\end{align*}
		\caption{Covering relation for a generic vertex $w$ with helicity $2\le h_w\le\deg(w)-2$. Here $\deg(u)+\deg(v)=\deg(w)+2$ and $h_u+h_v=h_w+1$.}
	\end{subfigure}
	\null
	\hfill
	\begin{subfigure}{0.49\textwidth}
		\begin{align*}
			\vcenter{\hbox{\includegraphics[scale=0.3]{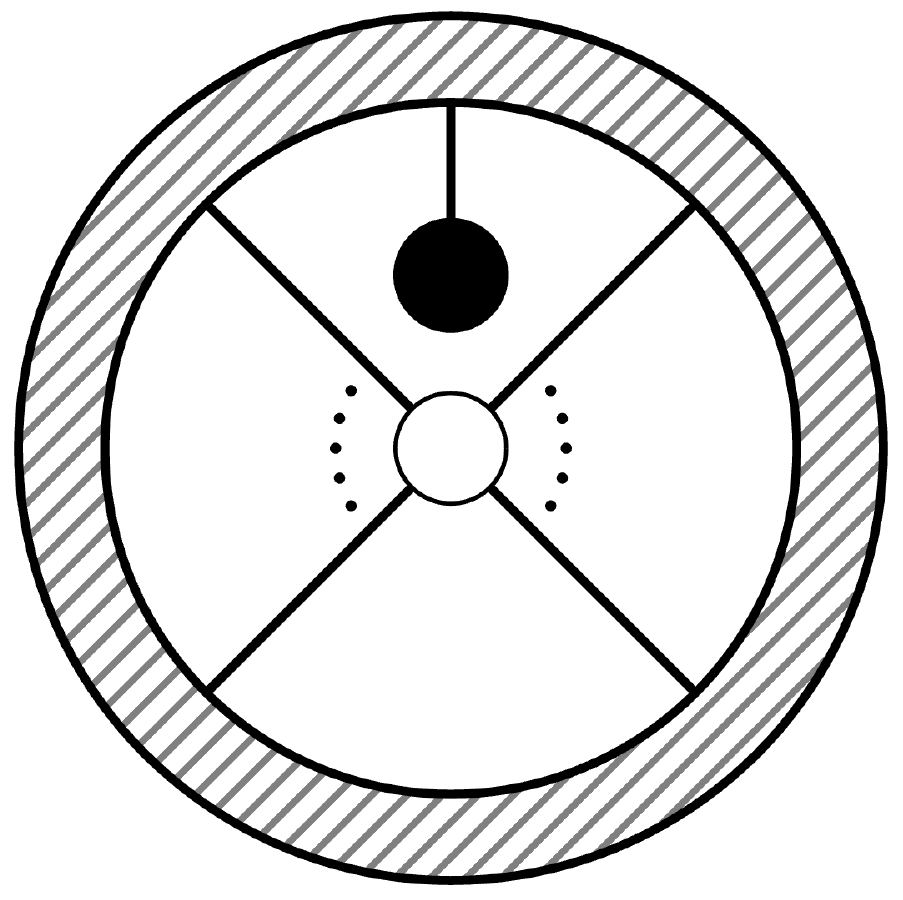}}}
			\precdot_{\symbolG}	
			\vcenter{\hbox{\includegraphics[scale=0.3]{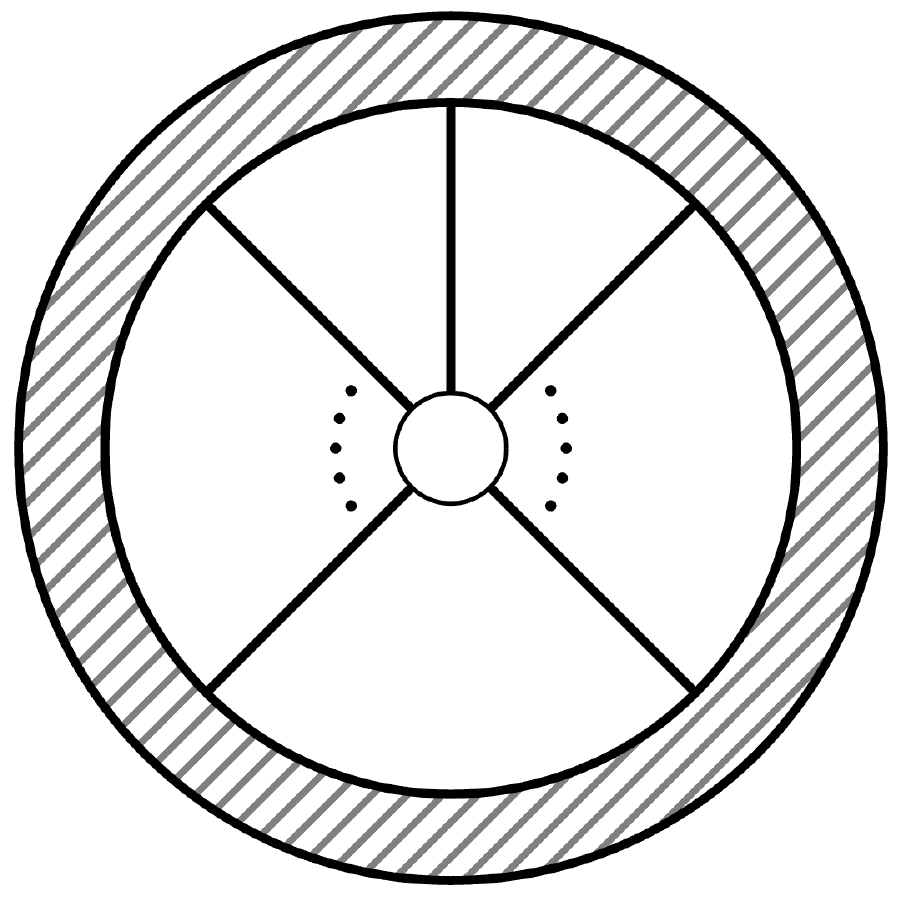}}}
		\end{align*}
		\caption{Covering relation for a white vertex.}
	\end{subfigure}
	\hfill
	\begin{subfigure}{0.49\textwidth}
		\begin{align*}
			\vcenter{\hbox{\includegraphics[scale=0.3]{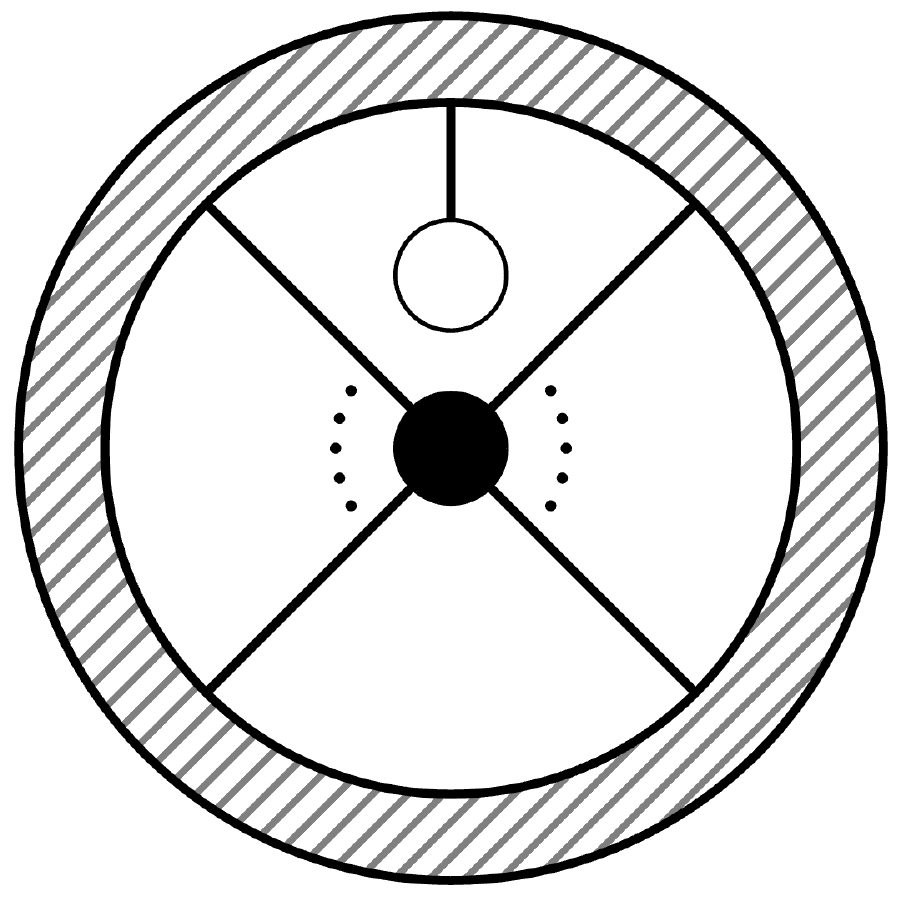}}}
			\precdot_{\symbolG}	
			\vcenter{\hbox{\includegraphics[scale=0.3]{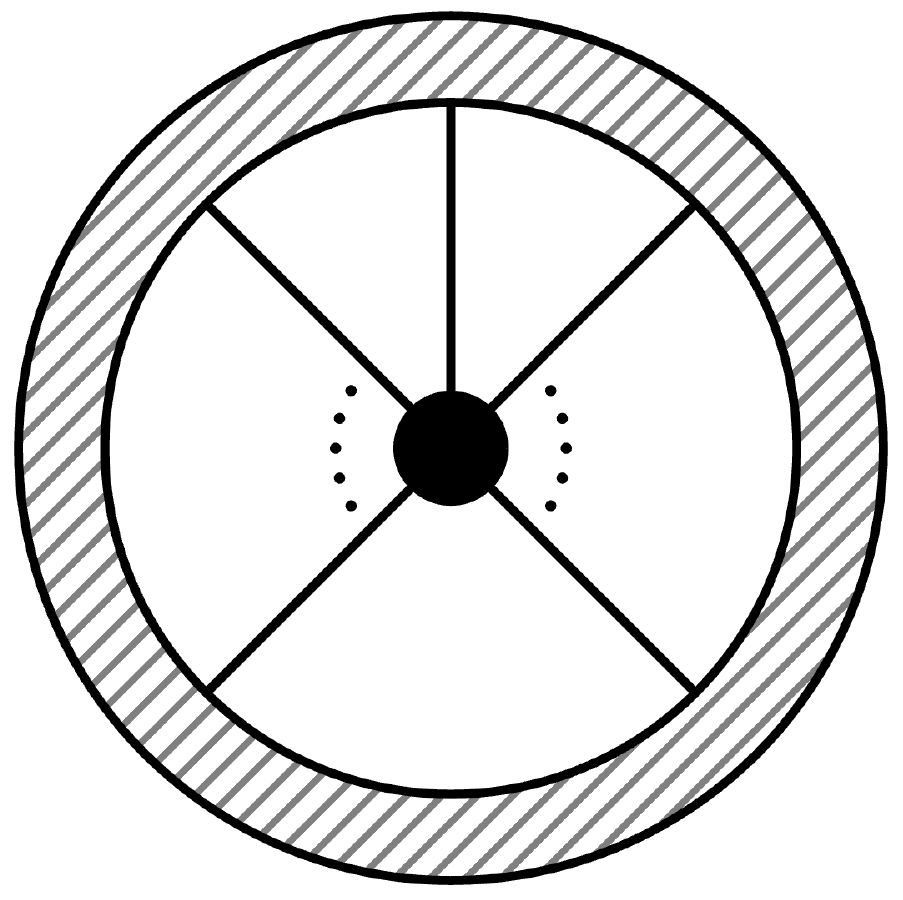}}}
		\end{align*}
		\caption{Covering relation for a black vertex.}
	\end{subfigure}
	\hfill
	\null
	\begin{subfigure}{\textwidth}
		\begin{align*}
			\vcenter{\hbox{\includegraphics[scale=0.3]{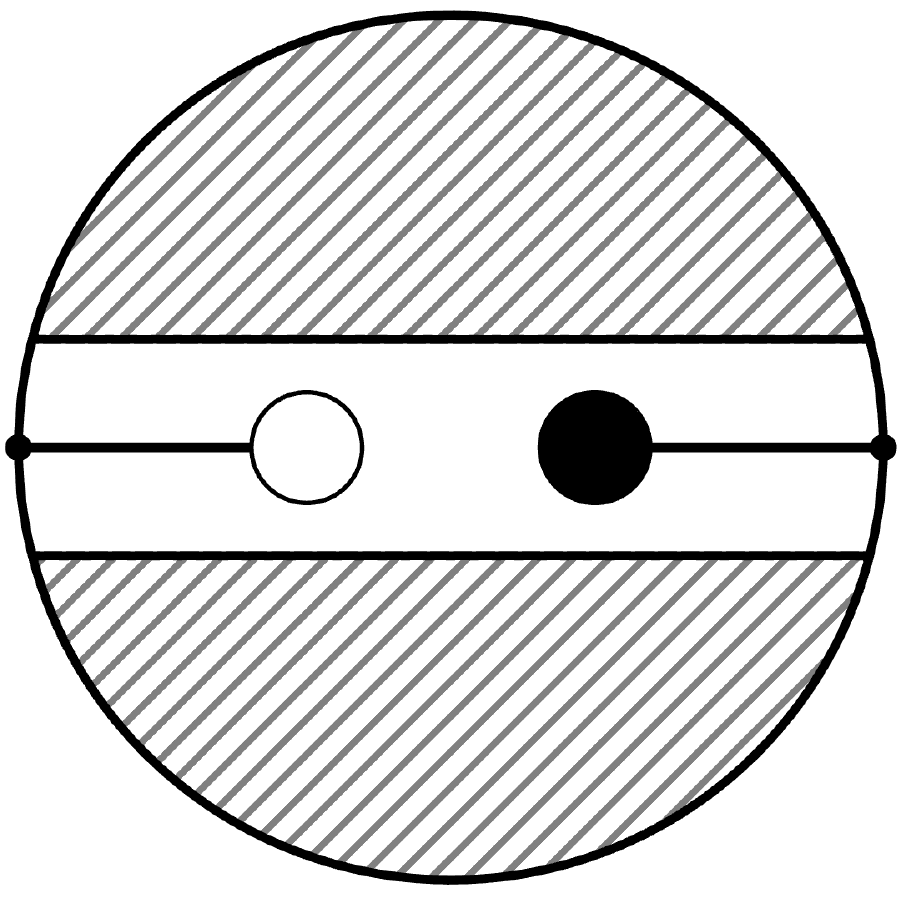}}}
			\precdot_{\symbolG}	
			\vcenter{\hbox{\includegraphics[scale=0.3]{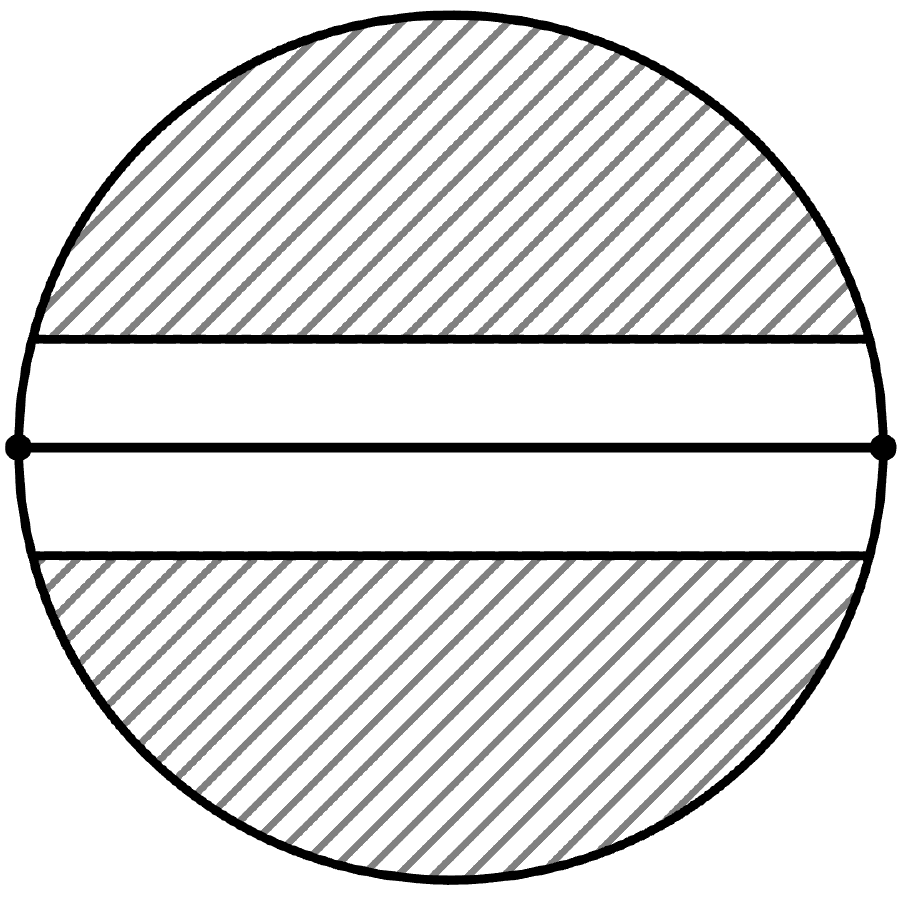}}}
		\end{align*}
		\caption{Covering relation for adjacent boundary vertices.}
	\end{subfigure}	
	\caption{Covering relations for $\Forest_{k,n}^{\symbolG}$ with respect to $\preceq_{\symbolG}$. Shaded regions represent unaffected graph components. Labels inside vertices denote helicities.}
	\label{fig:grass-covering}
\end{figure}

Let $\Graph_{k,n}^{\symbolG}$ be the set of refinement-equivalence classes of reduced Grassmannian graphs of type $(k,n)$. Since positroid cells of $\tnnG{k}{n}$ are in bijection with elements of $\Graph_{k,n}^{\symbolG}$, the partial order on closed positroid cells defined by set inclusion translates to a partial order on $\Graph_{k,n}^{\symbolG}$, which we denote by $\preceq_{\symbolG}$. Let $\Forest_{k,n}^{\symbolG}\subset\Graph_{k,n}^{\symbolG}$ be the set of refinement-equivalence classes of Grassmannian forests of type $(k,n)$. Then $(\Forest_{k,n}^{\symbolG},\preceq_{\symbolG})$ is an induced subposet of $(\Graph_{k,n}^{\symbolG},\preceq_{\symbolG})$. The covering relations for $(\Forest_{k,n}^{\symbolG},\preceq_{\symbolG})$ are given (without proof) in \cref{fig:grass-covering}.

\section{The Orthogonal Grassmannian}
\label{sec:ograss}

\subsection{Definition}
\label{sec:ograss-def}

Fix $k$ and $n$ to be positive integers for which $k\le\lfloor\frac{n}{2}\rfloor$ where $\lfloor\cdot\rfloor$ denotes the floor operation. Let $\lrangle{\cdot,\cdot}:\complex^n\times\complex^n\to\complex$ be a non-degenerate symmetric bilinear form. For $n>2k$, the \emph{orthogonal Grassmannian} $\OG{k}{n}$ is the subset of all $k$-dimensional subspaces of $\complex^n$ isotropic with respect to $\lrangle{\cdot,\cdot}$ \cite{Coskun:2013-lecture-notes}: 
\begin{align}\label{eq:ograss-k-n}
	\OG{k}{n}\coloneqq\left\{[C]\in\G{k}{n}:\langle C_a,C_b\rangle = 0~\text{for all}~a,b\in[k]\right\}.
\end{align}
Since $\lrangle{\cdot,\cdot}$ is symmetric, the isotropy condition produces $k(k+1)/2$ constraints. Consequently,
\begin{align}\label{eq:ograss-dim-k-n}
	\dim(\OG{k}{n}) = \dim(\G{k}{n}) - \frac{k(k+1)}{2} = \frac{k(2n-3k-1)}{2}\,.
\end{align}

This dissertation focusses on the case when $n=2k$. In this case, the algebraic set defined in \eqref{eq:ograss-k-n} has two irreducible isomorphic components, and $\OG{k}{2k}$ denotes one of them \cite{Coskun:2013-lecture-notes}. Let us begin by examining the two components of \eqref{eq:ograss-k-n}. One can always choose a basis $\{e_i\}_{i=1}^n$ for $\complex^n$ such that 
\begin{align}
	\eta_{ij}\coloneqq\lrangle{e_i,e_j}\,,
\end{align}
defines a diagonal matrix $\eta$ with diagonal entries either $\pm1$. Fix $p$ to be an integer such that $0\le p \le n$ and let $P\in\binom{[n]}{p}$. Assume that the diagonal entries of $\eta$ are given by
\begin{align}
	\eta_{ii} = \begin{cases}
		+1 & \text{if}~i\in P\\
		-1 & \text{otherwise}
	\end{cases},
\end{align} 
for each $i\in[n]$. Let $I=\{i_1,\ldots,i_k\}\in\binom{[n]}{k}$ be any ordered $k$-element subset of $[n]$ and let $I'=\{i'_1,\ldots,i'_k\}$ denote the complement $[n]\setminus I$ regarded as an ordered set. Then for each element  $[C]\in\OG{k}{2k}$, the isotropy condition $\langle C_a,C_b\rangle = 0$ reads 
\begin{align}\label{eq:ograss-isotropy-I}
	0 = \sum_{i\in I} C_{ai}\eta_{ii}C_{bi} + \sum_{i'\in I'}C_{ai'}\eta_{i'i'}C_{bi'}\,,
\end{align}
where $\eta_{ii'}=0$ for $i\in I$ and $i'\in I'$. By moving the summation over $I$ to the left-hand side of \eqref{eq:ograss-isotropy-I} and taking the determinant of both sides, we obtain 
\begin{align}\label{eq:ograss-isotropy-I-det}
	p_{I}(C)^2|{-\eta}|_{I,I} = p_{I'}(C)|\eta|_{I',I'}\,,
\end{align}
where the vertical bars $|\bullet|_{A,B}$ denote the minor in the row set $A$ and the column set $B$. Substituting
\begin{align}
	|{-\eta}|_{I,I} = (-1)^{|P\cap I|}\,,\qquad
	|\eta|_{I',I'} = (-1)^{|P'\cap I'|} = (-1)^{k-p+|P\cap I|}\,,
\end{align}
into 
\eqref{eq:ograss-isotropy-I-det} yields
\begin{align}\label{eq:ograss-isotropy-I-det-simplified-sqrt}
	p_{I}(C) = \pm e^{i\pi(k-p)}p_{I'}(C)\,.
\end{align}
The choice of sign in \eqref{eq:ograss-isotropy-I-det-simplified-sqrt} defines the two components of \eqref{eq:ograss-dim-k-n}. To extend Postnikov's definitions of positivity and non-negativity to the orthogonal Grassmannian, one must fix the sign together with $p$ so that all the maximal minors of $C$ can consistently be real and positive. These conditions are met by choosing the positive branch and setting $p=k$. For concreteness, we adopt the convention of \cite{Huang:2013owa,Huang:2014xza} and let $\eta=\text{diag}(+1,-1,\ldots,+1,-1)$. Consequently, the orthogonal Grassmannian $\OG{k}{2k}$ is defined as \cite{Galashin:2020jvd}
\begin{align}\label{eq:ograss-complex-k}
	\OG{k}{2k}\coloneqq\left\{V\in\G{k}{n}:p_I(V) = p_{[n]\setminus I}(V)~\text{for all}~I\in\binom{[2k]}{k}\right\},
\end{align}
and it has dimension
\begin{align}\label{eq:ograss-complex-dim-k}
	\dim(\OG{k}{2k}) = \frac{k(k-1)}{2}\,.
\end{align}
The positroid stratification of $\G{k}{2k}$ induces a decomposition of $\OG{k}{2k}$, the \emph{orthitroid stratification}\footnote{This terminology was coined in \cite{Lukowski:2021fkf}.}. It was first observed in \cite{Huang:2013owa,Huang:2014xza}, and latter proven in \cite{Galashin:2020jvd}, that the induced strata of $\OG{k}{2k}$ are labelled by matchings of $[2k]$ points on a circle. 
We define a \emph{matching permutation} on $[2k]$ to be a permutation on $[2k]$ expressible as the product of $k$ disjoint two-cycles. Matchings of points on a circle are naturally in bijection with matching permutations. Moreover, we define an \emph{matching affine permutation} on $[2k]$ to be an affine permutation on $[2k]$ which descends to a matching permutation on $[2k]$ modulo $2k$. Then the orthitroid stratification of $\OG{k}{2k}$ is given by
\begin{align}
	\OG{k}{2k}=\bigsqcup_{f}\OrthCirc{f}\,,
\end{align} 
where $\OrthCirc{f}\coloneqq\OG{k}{2k}\cap\PosCirc{f}$ denotes the subset of $\OG{k}{2k}$ labelled by the matching affine permutation $f$. Moreover, we define the \emph{orthitroid variety} $\Orth{f}$ to be the closure of $\OrthCirc{f}$ in $\OG{k}{2k}$.

\subsection{The Non-negative Orthogonal Grassmannian}
\label{sec:ograss-tnn}

Let $\OGReal{k}{2k}$ denote the  \emph{real orthogonal Grassmannian}, the real part of $\OG{k}{2k}$. The \emph{positive orthogonal Grassmannian} $\tpOG{k}{2k}$ is defined as 
\begin{align}\label{eq:ograss-tp}
	\tpOG{k}{2k}\coloneqq\OGReal{k}{2k}\cap\tpG{k}{2k}\,,
\end{align}
and the \emph{non-negative orthogonal Grassmannian} $\tnnOG{k}{2k}$ is defined as \cite{Huang:2013owa,Huang:2014xza,Galashin:2020jvd}
\begin{align}\label{eq:ograss-tnn}
	\tnnOG{k}{2k}\coloneqq\OGReal{k}{2k}\cap\tnnG{k}{2k}\,.
\end{align}
For every matching affine permutation $f$ on $[2k]$, the intersections
\begin{align}
	\begin{split}
		\tpOrth{f}\coloneqq\tpOG{k}{2k}\cap\OrthCirc{f}\,,
	\end{split}
	\begin{split}
		\tnnOrth{f}\coloneqq\tnnOG{k}{2k}\cap\Orth{f}\,,
	\end{split}
\end{align}
are called open and closed \emph{orthitroid cells}, respectively. We claim that $(\Orth{f},\tnnOrth{f})$ is a positive geometry. 

The non-negative orthogonal Grassmannian was first introduced in \cite{Huang:2013owa}, to describe amplitudes in $\mathcal{N}=6$ Aharony--Bergman--Jafferis--Maldacena (ABJM) theory. Its orthitroid stratification was studied in \cite{Kim:2014hva,Huang:2014xza}, and later in \cite{Galashin:2020jvd}. The non-negative orthogonal Grassmannian plays a central role in the construction of the orthogonal Momentum Amplituhedron in \cref{chp:omom}.

As a counterpart to \cite{Arkani-Hamed:2012zlh}, the authors of \cite{Kim:2014hva} conjecture a canonically positive parametrization for $\Orth{f}$ using \emph{``BCFW bridges''}. They devise a map $C_f:\real^d\to\Mat{k}{2k}{\real}$, where $d=\dim(\Orth{f})$ and $[C_f(\real^d)]\subset\Orth{f}$, which furnishes \emph{canonically positive coordinates} for $\tpOrth{f}$: $[C_f(t)]\in\tpOrth{f}$ if and only if $t\in\real_{>0}^d$. Their parametrization is reviewed in \cite{Lukowski:2021fkf} and implemented in the \Mathematica{} package \orthitroids{} \cite{Lukowski:2021fkf}. The expression for the canonical form $\Omega(\tnnOrth{f})$ in terms of canonically positive coordinates $t=(t_1,\ldots,t_d)$ is given by \cite{Kim:2014hva}
\begin{align}\label{eq:ograss-orthitroid-Omega}
	\Omega(\tnnOrth{f})=\bigwedge_{i=1}^dd\log(\tanh(t_i))\,.
\end{align}

\subsection{Orthogonal Grassmannian Graphs, Trees and Forest}
\label{sec:ograss-graph}

In analogy with the non-negative Grassmannian, the orthitroid cells of the non-negative orthogonal Grassmannian are in bijection with various combinatorial objects, including a special subset of refinement-equivalence classes of reduced Grassmannian graphs called orthogonal Grassmannian graphs or OG graphs \cite{Lukowski:2021fkf}. In this section, we introduce OG graphs together with OG trees and forests. The latter are essential to the analysis of \cref{chp:omom}.

The following definitions are taken from \cite{Lukowski:2021fkf}. An \emph{orthogonal Grassmannian graph} or \emph{OG graph} of type $\underline{k}$ is a Grassmannian graph $\Gamma$ of type $(k,2k)$ satisfying $\deg(v)=2h(v)>2$ for every internal vertex $v$. This restriction on internal vertices is compatible with \eqref{eq:grass-helicity-graph} since it implies $h(\Gamma) = \frac{n}{2} = k$.  Given a reduced OG graph $\Gamma$ of type $\underline{k}$, the associated decorated permutation $\pi_\Gamma$ on $[2k]$ is always a matching permutation and hence carries no decoration. Thus, we will refer to $\pi_\Gamma$ as the associated matching permutation. 

An \emph{orthogonal Grassmannian forest} or \emph{OG forest} is an acyclic OG graph and an \emph{orthogonal Grassmannian tree} or \emph{OG tree} is a connected OG forest. OG forests are automatically reduced because the same property holds for Grassmannian forests. Moreover, OG forests are unique with respect to the refinement order $\preceq_\text{ref}$ (defined for Grassmannian graphs). Consequently, each OG forest uniquely defines a matching permutation, and vice versa.

\begin{figure}[t]
	\centering
	\null
	\hfill
	\begin{subfigure}{0.49\textwidth}
		\begin{align*}
			\vcenter{\hbox{\includegraphics[scale=0.3]{G-covering-generic-after}}}
			\precdot_{\symbolG}	
			\vcenter{\hbox{\includegraphics[scale=0.3]{G-covering-generic-before}}}
		\end{align*}
		\caption{Covering relation for a vertex $w$ with helicity $h_w>2$. Here $h_u+h_v=h_w+1$.}
	\end{subfigure}
	\hfill
	\begin{subfigure}{0.49\textwidth}
		\begin{align*}
			\vcenter{\hbox{\includegraphics[scale=0.3]{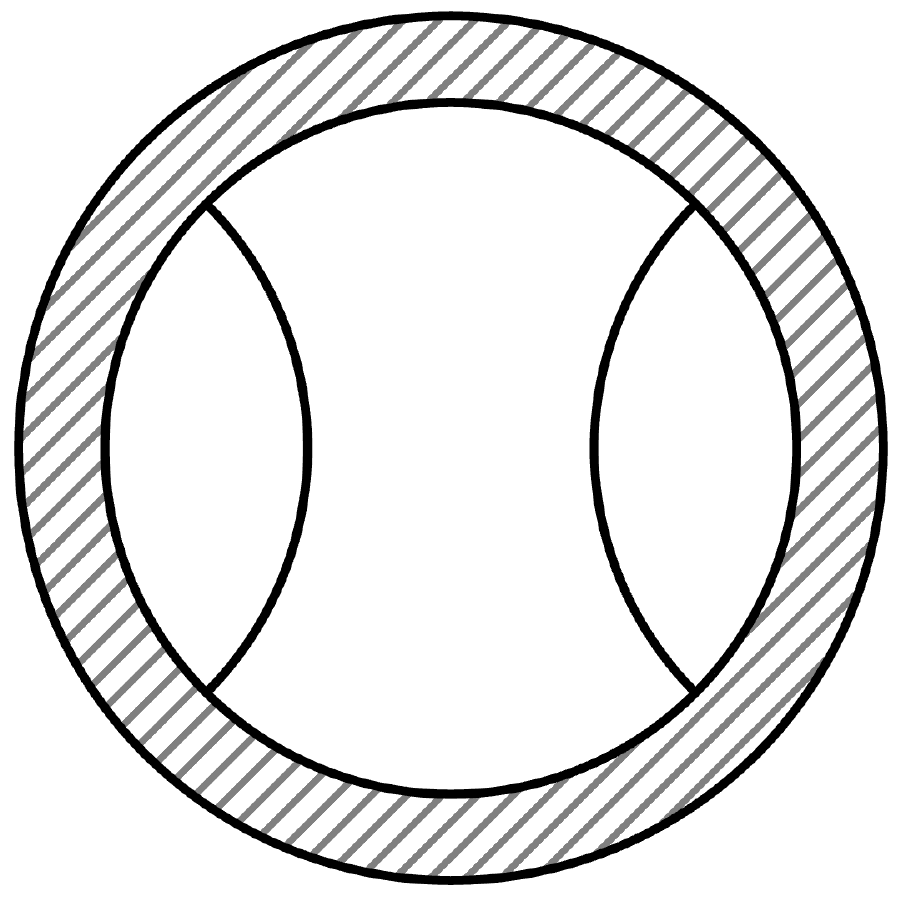}}}
			\precdot_{\symbolG}	
			\vcenter{\hbox{\includegraphics[scale=0.3]{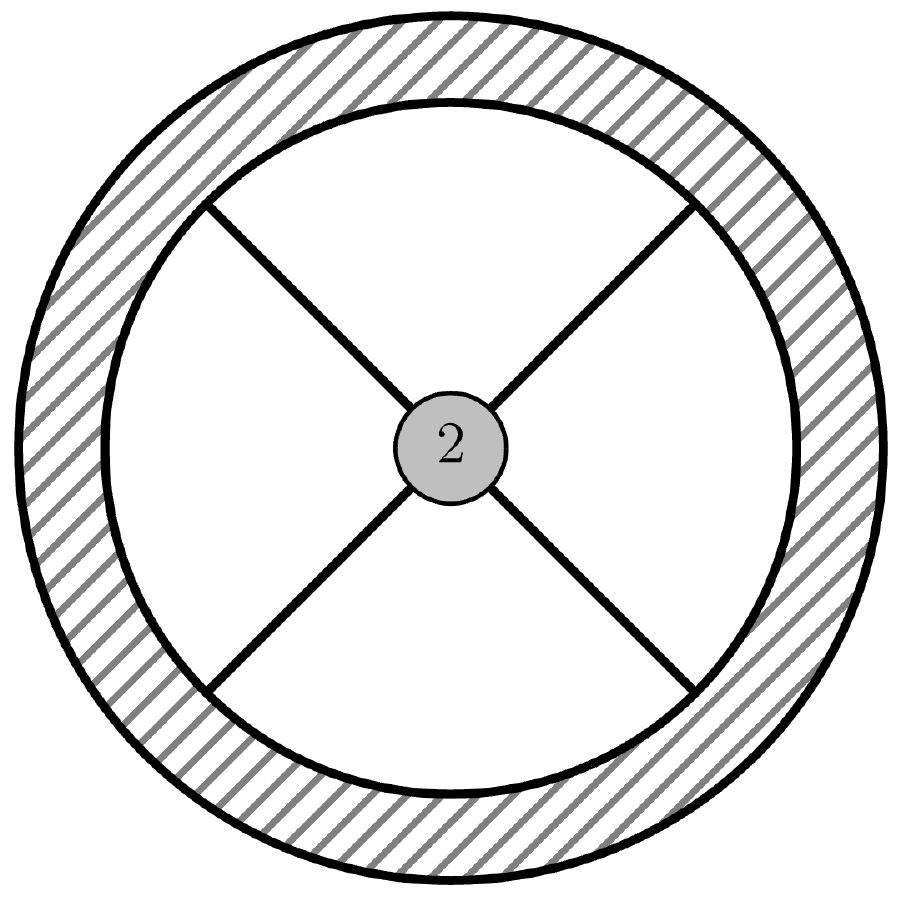}}}
		\end{align*}
		\caption{Covering relation for a vertex with helicity equal to $2$.}
	\end{subfigure}
	\hfill
	\null
	\caption{Covering relations for $\Forest_{k}^{\symbolOG}$ with respect to $\preceq_{\symbolOG}$. Shaded regions represent unaffected graph components. Labels inside vertices denote helicities.}
	\label{fig:ograss-covering}
\end{figure}

Let $\Graph_{k}^{\symbolOG}$ be the set of refinement-equivalence classes of reduced OG graphs of type $\underline{k}$. Since orthitroid cells of $\tnnOG{k}{2k}$ are in bijection with elements of $\Graph_{k}^{\symbolOG}$, the partial order on closed orthitroid cells defined by set inclusion translates to a partial order on $\Graph_{k}^{\symbolOG}$, which we denote by $\preceq_{\symbolOG}$. Let $\Forest_{k}^{\symbolOG}\subset\Graph_{k}^{\symbolOG}$ be the set of OG forests of type $\underline{k}$. Then $(\Forest_{k}^{\symbolOG},\preceq_{\symbolOG})$ is an induced subposet of $(\Graph_{k}^{\symbolOG},\preceq_{\symbolOG})$. The covering relations for $(\Forest_{k}^{\symbolOG},\preceq_{\symbolOG})$ are given (without proof) in \cref{fig:ograss-covering}. What is more, $(\Graph_{k}^{\symbolOG},\preceq_{\symbolOG})$ is an induced subposet of $(\Graph_{k,2k}^{\symbolG},\preceq_{\symbolG})$ and $(\Forest_{k}^{\symbolOG},\preceq_{\symbolOG})$ is an induced subposet of $(\Forest_{k,2k}^{\symbolG},\preceq_{\symbolG})$, where $\preceq_{\symbolOG}$ is understood to descend from $\preceq_{\symbolG}$ through the intersection defined in \eqref{eq:ograss-tnn}. 

\section{Summary}
In this chapter, we introduced two auxiliary positive geometries. They are the non-negative Grassmannian and the non-negative orthogonal Grassmannian. Their boundary stratifications are well-known \cite{Postnikov:2006kva,Postnikov:2018jfq,Huang:2013owa,Huang:2014xza,Galashin:2020jvd,Kim:2014hva} and classified in terms of Grassmannian graphs \cite{Arkani-Hamed:2012zlh,Postnikov:2018jfq} and orthogonal Grassmannian graphs \cite{Lukowski:2021fkf}. In \cref{chp:mom,chp:omom}, we will define new, physically-relevant positive geometries in terms of these positive geometries via morphisms. In particular, the (orthogonal) Momentum Amplituhedron is the image of the non-negative (orthogonal) Grassmannian. We will use the cell decomposition of the non-negative (orthogonal) Grassmannian and the outputs of \cref{chp:pos} (see \cref{conj:pos-boundaries}) to determine the boundaries of the (orthogonal) Momentum Amplituhedron. Anticipating the results of \cref{chp:mom,chp:omom}, we have focussed on (orthogonal) Grassmannian forests which are acyclic (orthogonal) Grassmannian graphs. They inherit a partial order with covering relations presented in \cref{fig:grass-covering,fig:ograss-covering}. These covering relations mimic the expected factorisation channels of tree-level amplitudes in $\mathcal{N}=4$ SYM and $\mathcal{N}=6$ ABJM. \cref{chp:mom,chp:omom} further discuss this connection in detail. 	
	\chapter{Pushforwards}
\label{chp:push}

\lettrine{I}{n \cref{chp:pos}}, we introduced morphisms between positive geometries and defined the pushforward of canonical forms via morphisms. The utility of this operation is that the pushforward of a canonical form is again a canonical form. This chapter addresses the problem of performing pushforwards in practice. Operationally, the pushforward involves finding the local inverses of a surjective meromorphic map. The pullback of each local inverse is then added together to give the pushforward. The local inverses are solutions to a system of polynomial equations, i.e.\ points in the zero-set of some polynomial ideal. In general, they are impossible to write down in a closed form, which prompts an important question: ``Is it possible to compute the pushforward of a rational form via a surjective meromorphic map without needing to determine its local inverses?'' This question is answered affirmatively in \cite{Lukowski:2022fwz}. In this chapter, we review the results of \cite{Lukowski:2022fwz}, presenting three methods based on ideas from computational algebraic geometry for calculating pushforwards. These tools allow for the efficient evaluation of pushforwards in \cref{chp:abhy,chp:mom,chp:omom,chp:meet}.

To illustrate the above-mentioned difficulty of finding local inverses, consider the following example. Let $X=Y=\complex$. Define $\tnn{X}=[a,b]$ and $\tnn{Y}=[a^2,b^2]$ as intervals in $\real$ where $a<b$ are positive real numbers. Both $(X,\tnn{X})$ and $(Y,\tnn{Y})$ are $1$-dimensional positive geometries with canonical forms given by
\begin{align}
\Omega(\tnn{X})=\frac{(b-a)dx}{(x-a)(x-b)}\,,\qquad
\Omega(\tnn{Y})=\frac{(b^2-a^2)dy}{(y-a^2)(y-b^2)}\,,
\end{align}
respectively. The two positive geometries are related via the rational map 
\begin{align}
	\Phi:X\dashrightarrow Y,x\mapsto y=x^2\,,
\end{align}
which defines a morphism. In particular, $\Phi(\tp{X}) = \tp{Y}$ and $\Phi|_{\tp{X}}:\tp{X}\to\tp{Y}$ is an orientation-preserving diffeomorphism. There are two local inverses,
\begin{align}
	\Xi^{(1)}:Y\to U,y\mapsto x = \sqrt{y}\,,\qquad \Xi^{(2)}:Y\to X\setminus U,y\mapsto x=-\sqrt{y}\,,
\end{align}
where $U\coloneqq\{x\in X:x=0~\text{or}~0\le\mathrm{Arg}(x)<\pi\}$ and $\tnn{X}\in U$. Notice that neither of the local inverses are meromorphic. The pushforward of $\Omega(\tnn{X})$ via $\Phi$, denoted by $\Phi_\ast\Omega(\tnn{X})$, is the sum of the pullbacks of $\Omega(\tnn{X})$ via the local inverses of $\Phi$:
\begin{align*}
	\Phi_\ast\Omega(\tnn{X}) &= \left(\Xi^{(1)}\right)^\ast\Omega(\tnn{X})+\left(\Xi^{(2)}\right)^\ast\Omega(\tnn{X})\\
	&= \frac{(b-a)d(\sqrt{y}))}{(\sqrt{y}-a)(\sqrt{y}-b)}+\frac{(b-a)d(-\sqrt{y})}{(-\sqrt{y}-a)(-\sqrt{y}-b)}\\
	&= \frac{(b^2-a^2)2\sqrt{y}d(\sqrt{y})}{(y-a^2)(y-b^2)}\\
	&= \frac{(b^2-a^2)dy}{(y-a^2)(y-b^2)} \\
	&= \Omega(\tnn{Y})\,.
\end{align*}
The square-roots appearing in intermediate steps of the above computation disappear from the final answer, demonstrating a general phenomenon: the pushforward of a rational form via a surjective meromorphic map is always a rational form, despite the non-meromorphicity of its local inverses.  

\section{Mathematical Preliminaries}

This chapter studies the problem of pushing forward rational forms using computational algebraic geometry. By reformulating morphisms as ideals inside polynomial rings over fraction fields, this problem is amenable to the powerful tools of Gr\"{o}bner bases and their applications to elimination theory. We will assume familiarity with basic notions relating to Gr\"{o}bner bases as presented in \cite{cox2013ideals}.

\subsection{Morphisms as Ideals}

Let $x\coloneqq(x_1,\ldots,x_m)$ and $y\coloneqq(y_1,\ldots,y_n)$ be tuples of indeterminates where $m$ and $n$ are positive integers. Designate by $\complex(y)[x]$ the ring of polynomials in $x$ whose coefficients are complex rational functions in $y$. Any polynomial $h\in\complex(y)[x]$ can be regarded as a map $h:\overbar{\complex(y)}^m\to\overbar{\complex(y)}$ defined by $x\mapsto h(x)$, where $\overbar{\complex(y)}$ denotes the algebraic closure of $\complex(y)$. Moreover, for each $x\in\overbar{\complex(y)}^m$, $h(x)\in\overbar{\complex(y)}$ defines a map $h(x):\complex^n\to\complex$ by $y\mapsto h(x)(y)$. We will often want to restrict $h$, specifically the coefficients of $h$, to some particular value for $y$. Given a point $y\in\complex^n$ for which all coefficients of $h$ are analytic, let $h(\bullet)(y)$ be the polynomial in $\complex[x]$ defined by $x\mapsto h(x)(y)$.

For our purposes, we can think of $x$ (resp., $y$) as coordinates in an affine chart of the embedding space $X$ (resp., $Y$) associated with some positive geometry $(X,\tnn{X})$ (resp., $(Y,\tnn{Y})$) of dimension $m$ (resp., $n$). For example, $x$ could represent coordinates for some worldsheet moduli space (see \cref{chp:abhy,chp:mom,chp:omom}) and $y$ could represent coordinates for some kinematic space. In this case, the two are related via some twistor-string map or version of the scattering equations. The graph of any morphism (restricted to a tile) can always be expressed as a \emph{complete intersection ideal} in $\complex(y)[x]$, i.e.\ an ideal generated by precisely $m$ polynomials in $\complex(y)[x]$. Since our ideal lives in $\complex(y)[x]$, we can always choose generators from $\complex[y][x]$: polynomials in $x$ whose coefficients are complex polynomials (as opposed to rational functions) in $y$. Suppose $f_1,\ldots,f_m\in\complex[y][x]$ generate the complete intersection ideal
\begin{align}\label{eq:push-ideal}
	\mathcal{I}\coloneqq\lrangle{f_1,\ldots,f_m}\subset\complex(y)[x]\,.
\end{align}
Let $F\coloneqq(f_1,\ldots,f_m)$ collect the generators of $\mathcal{I}$ into an $m$-tuple. For each $y\in\complex^n$, define
\begin{align}\label{eq:push-ideal-y}
	\mathcal{I}(y)\coloneqq\lrangle{f_1(\bullet)(y),\ldots, f_m(\bullet)(y)}\subset\complex[x]\,.
\end{align}
The \emph{zero-set} of $\mathcal{I}(y)$, denoted by $\zero(\mathcal{I}(y))$, is given by
\begin{align}\label{eq:push-ideal-y-zero}
	\zero(\mathcal{I}(y))\coloneqq\left\{\xi\in\complex^m:h(\xi)(y)=0~\text{for all}~h\in\mathcal{I}(y)\right\},
\end{align}
and the \emph{radical} of $\mathcal{I}(y)$, denoted by $\sqrt{\mathcal{I}(y)}$, is defined as
\begin{align}\label{eq:push-ideal-y-radical}
	\sqrt{\mathcal{I}(y)}\coloneqq\left\{h\in\complex[x]:h^r\in\mathcal{I}(y)~\text{for some}~r\in\integer_{>0}\right\}.
\end{align}
We say that $\mathcal{I}(y)$ is \emph{zero-dimensional} if $\zero(\mathcal{I}(y))$ consists of finitely many points, and \emph{radical} if $\mathcal{I}(y)=\sqrt{\mathcal{I}(y)}$. Let $\mathcal{Y}_F\subset\complex^n$ denote the subset of $y$-variables for which $\mathcal{I}(y)$ is both zero-dimensional and radical. We will refer to points in $\mathcal{Y}_F$ as generic $y$-variables. Suppose that $\mathcal{Y}_F$ is a non-empty Zariski open set. Then we will refer to $\mathcal{I}$ as a zero-dimensional radical ideal, meaning that $\mathcal{I}(y)$ is a zero-dimensional radical ideal for generic $y$-variables. 

The zero-set of $\mathcal{I}$, denoted by $\zero(\mathcal{I})$, is given by 
\begin{align}\label{eq:push-ideal-zero}
	\zero(\mathcal{I})\coloneqq\left\{\xi\in\overbar{\complex(y)}^{m}:h(\xi)=0~\text{for all}~h\in\mathcal{I}\right\}.
\end{align}
The points in $\zero(\mathcal{I})$ are functions of $y$ and non-singular for generic $y$-variables. We will assume (for generic $y$-variables) that these points are non-degenerate (i.e.\ they have multiplicity one) and enumerate them by $\zero(\mathcal{I})=\{\xi^{(\alpha)}\}_{\alpha=1}^d$.

\subsection{Notation and Conventions}

To streamline our presentation, let us collect a few definitions and establish notation. 

A \emph{monomial} in $x$ is an expression of the form $x^\alpha = x_1^{\alpha_1}\cdots x_m^{\alpha_m}$ where $\alpha\coloneqq(\alpha_1,\ldots,\alpha_m)\allowbreak\in\integer_{\ge0}^m$. A \emph{monomial ordering} $\succ$ on the polynomial ring $\complex(y)[x]$ is a total well-ordering on monomials in $x$, compatible with multiplication: if $x^\alpha\succ x^\beta$ then $x^{\alpha+\gamma}\succ x^{\beta+\gamma}$ for all $\gamma\in\integer_{\ge0}^n$. 

Recall that $\mathcal{I}$ is a zero-dimensional radical complete intersection ideal in $\complex(y)[x]$. The \emph{leading monomial} of a polynomial in $\mathcal{I}$ is the largest monomial with respect to $\succ$. Moreover, the \emph{leading coefficient} of a polynomial in $\mathcal{I}$ is the coefficient of the leading monomial and the \emph{leading term} is the term corresponding to the leading monomial. There are three monomial orderings which will be particularly useful later. To this end, declare $x_1\succ\cdots\succ x_m$ and let $\alpha,\beta\in\integer_{\ge0}^m$.
\begin{itemize}
	\item \emph{Lexicographic (lex)} order: $x^{\alpha}\succ_\text{lex}x^{\beta}$ if the leftmost non-zero entry of $\alpha-\beta$ is positive. Otherwise $x^{\beta}\succ_\text{lex}x^{\alpha}$.
	\item \emph{Graded lexicographic (grlex)} ordering: $x^\alpha\succ_\text{grlex}x^\beta$ if $\sum_{i=1}^m \alpha_i > \sum_{i=1}^m \beta_i$, or if $\sum_{i=1}^m \alpha_i = \sum_{i=1}^m \beta_i$ and the leftmost non-zero entry of $\alpha-\beta$ is positive. Otherwise $x^\beta\succ_\text{grlex}x^\alpha$.
	\item \emph{Graded reverse lexicographic (grevlex)} ordering: $x^\alpha\succ_\text{grevlex}x^\beta$ if $\sum_{i=1}^m \alpha_i > \sum_{i=1}^m \beta_i$, or if $\sum_{i=1}^n \alpha_i = \sum_{i=1}^n \beta_i$ and the rightmost non-zero entry of $\alpha-\beta$ is negative. Otherwise $x^\beta\succ_\text{grevlex}x^\alpha$.
\end{itemize}

A \emph{Gr\"{o}bner basis} for $\mathcal{I}$, with respect to a monomial ordering $\succ$, is a finite generating set for $\mathcal{I}$. It has a special property: the leading term of every (non-zero) polynomial in $\mathcal{I}$ is divisible by the leading term of some polynomial in the Gr\"{o}bner basis. We designate the \emph{unique reduced Gr\"{o}bner basis} for $\mathcal{I}$ with respect to $\succ$ by $\mathcal{G}_\succ(\mathcal{I})$ (or simply $\mathcal{G}$).

Since $|{\zero(\mathcal{I})}|=d$, the \emph{quotient ring} $\mathcal{Q}\coloneqq\complex(y)[x]/\mathcal{I}$ is a $d$-dimensional commutative algebra over $\complex(y)$. Let $\mathcal{B}_\succ(\mathcal{I})$ (or simply $\mathcal{B}$) denote the set of all monomials in $x$ that are indivisible by the leading terms of $\mathcal{G}$. This forms a basis for $\mathcal{Q}$ called the \emph{standard basis}: every element of $\mathcal{Q}$ can be represented as a $\complex(y)$-linear combination of monomials in $\mathcal{B}$. Moreover, every element of $\complex(y)[x]$ can be canonically assigned to an element of $\mathcal{Q}$ via the division algorithm with respect to $\mathcal{G}$. We enumerate the elements of the standard basis by $\mathcal{B}=\{e_\alpha\}_{\alpha=1}^{d}$. Importantly, they are monomials in $x$ and independent of $y$. Given a polynomial $h\in\complex(y)[x]$, let $\overbar{h}^{\mathcal{G}}$ denote the \emph{normal form} of $h$, i.e.\ the remainder of $h$ on division by $\mathcal{G}$. Then $\overbar{h}^{\mathcal{G}} = \sum_{\alpha=1}^d\overbar{h}_\alpha e_\alpha$ for some choice of coefficients $\overbar{h}_1,\ldots,\overbar{h}_d\in\complex(y)$. Alternatively, $\overbar{h}^\mathcal{G}=\overbar{h}\cdot e$ where $\overbar{h}\coloneqq(\overbar{h}_1,\ldots,\overbar{h}_d)$ and $e\coloneqq(e_1,\ldots,e_d)$. Lastly, $[h]=[\overbar{h}^\mathcal{G}]\in\mathcal{Q}$ for every $h\in\complex(y)[x]$.

\subsection{Pullbacks and Pushforwards}
With the above definitions and notation, let us review the pullback and pushforward operations in terms of $x$ and $y$. Fix $I\in\binom{[m]}{p}$ for some $0\le p\le m$ and let $\omega$ be a rational $p$-form
\begin{align}\label{eq:push-form}
	\omega(x)\coloneqq\underline{\omega}(x)\bigwedge_{i\in I}dx_i\,,
\end{align}
where $\underline{\omega}$ is some rational function in $\complex(x)$. Suppose $\psi:\complex^n\to\complex^m$ is a differentiable map from $y$-variables to $x$-variables. Then the pullback of $\omega$ via $\psi$ is the $p$-form defined by 
\begin{align}\label{eq:push-pullback-standard}
(\psi^\ast\omega)(y)\coloneqq \omega(\psi(y)) = \underline{\omega}(\psi(y))\bigwedge_{i\in I}d\psi_i(y) = \underline{\omega}(\psi(y))\sum_{J\in\binom{[n]}{p}}\left|\frac{\partial\psi}{\partial y}\right|_{I,J}\bigwedge_{j\in J}dy_j\,,
\end{align}
where ${\partial\psi}/{\partial y}$ is the $m\times n$ Jacobian matrix of partial derivatives ${\partial\psi_i}/{\partial y_j}$ (with $i\in[m]$ and $j\in[n]$) and $|{\partial\psi}/{\partial y}|_{I,J}$ denotes the minor of ${\partial\psi}/{\partial y}$ in the row set $I$ and the column set $J$.

We can recast the  above pullback as a pushforward via an ideal which implicitly defines a meromorphic map $\psi$. To this end, let $\mathcal{J}$ be a zero-dimensional radical complete intersection ideal in $\complex(y)[x]$ whose zero-set is a singleton. Then for generic $y$-variables, the Shape Lemma (see \cref{thm:shape}) implies that the Gr\"{o}bner basis $\mathcal{G}_\text{lex}(\mathcal{J})$ with respect to lex order has the shape
\begin{align}
	\mathcal{G}_\text{lex}(\mathcal{J})=\left\{x_1-\psi_1(y),\ldots,x_m-\psi_m(y)\right\},
\end{align} 
where $\psi\coloneqq(\psi_1,\ldots,\psi_m)$ is an $m$-tuple of rational functions in $\complex(y)$. Consequently, the zero-set of $\mathcal{J}$ is given by $\zero(\mathcal{J})=\{\psi\}$. This motivates the following definition. The pushforward of $\omega$ via $\mathcal{J}$, denoted by $\mathcal{J}_\ast\omega$, is defined by
\begin{align}\label{eq:push-pullback-ideal}
	(\mathcal{J}_\ast\omega)(y)\coloneqq(\psi^\ast\omega)(y) = \omega(\psi(y))\,,
\end{align}
and coincides with \eqref{eq:push-pullback-standard}.

\begin{figure}
	\centering
	\includegraphics[width=0.8\textwidth]{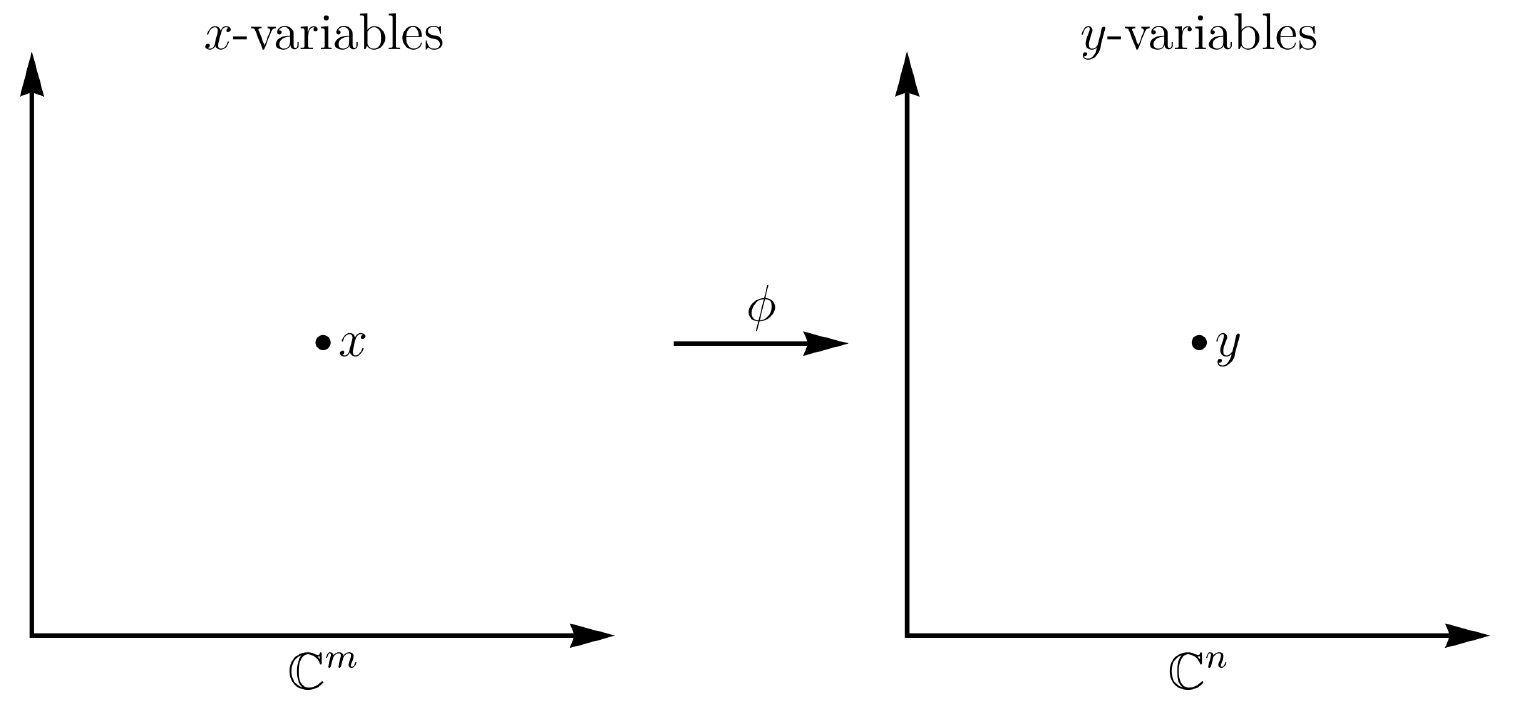}
	\caption{The map $\phi:\mathbb{C}^m\to\mathbb{C}^n$ takes $x$ to $y=\phi(x)$.}
	\label{fig:push-phi}
\end{figure}

\begin{figure}
	\centering
	\includegraphics[width=0.8\textwidth]{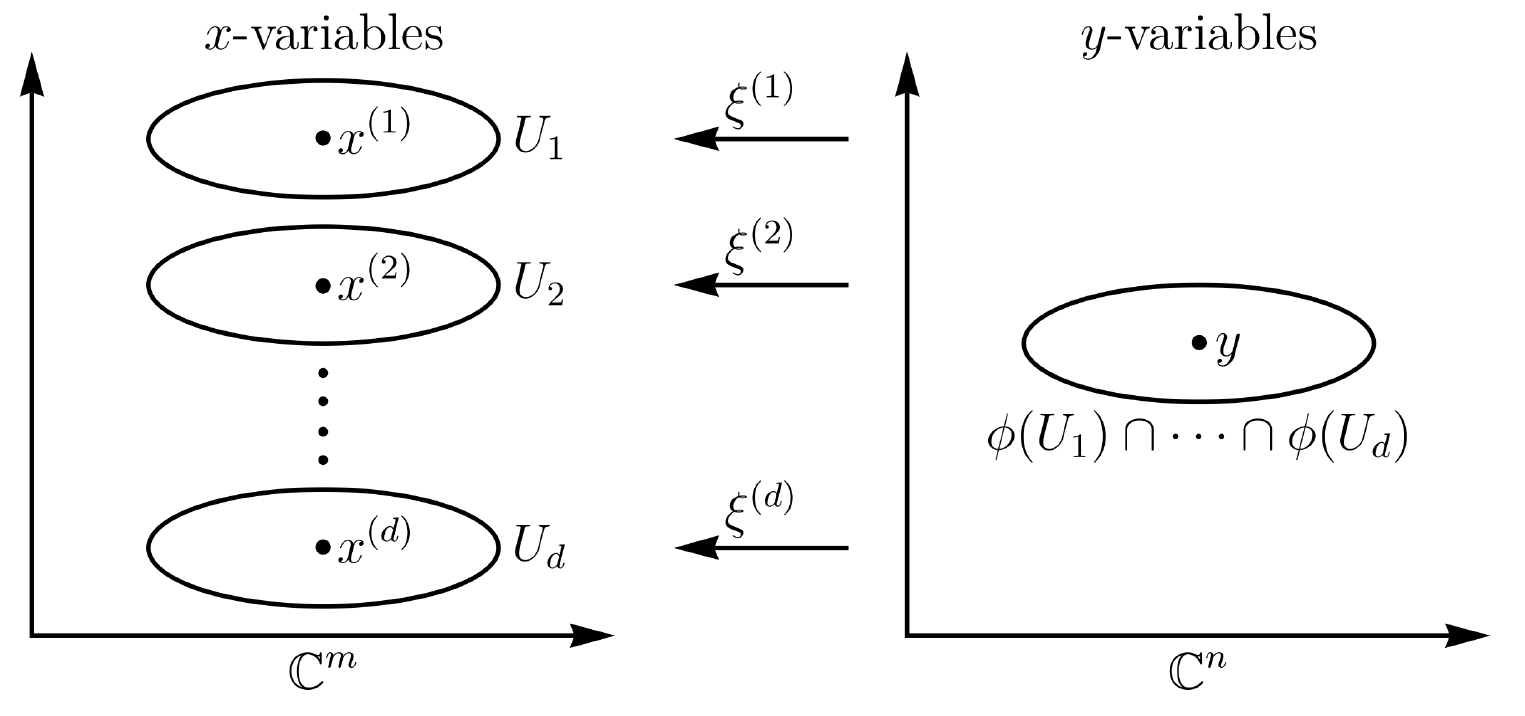}
	\caption{For each point $y$ in the image of $\phi$ there are $d$ local inverse maps $\xi^{(\alpha)}\coloneqq\phi|_{U_\alpha}^{-1}:\phi(U_\alpha)\to U_\alpha$ such that $x^{(\alpha)}=\xi^{(\alpha)}(y)\in U_\alpha$.}
	\label{fig:push-phi-inverses}
\end{figure}

Going in the direction opposite to $\psi$, let $\phi:\complex^m\to\complex^n$ be a meromorphic map of degree $d$ from $x$-variables to $y$-variables; see \cref{fig:push-phi}. Let $y$ be a point in the image of $\phi$. Then $y$ has $d$ preimages labelled $\phi^{-1}(\{y\}) = \{x^{(\alpha)}\}_{\alpha=1}^{d}$. For each $\alpha=1,\ldots,d$ there exists an open neighbourhood $U_\alpha$ containing $x^{(\alpha)}$ such that $\phi|_{U_\alpha}$ is invertible. Let $\xi^{(\alpha)}\coloneqq\phi|_{U_\alpha}^{-1}$. The local inverse maps are depicted in \cref{fig:push-phi-inverses}. Then the pushforward of $\omega$ via $\phi$ is a rational $p$-form defined by 
\begin{align}\label{eq:push-pushforward-standard}
	(\phi_\ast\omega)(y) = \sum_{\alpha=1}^{d} (\xi^{(\alpha)\ast}\omega)(y) =\sum_{J\in\binom{[n]}{p}} \sum_{\alpha=1}^d\left( \underline{\omega}(\xi^{(\alpha)}(y))\left|\frac{\partial\xi^{(\alpha)}}{\partial y}\right|_{I,J}\right)\bigwedge_{j\in J}dy_j \,,
\end{align}
where ${\partial\xi^{(\alpha)}}/{\partial y}$ is the $m\times n$ Jacobian matrix of mixed partial derivatives ${\partial\xi^{(\alpha)}_i}/{\partial y_j}$ (with $i\in[n]$ and $j\in[m]$). 

\begin{figure}
	\centering
	\includegraphics[width=0.8\textwidth]{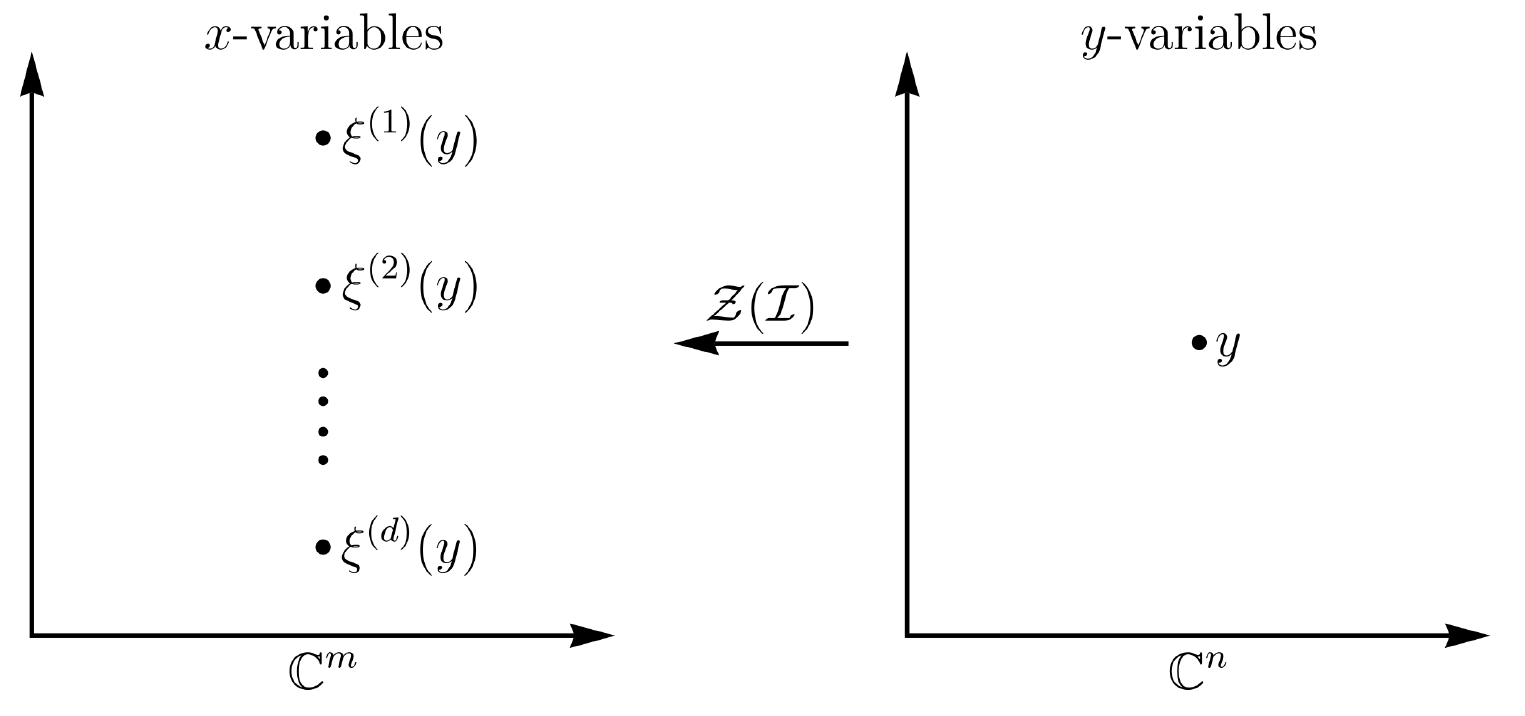}
	\caption{The zero-set $\mathcal{Z}(\mathcal{I})=\{\xi^{(\alpha)}\}_{\alpha=1}^d$ consists of $d$ maps in $\overbar{\mathbb{C}(y)}^m$ which take $y$ to $\xi^{(\alpha)}(y)$.}
	\label{fig:push-ideal}
\end{figure}

This chapter's primary focus is pushing forward rational forms via ideals obtained as the graph of some morphism between positive geometries. To this end, let $\mathcal{I}$ be the ideal in $\complex(y)[x]$ from before. The pushforward of $\omega$ via $\mathcal{I}$, denoted by $\mathcal{I}_\ast\omega$, is defined by
\begin{align}\label{eq:push-pushforward-ideal}
	(\mathcal{I}_\ast\omega)(y)\coloneqq\sum_{\xi\in\zero(\mathcal{I})}(\xi^\ast\omega)(y)\,,
\end{align}
where $\zero(\mathcal{I})$ is a finite subset of $\overbar{\complex(y)}^m$ consisting of maps from $y$-variables to $x$-variables; see \cref{fig:push-ideal}. Comparing \eqref{eq:push-pushforward-ideal} and \eqref{eq:push-pullback-ideal}, the pullback is a special case of the pushforward; in the former case, the zero-set contains a single point, while the zero-set in the latter case is a reducible algebraic set.

\cref{eq:push-pushforward-ideal} contains a sum over all points in the zero-set of $\mathcal{I}$. In general, it is impossible to determine this zero-set in closed form, thereby prompting the following analogue of an earlier question: ``Is it possible to compute the pushforward of a rational form via an ideal without needing to determine its zero-set?''

The special case of pushing forward rational functions (rational $0$-forms) has been studied in the scattering amplitudes literature in the context of the CHY formalism. At least two approaches can be generalised to rational forms \cite{Lukowski:2022fwz}. The approaches employ ideas from computational algebraic geometry; the first uses \emph{companion matrices} \cite{Huang:2015yka,Cardona:2015eba,Cardona:2015ouc} while the second uses the \emph{duality of global residues} \cite{Sogaard:2015dba}. By rephrasing the problem of pushing forward rational forms in terms of pushing forward rational functions, the above methods can be readily applied. To this end, let us reformulate \eqref{eq:push-pushforward-ideal}. Applying the chain rule to \eqref{eq:push-pushforward-ideal} yields
\begin{align}\label{eq:push-pushforward-ideal-chain-rule}
	(\mathcal{I}_\ast\omega)(y)=\sum_{J\in\binom{[n]}{p}}\left(\sum_{\xi\in\zero(\mathcal{I})}\underline{\omega}(\xi)\left|\frac{\partial\xi}{\partial y}\right|_{I,J}\right)\,\bigwedge_{j\in J}dy_j\,,
\end{align}
where ${\partial\xi}/{\partial y}$ is the $m\times n$ Jacobian matrix of mixed partial derivatives ${\partial\xi_i}/{\partial y_j}\in\overbar{\complex(y)}$ (with $i\in[m]$ and $j\in[n]$). The minors of ${\partial\xi}/{\partial y}$ can be judiciously rewritten using the following observation. Since $f_\ell(\xi(y))=0$ for each $\ell\in[m]$ and $\xi\in\zero(\mathcal{I})$, we have that
\begin{align}\label{eq:push-total-derivative}
	0 = \frac{df_{\ell}}{dy_j} = \frac{\partial f_\ell}{\partial y_j} + \frac{\partial f_\ell}{\partial \xi_i}\frac{\partial \xi_i}{\partial y_j} \implies \frac{\partial f_\ell}{\partial \xi_i}\frac{\partial \xi_i}{\partial y_j} = - \frac{\partial f_\ell}{\partial y_j}\,,
\end{align}
and hence
\begin{align}\label{eq:push-total-derivative-jacobian}
	\left|\frac{\partial\xi}{\partial y}\right|_{I,J}  = (-1)^p\, \left|\left(\frac{\partial F}{\partial x}(\xi)\right)^{-1}\frac{\partial F}{\partial y}(\xi)\right|_{I,J},
\end{align}
where ${\partial F}/{\partial x}$ is the $m\times m$ Jacobian matrix of partial derivatives ${\partial f_\ell}/{\partial x_i}\in\complex(y)[x]$ (with $\ell,i\in[m]$) and ${\partial F}/{\partial y}$ is the $m\times n$ Jacobian matrix of partial derivatives ${\partial f_\ell}/{\partial y_j}\in\complex(y)[x]$ (with $j\in[n]$). The matrix ${\partial F}/{\partial x}$ is assumed invertible for generic $y$. Substituting \eqref{eq:push-total-derivative-jacobian} into \eqref{eq:push-pushforward-ideal-chain-rule} leads to the master equation
\begin{align}\label{eq:push-master}
	(\mathcal{I}_\ast\omega)(y)=\sum_{J\in \binom{[n]}{p}}(\mathcal{I}_\ast\underline{\omega}_{I,J})(y)\,\bigwedge_{j\in J}dy_{j}\,,
\end{align}
where
\begin{align}\label{eq:push-master-coefficients}
	\underline{\omega}_{I,J}(x)\coloneqq (-1)^p\,\underline{\omega}(x)\left|\left(\frac{\partial F}{\partial x}\right)^{-1}\frac{\partial F}{\partial y}\right|_{I,J}\in\mathbb{C}(y)(x)\,.
\end{align}


In summary, we have reduced the problem pushing forward a rational $p$-form $\omega$ to pushing forward the rational functions $\underline{\omega}_{I,J}\in\complex(y)(x)$. In the next three sections, we present three methods for computing $\mathcal{I}_\ast\underline{\omega}_{I,J}$. To demonstrate the application of each method, let us establish the following running example. 

\begin{example}[Pushforwards via Brute Force]\label{ex:push-formal}
	Let $n=2$ and $m=3$. Let $\mathcal{I}$ be generated by $F\coloneqq(f_1,f_2)=(y_1x_1+y_2x_2,y_3x_2^2-1)$. Clearly $\mathcal{I}(y)$ is zero-dimensional provided $y_1$ and $y_3$ are non-zero. Hence,  $\mathcal{Y}_F=\complex^3\setminus\{y_1=0~\text{or}~y_3=0\}$. The zero-set of $\mathcal{I}$ is easily determined and given by 
	\begin{align}\label{eq:push-ex-ideal-zero}
		\zero(\mathcal{I})=\left\{\left( \frac{y_2}{y_1\sqrt{y_3}},-\frac{1}{\sqrt{y_3}}\right),\left( -\frac{y_2}{y_1\sqrt{y_3}},\frac{1}{\sqrt{y_3}}\right)\right\}.
	\end{align}
	The points in $\zero(\mathcal{I})$ are supported on $\mathcal{Y}_F$. Let us consider pushing forward the rational $2$-form
	\begin{align}\label{eq:push-ex-form}
		\omega(x)= \underline{\omega}(x)\,dx_1\wedge dx_2\,,
		\qquad
		\underline{\omega}(x)=\frac{1}{x_1x_2}\,.
	\end{align}
	In fact, $\omega$ is a $d\log$-form which can be thought of as the canonical form of the positive geometry $(X,\tnn{X})$ where $X=\complex^2$ and $\tnn{X}$ is the positive quadrant in $\real^2$. Since we know $\zero(\mathcal{I})$, we can directly push $\omega$ forward via $\mathcal{I}$: using \eqref{eq:push-pushforward-ideal} we obtain
	\begin{align}\label{eq:push-ex-pushforward-ideal}
		(\mathcal{I}_\ast\omega)(y) = \frac{dy_1\wedge dy_3}{y_1y_3} - \frac{dy_2\wedge dy_3}{y_2y_3}\,.
	\end{align}
	Let us verify this result using \eqref{eq:push-master}, the master equation. The Jacobian matrices relevant for this calculation are given by 
	\begin{align}\label{eq:push-ex-jacobians}
		\frac{\partial F}{\partial x}=
		\begin{bmatrix}
			y_1 & y_2\\
			0 & 2 y_3x_2
		\end{bmatrix},
		\qquad
		\frac{\partial F}{\partial y}=
		\begin{bmatrix}
			x_1 & x_2 & 0\\
			0 & 0 & x_2^2
		\end{bmatrix}.
	\end{align}
	Unsurprisingly, ${\partial F}/{\partial x}$ is invertible if and only if $y\in\mathcal{Y}_F$ because $\det(\frac{\partial F}{\partial x})=2y_1y_3x_2$. The coefficients $\underline{\omega}_{J}\coloneqq\underline{\omega}_{[2],J}$, calculated using \eqref{eq:push-master-coefficients}, are given by
	\begin{align}\label{eq:push-ex-master-coefficients}
		\underline{\omega}_{\{1,2\}}(x)=0\,,\qquad\underline{\omega}_{\{1,3\}}(x)=\frac{1}{2y_1y_3}\,,\qquad \underline{\omega}_{\{2,3\}}(x)=\frac{x_2}{2 y_1 y_3 x_1}\,.
	\end{align}
	Pushing each of them forward via $\mathcal{I}$, i.e.\ summing over all points in $\zero(\mathcal{I})$, we obtain
	\begin{align}\label{eq:push-ex-master-coefficients-pushforward}
		(\mathcal{I}_\ast\underline{\omega}_{\{1,2\}})(y)=0\,,\qquad (\mathcal{I}_\ast\underline{\omega}_{\{1,3\}})(y) = \frac{1}{y_1y_3}\,,\qquad (\mathcal{I}_\ast\underline{\omega}_{\{2,3\}})(y) = -\frac{1}{y_2y_3}\,,
	\end{align}
	in clear agreement with \eqref{eq:push-ex-pushforward-ideal}.
\end{example}

\section{Pushforwards via Companion Matrices}
\label{sec:push-companion}

Our first method uses companion matrices. Companion matrices help sidestep the problem of determining $\zero(\mathcal{I})$. This method was first employed (in the scattering amplitudes literature) in the context of the CHY formalism \cite{Huang:2015yka,Cardona:2015eba,Cardona:2015ouc}. Companion matrices represent polynomial multiplication as an endomorphism on the quotient ring $\mathcal{Q}=\complex(y)[x]/\mathcal{I}$. Let $\mathcal{G}$ be a Gr\"{o}bner basis for $\mathcal{I}$ and let $\mathcal{B}$ be the associated standard basis. For each $i\in[m]$, multiplication by $x_i$ in $\complex(y)[x]$ induces an endomorphism on $\mathcal{Q}$ via the division algorithm with respect to $\mathcal{G}$: 
\begin{align}
\begin{tikzcd}
	\complex(y)[x]\to\complex(y)[x], h\mapsto x_i h \arrow[d, "\text{modulo}~\mathcal{G}"]\\ \mathcal{Q}\to\mathcal{Q},[\overbar{h}^{\mathcal{G}}]\mapsto[\overbar{x_ih}^{\mathcal{G}}]
\end{tikzcd}
\end{align}
In the standard basis, this endomorphism is represented by a matrix $T_i\in\Mat{d}{d}{\complex(y)}$ whose components are given by 
\begin{align}
	\overbar{z_i e_\alpha}^{\mathcal{G}}=(T_i)_{\alpha\beta}e_\beta\,,
\end{align}
where repeated indices are implicitly summed. The matrix $T_i$ is called the $i$\textsuperscript{th} \emph{companion matrix} of $\mathcal{I}$ \cite{sturmfels2002solving}. The companion matrices of $\mathcal{I}$ mutually commute, because multiplication in the polynomial ring is commutative. Consequently, they generate a commutative subalgebra $\lrangle{T_1,\ldots,T_m}\subset\Mat{d}{d}{\complex(y)}$ isomorphic to $\mathcal{Q}$:
\begin{align}
	\lrangle{T_1,\ldots,T_m}\to \mathcal{Q},T_i\mapsto[\overbar{x_i}^{\mathcal{G}}]\,.
\end{align}
Stickelberger's Theorem (see \cref{thm:stickelberger}) informs us that (for generic $y$-variables) $\zero(\mathcal{I})$ is the set of vectors of simultaneous eigenvalues of the companion matrices of $\mathcal{I}$:
\begin{align}
	\zero(\mathcal{I})=\left\{\xi\in\overbar{\complex(y)}^m: (\exists~v\in\overbar{\complex(y)}^d\setminus\{0\})~(T_i)_{\alpha\beta} v_\beta = \xi_i v_{\alpha}~\text{for all}~i\in[m]\right\}.
\end{align}
Moreover, \cref{thm:simultaneous} states that (for generic $y$-variables) the companion matrices of $\mathcal{I}$ are simultaneously diagonalizable. Let $T\coloneqq(T_1,\ldots,T_m)$ collect the companion matrices of $\mathcal{I}$ into an $m$-tuple. Take any rational function $r\in\complex(y)(x)$. From the above considerations, $r(T)$ is well-defined and similar to a diagonal matrix with diagonal entries given by $r(\xi)$ for $\xi\in\zero(\mathcal{I})$. Consequently, 
\begin{align}\label{eq:push-companion-trace-rational}
	(\mathcal{I}_\ast r)(y)
	=\Tr\left(r(T(y))(y)\right).
\end{align}
In other words, provided we know the companion matrices of $\mathcal{I}$ (which requires a Gr\"{o}bner basis computation), we can push forward any rational function via $\mathcal{I}$ using \eqref{eq:push-companion-trace-rational}. In particular, 
\begin{align}\label{eq:push-companion-trace-coefficients}
	(\mathcal{I}_\ast\underline{\omega}_{I,J})(y) = \Tr\left(\underline{\omega}_{I,J}(T(y))(y)\right).
\end{align}
In this way, the problem of pushing forward $\omega$ via $\mathcal{I}$ is reduced to simple linear algebra.

\begin{example}[Pushforwards via Companion Matrices]\label{ex:push-companion}
	Let us re-derive \eqref{eq:push-ex-master-coefficients-pushforward} using \eqref{eq:push-companion-trace-coefficients}. Without loss of generality, assume lex-order with $x_1\succ x_2$. The Gr\"{o}bner basis\footnote{For this example, the Gr\"{o}bner basis is unchanged if we consider grevlex-order instead of lex-order.} for $\mathcal{I}$ is given by $\mathcal{G}=\{x_1+\frac{y_2}{y_1}x_2,x_2^2-\frac{1}{y_3}\}$ and the standard basis is given by $\mathcal{B}=\{x_2,1\}$. The companion matrices of $\mathcal{I}$ read
	\begin{align}\label{eq:ex-push-cm}
		T_1(y)=
		\begin{bmatrix}
			0 & -\frac{y_2}{y_1 y_3} \\
			-\frac{y_2}{y_1} & 0 \\
		\end{bmatrix},\qquad
		T_2(y)=
		\begin{bmatrix}
			0 & \frac{1}{y_3} \\
			1 & 0 \\
		\end{bmatrix}.
	\end{align}
	Evaluating \eqref{eq:push-companion-trace-coefficients} on \eqref{eq:push-ex-master-coefficients} and \eqref{eq:ex-push-cm} then gives
	\begin{align}
		(\mathcal{I}_\ast\underline{\omega}_{\{1,2\}})(y)&=\Tr[\underline{\omega}_{\{1,2\}}(T(y))(y)] = \vphantom{\frac11}0\,,\\
		(\mathcal{I}_\ast\underline{\omega}_{\{1,3\}})(y)&=\Tr[\underline{\omega}_{\{1,3\}}(T(y))(y)] = \frac{1}{2y_1y_3}\Tr[\mathbb{I}_{2\times 2}] = \frac{1}{y_1y_3}\,,\\
		(\mathcal{I}_\ast\underline{\omega}_{\{2,3\}})(y)&=\Tr[\underline{\omega}_{\{2,3\}}(T(y))(y)] = \frac{1}{2 y_1 y_3}\Tr[T_1^{-1}(y)\cdot T_2(y)] = -\frac{1}{y_2y_3}\,,
	\end{align}
	in clear agreement with \eqref{eq:push-ex-master-coefficients-pushforward}.
\end{example}
 
\section{Pushforwards via Derivatives of Companion Matrices}
\label{sec:push-companion-alt}

The previous method evaluates the rational functions $\underline{\omega}_{I,J}$ on companion matrices. The rational functions depend on the generators in $F$ through two Jacobian matrices. Consequently, $\underline{\omega}_{I,J}$ can be arbitrarily complicated, due to the multiplying minor in \eqref{eq:push-master-coefficients}, even if $\underline{\omega}$ is relatively simple, posing a significant bottleneck. In this section, we present a second method for pushing forward $\omega$ via $\mathcal{I}$. The method also uses companion matrices, but it avoids evaluating the multiplying minor in \eqref{eq:push-master-coefficients} on companion matrices. Instead, it uses derivatives of companion matrices. We also present a novel algorithm for efficiently determining companion matrix derivatives, developed with a view towards numerical sampling.

By comparing \eqref{eq:push-pushforward-ideal-chain-rule} with our master equation in \eqref{eq:push-master}, we arrive at
\begin{align}
	(\mathcal{I}_\ast\underline{\omega}_{I,J})(y) = \sum_{\xi\in\zero(\mathcal{I})}\underline{\omega}(\xi(y))\left|\frac{\partial\xi}{\partial y}\right|_{I,J}.
\end{align}
We can uplift this equation to one involving companion matrices in the obvious way:
\begin{align}\label{eq:push-companion-alt-trace-coefficients}
	(\mathcal{I}_\ast\underline{\omega}_{I,J})(y) = \Tr\left(\underline{\omega}(T(y))\sum_{\sigma\in S_p}\sign(\sigma)\frac{\partial T_{i_{\sigma(1)}}}{\partial y_{j_1}}\cdots\frac{\partial T_{i_{\sigma(p)}}}{\partial y_{j_p}}\right),
\end{align}	
where the sum is over the permutation group $S_p$, $\sign(\sigma)$ denotes the signature of $\sigma$, $I=\{i_1,\ldots,i_p\}$ with $i_1<\cdots<i_p$ and $J=\{j_1,\ldots,j_p\}$ with $j_1<\cdots<j_p$. Reference  \cite{Lukowski:2022fwz} proves \cref{eq:push-companion-alt-trace-coefficients}. The proof is non-trivial, because derivatives of companion matrices are not guaranteed to simultaneously commute.

\begin{example}[Pushforwards via Derivatives of Companion Matrices]\label{ex:push-companion-alt}
	Let us re-derive \eqref{eq:push-ex-master-coefficients-pushforward} using \eqref{eq:push-companion-alt-trace-coefficients}. Recall that the companion matrices, calculated in \cref{ex:push-companion}, are given by 
	\begin{align}\label{eq:ex-push-companion-alt-cm}
		T_1(y)=
		\begin{bmatrix}
			0 & -\frac{y_2}{y_1 y_3} \\
			-\frac{y_2}{y_1} & 0 \\
		\end{bmatrix},\qquad
		T_2(y)=
		\begin{bmatrix}
			0 & \frac{1}{y_3} \\
			1 & 0 \\
		\end{bmatrix}.
	\end{align}
	Their derivatives with respect to $y$-variables read
	\begin{equation}
		\label{eq:ex-push-companion-alt-cm-pd}
	\begin{alignedat}{3}
		\frac{\partial T_1}{\partial y_1} &= \begin{bmatrix}
			0 & \frac{y_2}{y_1^2y_3} \\
			\frac{y_2}{y_1^2} & 0
		\end{bmatrix}, \qquad& \frac{\partial T_1}{\partial y_2} &= \begin{bmatrix}
			0 & -\frac{1}{y_1y_3} \\
			-\frac{1}{y_1} & 0
		\end{bmatrix}, \qquad& \frac{\partial T_1}{\partial y_3} &=
		\begin{bmatrix}
			0 & \frac{y_2}{y_1y_3^2} \\
			0 & 0
		\end{bmatrix},\\
		\frac{\partial T_2}{\partial y_1} &= 0_{2\times 2}, \qquad& \frac{\partial T_2}{\partial y_2} &=
		0_{2\times 2}, \qquad& \frac{\partial T_2}{\partial y_3} &=
		\begin{bmatrix}
			0 & -\frac{1}{y_3^2} \\
			0 & 0
		\end{bmatrix}.
	\end{alignedat}
	\end{equation}
	Substituting \eqref{eq:ex-push-companion-alt-cm} and \eqref{eq:ex-push-companion-alt-cm-pd} into \eqref{eq:push-companion-alt-trace-coefficients} where $\underline{\omega}(x)=\frac{1}{x_1x_2}$, we find that
	\begin{align}
		(\mathcal{I}_\ast \omega_{\{1,2\}})(y) &= \Tr\left(\underline{\omega}(T(y))\left(\frac{\partial T_1}{\partial y_1}\frac{\partial T_2}{\partial y_2}-\frac{\partial T_2}{\partial y_1}\frac{\partial T_1}{\partial y_2}\right)\right) = 0\,,\\
		(\mathcal{I}_\ast \omega_{\{1,3\}})(y) &= \Tr\left(\underline{\omega}(T(y))\left(\frac{\partial T_1}{\partial y_1}\frac{\partial T_2}{\partial y_3}-\frac{\partial T_2}{\partial y_1}\frac{\partial T_1}{\partial y_3}\right)\right) = \frac{1}{y_1y_3}\,,\\
		(\mathcal{I}_\ast \omega_{\{2,3\}})(y) &= \Tr\left(\underline{\omega}(T(y))\left(\frac{\partial T_1}{\partial y_2}\frac{\partial T_2}{\partial y_3}-\frac{\partial T_2}{\partial y_2}\frac{\partial T_1}{\partial y_3}\right)\right) = - \frac{1}{y_2y_3}\,,
	\end{align}
	in clear agreement with \eqref{eq:push-ex-master-coefficients}.
\end{example}

In many instances, \eqref{eq:push-companion-alt-trace-coefficients} is more efficient to evaluate than \eqref{eq:push-companion-trace-coefficients}. In general, $\underline{\omega}$ is significantly simpler than $\underline{\omega}_{I,J}$, allowing for quicker evaluation on companion matrices. Moreover, once $\underline{\omega}(T(y))$ is known, no further matrix inversions are needed for \eqref{eq:push-companion-alt-trace-coefficients}. The major disadvantage of \eqref{eq:push-companion-alt-trace-coefficients} is its dependence on companion matrix derivatives. If the $y$-dependence of companion matrices is known, then such derivatives are straightforwardly calculated. However, it is often intractable to determine the $y$-dependence of companion matrices; often it is only possible to compute companion matrices for fixed numerical values of $y$ (i.e.\ sampling over finite fields). To remedy this limitation, we present an algorithm for working out exact derivatives of companion matrices, suitable for numerical sampling.

Fix $j\in[n]$ and $y\in\mathcal{Y}_F$. Regarding $\frac{\partial x}{\partial y_j}\coloneqq(\frac{\partial x_1}{\partial y_j},\ldots,\frac{\partial x_m}{\partial y_j})$ as an $m$-tuple of formal expressions, let $\complex[x,\frac{\partial x}{\partial y_j}]$ denote the ring of complex polynomials in $x$ and $\frac{\partial x}{\partial y_j}$. Furthermore, let $\mathcal{I}(y)$ and $\mathcal{I}_j(y)$ denote the polynomial ideals
\begin{align}
	\mathcal{I}(y)\coloneqq\lrangle{F(\bullet)(y)}
	\subset\complex[x]\,,\qquad
	\mathcal{I}_j(y)\coloneqq\lrangle{F(\bullet)(y),\frac{dF}{dy_j}(\bullet)(y)}
	\subset\complex[x,\frac{\partial x}{y_j}]\,,
\end{align}
where $\frac{dF}{dy_j}\coloneqq(\frac{\partial f_1}{\partial y_j} + \frac{\partial f_1}{\partial x_i}\frac{\partial x_i}{\partial y_j},\ldots,\frac{\partial f_m}{\partial y_j} + \frac{\partial f_m}{\partial x_i}\frac{\partial x_i}{\partial y_j})$. Let $\mathcal{G}(y)$ (resp., $\mathcal{G}_j(y)$) denote the unique reduced Gr\"{o}bner basis for $\mathcal{I}(y)$ (resp., $\mathcal{I}_j(y)$) with respect to lex-order where ${\partial x_1}/{\partial y_j}\succ\cdots\succ{\partial x_n}/{\partial y_j}\succ x_1\succ\cdots\succ x_n$. We designate that standard basis for $\complex[x]/\mathcal{I}(y)$ (resp., $\complex[x,\frac{\partial x}{\partial y_j}]/\mathcal{I}_j(y)$) by $\mathcal{B}(y)$ (resp., $\mathcal{B}_j(y)$). Suppose $\mathcal{B}_j(y)=\mathcal{B}(y)=\{e_\alpha\}_{\alpha=1}^d$. Then the $i$\textsuperscript{th} companion matrix of $\mathcal{I}$ evaluated at $y$ is given by
\begin{align}\label{eq:push-companion-alt-cm-components}
	\overbar{x_i e_\alpha}^{\mathcal{G}_j(y)} = T_i(y)_{\alpha\beta}e_{\beta}\,,
\end{align}
and its partial derivative with respect to $y_j$, evaluated at $y$, can be extracted via
\begin{align}\label{eq:push-companion-alt-cm-pd-components}
	\overbar{\frac{\partial(x_i e_\alpha)}{\partial y_j} - T_i(y)_{\alpha\beta}\frac{\partial e_\beta}{\partial y_j}}^{\mathcal{G}_j(y)} = \frac{\partial T_i(y)_{\alpha\beta}}{\partial y_j}e_\beta\,.
\end{align}
On the left-hand side of \eqref{eq:push-companion-alt-cm-pd-components}, $x$-variables and standard basis elements are regarded as implicit functions of $y_j$. Importantly, the dividends on the left-hand sides of \eqref{eq:push-companion-alt-cm-components} and \eqref{eq:push-companion-alt-cm-pd-components} are polynomials in $\complex[x,\frac{\partial x}{\partial y_j}]$ while the right-hand sides are polynomials in $\complex[x]$.

The validity of \eqref{eq:push-companion-alt-cm-components} and \eqref{eq:push-companion-alt-cm-pd-components} can be understood as follows. Each generator in $\frac{dF}{dy_j}$ defines a non-constant linear function in $\frac{\partial x}{\partial y_j}$. Consequently, the $m$\textsuperscript{th} elimination ideal of $\mathcal{I}_j(y)$ is given by $\mathcal{I}_j(y)\cap \complex[x] = \mathcal{I}(y)$ and the Elimination Theorem (see \cref{thm:elimination}) implies that $\mathcal{G}_j(y)\cap \complex[x] = \mathcal{G}(y)$. Since $x_i e_\alpha$ is a monomial in $\complex[x]$, division by $\mathcal{G}_j(y)$ coincides with division by $\mathcal{G}_j(y)\cap \complex[x] = \mathcal{G}(y)$. Therefore, \eqref{eq:push-companion-alt-cm-components} computes the $i$\textsuperscript{th} companion matrix of $\mathcal{I}(y)$. Moreover, since division by $\mathcal{G}_j(y)$ is a continuous operation, differentiating both sides of \eqref{eq:push-companion-alt-cm-components} with respect to $y_j$ produces \eqref{eq:push-companion-alt-cm-pd-components}. 

\begin{example}[Derivatives of Companion Matrices via the Elimination Theorem]
	Recall that the companion matrices, calculated in \cref{ex:push-companion}, read
	\begin{align*}
		T_1(y)=
		\begin{bmatrix}
			0 & -\frac{y_2}{y_1 y_3} \\
			-\frac{y_2}{y_1} & 0 \\
		\end{bmatrix},\qquad
		T_2(y)=
		\begin{bmatrix}
			0 & \frac{1}{y_3} \\
			1 & 0 \\
		\end{bmatrix}.
	\end{align*}
	Their derivatives with respect to $y$-variables, shown in  \cref{ex:push-companion-alt}, are given by
	\begin{alignat*}{3}
		\frac{\partial T_1}{\partial y_1} &= \begin{bmatrix}
			0 & \frac{y_2}{y_1^2y_3} \\
			\frac{y_2}{y_1^2} & 0
		\end{bmatrix}, \qquad& \frac{\partial T_1}{\partial y_2} &= \begin{bmatrix}
			0 & -\frac{1}{y_1y_3} \\
			-\frac{1}{y_1} & 0
		\end{bmatrix}, \qquad& \frac{\partial T_1}{\partial y_3} &=
		\begin{bmatrix}
			0 & \frac{y_2}{y_1y_3^2} \\
			0 & 0
		\end{bmatrix},\\
		\frac{\partial T_2}{\partial y_1} &= 0_{2\times 2}, \qquad& \frac{\partial T_2}{\partial y_2} &=
		0_{2\times 2}, \qquad& \frac{\partial T_2}{\partial y_3} &=
		\begin{bmatrix}
			0 & -\frac{1}{y_3^2} \\
			0 & 0
		\end{bmatrix}.
	\end{alignat*}
	Let us re-derive these derivatives using \eqref{eq:push-companion-alt-cm-pd-components}, beginning with $y_1$. In this case, the relevant ideal is
	\begin{align}
	\mathcal{I}_1 = \lrangle{F,\frac{dF}{dy_1}} =  \left\langle y_1x_1+y_2x_2,
	y_3x_2^2-1,
	x_1+y_1\frac{\partial x_1}{\partial y_1}+y_2\frac{\partial x_2}{\partial y_1},
	2 y_3 x_2\frac{\partial x_2}{\partial y_1}\right\rangle.
	\end{align}
	Its Gr\"{o}bner basis with respect to lex-order is given by 
	\begin{align}
		\mathcal{G}_1=\left\{
		\frac{\partial x_1}{\partial y_1} - \frac{y_2}{y_1^2}x_2,
		\frac{\partial x_2}{\partial y_1},
		x_1+\frac{y_2}{y_1}x_2,
		x_2^2-\frac{1}{y_3}
		\right\}.
	\end{align}
	Notice that $\mathcal{G}_1\cap\complex[x_1,x_2]=\mathcal{G}$ where $\mathcal{G}$ is the Gr\"{o}bner basis for $\mathcal{I}$ from \cref{ex:push-companion}. The standard basis for $\complex[x_1,x_2,\frac{\partial x_1}{\partial y_1},\frac{\partial x_2}{\partial y_1}]/\mathcal{I}_1$ associated with $\mathcal{G}_1$ is $\mathcal{B}_1=\{x_2,1\}$, identical to the standard basis $\mathcal{B}$ used in \cref{ex:push-companion}. With $e\coloneqq(x_2,1)$,
	\begin{align}
		\overbar{\frac{\partial(x_1 e)}{\partial y_1} - T_1\cdot\frac{\partial e}{\partial y_1}}^{\mathcal{G}_1} 
= \begin{bmatrix}
		\frac{y_2}{y_1^2 y_3}\\
		\frac{y_2 x_2}{y_1^2}
	\end{bmatrix},\qquad
		\overbar{\frac{\partial(x_2 e)}{\partial y_j} - T_2\cdot\frac{\partial e}{\partial y_1}}^{\mathcal{G}_1} 
= \begin{bmatrix}
	0\\0
\end{bmatrix},
	\end{align}
	from which we can be read off
	\begin{align}
		\frac{\partial T_1}{\partial y_1}= \begin{bmatrix}
			0 & \frac{y_2}{y_1^2y_3} \\
			\frac{y_2}{y_1^2} & 0
		\end{bmatrix}, \qquad\frac{\partial T_2}{\partial y_1}= 0_{2\times 2}\,,
	\end{align}
	in clear agreement with \eqref{eq:ex-push-companion-alt-cm-pd}. Derivatives of the companion matrices with respect to other $y$-variables are similarly confirmed. 
\end{example}

\section{Pushforwards via Global Duality Theorem}
\label{sec:global-residue}
Our third method for evaluating \eqref{eq:push-master} does not use companion matrices, but uses the duality of global residues. This involves constructing a dualizing inner product on $\mathcal{Q}=\complex(y)[x]/\mathcal{I}$. Our method uses machinery from \cite{cattani2005introduction} applied to the CHY scattering equations in \cite{Sogaard:2015dba}. 

\subsection{Multi-Dimensional Residues}
Let us begin by defining local and global residues. Consider any polynomial $h\in\complex(y)[x]$. Let $\xi\in\zero(\mathcal{I})$. Then $\xi(y)\in\zero(\mathcal{I}(y))$ for generic $y$. Let $U(y)\subset\complex^m$ be an open neighbourhood of $\xi(y)$ which excludes all other points in $\zero(\mathcal{I}(y))$. The \emph{local residue} of $h(\bullet)(y)$ at $\xi(y)$ with respect to the divisors in $F(\bullet)(y)$ is the multi-dimensional contour integral \cite{cattani2005introduction}
\begin{align}\label{eq:push-gr-local-residue}
	\Res_{F(\bullet)(y),\xi(y)}(h(\bullet)(y))\coloneqq \frac{1}{(2\pi i)^m}\oint_{\Gamma_\delta(\xi(y))}h(x)(y)\frac{dx_1\wedge\cdots\wedge dx_m}{f_1(x)(y)\cdots f_m(x)(y)}\,.
\end{align}
Here $\delta\coloneqq(\delta_1,\ldots,\delta_m)\in\real_{>0}^m$ is an $m$-tuple of parameters such that
\begin{align}
	\Gamma_\delta(\xi(y))\coloneqq\left\{x\in U(y):|f_\ell(x)(y)|=\delta_\ell~\text{for all}~\ell\in[m]\right\},
\end{align}
is a smooth real $m$-cycle, oriented by the $m$-form $d\arg(f_1)\wedge\cdots\wedge d\arg(f_m)$. \cref{eq:push-gr-local-residue} naturally extends to a definition for $\Res_{F,\xi}(h)$, the local residue of $h$ at $\xi\in\zero(\mathcal{I})$ with respect to the divisors in $F$. For generic $y$-variables, $\mathcal{I}$ is radical and the Jacobian matrix $\frac{\partial F}{\partial x}$ is non-degenerate on $\zero(\mathcal{I})$. Using Stokes' theorem, $\Res_{F,\xi}(h)$ evaluates to
\begin{align}\label{eq:push-gr-local-residue-cauchy}
	\Res_{F,\xi}(h)=\left|\frac{\partial F}{\partial x}(\xi)\right|^{-1}h(\xi)\,.
\end{align}
The \emph{global residue} of $h$ with respect to the divisors in $F$ is the sum of all local residues  \cite{cattani2005introduction}
\begin{align}\label{eq:push-gr-global-residue}
	\Res_F(h)\coloneqq \sum_{\xi\in\zero(\mathcal{I})}\Res_{F,\xi}(h)\,.
\end{align}
By combining \eqref{eq:push-gr-global-residue} with \eqref{eq:push-gr-local-residue-cauchy}, the pushforward of $h$ via $\mathcal{I}$ can be expressed as
\begin{align}\label{eq:push-gr-pushforward-ideal}
	\mathcal{I}_\ast h \coloneqq \Res_F\left(\left|\frac{\partial F}{\partial x}\right|h\right).
\end{align}

\subsection{Global Duality Theorem}
We can evaluate \eqref{eq:push-gr-pushforward-ideal} by exploiting the Global Duality Theorem \cite{cattani2005introduction}. To this end, let $\mathcal{G}$ be a Gr\"{o}bner basis for $\mathcal{I}$ and let $\mathcal{B}=\{e_\alpha\}_{\alpha=1}^d$ be the associated standard basis. Then the normal form of $h$, denoted by $\overbar{h}^{\mathcal{G}}$, provides a canonical representative for $[h]=[\overbar{h}^{\mathcal{G}}]$ in $\mathcal{Q}$. On the support of $\zero(\mathcal{I})$, $h$ and $\overbar{h}^{\mathcal{G}}$ coincide:
\begin{align}
	h(\xi(y))(y) = \overbar{h}^{\mathcal{G}}(\xi(y))(y)~\text{for all}~\xi\in\zero(\mathcal{I})\,.
\end{align}
This observation motivates the definition of the following symmetric inner-product on $\mathcal{Q}$:
\begin{align}
	\lrangle{\bullet,\bullet}:\mathcal{Q}\times \mathcal{Q}\to\complex(y);([h_1],[h_2])\mapsto\lrangle{[h_1],[h_2]}\coloneqq\Res_F(h_1h_2)\,.
\end{align}
For generic $y$-variables, $\lrangle{\bullet,\bullet}$ is non-degenerate by the Global Duality Theorem (see \cref{thm:global-duality}). Non-degeneracy guarantees a \emph{dual basis} $\mathcal{B}^\vee=\{\Delta_\alpha\}_{\alpha=1}^d$ for $\mathcal{Q}$ satisfying
\begin{align}
	\lrangle{e_\alpha,\Delta_\beta}=\delta_{\alpha\beta}\,.
\end{align}
Since $1\in \mathcal{Q}$, it can be decomposed in the dual basis: 
\begin{align}
	1=\sum_{\alpha=1}^{d}\mu_\alpha\Delta_\alpha\,,
\end{align}
for some choice of rational function coefficients $\mu_1,\ldots,\mu_d\in\complex(y)$. Once these coefficients are known, the pushforward of $h$ via $\mathcal{I}$ can be calculated as 
\begin{align}\label{eq:push-gr-pushforward-ideal-sum-p}
	\mathcal{I}_\ast h 
	=\lrangle{1,\overbar{\left(\left|\frac{\partial F}{\partial x}\right|h\right)}^{\mathcal{G}}} = \sum_{\alpha=1}^d\mu_\alpha\overbar{\left(\left|\frac{\partial F}{\partial x}\right|h\right)}_\alpha,
\end{align}
where $\overbar{(|{\partial F}/{\partial x}|h)}_1,\ldots,\overbar{(|{\partial F}/{\partial x}|h)}_d\in\complex(y)$ are the coefficients of $\overbar{(|{\partial F}/{\partial x}|h)}^{\mathcal{G}}$ in the standard basis. 

\subsection{Bezoutian Matrix}
To decompose $1$ in the dual basis, we need to know the dual basis. So, in this subsection, we outline a procedure for determining the dual basis.

Let $\mathcal{G}$ be a Gr\"{o}bner basis for $\mathcal{I}$ with respect to any graded monomial ordering (grlex, grevlex, etc.) and let $\mathcal{B}=\{e_\alpha\}_{\alpha=1}^d$ be the associated standard basis. Let $\tilde{x}\coloneqq(\tilde{x}_1,\ldots,\tilde{x}_m)$ be an $m$-tuple of auxiliary variables. The \emph{Bezoutian matrix} $B$ for $\mathcal{I}$ is an $m\times m$ matrix with components given by 
\begin{align}\label{eq:push-bezoutian}
	B_{i,j}\coloneqq\frac{f_i(\tilde{x}_1,\ldots,\tilde{x}_{j-1},x_j,\ldots,x_m)-f_i(\tilde{x}_1,\ldots,\tilde{x}_j,x_{j+1},\ldots,x_m)}{x_j-\tilde{x}_j}\in\complex(y)[x,\tilde{x}]\,.
\end{align}
Let $\tilde{\mathcal{G}}$ denote the set obtained from $\mathcal{G}$ by replacing $x$ with $\tilde{x}$. The remainder of $\det(B)\in\complex(y)[x,\tilde{x}]$ on division by $\mathcal{G}\cup\tilde{\mathcal{G}}$ is given by \cite{cattani2005introduction}
\begin{align}\label{eq:push-bezoutian-determinant}
\overbar{\det(B)}^{\mathcal{G}\cup\tilde{\mathcal{G}}} = \sum_{\alpha=1}^{d}e_\alpha(x)\Delta_\alpha(\tilde{x})\,,
\end{align}
from which we can read off the dual basis elements $\mathcal{B}^\vee=\{\Delta_\alpha\}_{\alpha=1}^d$. The dual basis for our running example is calculated below.

\begin{example}[Dual Bases]\label{ex:push-dual-basis}
	From \cref{ex:push-companion}: $\mathcal{G}= \{x_1+\frac{y_2}{y_1}x_2,x_2^2-\frac{1}{y_3}\}$ and $\mathcal{B}=\{e_1,e_2\}=\{x_2,1\}$. Introducing auxiliary variables $\tilde{x}=(\tilde{x}_1,\tilde{x}_2)$, the Bezoutian matrix $B$ for $\mathcal{I}$, calculated according to \eqref{eq:push-bezoutian}, is given by 
	\begin{align}
		B=\begin{bmatrix}
			y_1 & y_2 \\
			0 & y_3(x_2+\tilde{x}_2)
		\end{bmatrix}.
	\end{align}
	The remainder of $\det B$ on division by $\mathcal{G}\cup\tilde{\mathcal{G}}$, expanded in the standard basis, is given by
	\begin{align}
		\overbar{\det B}^{\mathcal{G}\cup\tilde{\mathcal{G}}}=y_1y_3(x_2+\tilde{x}_2) = (y_1y_3)e_1+(y_1y_3\tilde{x}_2)e_2\,,
	\end{align}
    from which we read off the dual basis: $\mathcal{B}^\vee=\{\Delta_1,\Delta_2\}\coloneqq \{y_1y_3,y_1y_3x_2\}$. Since $\Delta_1=y_1y_3$ is independent of $x$, $1$ decomposes as
	\begin{align}
		1 = \mu_1\Delta_1+\mu_2\Delta_2=(y_1y_3\mu_2)x_2+(y_1y_3\mu_1)\,,
	\end{align}
	where $\mu_1(y)=\frac{1}{y_1y_3}$ and $\mu_2(y)=0$.
\end{example}

\subsection{Polynomial Inverses}

To extend our analysis to rational functions we need to find representatives for rational functions in $\mathcal{Q}$. This brings us to the topic of polynomial inverses. Suppose $q\in\complex(y)[x]$ is a polynomial with no zeros in common with $\mathcal{I}$, i.e.\ $\zero(\lrangle{q}+\mathcal{I})=\emptyset$. By Hilbert's Weak Nullstellensatz (see \cref{thm:nullstellensatz}), there exists polynomials $\tilde{f}_1,\ldots,\tilde{f}_m$ and $q_\text{inv}$ in $\complex(y)[x]$ such that
\begin{align}
	f_1\tilde{f}_1 + \ldots + f_m\tilde{f}_m + qq_\text{inv} = 1\,.
\end{align}
Here $q_\text{inv}$ (resp., $q$) is called the \emph{polynomial inverse} for $q$ (resp., $q_\text{inv}$). The polynomial inverse for $q$ can be determined via a Gr\"{o}bner basis computation; see \cite[Algortihm 2]{Sogaard:2015dba}. Consider the ideal $\mathcal{J}=\lrangle{F,wq-1}\in\complex(y)[w,x]$ where $w$ is a single auxiliary variable. Define a block monomial ordering $\succ$ which first compares the degrees of $w$, and breaks ties using any graded monomial ordering (grlex, grevlex, etc.) on $x$. Then the Gr\"{o}bner basis $\mathcal{G}_\succ(\mathcal{J})$ contains a polynomial of the form $w-q_\text{inv}(x)$ where $q_\text{inv}$ is the polynomial inverse for $q$ in $\mathcal{Q}$.

Let $r\in\complex(y)(x)$ be a rational function where $r(x)=\frac{h(x)}{q(x)}$ for some polynomials $h,q\in\complex(y)[x]$ and suppose $q$ has no zeros in common with the generators of $\mathcal{I}$. Then the pushforward of $r$ via $\mathcal{I}$ can be evaluated as 
\begin{align}\label{eq:push-gr-pushforward-ideal-sum-r}
	\mathcal{I}_\ast r 
	=\lrangle{1,\overbar{\left(\left|\frac{\partial F}{\partial x}\right|hq_\text{inv}\right)}^{\mathcal{G}}} = \sum_{\alpha=1}^d\mu_\alpha\overbar{\left(\left|\frac{\partial F}{\partial x}\right|hq_\text{inv}\right)}_\alpha.
\end{align}
where $\overbar{(|\frac{\partial F}{\partial x}|hq_\text{inv})}_1,\ldots,\overbar{(|\frac{\partial F}{\partial x}|hq_\text{inv})}_d\in\complex(y)$ are the coefficients of $\overbar{(|\frac{\partial F}{\partial x}|hq_\text{inv})}^{\mathcal{G}}$ in the standard basis. 

\begin{example}[Polynomial Inverses]\label{ex:push-polynomial-inverses}
	Let $\mathcal{I}$ be the polynomial ideal defined in \cref{ex:push-formal}. Suppose we want to know the polynomial inverse of $q(x)=x_1x_2$ in $\mathcal{Q}=\complex(y_1,y_2,y_3)[x_1,x_2]/\mathcal{I}$ where $q(x)$ is the denominator of the rational function $\underline{\omega}(x)=\frac{1}{x_1x_2}$. Consider the ideal  $\mathcal{J}=\lrangle{f_1,f_2,wq-1}\subset\complex(y_1,y_2,y_3)[w,x_1,x_2]$. Let $\succ$ be the block order that first compares the degrees of $w$ and breaks ties using grevlex order on $x$. Then the Gr\"{o}bner basis $\mathcal{G}_\succ(\mathcal{J})$ contains the polynomial $w+\frac{y_1y_3}{y_2}$. Hence $q_\text{inv}(x)=-\frac{y_1y_3}{y_2}$ is the polynomial inverse of $q(x)$ in $\mathcal{Q}$.
\end{example}

\subsection{Pushforwards} 
\cref{eq:push-gr-pushforward-ideal-sum-r} provides an alternative to companion matrix methods for pushing forward $\underline{\omega}_{I,J}$ via $\mathcal{I}$. There are some simplifications that occur for top-dimensional rational forms which are worth noting. To this end, let $p=m$ and define $\underline{\omega}_J\coloneqq\underline{\omega}_{[m],J}$ where $J\in\binom{[n]}{m}$. Then using \eqref{eq:push-master-coefficients}, 
\begin{align}\label{eq:push-master-coefficients-p=m}
	\left|\frac{\partial F}{\partial x}\right|\underline{\omega}_J = (-1)^m\underline{\omega}(x)\left|\frac{\partial F}{\partial y}\right|_{[m],J},
\end{align}
where $|\partial F/\partial y|_{[m],J}$ denotes the minor of $\partial F/\partial y$ in the column set $J$ and the determinant $|\partial F/\partial x|$ disappears in the final result. Let $\underline{\omega}(x)=h(x)/q(x)$ where $h$ and $q$ are polynomials. Assume $q$ has no zeros in common with $\mathcal{I}$, i.e.\ $\zero(\lrangle{q}+\mathcal{I})=\emptyset$. By combining \eqref{eq:push-master-coefficients-p=m} with \eqref{eq:push-gr-pushforward-ideal-sum-r}, we have the following formula:
\begin{align}\label{eq:push-gr-pushforward-ideal-sum-top}
	\mathcal{I}_\ast\underline{\omega}_{J}=  (-1)^m\sum_{\alpha=1}^{d}\mu_\alpha\overbar{\left(\left|\frac{\partial F}{\partial y}\right|_J hq_\text{inv}\right)}_\alpha.
\end{align}

\begin{example}[Pushforwards via the Global Duality Theorem]\label{ex:push-gr}
	Let us re-derive \eqref{eq:push-ex-master-coefficients-pushforward} using \eqref{eq:push-gr-pushforward-ideal-sum-top}. The relevant Jacobian matrices from \cref{ex:push-formal} read
		\begin{align*}
		\frac{\partial F}{\partial x}=
		\begin{bmatrix}
			y_1 & y_2\\
			0 & 2 y_3x_2
		\end{bmatrix},
		\qquad
		\frac{\partial F}{\partial y}=
		\begin{bmatrix}
			x_1 & x_2 & 0\\
			0 & 0 & x_2^2
		\end{bmatrix}.
	\end{align*}
	In \cref{ex:push-dual-basis}, we showed that $\mu_1(y)=\frac{1}{y_1y_3}$ and $\mu_2(y)=0$, while in \cref{ex:push-polynomial-inverses}, we found that $q_\text{inv}(x)=-\frac{y_1y_3}{y_2}$ is the polynomial inverse of $q(x)=x_1x_2$. Using \eqref{eq:push-gr-pushforward-ideal-sum-top} we obtain
	\begin{align}
		(\mathcal{I}_\ast\underline{\omega}_{\{1,2\}})(y) &=\mu_1\overbar{\left(\left|\frac{\partial F}{\partial y}\right|_{\{1,2\}}q_\text{inv}\right)}_1 =-\frac{1}{y_2}\overbar{(0)}_1=0\,,\\
		(\mathcal{I}_\ast\underline{\omega}_{\{1,3\}})(y) & =\mu_1\overbar{\left(\left|\frac{\partial F}{\partial y}\right|_{\{1,3\}}q_\text{inv}\right)}_1 =-\frac{1}{y_2}\overbar{(x_1x_2^2)}_1 = \frac{1}{y_1y_3}\,,\\
		(\mathcal{I}_\ast\underline{\omega}_{\{2,3\}})(y) & =\mu_1\overbar{\left(\left|\frac{\partial F}{\partial y}\right|_{\{2,3\}}q_\text{inv}\right)}_1 =-\frac{1}{y_2}\overbar{(x_2^3)}_1 = -\frac{1}{y_2y_3}\,,
	\end{align}
	in clear agreement with \eqref{eq:push-ex-master-coefficients}.
\end{example}

\section{Summary}
We have shown three methods for pushing forward rational forms via ideals. These methods use elements of computational algebraic geometry and sidestep the problem of determining zero-sets. Reference \cite{Lukowski:2022fwz} discusses the relative pros and cons of each approach. Together these methods provide novel conceptual and practical resources for calculating pushforwards. 

This work's motivation comes from its application to positive geometries, specifically for pushing forward canonical forms via morphisms. The results of this chapter apply to canonical forms of old positive geoemtries (see \cref{chp:abhy,chp:mom,chp:omom}) and new positive geometries, including the conjectured ``symplectic Momentum Amplituhedron'' of \cite{He:2021llb}. \cref{chp:meet} gives further details regarding this conjectured positive geometry.

Lastly, all three methods require a Gr\"{o}bner basis computation which poses a significant bottleneck for high-multiplicity examples or when working over fraction fields (as opposed to sampling over finite fields). Improving the efficiency of Gr\"{o}bner basis computations for fraction fields would significantly strengthen our results. We leave this problem to future research. It is also worthwhile investigating alternatives which do not depend on Gr\"{o}bner bases.

	\chapter{The ABHY Associahedron}
\label{chp:abhy}

\lettrine{T}{he ABHY associahedron} is a positive geometry that captures the tree-level S-matrix of bi-adjoint scalar theory. It is the simplest of the three positive geometries studied in this dissertation. Its boundaries are in bijection with series-reduced planar trees on $n$-leaves and capture all physical singularities of the corresponding S-matrix. The ABHY associahedron relates to the worldsheet associahedron via the CHY scattering equations. This connection between the CHY formalism and positive geometries is the first of three examples explored in this dissertation. This chapter introduces the ABHY associahedron, the worldsheet associahedron, and the CHY scattering equations. These objects are also relevant for \cref{chp:mom,chp:omom,chp:meet}. This chapter also applies the results of \cref{chp:enum} to determine the rank generating function for the boundaries of the ABHY associahedron. While this generating function is known, its derivation serves as a warm-up exercise for the more complicated generating functions derived in \cref{chp:mom,chp:omom}. In \cref{chp:meet}, we demonstrate a web of relationships between the canonical form of the ABHY associahedron (restricted to particular spacetime dimensions) and the canonical forms of the Momentum Amplituhedron and orthogonal Momentum Amplituhedron. 

\section{The ABHY Associahedron}
\label{sec:abhy-abhy}
The \emph{Arkani-Hamed--Bai--He--Yan (ABHY) associahedron} is a positive geometry whose canonical form encodes tree-level amplitudes in general $d$-dimensions for \emph{bi-adjoint scalar (BAS) theory} \cite{Arkani-Hamed:2017mur}. First introduced in \cite{Cachazo:2013iea}, BAS is a theory of massless scalars, transforming in the adjoint representation of the product of two different colour groups $\U{N}\times\U{\tilde{N}}$, and it contains a single (cubic) interaction term. The $n$-particle tree-level amplitude $m_n^\text{tree}$ can be decomposed into double partial amplitudes $m_n^\text{tree}[\alpha|\beta]$ where $\alpha,\beta$ are two colour orderings (permutations of $[n]$). In this chapter, we specialise to the case where $\alpha=(1,\ldots,n)=\beta$ and we define $m_n^\text{tree}[1,\ldots,n]\coloneqq m_n^\text{tree}[1,\ldots,n|1,\ldots,n]$. This double partial amplitude receives contributions from all $n$-particle planar tree Feynman diagrams (i.e.\ series-reduced planar trees on $n$ leaves with trivalent internal vertices), where each diagram contributes the product of its propagators (the Mandelstam variables assigned to each internal edge). Therefore, $m_n^\text{tree}[1,\ldots,n]$ is a rational function of $n$-particle Mandelstam variables.  

\begin{figure}
	\centering
\begin{align*}
	\vcenter{\hbox{\includegraphics[scale=0.45]{enum-trees}}}
	\iff
	\vcenter{\hbox{\includegraphics[scale=0.45]{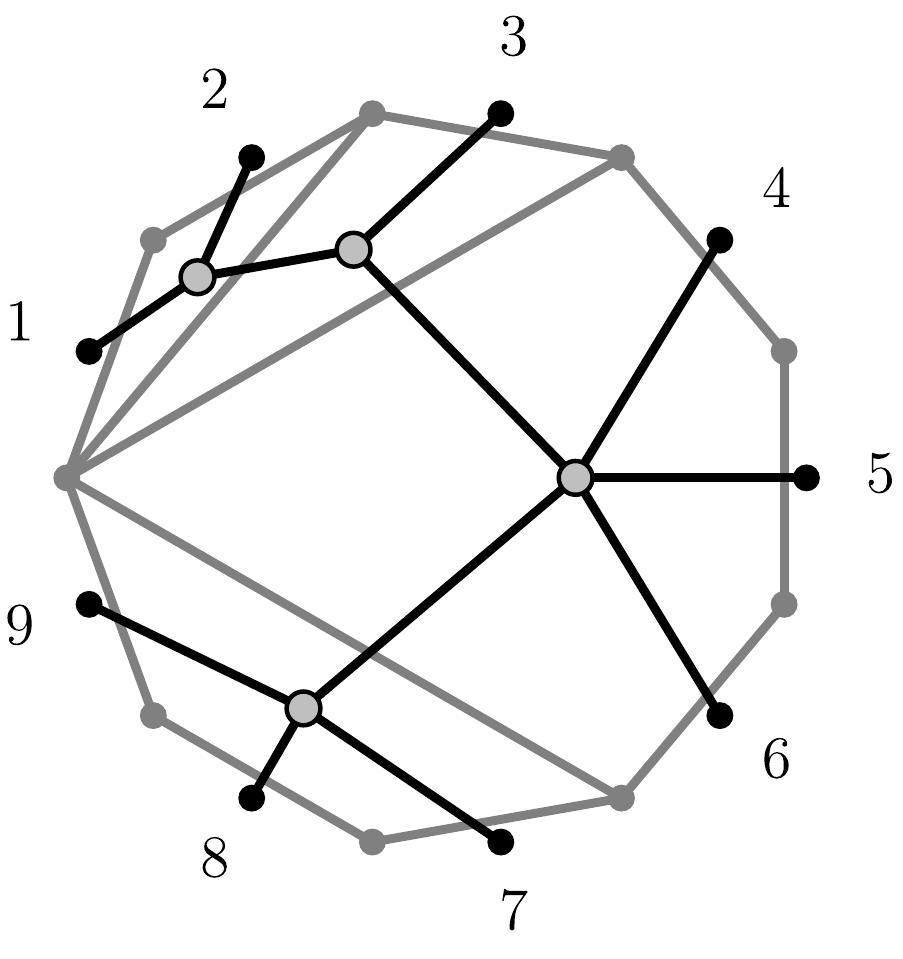}}}
\end{align*}
	\caption{Bijection between series-reduced planar trees on $n$ leaves and dissections of a regular $n$-gon. The labelling of leaves on the left corresponds to the labelling of polygon edges on the right.}
	\label{fig:abhy-planar}
\end{figure}

The kinematic space of BAS is the space of (complex) Mandelstam variables for $n$-particle scattering, which we denote by $\Mandelstam{n}$. It is spanned by the two-particle Mandelstam variables $s_{i,j}\coloneqq(p_i+p_j)^2$ subject to $n$ constraints imposed by momentum conservation:
$\sum_{j\ne i}s_{i,j}=0$ for all $i\in[n]$. Consequently, it has dimension
\begin{align}\label{eq:abhy-kin}
	\dim(\Mandelstam{n}) = \binom{n}{2}-n = \frac{n(n-3)}{2}\,,
\end{align}
for general $d$-dimensions. For $d<n-1$, there are additional constraints which we consider in \cref{chp:meet}. This chapter assumes $d\ge n-1$. A natural basis for $\Mandelstam{n}$ is given by the \emph{planar Mandelstam variables}, multi-particle Mandelstam variables defined by the momenta of consecutive particles:
\begin{align}\label{eq:abhy-planar}
	X_{i,j}\coloneqq s_{i,i+1,\ldots,j-1} = (p_i+p_{i+1}+\ldots+p_{j-1})^2\,,
\end{align}
where $1\le i<j\le n-\delta_{1,i}$. Given a regular $n$-gon with vertices labelled $\{1,2,\ldots,n\}$ in clockwise order, one can visualise $X_{i,j}$ as the diagonal between vertices $i$ and $j$ as depicted in \cref{fig:abhy-planar}. Let $\setMandelstam{n}$ denote the set of all planar Mandelstam variables for $n$-particle scattering.

The ABHY associahedron is defined as the intersection of two spaces: an affine hyperplane and a non-negative region. The affine hyperplane $H_n\subset\Mandelstam{n}$ is defined by
\begin{align}\label{eq:abhy-H}
	X_{i,j}+X_{i+1,j+1}-X_{i,j+1}-X_{i+1,j}=c_{i,j}~\text{for all}~1\le i<j\le n-1\,,
\end{align}
where $c_{i,j}$ are positive constants. Equivalently, \eqref{eq:abhy-H} requires $s_{i,j}=-c_{i,j}$ to be a negative constant for $1\le i<j\le n-1$. Counting the number of constraints, we find that the dimension of $H_n$ is
\begin{align}\label{eq:abhy-H-dim}
	\dim(H_n) = \dim(\Mandelstam{n})-\binom{n-2}{2} = n-3\,.
\end{align}
The non-negative region $\Delta_n\subset\Mandelstam{n}(\real)$ is the semi-algebraic set cut out by 
\begin{align}\label{eq:abhy-Delta}
X_{i,j}\ge 0~\text{for}~1\le i<j\le n-\delta_{1,i}\,.
\end{align}
The ABHY associahedron is the intersection $\Abhy{n}\coloneqq H_n\cap\Delta_n$. It is an $(n-3)$-dimensional \emph{associahedron} or \emph{Stasheff polytope} \cite{10.2307/1993608,10.2307/1993609}. Its boundaries are enumerated by series-reduced planar trees on $n$ leaves. The $0$-dimensional boundaries correspond to $n$-particle planar tree Feynman diagrams where every internal vertex is trivalent.

The \emph{planar scattering form} of rank $n-3$ is defined as 
\begin{align}\label{eq:abhy-omega}
	\omega_n^\text{ABHY}\coloneqq\sum_{T}\sign(T)\bigwedge_{a=1}^{n}d\log X_{i_a,j_a}\,,
\end{align}
where the sum is over all $n$-particle planar tree Feynman diagrams and $\sign(T)\in\{\pm1\}$ is chosen separately for each $T$ such that $\omega_n$ is projectively invariant. The pair $(H_n,\Abhy{n})$ is a positive geometry whose canonical form $\Omega(\Abhy{n})$ is the pullback of $\omega_n^\text{ABHY}$ to $H_n$. Moreover
\begin{align}\label{eq:abhy-Omega}
\Omega(\Abhy{n})=\underline{\Omega}(\Abhy{n})d^{n-3}X\,,\qquad\underline{\Omega}(\Abhy{n})=m_n^\text{tree}[1,\ldots,n]\,,
\end{align}
where $d^{n-3}X=dX_{1,3}\wedge\cdots\wedge dX_{1,n-1}$ and $\underline{\Omega}(\Abhy{n})$, the \emph{canonical function} of $\Abhy{n}$, is precisely the double partial amplitude $m_n^\text{tree}[1,\ldots,n]$. We refer the reader to \cite{Arkani-Hamed:2017mur} for further details.

\begin{figure}
	\centering
	\includegraphics[width=0.5\textwidth]{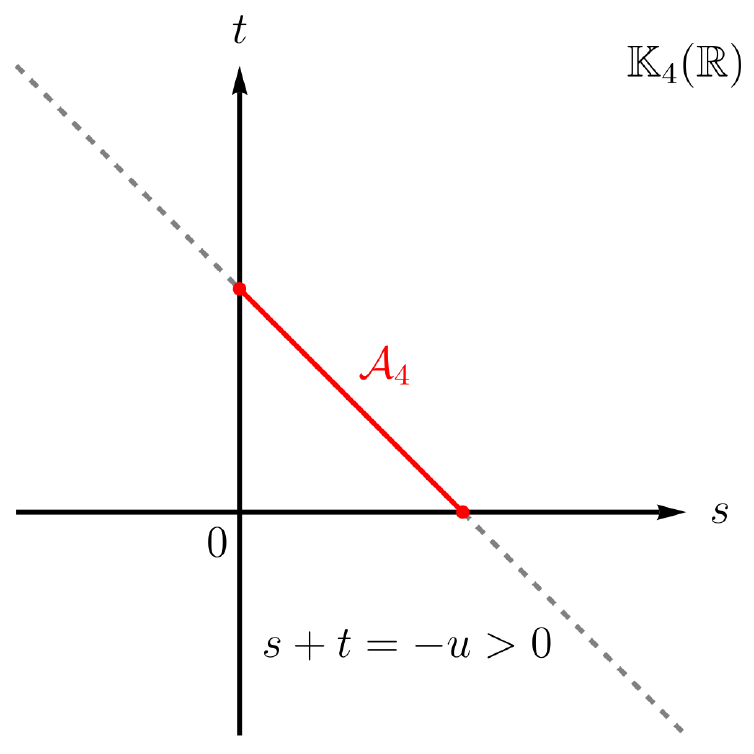}
	\caption{The ABHY associahedron $\Abhy{4}$ for four-particle scattering as a red line segment.}
	\label{fig:abhy-4}
\end{figure}

To clarify the above discussion, let us consider the simplest possible example, namely four-particle scattering. In this case $\Mandelstam{4}$ is two-dimensional and spanned by the Mandelstams $X_{1,3}=s=(p_1+p_2)^2$ and $X_{2,4}= t=(p_2+p_3)^2$. The affine hyperplane $H_4\subset\Mandelstam{4}$ is the line $s+t=-u$ where $u$ is a negative constant. Intersecting $H_4$ with the non-negative region $\Delta_4\subset\Mandelstam{4}(\real)$ given by $s,t\ge0$ produces $\Abhy{4}$, a line segment, as depicted in \cref{fig:abhy-4}.	The four-particle planar scattering form is given by
\begin{align}
	\omega_4^\text{ABHY} = d\log\frac st = \frac{ds}s - \frac{dt}t\,.
\end{align}
Pulling $\omega_4^\text{ABHY}$ back to $H_4$ (i.e.\ writing $t =-s-u$ where $u$ is a negative constant) yields the canonical form
\begin{align}
	\Omega(\Abhy{4}) = \frac{ds}s - \frac{d(-s-u)}{t} = \left(\frac 1s + \frac 1t\right)ds\,,
\end{align}
from which we read off the four-particle double partial amplitude $m_4^\text{tree}[1,2,3,4]=\frac 1s + \frac 1t$.

\section{The Worldsheet Associahedron}
\label{sec:abhy-world}

The \emph{genus zero moduli space} $\Moduli{0}{n}$ is the configuration space of $n$ distinct punctures on the Riemann sphere $\CP^1$ modulo $\SL{2}{\complex}$. Its real part $\ModuliReal{0}{n}$ is the \emph{open string moduli space}, consisting of $n$ distinct punctures on the real projective line $\RP^1$ modulo $\SL{2}{\real}$. The \emph{positive moduli space} is the positive part of $\ModuliReal{0}{n}$ given by
\begin{align}\label{eq:abhy-positive}
	\tpModuli{0}{n}\coloneqq\{(\sigma_1,\ldots,\sigma_n)\in\real^n:\sigma_1<\ldots<\sigma_n\}/\SL{2}{\real}\,.
\end{align}
Let $\ModuliCompact{0}{n}$ denote the well-known \emph{Deligne-Knudsen-Mumford compactification} of $\ModuliReal{0}{n}$ \cite{deligne1969irreducibility}. The \emph{worldsheet associahedron} $\tnnModuli{0}{n}$  is an intermediate partial compactification satisfying \cite{Arkani-Hamed:2019mrd}
\begin{align}\label{eq:abhy-world}
	\ModuliReal{0}{n}\subsetneq\tnnModuli{0}{n}\subsetneq\ModuliCompact{0}{n}\,,
\end{align}
whose boundary stratification is combinatorially equivalent to $\Abhy{n}$ \cite{Brown:2009qja}. The pair $(\Moduli{0}{n},\allowbreak\tnnModuli{0}{n})$ is a positive geometry \cite{Arkani-Hamed:2017mur} whose canonical form $\Omega(\tnnModuli{0}{n})$ is the \emph{Parke-Taylor form}, given by 
\begin{align}\label{eq:abhy-world-Omega}
	\omega_n^\text{WS}\coloneqq\frac{1}{\vol(\SL{2}{\real})}\bigwedge_{i=1}^n\frac{d\sigma_i}{\sigma_{i,i+1}}\,,
\end{align}
where $\sigma_{i,j}\coloneqq\sigma_{i}-\sigma_{j}$. 
One can ``gauge-fix'' the $\SL{2}{\real}$ redundancy in the definition of $\tpModuli{0}{n}$ by fixing the values of any three punctures. In the standard approach where $(\sigma_1,\sigma_{n-1},\sigma_n)=(0,1,\infty)$, the Parke-Taylor form reads
\begin{align}
	\omega_n^\text{WS}=-\frac{d\sigma_2\wedge\cdots\wedge d\sigma_{n-2}}{\sigma_{1,2}\sigma_{2,3}\cdots\sigma_{n-2,n-1}}\,.
\end{align}
Alternatively, $\tpModuli{0}{n}$ can be regarded as the \emph{positive configuration space} $\tpConf{2}{n}$ \cite{Arkani-Hamed:2020cig} --- the positive Grassmannian $\tpG{2}{n}$ modded out by the torus action $\real_{>0}^n$ --- parametrised by matrices of the form
\begin{align}\label{eq:abhy-tpModuli-0-n}
	\begin{pmatrix}
		1 & 1 & \cdots & 1\\
		\sigma_1 & \sigma_2 & \cdots & \sigma_n
	\end{pmatrix},
\end{align}
where the punctures $\sigma=(\sigma_1,\ldots,\sigma_n)\in\real^n$ are ordered according to $\sigma_1<\ldots<\sigma_n$. In this case, the Parke-Taylor form can be expressed as
\begin{align}
	\omega_n^\text{WS}=\frac{1}{\vol(\SL{2}{\real}\times\GL{1}{\real}^n)}\bigwedge_{i=1}^n\frac{d^2x_i}{(ii+1)}\,,
\end{align}
where $X$ is a $2\times n$ real matrix representing a point in $\tpG{2}{n}$, the columns of $X$ are labelled by $x_1,\ldots,x_n\in\real^2$, and $(ii+1)$ denotes the maximal minor of $X$ located in the column set $\{i,i+1\}$.

\section{The Scattering Equations}
\label{sec:abhy-SE}

The \emph{scattering equations} govern the scattering of $n$ massless particles, connecting the kinematic space $\Mandelstam{n}$ to the genus zero moduli space $\Moduli{0}{n}$. They were originally presented as a system of rational equations \cite{Cachazo:2013gna}
\begin{align}\label{eq:abhy-SE-rational}
	E_i\coloneqq\sum_{j\ne i}\frac{s_{i,j}}{\sigma_{i,j}}=0~\text{for}~1\le i \le n\,.
\end{align}
Due to momentum conservation $\sum_{j\ne i}s_{i,j}=0$, the scattering equations (of which there are $n$) are M\"{o}bius invariant, so only $n-3$ of them are linearly independent. 

The CHY formalism expresses the double partial amplitude $m_n^\text{tree}[1,\ldots,n]$ as an integral over $\Moduli{0}{n}$  which localises on solutions to the scattering equations \cite{Cachazo:2013iea}:
\begin{align}\label{eq:abhy-chy}
	m_n^\text{tree}[1,\ldots,n]\coloneqq \int \omega_n^\text{WS}\left(\frac{1}{\prod_{i=1}^n\sigma_{i,i+1}}\prod_{\ell=1}^n{'}\delta(E_\ell)\right),
\end{align}
where 
\begin{align}
	\prod_{\ell=1}^n{'}\delta(E_\ell)\coloneqq \sigma_{i,j}\sigma_{j,k}\sigma_{k,i}\prod_{\ell\in[n]\setminus\{i,j,k\}}\delta(E_\ell)\,,
\end{align}
is independent of the choice of $i,j,k$ \cite{Cachazo:2013gna}.

In \Cite{Dolan:2015iln}, Dolan and Goddard recast the scattering equations as a 
system of polynomial equations
\begin{align}\label{eq:abhy-SE-polynomial}
h_\ell\coloneqq\sum_{I\in\binom{[n]}{\ell+1}}s_I\sigma_I=0~\text{for}~1\le\ell\le n-3\,,
\end{align}
where $\sigma_I\coloneqq\prod_{i\in I}\sigma_i$. Applying the standard gauge where $(\sigma_1,\sigma_{n-1},\sigma_n)=(0,1,\infty)$, \eqref{eq:abhy-SE-polynomial} is equivalent to
\begin{align}
	f_\ell\coloneqq\lim_{\sigma_1\to0}\lim_{\sigma_{n-1}\to1}\lim_{\sigma_n\to\infty}\frac{h_\ell}{\sigma_n} = 0~\text{for}~1\le\ell\le n-3\,,
\end{align}
a system of $n-3$ polynomial equations for $n-3$ punctures. In this form, the scattering equations define a zero-dimensional complete intersection ideal $\mathcal{I}_n\coloneqq\lrangle{f_1,\ldots,f_{n}}\subset\complex[\setMandelstam{n}][\sigma_2,\ldots,\sigma_{n-2}]$ whose zero-set contains $(n-3)!$ distinct points in $\Moduli{0}{n}$ \cite{Cachazo:2013gna,Dolan:2015iln}. 

Remarkably, the interiors of the ABHY associahedron and the worldsheet associahedron are diffeomorphic via the scattering equations \cite{Arkani-Hamed:2017mur}. Evaluating the scattering equations on the affine hyperplane $H_n$ defines the \emph{scattering equation map} $\Moduli{0}{n}\to H_n$ which (conjecturally) restricts to a diffeomorphism $\tpModuli{0}{n}\to \tpAbhy{n}$ \cite{Arkani-Hamed:2017mur}. Consequently, the scattering equations define a morphism between the positive geometries $(\Moduli{0}{n},\tnnModuli{0}{n})$ and $(H_n,\Abhy{n})$. Moreover, the pushforward of $\omega_n^\text{WS}$ via the scattering equation map equals $\omega_n^\text{ABHY}$ \cite{Arkani-Hamed:2017mur}:
\begin{align}\label{eq:abhy-SE-push}
	(\mathcal{I}_n)_\ast\omega_n^\text{WS} \coloneqq \sum_{\sigma\in\zero(\mathcal{I}_n)}\omega_n^\text{WS}(\sigma)= \omega_n^\text{ABHY}\,.
\end{align}
Using methods discussed in \cref{chp:push}, one can evaluate \eqref{eq:abhy-SE-push} without needing to solve the scattering equations explicitly.

\section{Rank Generating Function}

We conclude this chapter with a derivation of a rank generating function for the boundaries of the ABHY associahedron. Although the result is already known, its derivation serves as a warm-up exercise for the generating functions encountered in \cref{chp:mom,chp:omom}.

Recall that the boundaries of the $(n-3)$-dimensional Stasheff polytope are enumerated by series-reduced planar trees on $n$ leaves. The dimension of a boundary can be calculated as a function of corresponding tree as follows. Let $T$ be a series-reduced planar tree on $n$ leaves. We denote by $\Vertices_\text{int}(T)$ the set of internal vertices of $T$. The \emph{associahedron dimension} or \emph{$\symbolAbhy$-dimension} of $T$, denoted by $\dim_{\symbolAbhy}(T)$, is defined as
\begin{align}\label{eq:abhy-dim-T}
	\dim_{\symbolAbhy}(T)\coloneqq n-2 - |\Vertices_\text{int}(T)| = \sum_{v\in\Vertices_\text{int}(T)}(\deg(v)-3)\,,
\end{align}
where
\begin{align}\label{eq:abhy-n-T}
	n = 2 + \sum_{v\in\Vertices_\text{int}(T)}(\deg(v)-2)\,.
\end{align}
In particular, $\dim_{\symbolAbhy}(T)$ depends only on the degrees of the internal vertices of $T$. Consequently, we can use the series-reduced planar tree analogue of the Exponential formula (see \cref{thm:enum-tree}) to derive a rank generating function for the ABHY associahedron $\Abhy{n}$. To this end, let $r$ be the $\symbolAbhy$-dimension of some $T\in\Tree_n$ and let $q$ be an auxiliary variable for keeping track of $r$ in our rank generating function. Using \eqref{eq:abhy-dim-T}, $q^r$ is given by
\begin{align}\label{eq:abhy-q-r}
	q^r = \prod_{v\in\Vertices_\text{int}(T)}q^{\deg(v)-3}\,.
\end{align}
Comparing the right-hand side of \eqref{eq:abhy-q-r} with the definition for $h(n)$ in \eqref{eq:enum-tree-h}, we are motivated to define the following function $f:\integer_{\ge3}\to\rational(q), d\mapsto q^{d-3}$. Let $h:\integer_{\ge3}\to\rational(y)$ be defined in terms of $f$ as in \cref{thm:enum-tree}. Furthermore, let $F(x)=\sum_{d\ge3}f(d)x^d$ and $H_{\tree}(x)=x^2+\sum_{n\ge 3}h(n)x^n$. It is easy to verify that $F(x)$ is given by
\begin{align}
	F(x) = \frac{x^3}{1-xq}\,.
\end{align}
Let $\mathcal{G}_{\symbolAbhy}(x,q)$ denote the rank generating function $H_{\tree}(x)$, computed in terms of $F(x)$ according to \cref{thm:enum-tree}. By solving the quadratic equation satisfied by the inverse of $x-\frac{1}{x}F(x)$ with respect to $x$, we find the following expression for $\mathcal{G}_{\symbolAbhy}(x,q)$:
\begin{align}\label{eq:abhy-GF-T}
\mathcal{G}_{\symbolAbhy}(x,q) = \frac{x}{2 (q+1)} \left(1+xq-\sqrt{(1-xq)^2-4 x}\right).
\end{align}
Therefore, the number of series-reduced planar trees on $n$ leaves with $\symbolAbhy$-dimension $r$ is equal to the coefficient $[x^nq^r]\mathcal{G}_{\symbolAbhy}(x,q)$. \cref{eq:abhy-GF-T} implies that $\mathcal{G}_{\symbolAbhy}(x,q)$ is an algebraic generating function of degree $2$, satisfying the following quadratic equation
\begin{align}
	(1+q)\mathcal{G}_{\symbolAbhy}^2-x(1+xq)\mathcal{G}_{\symbolAbhy}+x^3=0\,.
\end{align}

Having derived a rank generating function for $\Abhy{n}$, we can verify that its Euler characteristic, the coefficient of $x^n$ in $\mathcal{G}_{\symbolAbhy}(x,-1)$, is one. Assuming $1-x>0$, we can evaluate \eqref{eq:abhy-GF-T} at $q=-1$ using L'H\^{o}pital's rule
\begin{align}
	\mathcal{G}_{\symbolAbhy}(x,-1) = \frac{x}{2}\lim\limits_{q\to-1}\frac{\partial}{\partial q}\left(1+xq-\sqrt{(1-xq)^2-4 x}\right) = \frac{x^3}{1-x} = \sum_{3\le n}x^n\,,
\end{align}
and every coefficient is equal to one.

\section{Summary}
While our focus in this chapter was the ABHY associahedron, there are three themes which will carry over to \cref{chp:mom,chp:omom}, which we highlight now. Firstly, the ABHY associahedron is a positive geometry defined in the on-shell momentum space relevant to BAS. In particular, simple inequalities cut it out without any reference to BAS. Secondly, this semi-algebraic set is the image of the worldsheet associahedron via the CHY scattering equations. Said differently, it is the image of a positive geometry – the most obvious choice – living in the CHY moduli space for BAS. This connection suggests that the CHY formalism for other theories provides a natural point of departure for identifying (suitable generalisations of) positive geometries in that theory’s appropriate kinematic space. Thirdly, the boundary stratification captures the combinatorics of tree-level amplitude singularities for BAS. Series-reduced planar trees on $n$ leaves are well-known labels for associahedron boundaries, so this last point is not as pertinent here as it will be in \cref{chp:mom,chp:omom}. However, these combinatorial labels allowed us to determine an algebraic rank generating function using the results of \cref{chp:enum}. Additionally, we were able to prove that the associahedron has an Euler characteristic of one, as expected. In \cref{chp:mom,chp:omom}, we will again see how, with suitable combinatorial labels for boundaries, we can derive algebraic rank generating functions for the Momentum Amplituhedron and the orthogonal Momentum Amplituhedron.

	\chapter{The Momentum Amplituhedron}
\label{chp:mom}

\lettrine{T}{he Momentum Amplituhedron} is this chapter’s focus. It is (conjecturally) a positive geometry whose canonical form captures the tree-level $S$-matrix of $\mathcal{N}=4$ SYM. Unlike the Amplituhedron, which uses momentum twistors \cite{Arkani-Hamed:2013jha,Arkani-Hamed:2017vfh}, the Momentum Amplituhedron encodes amplitudes using spinor-helicity variables \cite{Damgaard:2019ztj}. It is rich with physical information and mathematical detail. In this chapter, we give special attention to the combinatorial structure of its boundaries, highlighting the results of \cite{Ferro:2020lgp,Moerman:2021cjg}. The Momentum Amplituhedron’s boundaries encode all singularities of the tree-level S-matrix for SYM. Remarkably, the boundary stratification forms an induced subposet of the positroid stratification. Recall that the positroid stratification is the boundary stratification of the non-negative Grassmannian. In particular, Momentum Amplituhedron boundaries are labelled by contracted Grassmannian forests, a special subset of a Grassmannian graphs. Using this labelling, we derive an algebraic rank generating function from which we prove that the Momentum Amplituhedron’s Euler characteristic is one. In addition to studying the Momentum Amplituhedron’s boundaries, we also review its connection to the worldsheet associahedron as investigated in \cite{He:2021llb}. In this case, the Momentum Amplituhedron is the image of a positive geometry in the appropriate CHY moduli space via the four-dimensional scattering equations. 

\section{Maximally Supersymmetric Yang-Mills Theory}
\label{sec:mom-theory}

To maintain manifest locality and unitarity, the standard description of quantum field theory introduces off-shell redundancy, causing computational complexity to cascade. This drawback of Feynman diagrams obscures the inherent simplicity and structure of the final result as famously demonstrated by Parke and Taylor \cite{Parke:1985ax,Parke:1986gb}. These issues motivated the study of amplitudes in terms of on-shell building blocks; 
for two extensive reviews, see \cite{Elvang:2013cua,Henn:2014yza}. 

The planar limit of \emph{maximally supersymmetric Yang-Mills theory} (or $\mathcal{N}=4$ SYM) has been a prime laboratory for the development of modern on-shell methods \cite{Elvang:2013cua,Henn:2014yza}. Although supersymmetric, this theory has many important phenomenological applications. For example, its tree-level $S$-matrix contains that of pure Yang-Mills theory which describes gluons in the standard model of particle physics.

The spectrum of $\mathcal{N}=4$ SYM consists of a CPT self-dual supermultiplet containing $16$ massless states. They transform as representations of the $R$-symmetry group $\SU{4}$ where capitalized Roman indices take values $1,2,3,4$. The supermultiplet can be conveniently collected into a single chiral on-shell superfield $\ket{\tilde\eta}$ by introducing four Grassmann variables $\tilde\eta^I$ transforming in the anti-fundamental representation:
\begin{align}\label{eq:mom-ket-chiral}
	\ket{\tilde\eta} = \ket{1}+\tilde\eta^I\ket{\textstyle\frac{1}{2}}_I + \frac{1}{2!}\tilde\eta^I\tilde\eta^J\ket{0}_{I,J} + \frac{1}{3!}\tilde\eta^I\tilde\eta^J\tilde\eta^K\varepsilon_{IJKL}\ket{\textstyle-\frac{1}{2}}^L+ \frac{1}{4!}\tilde\eta^I\tilde\eta^J\tilde\eta^K\tilde\eta^L\varepsilon_{IJKL}\ket{-1}\,,
\end{align}
where $\ket{\pm1}$ is a gluon with helicity $\pm 1$ in the singlet representation, $\ket{+\frac{1}{2}}_I$ (resp., $\ket{-\frac{1}{2}}^L$) labels four gluinos (resp., anti-gluinos) with helicity $+\frac{1}{2}$ (resp., $-\frac{1}{2}$) transforming in the fundamental (resp., anti-fundamental) representation, and $\ket{0}_{I,J} $ transforms in the anti-symmetric representation and denotes six scalars.

An $n$-particle superamplitude $A_n=A_n(\ket{\tilde\eta_1},\ldots,\ket{\tilde\eta_n})$ describes the scattering on $n$ chiral on-shell superfields $\ket{\tilde\eta_1},\ldots,\ket{\tilde\eta_n}$. It can be expanded in terms of helicity sectors as follows:
\begin{align}
	A_n = \begin{cases}
A_{3,1}+A_{3,2} & \text{if}~n=3\\
\sum_{k=2}^{n-2}A_{n,k} & \text{if}~n\ge 4
	\end{cases},
\end{align}
where $A_{n,k}$ has Grassmann degree $4k$ and helicity $k$. Each amplitude $A_{n,k}$ is given a unique designation: $A_{n,2}$ is the \emph{maximally-helicity-violating (MHV)} amplitude, $A_{n,3}$ is the \emph{next-to-maximally-helicity-violating (NMHV)} amplitude, etc., with the exception of $A_{n,n-2}$ which is referred to as the \emph{googly} or $\overline{\text{MHV}}$ amplitude. More generally, $A_{n,k}$ denotes the N\textsuperscript{$k-2$}MHV amplitude.

In planar $\mathcal{N}=4$ SYM, each amplitude $A_{n,k}$ can be further expanded in the t'Hooft coupling constant $\lambda$ as follows:
\begin{align}
	A_{n,k}=\sum_{\ell=0}^{\infty}\lambda^\ell A_{n,k}^{(\ell)}\,,
\end{align}
where $A_{n,k}^\text{tree}\coloneqq A_{n,k}^{(0)}$ is the tree-level amplitude and $A_{n,k}^{(\ell)}$ is the $\ell$-loop amplitude for $\ell>0$. Each tree-level amplitude $A_{n,k}^\text{tree}$ can be further decomposed into partial amplitudes for different colour orderings \cite{Dixon:1996wi}. We will denote by $A_{n,k}^\text{tree}[1,\ldots,n]$ the partial amplitude associated with the standard colour ordering.

The predominant precursor to the development of positive geometries --- the Amplituhedron \cite{Arkani-Hamed:2013jha,Arkani-Hamed:2017vfh}, the Momentum Amplituhedron \cite{Damgaard:2019ztj} and the loop Momentum Amplituhedron \cite{Ferro:2022abq} --- is the Grassmannian description of amplitudes in planar $\mathcal{N}=4$ SYM. This picture builds on the \emph{Britto--Cachazo--Feng--Witten (BCFW) recursions relations} for tree-level amplitudes \cite{Britto:2004ap,Britto:2005fq}, and their loop-level generalization \cite{Arkani-Hamed:2010zjl,Boels:2010nw}. There are Grassmannian formulae for (the leading singularities of) amplitudes in planar $\mathcal{N}=4$ SYM with respect to different kinematic spaces, including twistor variables \cite{Arkani-Hamed:2009ljj,Bourjaily:2010kw}, momentum twistor variables \cite{Mason:2009qx,Arkani-Hamed:2009nll}, and spinor-helicity variables \cite{Arkani-Hamed:2012zlh,He:2018okq}. These Grassmannian formulae are given by contour integrals over the (complex) Grassmannian for some suitable choice of integration contour. The contour integral evaluates to a sum of residues on particular positroid cells. Each residue, referred to as an \emph{on-shell function} and labelled by an \emph{on-shell diagram}, corresponds to an individual term in a solution to the BCFW recursion relations. Different solutions to the BCFW recursion relations yield different representations for (the leading singularities of) amplitudes. They correspond to different choices for the integration contour and their equivalence in the Grassmannian picture stems from the global residue theorem. The equivalence of different representations inspired Hodges to conceptualise the NMHV tree-level amplitude as the volume of a polytope in momentum twistor space where different solutions correspond to different triangulations \cite{Hodges:2009hk}. The Amplituhedron and other Grassmannian-related positive geometries generalize this idea with different solutions corresponding to different positroid tilings.

In \emph{chiral on-shell superspace} $(\lambda_i^a,\tilde{\lambda}_i^{\dot{a}}|\tilde{\eta}_{i}^A)$, the chiral supersymmetric extension of $D=4$ spinor-helicity space, the Grassmannian formula for the partial amplitude $A_{n,k}^\text{tree}[1,\ldots,n]$ reads \cite{Arkani-Hamed:2012zlh}
\begin{align}\label{eq:mom-A-chiral}
	A_{n,k}^\text{tree}[1,\ldots,n]=\oint_{\gamma}\frac{d^{k\times n}C\delta^{k\times2}(C\tilde{\lambda}^T)\delta^{(n-k)\times2}(C^\perp\lambda^T)\delta^{k\times4}(C\tilde{\eta}^T)}{\vol{\GL{k}{\complex}}\prod_{i=1}^n(i,\ldots,i+k-1)}\,,
\end{align}
where the integration contour $\gamma$ can be chosen using the BCFW recursion relations, for example. To define the Momentum Amplituhedron, and to uplift partial amplitudes from rational functions to rational forms \cite{He:2018okq}, we need \emph{non-chiral on-shell superspace} $(\lambda_i^a,\tilde{\lambda}_i^{\dot{a}}|\eta_i^a,\tilde{\eta}_i^{\dot a})$. One can map the chiral description to the non-chiral setting by performing a Fourier transform on half of the Grassmann variables, say from $\tilde{\eta}_i^3,\tilde{\eta}_i^4$ to $\eta_i^1,\eta_i^2$ \cite{Huang:2011um}. This breaks manifest $R$-symmetry, mapping \eqref{eq:mom-ket-chiral} to a non-chiral on-shell superfield $\ket{\eta,\tilde{\eta}}$:
\begin{align}
	\begin{split}
\ket{\eta,\tilde{\eta}}=
(\eta)^2\ket{+1}
-(\tilde{\eta})^2\ket{-1}
+(\eta)^2(\tilde{\eta})^2\ket{0}_{1,2}
+\eta^a\tilde{\eta}^{\dot{a}}\ket{0}_{5-a,\dot{a}}
+\ket{0}_{3,4}\\
+(\eta)^2\tilde{\eta}^{\dot{a}}\ket{\textstyle+\frac{1}{2}}_{\dot{a}}
+\eta^a\ket{\textstyle+\frac{1}{2}}_{5-a}
-(\tilde{\eta})^2\eta^{a}\varepsilon_{ab}\ket{\textstyle-\frac{1}{2}}^{5-b}
-\tilde{\eta}^{\dot{a}}\eta^{a}\varepsilon_{\dot{a}\dot{b}}\ket{\textstyle-\frac{1}{2}}^{\dot{a}}\,.
	\end{split}	
\end{align}
In this non-chiral arena, \eqref{eq:mom-A-chiral} reads \cite{He:2018okq}
\begin{align}\label{eq:mom-A-nonchiral}
	A_{n,k}^\text{tree}[1,\ldots,n]=\oint_{\gamma}\frac{d^{k\times n}C\delta^{k\times2}(C\tilde{\lambda}^T)\delta^{(n-k)\times2}(C^\perp\lambda^T)\delta^{k\times2}(C\tilde{\eta}^T)\delta^{(n-k)\times2}(C^\perp\eta^T)}{\vol{\GL{k}{\complex}}\prod_{i=1}^n(i,\ldots,i+k-1)}\,.
\end{align}
Via the replacement
\begin{align}
	\eta_i^a\mapsto d\lambda_i^a\,,\qquad\tilde{\eta}_i^{\dot{a}}\mapsto d\tilde{\lambda}_i^{\dot{a}}\,,
\end{align}
the partial amplitude $A_{n,k}^\text{tree}[1,\ldots,n]$ uplifts to rational form of degree
$(2(n-k), 2k)$ in $(d\lambda,d\tilde{\lambda})$ which vanishes due to supersymmetric Ward identities \cite{He:2018okq}. However, by stripping off $d^4q$ or $d^4\tilde{q}$, where $q^{a,\dot{a}}\coloneqq\sum_{i=1}^n\lambda_i^a(d\tilde{\lambda}_i^{\dot{a}})$ and $\tilde{q}^{a,\dot{a}}\coloneqq\sum_{i=1}^n(d\lambda_i^a)\tilde{\lambda}_i^{\dot{a}}$, one obtains a non-vanishing rational form of degree $2n-4$. It is this rational form that becomes the canonical form of the Momentum Amplituhedron. 

Before defining the Momentum Amplituhedron, let us first discuss the singularities of partial amplitudes in $\mathcal{N}=4$ SYM. This discussion will help us to identify these singularities with the boundaries of the Momentum Amplituhedron.

\section{Amplitude Singularities}
\label{sec:mom-sing}

The behaviour of partial amplitudes in gauge theories under collinear (resp., soft) limits is governed by universal functions which depend only on the particles which become collinear (resp., the nearest neighbours of the soft particle); for an exposition on collinear and soft gluons in Yang-Mills theory see \cite{Henn:2014yza} and references therein. In this section, we present the super-splitting and the super-soft functions for $\mathcal{N}=4$ SYM, derived in \cite{Ferro:2020lgp}. There are two types of collinear limits: helicity-preserving and helicity-reducing types. For soft limits, we confirm that only the gluons in the superfield have divergent behaviour when becoming soft. The two types of soft limits arise from two consecutive collinear limits of the same type. We will work in chiral on-shell superspace, and translate our results to the non-chiral setup at the very end.

\begin{figure}
	\centering
	\null
	\hfill
	\begin{subfigure}[b]{0.3\textwidth}
		\centering
		\includegraphics[scale=0.45]{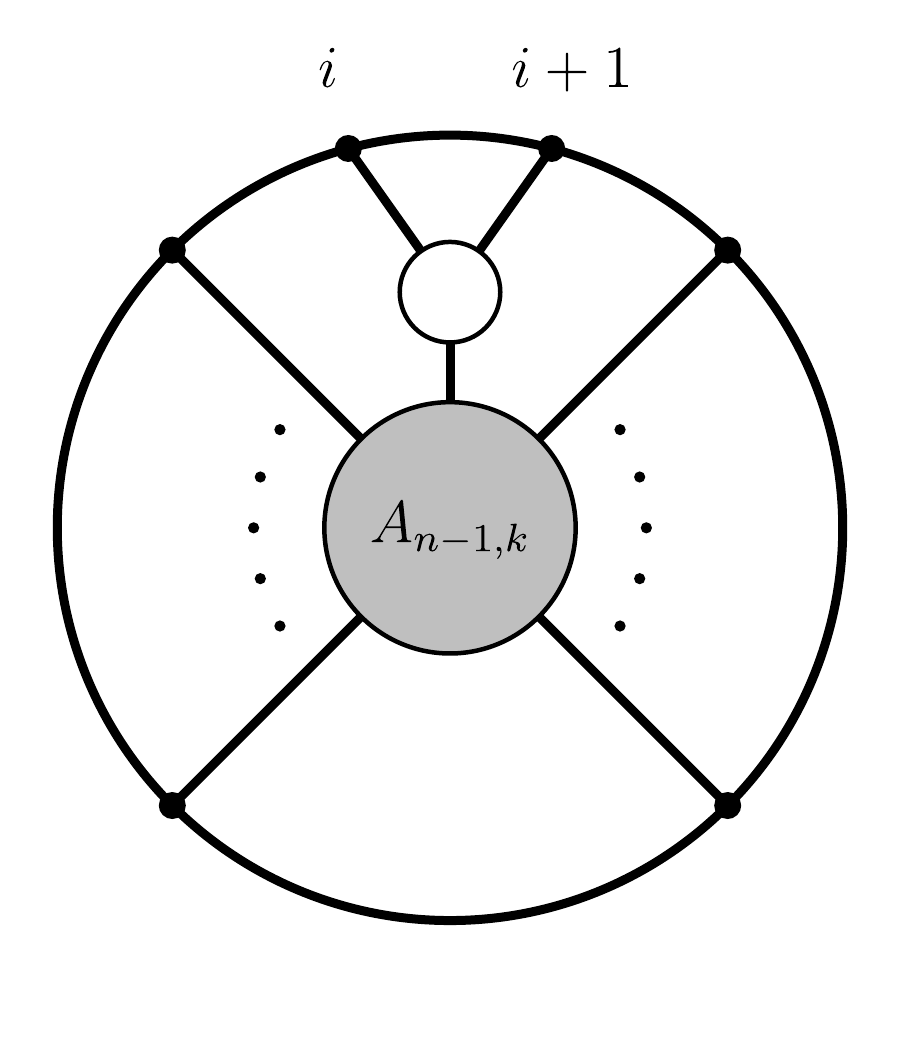}
		\caption{$\lrangle{ii+1}=0$}
		\label{fig:mom-collinear-preserving}
	\end{subfigure}
	\hfill
	\begin{subfigure}[b]{0.3\textwidth}
		\centering
		\includegraphics[scale=0.45]{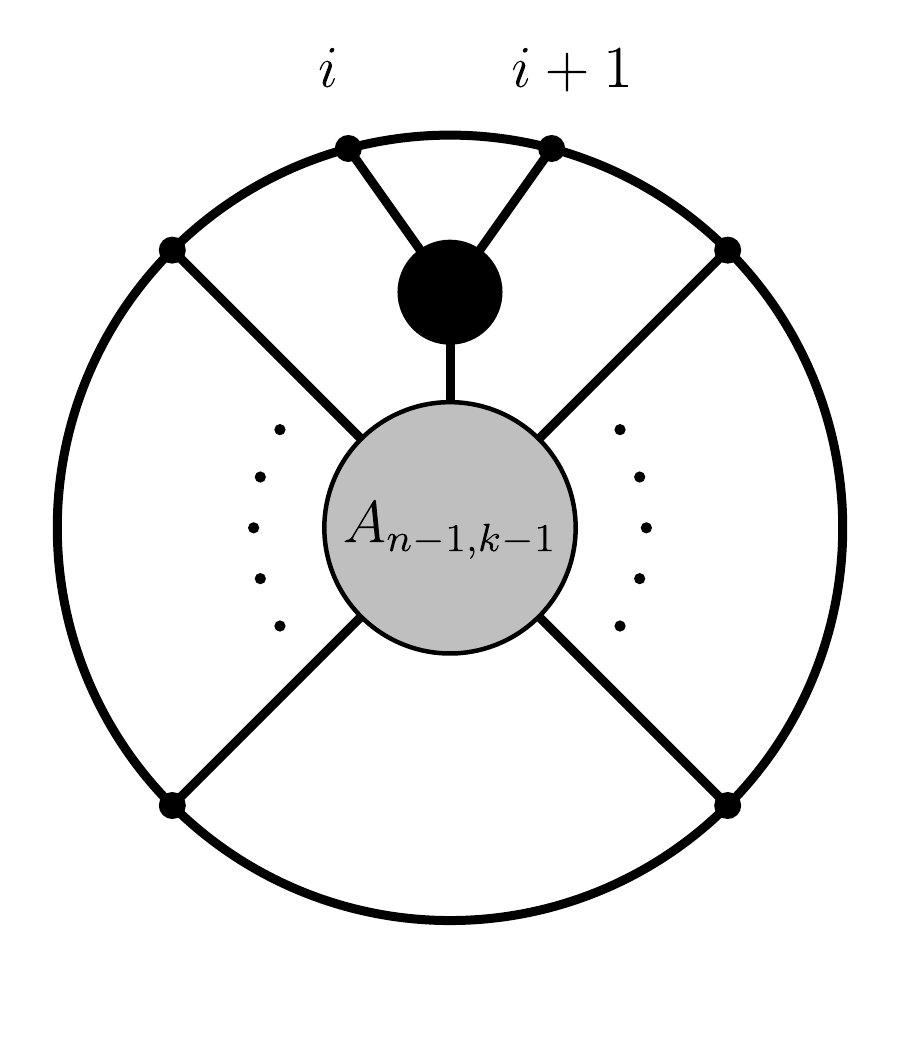}
		\caption{$[ii + 1] = 0$}
		\label{fig:mom-collinear-reducing}
	\end{subfigure}	
	\hfill
	\null
	\caption{Collinear limits: a white trivalent vertex represents $\text{Split}^\text{tree}_{0}$ in \cref{fig:mom-collinear-preserving} and a black trivalent vertex denotes $\text{Split}^\text{tree}_{-1}$ in \cref{fig:mom-collinear-reducing}.}
	\label{fig:mom-collinear}
\end{figure}

Partial amplitudes become singular when the sum of adjacent external momenta goes on-shell, i.e.\ $P_{i,j}^2\coloneqq (p_i+p_{i+1}+\ldots+p_j)^2\rightarrow 0$. On these multi-particle poles, partial amplitudes factorise according to \cite{Elvang:2013cua} 
\begin{align}\label{eq:mom-factorization}
	\begin{split}
		&A_{n,k}^\text{tree}[1,\ldots,n]\xrightarrow{P_{i,j}^2\to 0}\\
		&\sum_{k'=1}^{k}\int d^4\tilde{\eta}_{P_{i,j}} A_{j-i+2,k'}^\text{tree}[i,\ldots,j,P_{i,j}]\frac{1}{P_{i,j}^2}A_{n-j+i,k-k'+1}^\text{tree}[-P_{i,j},j+1,\dots,i-1]\,.
	\end{split}
\end{align}
Collinear singularities are a special case of \eqref{eq:mom-factorization} where the momenta of two adjacent particles become proportional. Without loss of generality, suppose particles $1$ and $2$ become collinear. Then the momentum $P_{1,2} =p_1+p_2$ goes on-shell and \eqref{eq:mom-factorization} becomes
\begin{align}\label{eq:mom-collinear}
	A_{n,k}^\text{tree}[1,\ldots,n]\xrightarrow{P_{1,2}^2\to 0}\sum_{k'=1}^{2}\int d^4\tilde{\eta}_{P_{1,2}} A_{3,k'}^\text{tree}[1,2,P_{1,2}]\frac{1}{P_{1,2}^2}A_{n-1,k-k'+1}^\text{tree}[-P_{1,2},3,\dots,n]\,.
\end{align}
The sum runs over two contributions: the helicity of the remaining amplitude is preserved when $k'=1$, whereas for $k'=2$ the helicity is reduced by one. The integration over $\tilde{\eta}_{P_{1,2}}$ aggregates all states exchanged in the factorisation channel. The collinearity condition
\begin{align}\label{eq:mom-super-collinear-momenta}
	P_{1,2}^2=(p_1+p_2)^2=2 p_1.p_2=\lrangle{12}[12]=0\,,
\end{align}
can be satisfied by taking $\lrangle{12}\to0$ or $[12]\to0$. The independence of these two limits is a consequence of three-particle special kinematics: on-shell three-particle amplitudes are non-vanishing provided that angle and square brackets are independent. The independence of spinor-helicity brackets is accomplished by considering complex momenta or changing the spacetime signature to $(2,2)$. In particular, $\lrangle{12}\to0$ parametrises the helicity-preserving collinear limit, while $[12]\to0$ parametrises the helicity-reducing one. Let $0\le z\le 1$ denote the fraction of the total momentum $P_{1,2} = p_1 + p_2$ carried by $p_1$. The analysis of \cite{Ferro:2020lgp} concludes that
\begin{align}
	A_{n,k}^\text{tree}[1,\ldots,n]\xrightarrow{\lrangle{12}\to 0}\int d^4\tilde{\eta}_{P_{1,2}}\text{Split}^\text{tree}_{0}(z;\tilde{\eta}_1,\tilde{\eta}_2,\tilde{\eta}_{P_{1,2}})A_{n-1,k}^\text{tree}[-P_{1,2},3,\dots,n]\,,
\end{align}
where 
\begin{align}\label{eq:mom-collinear-preserving-function} 
	\text{Split}^\text{tree}_{0}(z;\tilde{\eta}_1,\tilde{\eta}_2,\tilde{\eta}_3)\coloneqq \frac{1}{\sqrt{z(1-z)}}\frac{1}{\lrangle{12}}\prod_{I=1}^{4}(\tilde{\eta}_{3}^I-\sqrt{z}\tilde{\eta}_{1}^I-\sqrt{1-z}\tilde{\eta}_{2}^I)\,,
\end{align}
is the \emph{helicity-preserving super-splitting function}. One can obtain the Yang-Mills theory splitting functions by expanding \eqref{eq:mom-collinear-preserving-function}, focusing on terms proportional to say $(\tilde{\eta}_1)^4$. Similarly 
\begin{align}
	A_{n,k}^\text{tree}[1,\ldots,n]\xrightarrow{[12]\to 0}\int d^4\tilde{\eta}_{P_{1,2}}\text{Split}^\text{tree}_{-1}(z;\tilde{\eta}_1,\tilde{\eta}_2,\tilde{\eta}_{P_{1,2}})A_{n-1,k-1}^\text{tree}[-P_{1,2},3,\dots,n]\,,
\end{align}
where
\begin{align}\label{eq:mom-collinear-reducing-function}
	\text{Split}^\text{tree}_{-1}(z;\tilde{\eta}_1,\tilde{\eta}_2,\tilde{\eta}_3)\coloneqq \frac{1}{\sqrt{z(1-z)}}\frac{1}{[12]}\prod_{I=1}^{4}(\tilde{\eta}_{1}^I\tilde{\eta}_{2}^I-\sqrt{z}\tilde{\eta}_{2}^I\tilde{\eta}_{3}^I-\sqrt{1-z}\tilde{\eta}_{3}^I\tilde{\eta}_{1}^I)\,,
\end{align}
is the \emph{helicity-reducing super-splitting function}. Again, one recovers the Yang-Mills splitting functions by expanding \eqref{eq:mom-collinear-reducing-function}, focusing on terms proportional to say $(\tilde{\eta}_1)^4 (\tilde{\eta}_2)^4$. These super-splitting functions, parametrised by $\lrangle{12}\to0$ and $[12]\to0$, represent two physically distinct processes, depicted in \cref{fig:mom-collinear}. 

\begin{figure}
	\centering
	\null
	\hfill
	\begin{subfigure}[b]{0.3\textwidth}
		\centering
		\includegraphics[scale=0.45]{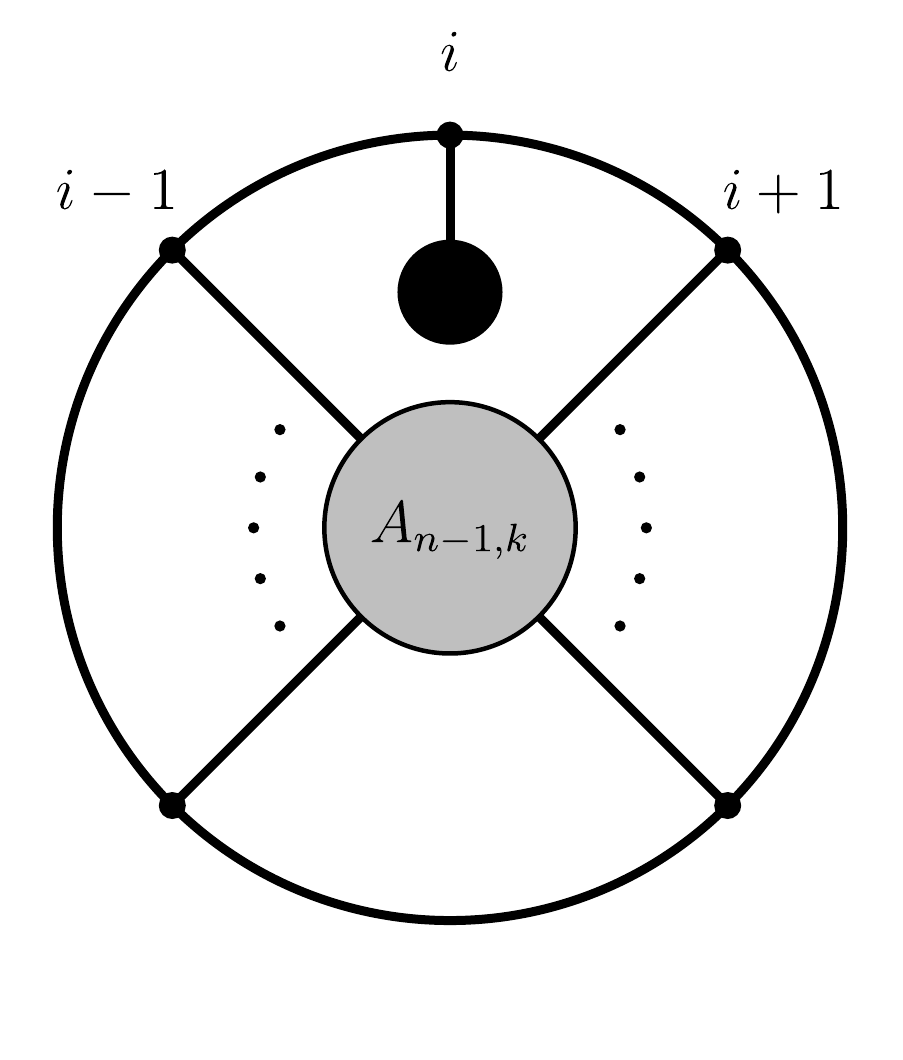}
		\caption{$\lrangle{ii+1}=0=\lrangle{ii-1}$}
		\label{fig:mom-soft-preserving}
	\end{subfigure}
	\hfill
	\begin{subfigure}[b]{0.3\textwidth}
		\centering
		\includegraphics[scale=0.45]{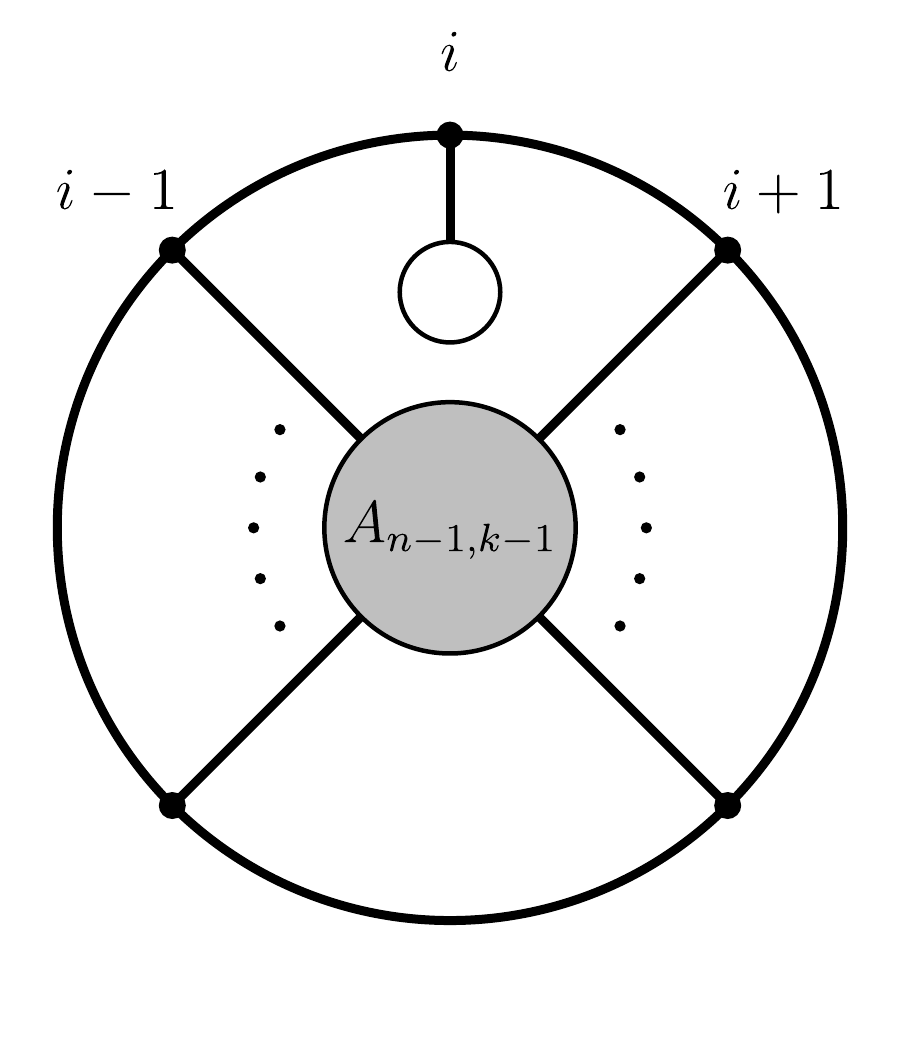}
		\caption{$[ii + 1] = 0 = [ii-1]$}
		\label{fig:mom-soft-reducing}
	\end{subfigure}
	\hfill
	\null
	\caption{Soft limits: a black or white boundary leaf depicts a soft particle arising from two consecutive collinear limits involving angle or square brackets, respectively.}
	\label{fig:mom-soft}
\end{figure}

As demonstrated, collinear limits are a special case of factorisation on two-particle poles. Taking two consecutive collinear limits of the same type produces a soft limit. In particular, 
 the \emph{helicity-preserving soft limit} is given by
\begin{equation}\label{eq:mom-soft-preserving}
	A_{n,k}^\text{tree}[1,2,3,\ldots,n]\xrightarrow{\lrangle{12},\lrangle{23}\to 0} \frac{\lrangle{13}}{\lrangle{12}\lrangle{23}}A_{n-1,k}^\text{tree}[1,3,\ldots,n]\,,
\end{equation}
while the \emph{helicity-reducing soft limit} is given by
\begin{align}\label{eq:mom-soft-reducing}
	A_{n,k}^\text{tree}[1,2,3,\ldots,n]\xrightarrow{[12],[23]\to 0} (\tilde{\eta}_2)^4 \frac{[13]}{[12][23]} A_{n-1,k-1}^\text{tree}[1,3,\ldots,n]\,.
\end{align}
In both cases, the only divergent contribution comes from soft gluons: \eqref{eq:mom-soft-preserving} comes solely from the positive helicity gluon, while only the negative helicity gluon contributes to \eqref{eq:mom-soft-reducing}. We depict these soft limits in \cref{fig:mom-soft}.

To translate our results to non-chiral on-shell superspace, we perform a Fourier transform on half of the Grassmann variables, from $\tilde{\eta}_i^3,\tilde{\eta}_i^4$ to $\eta_i^1,\eta_i^2$, producing the super-splitting functions
\begin{align}
	&\text{Split}^\text{tree}_{0}(z;\eta_1,\eta_2,\eta_3,\tilde\eta_1,\tilde\eta_2,\tilde\eta_3)=\\*
	\nonumber&\quad\frac{1}{\sqrt{z(1-z)}}\frac{1}{\langle 12\rangle}\prod_{r=1}^{2}(\eta_{1}^{r}\eta_{2}^{r}-\sqrt{z}\eta_{2}^{r}\eta_{3}^{r}-\sqrt{1-z}\eta_{3}^{r}\eta_{1}^{r})\prod_{\dot{r}=1}^{2}(\tilde\eta_{3}^{\dot{r}}-\sqrt{z}\tilde\eta_{1}^{\dot{r}}-\sqrt{1-z}\tilde\eta_{2}^{\dot{r}}) \,,\\
	&\text{Split}^\text{tree}_{-1}(z;\eta_1,\eta_2,\eta_3,\tilde\eta_1,\tilde\eta_2,\tilde\eta_3)=\\*
	\nonumber&\quad\frac{1}{\sqrt{z(1-z)}}\frac{1}{[12]}\prod_{\dot{r}=1}^{2}(\tilde\eta_{1}^{\dot{r}}\tilde\eta_{2}^{\dot{r}}-\sqrt{z}\tilde\eta_{2}^{\dot{r}}\tilde\eta_{3}^{\dot{r}}-\sqrt{1-z}\tilde\eta_{3}^{\dot{r}}\tilde\eta_{1}^{\dot{r}})\prod_{r=1}^{2}(\eta_{3}^{{r}}-\sqrt{z}\eta_{1}^{{r}}-\sqrt{1-z}\eta_{2}^{{r}}) \,,
\end{align}
and the soft limits
\begin{align}
	&A_{n,k}^\text{tree}[1,2,3,\ldots,n]\xrightarrow{\lrangle{12},\lrangle{23}\to0}(\eta_2)^2 \frac{\lrangle{13}}{\lrangle{12}\lrangle{23}} A_{n-1,k}^\text{tree}[1,3,\ldots,n] \,,\\
	&A_{n,k}^\text{tree}[1,2,3,\ldots,n]\xrightarrow{[12],[23]\to 0}(\tilde\eta_{2})^{2} \frac{[13]}{[12][23]}A_{n-1,k-1}^\text{tree}[1,3,\ldots,n] \,.	
\end{align}  

\section{The Momentum Amplituhedron}
\label{sec:mom-def}

\begin{figure}
	\centering
     \includegraphics{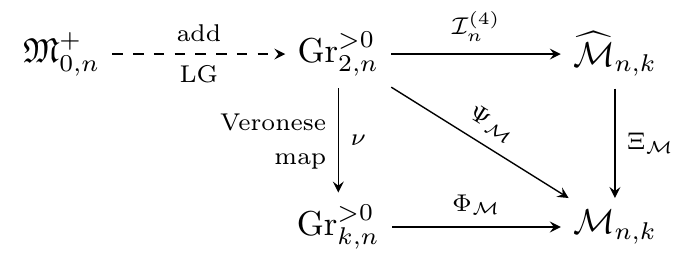}
     \caption{Web of connections between spaces associated with the Momentum Amplituhedron.}
     \label{fig:mom-web-spaces}
\end{figure}

As is the case for the Amplituhedron, the Momentum Amplituhedron admits two equivalent definitions: either using an auxiliary Grassmannian \cite{Damgaard:2019ztj} or directly in the kinematic space of four-dimensional spinor-helicity variables \cite{Ferro:2020ygk}. The latter is the image of the former in an affine chart. In this section, we present both definitions. We also relate the Momentum Amplituhedron to $\tpG{2}{n}\cong\tpModuli{0}{n}\times\projective(\real_{>0}^n)$ via the four-dimensional twistor-string map of \cite{He:2021llb}. This web of connections in depicted in \cref{fig:mom-web-spaces}. 
Throughout this section, we will fix $k$ and $n$ to be positive integers for which $2\le k\le n-2$.

\begin{figure}
	\centering
	\includegraphics{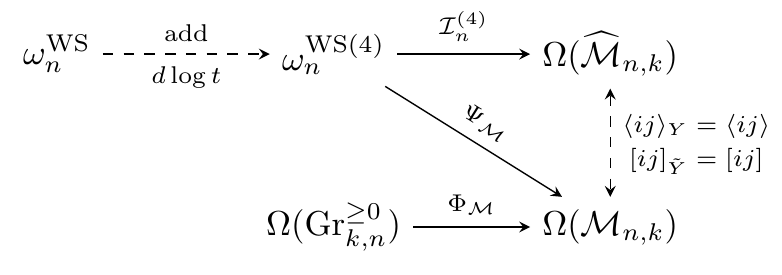}
	\caption{Web of connections between rational forms associated with the Momentum Amplituhedron. Solid lines designate pushforwards.}
	\label{fig:mom-web-forms}
\end{figure}

\subsection{Grassmannian Definition}
\label{sec:mom-grass}

Let us begin by establishing some definitions. Suppose $a$ and $b$ are integers such that $a\le b$. A real $b\times a$ matrix is called \emph{positive} if all of its maximal minors are positive. We denote by $\tpMat{b}{a}$ the set of all positive matrices in $\Mat{b}{a}{\real}$. A matrix $X\in\Mat{b}{a}{\real}$ is called \emph{twisted positive} if $\epsilon_{I,[b]\setminus I}\,p_{[b]\setminus I}(X)>0$ for each $I\in\binom{[b]}{b-a}$, where $\epsilon_{A,B} = (-1)^{\text{inv}(A,B)}$ and $\text{inv}(A,B)=\#\{(a,b)\in A\times B:a>b\}$ is the inversion number. The set of all twisted positive matrices in $\Mat{b}{a}{\real}$ is designated by $\tpTwistedMat{b}{a}$. 

Both positive and twisted positive matrices possess a \emph{twisted cyclic symmetry}. If $\tilde{X}\in\tpMat{b}{a}$ has rows $\tilde{X}_1,\ldots,\tilde{X}_b$ and if we define $\tilde{X}_{\hat{i}}\coloneqq(-1)^{a-1}\tilde{X}_i$, then the matrix with rows $\tilde{X}_2,\ldots,\tilde{X}_b,\tilde{X}_{\hat{1}}$ is also positive. Similarly, if $X\in\tpTwistedMat{b}{a}$ has rows $X_1,\ldots,X_b$ and if we define $X_{\hat{i}}\coloneqq(-1)^{a-1}X_i$ (as before), then the matrix with rows $X_2,\ldots,X_b,X_{\hat{1}}$ is also twisted positive.

Lastly, given a point $V$ in $\G{k}{n}$ with Pl\"{u}cker coordinates $p_I(V)$ for $I\in\binom{[n]}{k}$, its orthogonal complement $V^{\perp}$ is a point in $\G{n-k}{k}$ with Pl\"{u}cker coordinates $p_{[n]\setminus I}(V^{\perp}) = \epsilon_{[n]\setminus I,I}\,p_I(V)$ for $I\in\binom{[n]}{k}$ \cite{karp2017sign}.

Towards defining the Momentum Amplituhedron, let $\Lambda\in\tpTwistedMat{n}{n-k+2}$ be a twisted positive matrix and let $\tilde{\Lambda}\in\tpMat{n}{k+2}$ be a positive matrix. We refer to the pair $(\Lambda,\tilde{\Lambda})$ as the \emph{kinematic data}. Given a point $[C]\in\G{k}{n}$, and its orthogonal complement $[C^\perp]\in\G{n-k}{n}$, define $Y \coloneqq C^\perp\Lambda$ and $\tilde{Y}\coloneqq C\tilde{\Lambda}$. The \emph{Momentum Amplituhedron} $\tnnMom{k}{n}\coloneqq\MomMap{\tnnG{k}{n}}$ is the image of $\tnnG{k}{n}$ through the \emph{Momentum Amplituhedron map} \cite{Damgaard:2019ztj}
\begin{align}\label{eq:mom-map}
\symbolMomMap:\G{k}{n}\to\G{n-k}{n-k+2}\times\G{k}{k+2}, [C]\mapsto([Y],[\tilde{Y}])\,.
\end{align} 
Equivalently, the Momentum Amplituhedron $\tnnMom{k}{n}$ can be regarded as the closure of $\tpMom{k}{n}\coloneqq\MomMap{\tpG{k}{n}}$ in $\MomReal{k}{n}$ where $\tpMom{k}{n}$ is the interior of $\tnnMom{k}{n}$. From the above definition, it immediately follows that $\tnnMom{2}{n}$ and $\tnnMom{n-2}{n}$ are diffeomorphic to $\tnnG{2}{n}$. Consequently, their boundary stratifications are combinatorially equivalent. 

There are analogues of four-dimensional spinor-helicity brackets which define natural coordinates for the Momentum Amplituhedron. Let $([Y],[\tilde{Y}])$ be a point in $\G{n-k}{n-k+2}\times\G{k}{k+2}$. For each $i,j\in[n]$, we define the \emph{twistor coordinate}
\begin{align}\label{eq:mom-twistor-coodinate}
	[ij]_{\tilde{Y}}\coloneqq\det(\tilde{Y}_1,\ldots,\tilde{Y}_k,\allowbreak\tilde{\Lambda}_{i},\tilde{\Lambda}_{j})\,,
\end{align}
to be the determinant of the $(k+2)\times(k+2)$ matrix with rows  $\tilde{Y}_1,\ldots,\tilde{Y}_k,\tilde{\Lambda}_{i},\tilde{\Lambda}_{j}$ and we define the \emph{(dual) twistor coordinate}
\begin{align}\label{eq:mom-dual-twistor-coodinate}
	\lrangle{ij}_{Y}\coloneqq\det(Y_1,\ldots,Y_{n-k},\Lambda_{i},\Lambda_{j})\,,
\end{align}
to be the determinant of the $(n-k+2)\times(n-k+2)$ matrix with rows $Y_1,\ldots,Y_{n-k},\Lambda_{i},\Lambda_{j}$. Moreover, for each $I\subset[n]$, we define the \emph{Mandelstam variable} $S_I(Y,\tilde{Y})$ to be the following quadratic polynomial in twistor coordinates:
\begin{align}\label{eq:mom-mandelstam}
	S_I(Y,\tilde{Y})\coloneqq\sum_{\{i,j\}\in \binom{I}{2}}\lrangle{ij}_{Y}[ij]_{\tilde{Y}}\,.
\end{align}

A subset $I\subset[n]$ is said to be \emph{cyclically consecutive (CC)} if its elements, or the elements of its complement in $[n]$, denoted by $I'=[n]\setminus I$, are consecutive.
The kinematic data $(\Lambda,\tilde{\Lambda})$ must satisfy 
\begin{align}\label{eq:mom-kinematic-data}
S_I(Y,\tilde{Y})>0~&\text{for all $([Y],[\tilde{Y}])\in\tpMom{k}{n}$}\,,\\
\nonumber &\text{for all CC subsets $I\subset[n]$ with $2\le|I|\le n-2$}\,.
\end{align}
This additional requirement ensures that the codimension-one boundaries of $\tnnMom{k}{n}$ encode all factorisation channels of $A_{n,k}^\text{tree}[1,\ldots,n]$ \cite{Damgaard:2019ztj}. Using \Mathematica{} it was checked in \cite{Damgaard:2019ztj} for $n\le10$ that the following kinematic data satisfies \eqref{eq:mom-kinematic-data}:
\begin{align}
	(\Lambda^\perp)_{i,A'} = i^{A'-1}\,,\qquad
	\tilde{\Lambda}_{i,\dot{A}} = i^{\dot{A}-1}\,,
\end{align} 
where $i\in[n]$, $A'\in[k-2]$ and $\dot{A}\in[k+2]$. Note that rows of $\Lambda^\perp$ and $\tilde{\Lambda}$ are vectors lying on moment curves. Since the combinatorial properties of $\tnnMom{k}{n}$ are conjecturally independent of our choice of kinematic data, we omit explicit dependence on $(\Lambda,\tilde{\Lambda})$ throughout.
  
The twistor coordinates on $\G{k}{k+2}\times\G{n-k}{n-k+2}$ are projectively well-defined, embedding $\G{n-k}{n-k+2}\times\G{k}{k+2}$ into $\CP^{\binom{n}{2}-1}\times\CP^{\binom{n}{2}-1}$ where
\begin{align}
	\dim\left(\CP^{\binom{n}{2}-1}\times\CP^{\binom{n}{2}-1}\right) = (n-2)(n+1)\,,
\end{align}
is strictly larger than 
\begin{align}
	\dim\left(\G{k}{k+2}\times\G{n-k}{n-k+2}\right) = 2n\,.
\end{align}
Moreover, given a point $V\in\G{k}{n}$ mapped to a point $([Y],[\tilde{Y}])=\MomMap{V}\in\Mom{k}{n}$, it follows from the Cauchy-Binet formula that 
\begin{align}\label{eq:mom-cauchy-binet}
	[ij]_{\tilde{Y}} = \sum_{\mathclap{I\in\binom{[n]}{k}}}p_{I}(V)\,p_{I\cup\{i,j\}}(\tilde{\Lambda})\,,\qquad
	\lrangle{ij}_{Y} = \sum_{\mathclap{J\in\binom{[n]}{n-k}}}\epsilon_{J,[n]\setminus J}\,p_{[n]\setminus J}(V)\,p_{J\cup\{i,j\}}(\Lambda)\,.
\end{align}

For each point $([Y],[\tilde{Y}])\in\G{n-k}{n-k+2}\times\G{k}{k+2}$, let $Y^\perp$ and $\tilde{Y}^\perp$ represent the orthogonal complements of $[Y]$ and $[\tilde{Y}]$, respectively. The \emph{hypersurface of momentum conservation} is a codimension-$4$ algebraic subset of $\G{n-k}{n-k+2}\times\G{k}{k+2}$ defined as $P=\mathbbb{0}_{2\times 2}$ where
\begin{align}\label{eq:mom-P}
	P\coloneqq (Y^\perp\Lambda^T)(\tilde{Y}^\perp\tilde{\Lambda}^T)^T\,.
\end{align}
The identity 
\begin{align}
	\mathbbb{0}_{2,n-k}=Y^\perp Y^T = (Y^\perp\Lambda^T)(C^\perp)^T\,,
\end{align}
implies that $[(C^\perp)^\perp]=[C]$ contains the row span of $Y^\perp\Lambda^T$. Similarly, $[C^\perp]$ contains the row span of $\tilde{Y}^\perp\tilde{\Lambda}^T$ as demonstrated by the identity
\begin{align}
	\mathbbb{0}_{2,k}=\tilde{Y}^\perp \tilde{Y}^T = (\tilde{Y}^\perp\tilde{\Lambda}^T)(C)^T\,.
\end{align}
Since $Y^\perp\Lambda^T$ and $\tilde{Y}^\perp\tilde{\Lambda}^T$ span orthogonal subspaces, the Momentum Amplituhedron lies on the hypersurface of momentum conservation \cite{Damgaard:2019ztj}. 

Conjecturally, the Momentum Amplituhedron $\tnnMom{k}{n}$ is a $(2n-4)$-dimensional positive geometry whose canonical form $\Omega(\tnnMom{k}{n})$ encodes the partial amplitude $A^\text{tree}_{n,k}[1,\ldots,n]$ \cite{Damgaard:2019ztj}. Moreover, the Momentum Amplituhedron map $\symbolMomMap$ is (conjecturally) a morphism between $(\G{k}{n},\tnnG{k}{n})$ and $(\Mom{k}{n},\tnnMom{k}{n})$. Consequently, $\Omega(\tnnMom{k}{n})$ is the pushforward of $\Omega(\tnnG{k}{n})$ via $\symbolMomMap$. Let $\mathcal{C}=\{\tnnPos{f}\}_{f\in F}$ be a $\symbolMomMap$-induced tiling of $\tnnMom{k}{n}$ for some indexing set $F$ of affine permutations. For each $f\in F$, $\symbolMomMap|_{\tpPos{f}}:\tpPos{f}\to\symbolMomMap(\tpPos{f})$ is a diffeomorphism and $\Omega(\tnnPos{f})=\bigwedge_{i=1}^{2n-4}d\log\alpha_i$ as per \eqref{eq:grass-positroid-Omega}. Then
\begin{align}
	\Omega(\tnnMom{k}{n}) = 
	\sign(\mathcal{C})\sum_{f\in F}\symbolMomMap_\ast\Omega(\tnnPos{f})\,,
\end{align} 
where $\sign(\mathcal{C})=1$ (resp., $\sign(\mathcal{C})=-1$) if $\mathcal{C}$ is orientation-preserving (resp., orientation-reversing).

Wedging the canonical form $\Omega(\tnnMom{k}{n})$ of degree $(2n-4)$ with $d^4P\delta^4(P)$ produces a top-form on $\G{n-k}{n-k+2}\times\G{k}{k+2}$ multiplied by the \emph{canonical function} $\underline{\Omega}(\tnnMom{k}{n})$:
\begin{align}
	\Omega(\tnnMom{k}{n})\wedge d^4P\,\delta^4(P) = \bigwedge_{\alpha=1}^{n-k}\det(Y,d^2Y_{\alpha})\bigwedge_{\dot{\alpha}=1}^{k}\det(\tilde{Y},d^2\tilde{Y}_{\dot{\alpha}})\,\delta^4(P)\,\underline{\Omega}(\tnnMom{k}{n})\,.
\end{align}
To extract the partial amplitude $A_{n,k}^\text{tree}[1,\ldots,n]$ from $\underline{\Omega}(\tnnMom{k}{n})$ we follow the prescription given in \cite{Damgaard:2019ztj,Ferro:2020ygk}. Introduce $2n$ auxiliary Grassmann parameters $\phi^\alpha_a$ and $\tilde{\phi}^{\tilde{\alpha}}_{\dot{a}}$ where $a,\dot{a}\in[2]$, $\alpha\in[n-k]$ and $\dot{\alpha}\in[k]$ and define
\begin{align}
	(\Lambda^T_\ast)_{A,i} = 
	\begin{pmatrix}
		\lambda_i^a \\
		\phi^\alpha_b\eta_i^b
	\end{pmatrix},
	\qquad
	(\tilde{\Lambda}^T_\ast)_{\dot{A},i} = 
	\begin{pmatrix}
		\tilde{\lambda}_i^{\dot{a}} \\
		\tilde{\phi}^{\dot{\alpha}}_{\dot{b}}\tilde{\eta}_i^{\dot{b}}
	\end{pmatrix},
\end{align}
where $A=(a,\alpha)\in[n-k+2]$ and $\dot{A}\in[k+2]$. Choose the reference points
\begin{align}
	Y_\ast \coloneqq
	\begin{pmatrix}
		\mathbbb{0}_{(n-k)\times 2} & \mathbbb{1}_{(n-k)\times(n-k)}
	\end{pmatrix},
	\qquad
	\tilde{Y}_\ast \coloneqq
	\begin{pmatrix}
		\mathbbb{0}_{k\times 2} & \mathbbb{1}_{k\times k}
	\end{pmatrix}.
\end{align}
Then the partial amplitude $A_{n,k}^\text{tree}[1,\ldots,n]$ is extracted via the following integral
\begin{align}
	A_{n,k}^\text{tree}[1,\ldots,n]=\int d^{2(n-k)}\phi\wedge d^{2k}\tilde{\phi}\,\delta^4(P)\,\underline{\Omega}(\tnnMom{k}{n})\Big|_{(Y,\tilde{Y},\Lambda,\tilde{\Lambda})\to(Y_\ast,\tilde{Y}_\ast,\Lambda_\ast,\tilde{\Lambda}_\ast)}\,.
\end{align}

Alternatively, we can define the Momentum Amplituhedron as the intersection to two spaces. Let $\MomV{k}{n}$ denote the set of points $([Y],[\tilde{Y}])$ in $\G{n-k}{n-k+2}\times\G{k}{k+2}$ which lie on the hypersurface of momentum conservation $P=\mathbbb{0}_{2\times 2}$. Its real part is designated by $\MomVReal{k}{n}$. Define the \emph{winding space} $\MomW{k}{n}$ as the set of points $([Y],[\tilde{Y}])$ in $\GReal{n-k}{n-k+2}\times\GReal{k}{k+2}$ for which 
\begin{subequations}\label{eq:mom-winding}
	\begin{align}
		&\lrangle{ii+1}_Y>0~\text{for}~i\in[n-1]~\text{and}~\lrangle{n\hat{1}}_Y>0\,,\\
		&\var(\lrangle{12}_Y,\lrangle{13}_Y,\ldots,\lrangle{1n}_Y)=k-2\,,\\
		&[ii+1]_{\tilde{Y}}>0~\text{for}~i\in[n-1]~\text{and}~[n\hat{1}]_{\tilde{Y}}>0\,,\\
		&\var([12]_{\tilde{Y}},[13]_{\tilde{Y}},\ldots,[1n]_{\tilde{Y}})=k\,,\\
		&S_I(\tilde{Y},Y) > 0~\text{for all CC subsets $I\subset[n]$ with $2\le|I|\le n-2$}\,,
	\end{align}
\end{subequations}
where $\var$ counts the number of sign flips. It is conjectured that $\Mom{k}{n}$ equals $\MomV{k}{n}$, $\tpMom{k}{n}$ equals $\MomV{k}{n}\cap\MomW{k}{n}$, and that $\tnnMom{k}{n}$ equals the closure of $\MomV{k}{n}\cap\MomW{k}{n}$ in $\MomVReal{k}{n}$. We have already established $\Mom{k}{n}\subset\MomV{k}{n}$ while $\tpMom{k}{n}\subset\MomV{k}{n}\cap\MomW{k}{n}$ is proven in \cite{Damgaard:2019ztj}.

Let us illustrate some of the preceding discussion by considering the simplest possible example, namely four-particle MHV scattering. Fix the kinematic data $(\Lambda,\tilde{\Lambda})$ where $\Lambda$ is a $4\times 4$ twisted positive matrix and $\tilde{\Lambda}$ is a $4\times 4$ positive matrix. The top-cell of $\tnnG{2}{4}$ admits the following canonically positive parametrisation \cite{Damgaard:2019ztj} 
\begin{align}
	C = \begin{pmatrix}
		1 & \alpha_2 & 0 & -\alpha_3\\
		0 & \alpha_1 & 1 & \alpha_4
	\end{pmatrix}.
\end{align}
In this coordinate patch, the canonical form for $\tnnG{2}{4}$ is given by
\begin{align}
	\Omega(\tnnG{2}{4})=d\log\alpha_1\wedge d\log\alpha_2\wedge d\log\alpha_3\wedge d\log\alpha_4\,.
\end{align}
Since $\dim(\tnnG{2}{4}) = 4 = \dim(\tnnMom{2}{4})$, the trivial dissection of $\tnnMom{2}{4}$ (i.e.\ the collection containing only the top-cell of $\tnnG{2}{4}$) is the only tiling of $\tnnMom{2}{4}$. There are various ways to express the canonically positive parameters in terms of twistor coordinates. For example, using \eqref{eq:mom-cauchy-binet} we have that
\begin{align}
	\alpha_1=\frac{\lrangle{12}_{Y}}{\lrangle{13}_{Y}},\quad
	\alpha_2=\frac{\lrangle{23}_{Y}}{\lrangle{13}_{Y}},\quad
	\alpha_3=\frac{\lrangle{34}_{Y}}{\lrangle{13}_{Y}},\quad
	\alpha_4=\frac{\lrangle{14}_{Y}}{\lrangle{13}_{Y}}.
\end{align}
Hence, the pushforward of $\Omega(\tnnG{2}{4})$ via $\symbolMomMap$ yields an expression involving only $Y$:
\begin{align}\label{eq:mom-omega-4-2}
	\Omega(\tnnMom{2}{4}) = d\log\frac{\lrangle{12}_{Y}}{\lrangle{13}_{Y}}\wedge d\log\frac{\lrangle{23}_{Y}}{\lrangle{13}_{Y}}\wedge d\log\frac{\lrangle{34}_{Y}}{\lrangle{13}_{Y}}\wedge d\log\frac{\lrangle{14}_{Y}}{\lrangle{13}_{Y}}.
\end{align}
Equivalently, one can derive an expression involving only $\tilde{Y}$ which amounts to replacing $\lrangle{ij}_Y$ with $[ij]_{\tilde{Y}}$ in \eqref{eq:mom-omega-4-2}. We refer the reader to \cite{Damgaard:2019ztj} for further details.

\subsection{Kinematic Definition}
\label{sec:mom-kinematic}

The above definition can, however, be circumvented, and one can define the Momentum Amplituhedron directly in the kinematic space of four-dimensional spinor-helicity variables. Towards this goal, assume that the spacetime signature is $(2,2)$. In this case, spinor-helicity variables are real and independent, and the Lorentz group is $\Spin{2}{2}=\SL{2}{\real}\times\SL{2}{\real}$. Let $\OnShell{4}{n}$ denote the kinematic space of $n$-particle spinor-helicity variables in four dimensions. It is defined as
\begin{align}
	\OnShell{4}{n}\coloneqq\left\{([\lambda],[\tilde\lambda])\in\frac{\Mat{2}{n}{\real}}{\SL{2}{\real}}\times\frac{\Mat{2}{n}{\real}}{\SL{2}{\real}}:\sum_{i=1}^n\lambda_i^a\tilde\lambda_i^{\dot{a}} = 0\right\},
\end{align}
where $\SL{2}{\real}$ acts on $\Mat{2}{n}{\real}$ by left multiplication. Clearly $\OnShell{4}{n}$ has dimension
\begin{align}
	\dim(\OnShell{4}{n}) = 2(\dim(\Mat{2}{n}{\real})-\dim(\SL{2}{\real})) - 4 = 4n -10\,.
\end{align}
It can be regarded as an algebraic subset of $\real^{4n}$ since it can be equivalently defined as the zero set of a polynomial ideal in $\real[\lambda,\tilde\lambda]\coloneqq\real[\lambda_1^1,\lambda_1^2,\ldots,\lambda_n^1,\lambda_n^2,\tilde\lambda_1^1,\tilde\lambda_1^2,\ldots,\tilde\lambda_n^1,\tilde\lambda_n^2]$ generated by the \emph{Schouten identities}
\begin{align}
	\begin{split}
		\lrangle{ij}\lrangle{kl} + \lrangle{il}\lrangle{jk} + \lrangle{ik}\lrangle{lj} = 0\,,
	\end{split}
	\begin{split}
		[ij][kl] + [il][jk] + [ik][lj] = 0\,,
	\end{split}
\end{align}
together with the momentum conservation constraint $\sum_{i=1}^n\lambda_i^a\tilde\lambda_i^{\dot{a}}=0$ for $a,\dot{a}\in[2]$. Let $\OnShellComplex{4}{n}$ denote the complexification of $\OnShell{4}{n}$.

The Grassmannian is naturally covered by a family of (Zariski) open sets, each defined as the set of all points for which a particular Pl\"{u}cker coordinate is non-vanishing. To motivate the definition of the Momentum Amplituhedron in $\OnShell{4}{n}$, consider the open set $U\times\tilde{U}\subset\G{n-k}{n-k+2}\times\G{k}{k+2}$ where 
\begin{align}
	U\coloneqq\{V\in\G{n-k}{n-k+2}: p_{[n-k]+2}(V)\ne0\}\,,\qquad
	\tilde{U}\coloneqq\{\tilde{V}\in\G{k}{k}: p_{[k]+2}(\tilde{V})\ne0\}\,.
\end{align}
Every point $(V,\tilde{V})$ in $U\times\tilde{U}$ admits a representative $([Y],[\tilde{Y}])$ of the form
\begin{align}\label{eq:mom-chart}
	Y\coloneqq
	\begin{pmatrix}
		-y^T& \mathbbb{1}_{(n-k)\times (n-k)}
	\end{pmatrix},
	\qquad
	\tilde{Y}\coloneqq
	\begin{pmatrix}
		-\tilde{y}^T& \mathbbb{1}_{k\times k}
	\end{pmatrix},
\end{align}
for some $y\in\Mat{2}{n-k}{\complex}$ and $\tilde{y}\in\Mat{2}{k}{\complex}$. In fact, \eqref{eq:mom-chart} furnishes a smooth affine chart for $U\times\tilde{U}$. Moreover, there is a canonical choice of representative $([Y^\perp],[\tilde{Y}^\perp])$ for the orthogonal complement $(V^\perp,\tilde{V}^\perp)$ in $U^\perp\times\tilde{U}^\perp$ given by
\begin{align}\label{eq:mom-chart-perp}
	Y^\perp\coloneqq
	\begin{pmatrix}
		\mathbbb{1}_{2\times 2} & y
	\end{pmatrix},
	\qquad
	\tilde{Y}^\perp\coloneqq
	\begin{pmatrix}
		\mathbbb{1}_{2\times 2} & \tilde{y}
	\end{pmatrix},
\end{align}
which defines a smooth affine chart for $U^\perp\times\tilde{U}^\perp$. This choice in \eqref{eq:mom-chart-perp} is canonical in the sense that it relates the maximal minors of $Y^\perp$ and $\tilde{Y}^\perp$ to those of $Y$ and $\tilde{Y}$, respectively, as follows: 
\begin{align}\label{eq:mom-minors}
	p_{J}(Y^\perp) = \epsilon_{J,[n-k+2]\setminus J}\,p_{[n-k+2]\setminus J}(Y)\,,
	\qquad
	p_{\dot{J}}(\tilde{Y}^\perp) = \epsilon_{\dot{J},[k+2]\setminus\dot{J}}\,p_{[k+2]\setminus\dot{J}}(\tilde{Y})\,,
\end{align}
for all $J\in\binom{n-k+2}{2}$ and $\dot{J}\in\binom{k+2}{2}$. 
Given a point $(V,\tilde{V})\in U\times\tilde{U}$, define $\lambda\coloneqq Y^\perp\Lambda^T$ and $\tilde{\lambda}\coloneqq \tilde{Y}^\perp\tilde{\Lambda}^T$ where $\lambda$ is a function of $y\in\Mat{2}{n-k}{\complex}$ and $\tilde{\lambda}$ is a function of $\tilde{y}\in\Mat{2}{k}{\complex}$. This establishes the map
\begin{align}
	\symbolMomHatMap: U\times\tilde{U}\to\Mat{2}{n}{\complex}\times\Mat{2}{n}{\complex}, (V,\tilde{V})\mapsto(\lambda,\tilde{\lambda})\,.
\end{align}

Finally we can define an analogue of the Momentum Amplituhedron in (an affine chart of) $\OnShell{4}{n}$ as per \cite{Ferro:2020ygk}. Let $\MomHatV{k}{n}$ be the image of $\MomV{k}{n}\cap(U\times\tilde{U})$ under $\symbolMomHatMap$ and designate its real part by $\MomHatVReal{k}{n}$. Define the \emph{winding space} $\MomHatW{k}{n}$ as the image of $\MomW{k}{n}$ under $\symbolMomHatMap$. Then the \emph{Momentum Amplituhedron} $\tnnMomHat{k}{n}$ is the closure of $\tpMomHat{k}{n}\coloneqq\MomHatV{k}{n}\cap\MomHatW{k}{n}$ in $\MomHatVReal{k}{n}$ where $\tpMomHat{k}{n}$ is the interior of $\tnnMomHat{k}{n}$. Every point $(\lambda,\tilde{\lambda})\in\MomHatV{k}{n}$ satisfies momentum conservation and hence represents a point $([\lambda],[\tilde{\lambda}])$ in $\OnShellComplex{4}{n}$ where $[\lambda]=\SL{2}{\complex}\lambda$ and $[\tilde{\lambda}]=\SL{2}{\complex}\tilde{\lambda}$. Moreover, for a point $(V,\tilde{V})\in U\times\tilde{U}$ mapped to a point $(\lambda,\tilde{\lambda})=\MomHatMap{V,\tilde{V}}$, we have the following identification between twistor coordinates and spinor-helicity brackets \cite{Arkani-Hamed:2017vfh}
\begin{align}
	\lrangle{ij}_{Y} = \lrangle{ij}\,,\qquad[ij]_{\tilde{Y}}=[ij]\,,
\end{align}
for all $i,j\in[n]$ where $Y$ and $\tilde{Y}$ are defined as per \eqref{eq:mom-chart}. Consequently, the winding space $\MomHatW{k}{n}$ can be defined in analogy with \eqref{eq:mom-winding}: it is the set of points $(\lambda,\tilde{\lambda})$ in $\MomHatMap{U(\real)\times\tilde{U}(\real)}$ for which
\begin{subequations}\label{eq:mom-hat-winding}
	\begin{align}
		&\lrangle{ii+1}>0~\text{for}~i\in[n-1]~\text{and}~\lrangle{n\hat{1}}>0\,,\\
		&\var(\lrangle{12},\lrangle{13},\ldots,\lrangle{1n})=k-2\,,\\
		&[ii+1]>0~\text{for}~i\in[n-1]~\text{and}~[n\hat{1}]>0\,,\\
		&\var([12],[13],\ldots,[1n])=k\,,\\
		&s_I > 0~\text{for all CC subsets $I\subset[n]$ with $2\le|I|\le n-2$}\,,
	\end{align}
\end{subequations}
where for each $I\subset[n]$
\begin{align}
	s_I\coloneqq\sum_{\{i,j\}\in\binom{I}{2}}\lrangle{ij}[ij]\,,
\end{align}
and $\var$ counts the number of sign flips. The canonical form $\Omega(\tnnMomHat{k}{n})$ can be determined using one of two approaches: either via the replacement
\begin{align}\label{eq:mom-hat-omega}
	\Omega(\tnnMomHat{k}{n})=\Omega(\tnnMom{k}{n})\Big|_{\lrangle{ij}_{Y}\to\lrangle{ij},[ij]_{\tilde{Y}}\to[ij]}\,,
\end{align}
or using the \emph{inverse-soft construction} of \cite{He:2018okq}. In this setting, the partial amplitude $A_{n,k}^\text{tree}[1,\ldots,n]$ can be obtained directly from $\Omega(\tnnMomHat{k}{n})$ via \cite{Ferro:2020ygk}
\begin{align}\label{eq:mom-hat-amp}
	A_{n,k}^\text{tree}[1,\ldots,n]=\delta^4(p)\left(\Omega(\tnnMomHat{k}{n})\wedge d^4p\Big|_{d\lambda_i^a\to\eta_i^a,d\tilde{\lambda}_i^{\dot{a}}\to\tilde{\eta}_i^{\dot{a}}}\right),
\end{align}
where $p^{a,\dot{a}}\coloneqq\sum_{i=1}^n\lambda_i^a\tilde{\lambda}_i^{\dot{a}}$.

Let us return to our four-particle MHV example. Applying \eqref{eq:mom-hat-omega} to \eqref{eq:mom-omega-4-2} yields
\begin{align}
\Omega(\tnnMomHat{2}{4}) = d\log\frac{\lrangle{12}}{\lrangle{13}}\wedge d\log\frac{\lrangle{23}}{\lrangle{13}}\wedge d\log\frac{\lrangle{34}}{\lrangle{13}}\wedge d\log\frac{\lrangle{14}}{\lrangle{13}}.
\end{align}
As written, $\Omega(\tnnMomHat{2}{4})$ is a $4$-form in $d\lambda$. Moreover, $\lambda$ defines a $4$-dimensional hypersurface inside $\Mat{2}{4}{\real}$ (i.e.\ $\lambda$ is a function of $y\in\Mat{2}{2}{\real}$). Consequently wedging $\Omega(\tnnMomHat{2}{4})$ with $d^4p=(\tilde{\lambda}d\lambda^T + \lambda d\tilde{\lambda}^T)^4$ is equivalent to wedging it with $(\lambda d\tilde{\lambda}^T)^4$. After performing the replacement in \cref{eq:mom-hat-amp} and enforcing momentum conservation, one finds \cite{Damgaard:2019ztj}
\begin{align}
A_{n,k}^\text{tree}[1,2,3,4] = \frac{\delta^4(p)\delta^4(q)\delta^4(\tilde{q})}{\lrangle{12}\lrangle{23}[12][23]},
\end{align}
where $q^{a,\dot{a}}\coloneqq\sum_{i=1}^n\lambda_i^a\tilde{\eta}_i^{\dot{a}}$ and $\tilde{q}^{a,\dot{a}}\coloneqq\sum_{i=1}^n\eta_i^a\tilde{\lambda}_i^{\dot{a}}$.

\subsection{\texorpdfstring{$D=4$}{Four-dimensional} Twistor-String Map}
\label{sec:mom-string}

One can also regard the Momentum Amplituhedron as the image of the non-negative Grassmannian $\tnnG{2}{n}$ via the four-dimensional twistor-string map of \cite{He:2021llb}. Let $\tnzG{k}{n}$ denote the subset of $\G{k}{n}$ for which all Pl\"{u}cker coordinates are non-zero. The map of interest is defined on $\tnzG{2}{n}$. Given a point in $\tnzG{2}{n}$ one can always choose a representative with entries
\begin{align}\label{eq:mom-tnzGr-2-n}
	X(t,\sigma)=
	\begin{pmatrix}
		t_1&t_2&\cdots&t_n\\
		t_1\sigma_1&t_2\sigma_2&\cdots&t_n\sigma_n
	\end{pmatrix},
\end{align}
where $t=(t_1,\ldots,t_n)\in\complex_{\ne0}^n$ and $\sigma=(\sigma_1,\ldots,\sigma_n)\in\complex^n$ is a tuple of distinct points. Let us denote the Pl\"{u}cker coordinates of the matrix in \eqref{eq:mom-tnzGr-2-n} by $(ij)=t_it_j\sigma_{j,i}$.

The \emph{Veronese map} embeds $\tnzG{2}{n}$ in $\tnzG{k}{n}$ as follows. Let 
$C_\nu(t,\sigma)\in\Mat{k}{n}{\complex}$ and $C^\perp_\nu(t,\sigma)\in\Mat{n-k}{n}{\complex}$ be matrices with entries\footnote{Our definition for $C_\nu(t,\sigma)$ differs from \cite{He:2021llb} and coincides with \cite{Arkani-Hamed:2009kmp} so that $[C_\nu(t,\sigma)]$ is independent of the choice of representative for $C_\nu(t,\sigma)$.} given by
\begin{align}
	C_\nu(t,\sigma)_{\alpha,i} = t_i^{k-1}\sigma_i^{\alpha-1}\,,
	\qquad
	C^\perp_\nu(t,\sigma)_{\tilde{\alpha},i} = \tilde{t}_i^{n-k-1}\sigma_i^{\tilde{\alpha}-1}\,,
\end{align}
where
\begin{align}
	\tilde{t}_i^{n-k-1}\coloneqq
	\frac{\gamma t_i^{n-k-1}}{\prod_{\ell\in[n]\setminus\{i\}}(i\ell)}\,,\qquad
	\gamma^{n-k}\coloneqq (-1)^{\binom{n-k}{2}}\prod_{i<j\in[n]}(ij)\,.
\end{align}
For each $I\in\binom{[n]}{k}$
\begin{align}\label{eq:mom-veronese-pluckers}
	p_I(C_\nu(t,\sigma)) = \textstyle\prod_{i<j\in I}(ij)\,,\qquad
	p_{[n]\setminus I}(C^\perp_\nu(t,\sigma)) = \epsilon_{[n]\setminus I,I}\,p_I(C_\nu(t,\sigma))\,.
\end{align}
The first identity in \eqref{eq:mom-veronese-pluckers} is proven as follows:
\begin{align*}
	p_I(C_\nu(t,\sigma)) = \prod_{i\in I}t_i^{k-1}\prod_{i<j\in I}\sigma_{j,i}=  \prod_{i<j\in I}t_i t_j\sigma_{j,i}= \prod_{i<j\in I}(ij)\,,
\end{align*}
where the first equality follows from the well-known formula for the determinant of the Vandemonde matrix of order $k$. To prove the second identity in \eqref{eq:mom-veronese-pluckers}, let $I'=[n]\setminus I$ and consider the following equality chain: 
\begin{align*}
	p_{I'}(C^\perp_\nu(t,\sigma))= \prod_{i\in I'}\tilde{t}_i^{n-k-1}\prod_{i<j\in I'}\sigma_{j,i}=\prod_{i\in I'}\frac{\tilde{t}_i^{n-k-1}}{t_i^{n-k-1}}\prod_{i<j\in I'}t_i t_j\sigma_{j,i}=
	\frac{\gamma^{n-k}\prod_{i<j\in I'}(ij)}{\prod_{i\in I'}\prod_{\ell\in[n]\setminus\{i\}}(i\ell)}\,.
\end{align*}
Rewriting the denominator as
\begin{align*}
	\prod_{i\in I'}\prod_{\ell\in[n]\setminus\{i\}}(i\ell) = (-1)^{\binom{n-k}{2}}\prod_{i<j\in I'}(ij)^2\prod_{i\in I',j\in I}(ij)\,,
\end{align*}
and substituting $\gamma^{n-k}$ into our expression for $p_{I'}(C^\perp_\nu(t,\sigma))$ yields the desired outcome. Consequently, the Pl\"{u}cker coordinates of $C_\nu(t,\sigma)$ and $C^\perp_\nu(t,\sigma)$ are well-defined. Then  $\nu:\G{2}{n}\to\G{k}{n},[X(t,\sigma)]\mapsto[C_\nu(t,\sigma)]$ and $\nu^\perp:\G{2}{n}\to\G{n-k}{k},[X(t,\sigma)]\mapsto[C_\nu^\perp(t,\sigma)]$ define rational maps sharing $\tnzG{2}{n}$ as their domain. They satisfy
\begin{subequations}
\begin{alignat}{2}
	&\nu(\tnzG{2}{n})\subset\tnzG{k}{n}\,,\qquad&&\nu^\perp(\tnzG{2}{n})\subset\tnzG{n-k}{n}\,,\\
	&\nu(\tpG{2}{n})\subset\tpG{k}{n}\,,&&
		\nu^\perp(\tpG{2}{n})\subset\tpG{n-k}{n}\,.
\end{alignat}
\end{subequations}
The former is called the \emph{Veronese map} \cite{Arkani-Hamed:2009kmp}. Moreover, $C_\nu(t,\sigma)C^\perp_\nu(t,\sigma)^T = \mathbbb{0}_{k\times(n-k)}$. 

It is natural to decompose $\tpG{2}{n}$ as $\tpModuli{0}{n}\times\projective(\real_{>0}^n)$ where the positive moduli space $\tpModuli{0}{n}$ is parametrised by $\sigma=(\sigma_1,\ldots,\sigma_n)\in\real^n$ with $\sigma_1<\ldots<\sigma_n$ via \eqref{eq:abhy-tpModuli-0-n} and $\projective(\real_{>0}^n)$ is parametrised by the homogenous coordinates $[t]=[t_1:\ldots:t_n]$. These homogeneous coordinates can be regarded \emph{little group} variables \cite{Damgaard:2020eox}. In this setting, the canonical form $\Omega(\tnnG{2}{n})$ can be expressed as
\begin{align}
	\omega_n^{\text{WS}(4)} \coloneqq \omega_n^\text{WS}\wedge\Omega(\projective(\real_{\ge0}^n))=\Omega(\tnnG{2}{n})\,,
\end{align}
where $\omega_n^\text{WS}=\Omega(\tnnModuli{0}{n})$ is the Parke-Taylor form given by \eqref{eq:abhy-world-Omega} and $\Omega(\projective(\real_{\ge0}^n))=\sum_{i=1}^n(-1)^{i+1}\bigwedge_{j\ne i}\frac{dt_j}{t_j}$.

Finally, let $Y(t,\sigma)\coloneqq C_\nu^\perp(t,\sigma)\Lambda$ and $\tilde{Y}(t,\sigma)\coloneqq C_\nu(t,\sigma)\tilde{\Lambda}$. The \emph{$D=4$ twistor-string map} is the rational map
\begin{align}
	\symbolMomTwistorMap:\G{2}{n}\to\G{n-k}{n-k+2}\times\G{k}{k+2},[X(t,\sigma)]\mapsto([Y(t,\sigma)],[\tilde{Y}(t,\sigma)])\,,
\end{align}
with domain $\tnzG{2}{n}$. Composing $\symbolMomTwistorMap$ with $\symbolMomHatMap$ produces the \emph{$D=4$ scattering equations} of \cite{Geyer:2014fka}: the identities $\tilde{Y}^\perp(t,\sigma) \tilde{Y}^T(t,\sigma)=\mathbbb{0}_{2,k}$ and $Y^\perp(t,\sigma)Y^T(t,\sigma)=\mathbbb{0}_{2,n-k}$ imply
\begin{align}\label{eq:mom-SE}
	C_\nu(t,\sigma)\tilde{\lambda}^T=0\,,\qquad
	C_\nu^\perp(t,\sigma)\lambda^T=0\,.
\end{align}
where $\lambda$ is a function of $y\in\Mat{2}{n-k}{\complex}$ and $\tilde{\lambda}$ is a function of $\tilde{y}\in\Mat{2}{k}{\complex}$. Momentum conservation $\lambda\tilde{\lambda}^T=\mathbbb{0}_{2\times2}$ implies that \eqref{eq:mom-SE} contains only $2n-4$ independent equations. By gauge-fixing any three $\sigma$-variables (e.g.\ $(\sigma_1,\sigma_{n-1},\sigma_n)=(0,1,\infty)$) and any $t$-variable (e.g.\ $t_n=1$), \eqref{eq:mom-SE} defines a zero-dimensional complete intersection ideal $\mathcal{I}_n^{(4)}\subset\complex(y,\tilde{y})[\sigma,t]$ whose zero-set contains $E_{n-3,k-2}$ distinct points in $\tnzG{2}{n}$. Here $E_{n-3,k-2}$ is an \emph{Eulerian number} \cite{Cachazo:2013iaa}. The restriction $\symbolMomTwistorMap|_{\tpG{2}{n}}:\tpG{2}{n}\to\tpMom{k}{n}$ is a bijection for $k=2$ and $k=n-2$, and there is strong numerical evidence to support this claim also for $2<k<n-2$ \cite{He:2021llb}; the authors of \cite{He:2021llb} have numerically checked that for each point $(\lambda,\tilde{\lambda})$ in $\tpMomHat{k}{n}$, out of the $E_{n-3,k-2}$ (generally complex) solutions to the $D=4$ scattering equations, there is always a unique point in $\tpG{2}{n}$. Thus, the $D=4$ twistor-string map $\symbolMomTwistorMap$ is (conjecturally) a morphism between $(\G{2}{n},\tnnG{2}{n})$ and $(\Mom{k}{n},\tnnMom{k}{n})$. Consequently, the canonical form $\Omega(\tnnMom{k}{n})$ can be computed as the pushforward of $\omega_n^{\text{WS}(4)}$ via the $D=4$ twistor-string map $\symbolMomTwistorMap$
\begin{align}\label{eq:mom-pushforward-twistor-string}
	\Omega(\tnnMom{k}{n}) = \symbolMomTwistorMap_\ast\omega_n^{\text{WS}(4)}\,.
\end{align}
Equivalently, the canonical form $\Omega(\tnnMomHat{k}{n})$ can be computed as the pushforward of $\omega_n^{\text{WS}(4)}$ via the $D=4$ scattering equations
\begin{align}\label{eq:mom-pushforward-scattering-equations}
	\Omega(\tnnMomHat{k}{n}) = (\mathcal{I}_n^{(4)})_\ast\omega_n^{\text{WS}(4)}\,.
\end{align}

\section{Boundary Stratification and Rank Generating Function}
\label{sec:mom-boundaries}

The boundaries of the Momentum Amplituhedron $\tnnMom{k}{n}$ correspond to the singularities of the partial amplitude $A^\text{tree}_{n,k}[1,\ldots,n]$ \cite{Damgaard:2019ztj}. In this section, we explore the boundary stratification of $\tnnMom{k}{n}$, summarising the results of \cite{Ferro:2020lgp,Moerman:2021cjg}; we characterise this boundary stratification, derive a rank generating function for it, and prove that $\tnnMom{k}{n}$ has Euler characteristic one. In addition, we supply the covering relations for the Momentum Amplituhedron.

Since the Momentum Amplituhedron map $\symbolMomMap$ is (conjecturally) a morphism of positive geometries, we speculate that the $\symbolMomMap$-induced boundary stratification of $\tnnMom{k}{n}$ is the true boundary stratification. Thus, we will drop the epithet ``$\symbolMomMap$-induced''. Moreover, since the boundaries of $\tnnMom{k}{n}$ are images of positroid cells labelled by Grassmannian graphs, we will enumerate the boundaries of $\tnnMom{k}{n}$ accordingly.

\begin{figure}
	\centering
	\null
	\hfill
	\begin{subfigure}[t]{0.3\textwidth}
		\centering
		\includegraphics[scale=0.45]{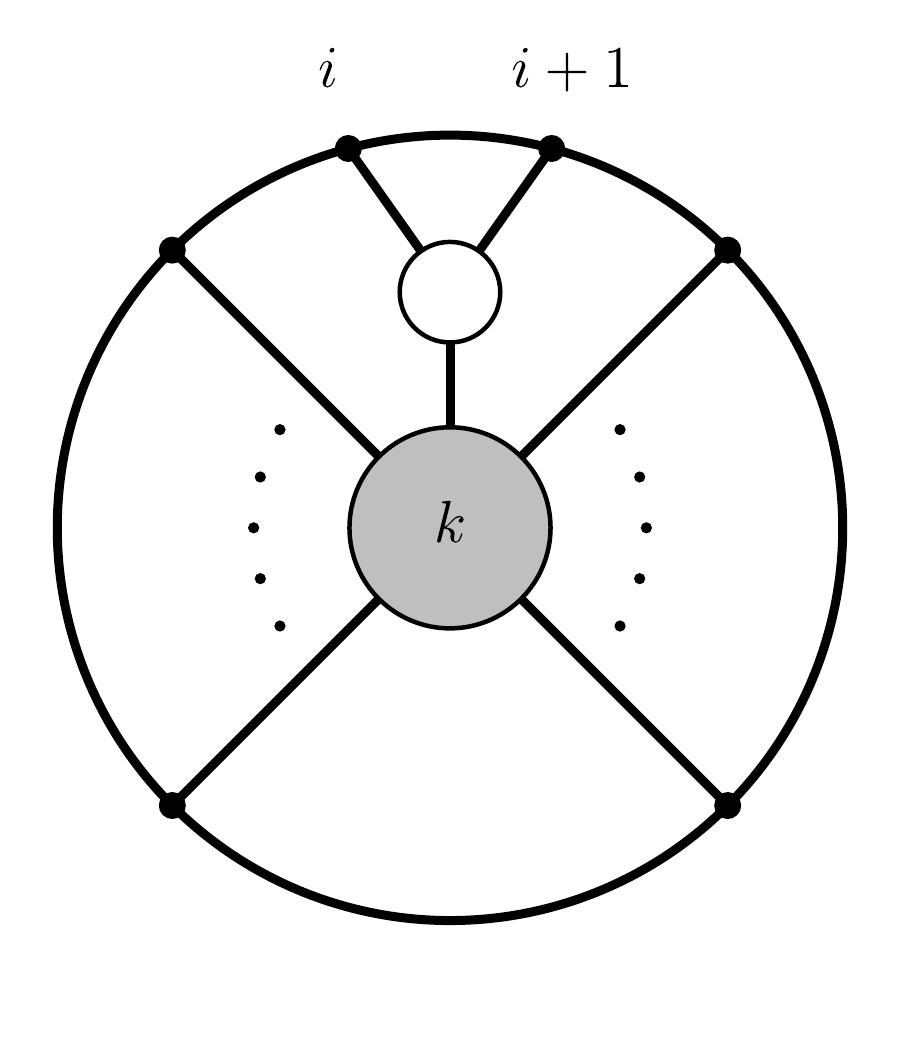}
		\caption{$\lrangle{ii+1}_Y=0$}
		\label{fig:mom-facet-collinear-preserving}
	\end{subfigure}
	\hfill
	\begin{subfigure}[t]{0.3\textwidth}
		\centering
		\includegraphics[scale=0.45]{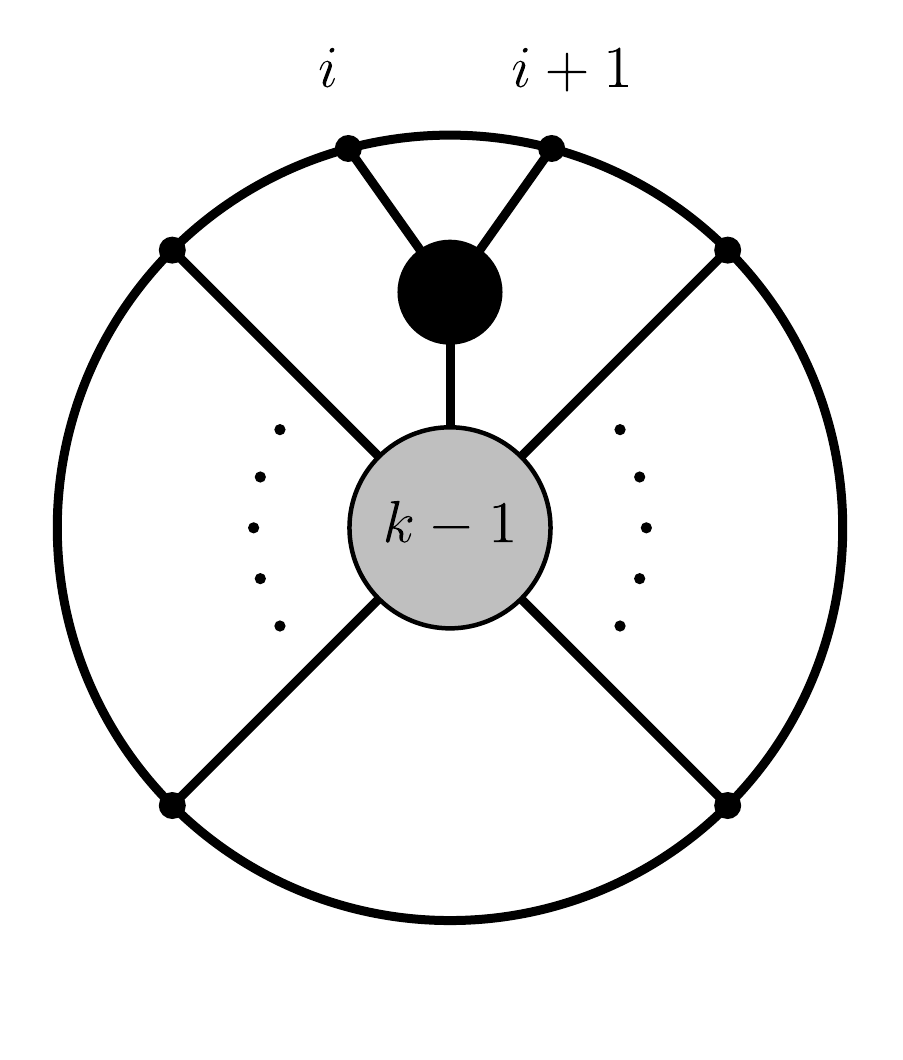}
		\caption{$[ii + 1]_{\tilde{Y}} = 0$}
		\label{fig:mom-facet-collinear-reducing}
	\end{subfigure}	
	\hfill
	\begin{subfigure}[t]{0.3\textwidth}
		\centering
		\includegraphics[scale=0.45]{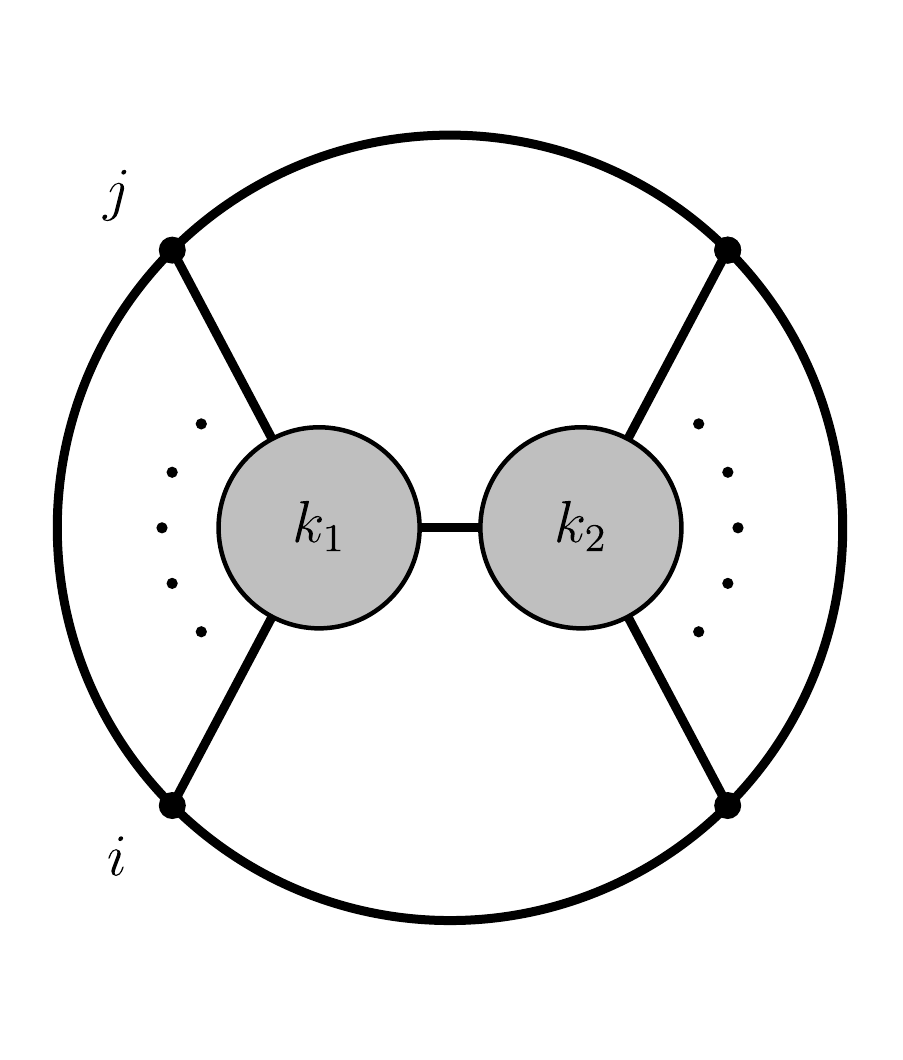}
		\caption{$S_I(\tilde{Y},Y)=0$}
		\label{fig:mom-facet-factorization}
	\end{subfigure}
	\hfill
	\null
	\caption{Grassmannian graph labels for facets of the Momentum Amplituhedron. Labels inside vertices denote helicities. In \cref{fig:mom-facet-factorization}, $I = \{i,i+1,\ldots,j\}$ is a cyclically consecutive subset of $[n]$ with $2<|I|<n-2$ and $k_1+k_2=k+1$.}
	\label{fig:mom-facets}
\end{figure}

The facets of the Momentum Amplituhedron were first studied in \cite{Damgaard:2019ztj}. They fall into three classes: $\lrangle{ii+1}_Y=0$ and $[ii+1]_{\tilde{Y}}=0$ parametrise helicity-preserving and helicity-reducing collinear limits, respectively, while $S_I(\tilde{Y},Y)=0$ denotes a factorisation channel for each cyclically consecutive $I\subset[n]$ with $2<|I|<n-2$. These three classes are the analogues of the factorisation channels and the two collinear limit types discussed in \cref{sec:mom-sing}. All three are present for $2<k<n-2$. For $k=2$ (resp., $k=n-2$) only facets corresponding to helicity-preserving (resp., helicity-reducing) collinear limits exist. To obtain the facets of the Momentum Amplituhedron, we apply the first algorithm of \cref{sec:pos-boundaries} to any tiling\footnote{In the physics literature, a tiling is often referred to as a ``triangulation''.}. (The BCFW recursion relations generate a class of tilings called \emph{BCFW tilings}.) These facets are labelled by the Grassmannian graphs depicted in \cref{fig:mom-facets} \cite{Ferro:2020lgp}. Unsurprisingly, the Grassmannian graphs depicted in \cref{fig:mom-facets} are precisely the on-shell diagrams for the singularities of the corresponding facets.

Higher codimension boundaries of the Momentum Amplituhedron can be generated using the second algorithm of \cref{sec:pos-boundaries} which is implemented in the \Mathematica{} package \amplituhedronBoundaries{} \cite{Lukowski:2020bya}. Using this package, the boundary stratification of $\tnnMom{k}{n}$ for all $n\le 9$ was synthesised in \cite{Ferro:2020lgp}. After careful analysis, the authors of \cite{Ferro:2020lgp} suggested a pictorial characterisation for all boundaries. Their pictorial characterisation was later reformulated in \cite{Moerman:2021cjg} as follows: $\symbolMomMap(\tnnPos{\Gamma})$ is a boundary of $\tnnMom{k}{n}$ if and only if $\Gamma$ is a Grassmannian forest of type $(k,n)$. Moreover, the boundary stratification of $\tnnMom{k}{n}$ is an induced subposet of the postroid stratification of $\tnnG{k}{n}$. Its covering relations are depicted in \cref{fig:grass-covering}.

\begin{figure}
	\centering
	\null
	\hfill
	\begin{subfigure}[b]{0.3\textwidth}
		\centering
		\includegraphics[scale=0.45]{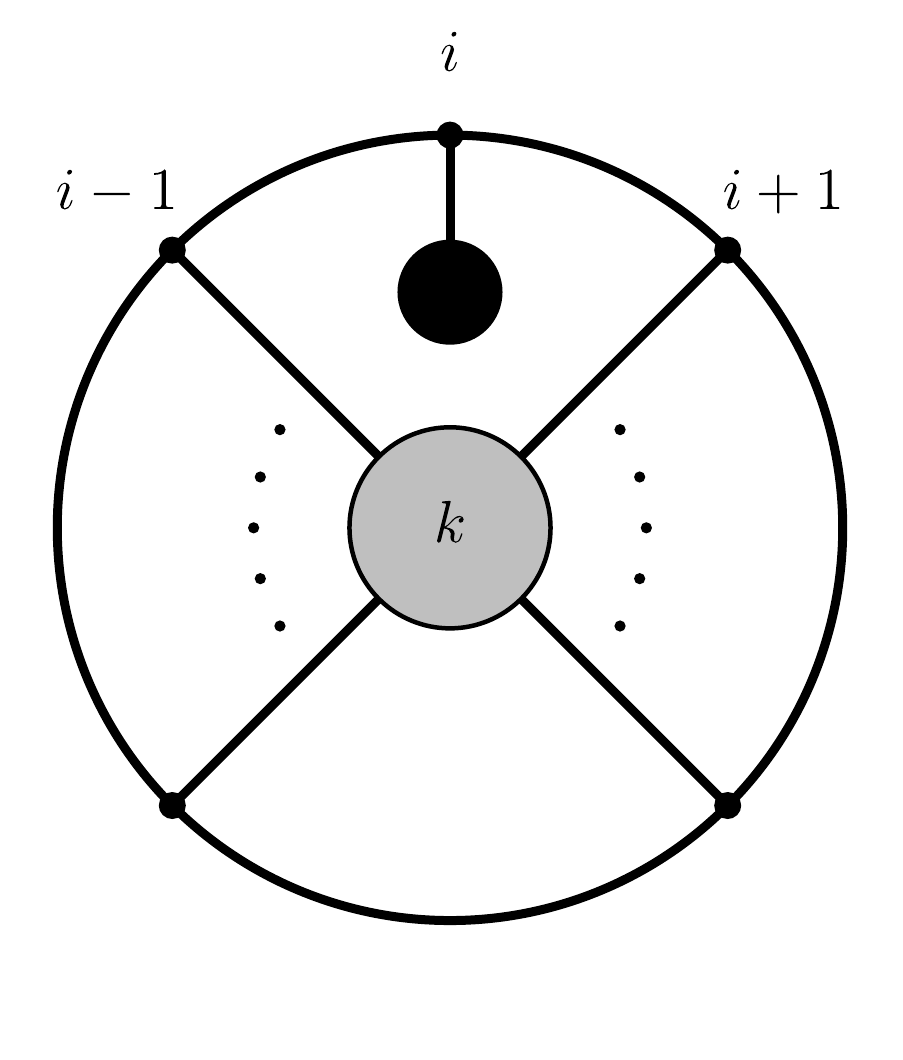}
		\caption{$\lrangle{ii+1}_Y=0=\lrangle{ii-1}_Y$}
		\label{fig:mom-boundary-soft-preserving}
	\end{subfigure}
	\hfill
	\begin{subfigure}[b]{0.3\textwidth}
		\centering
		\includegraphics[scale=0.45]{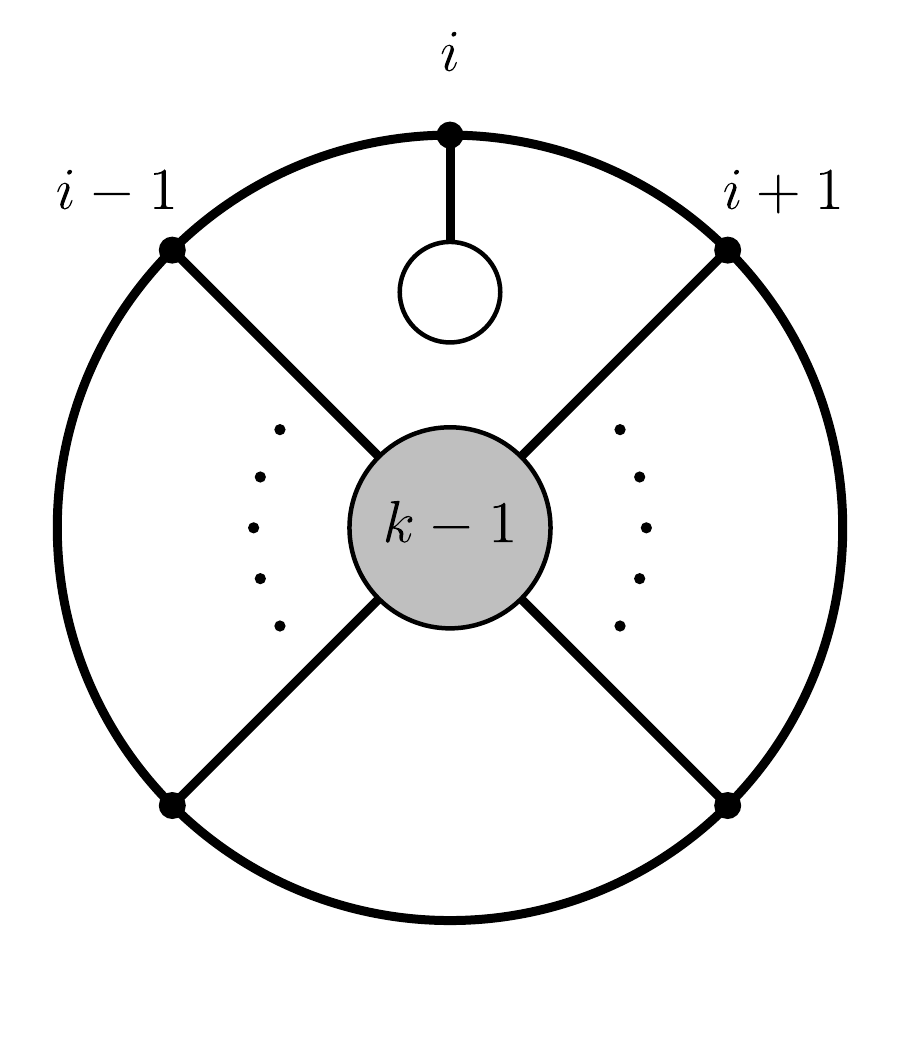}
		\caption{$[ii + 1]_{\tilde{Y}} = 0 = [i i-1]_{\tilde{Y}}$}
		\label{fig:mom-boundary-soft-reducing}
	\end{subfigure}
	\hfill
	\null
	\caption{Grassmannian graphs labels for codimension-two boundaries of the Momentum Amplituhedron corresponding to soft limits. Labels inside vertices denote helicities.}
	\label{fig:mom-boundary-soft}
\end{figure}

At codimension two, one finds additional factorisations of complete subgraphs of facet-labelling Grassmannian graphs as well as further collinear limits. A novel physical singularity emerges at the intersection of facets corresponding to two consecutive collinear limits of the same type. In this case, the Grassmannian graph contains a boundary leaf, an internal vertex of degree $1$ connected to a boundary vertex, detached from a complete subgraph, see \cref{fig:mom-boundary-soft}. The boundary leaf is coloured white (resp., black) in the helicity-preserving (resp., helicity-reducing) case. These are exactly the soft limits described in \cref{sec:mom-sing}. 

We conclude this section by deriving a rank generating function for enumerating the boundaries of the Momentum Amplituhedron according to their dimension. The following definitions, conjectures and results are taken from \cite{Moerman:2021cjg}, unless otherwise stated.

One can associate a dimension to each refinement-equivalence class of Grassmannian forests \cite{Ferro:2020lgp,Moerman:2021cjg} as follows. For each internal vertex $v$ in a Grassmannian forest, define the statistic
\begin{align}\label{eq:mom-dimension-m}
	m(v)\coloneqq
	\begin{cases}
		2\deg(v)-4\,,& \text{if}~v~\text{is generic}\\
		\deg(v)-1\,,& \text{otherwise}
	\end{cases}.
\end{align}
Then the \emph{Momentum Amplituhedron dimension} or \emph{$\symbolMom$-dimension} of a Grassmannian forest $F$, denoted by $\dim_{\symbolMom}(F)$, is defined as
\begin{align}\label{eq:mom-dimension-F}
	\dim_{\symbolMom}(F)\coloneqq\sum_{T\in\Trees(F)}\dim_{\symbolMom}(T)\,.
\end{align}
where $\dim_{\symbolMom}(T)$ is the $\symbolMom$-dimension of a Grassmannian tree $T$ given by
\begin{align}\label{eq:mom-dimension-T}
	\dim_{\symbolMom}(T)\coloneqq
	\begin{cases}
		n-1\,,&\text{if}~n\le2\\
		1+\sum_{v\in\Vertices_\text{int}(T)}(m(v)-1)\,,&\text{otherwise}
	\end{cases}.
\end{align}
Given a Grassmannian tree $T$ of type $(k,n)$, if we replace the helicity of each internal vertex $v$ by $\deg(v)-h(v)$, we obtain a Grassmannian tree of type $(n-k,n)$ by \eqref{eq:grass-helicity-T} and \eqref{eq:grass-multiplicity-T}. Consequently, this map gives a $\symbolMom$-dimension-preserving bijection between the Grassmannian forests of type $(k,n)$ and type $(n-k,n)$. Moreover, $\symbolMom$-dimension is an invariant of refinement-equivalence classes of Grassmannian forests. To prove this, it is sufficient to show that the $\symbolMom$-dimensions of refinement-equivalent Grassmannian trees are the same. Without loss of generality, let $T$ and $T'$ be refinement-equivalent Grassmannian trees where $T'$ is obtained from $T$ by applying a single vertex contraction move, resulting in some non-generic internal vertex $v_\ast\in\Vertices_\text{int}(T')$. Let $\Vertices_\text{int}(T)\setminus\Vertices_\text{int}(T')=\{v_1,v_2\}$ be the two non-generic internal vertices whose contraction results in $v_\ast$. Clearly $\deg(v_\ast)=\deg(v_1)+\deg(v_2)-2$. Then
\begin{align*}
	1+(m(v_1)-1)+(m(v_2)-1)&= m(v_1)+m(v_2)-1 \\
	&=(\deg(v_1)-1)+(\deg(v_2)-1) -1 \\
	&=(\deg(v_1)+\deg(v_2)-2) - 1 \\
	&=\deg(v_\ast)-1=m(v_\ast)=1+(m(v_\ast)-1)\,,
\end{align*}
from which it follows that $\dim_{\symbolMom}(T)=\dim_{\symbolMom}(T')$. Finally, we conjecture that 
\begin{align}
	\dim(\Phi(\tpPos{F})) = \dim_{\symbolMom}(F)\,,
\end{align}
for each Grassmannian forest $F$ of type $(k,n)$ \cite{Ferro:2020lgp,Moerman:2021cjg}.

A rank generating function for the Momentum Amplituhedron was derived in \cite{Moerman:2021cjg} using the results presented in \cref{chp:enum}. We quote the result below; its proof can be found in the original paper.	The number of contracted Grassmannian trees of type $(k,n)$ with $\symbolMom$-dimension $r$ is given by $[x^ny^kq^r] \mathcal{G}^{\symbolMom}_{\tree}(x,y,q)$ 
where 
\begin{align}\label{eq:mom-trees-GF}
	\mathcal{G}^{\symbolMom}_{\tree}(x,y,q)&=x\left(1+y+yq\,C^{\lrangle{-1}}_x(x,y,q)\right),
\end{align}
with
\begin{align}\label{eq:mom-trees-C}
	C(x,y,q)&=\frac{x(1-x (1+y) q^2-x^2 y q^2 (1+q-q^2)-x^4 y^2 q^5 (1+q))}{(1+xq)(1+xyq)(1-xq^2)(1-xyq^2)},
\end{align}
and the compositional inverse is with respect to the variable $x$. Moreover, the number of contracted Grassmannian forests of type $(k,n)$ with $\symbolMom$-dimension $r$ is given by $[x^ny^kq^r]\mathcal{G}^{\symbolMom}_{\forest}(x,y,q)$ where
\begin{align}\label{eq:mom-forest-GF-1}
	x\, \mathcal{G}^{\symbolMom}_{\forest}(x,y,q)=\left(\frac{x}{1+\mathcal{G}^{\symbolMom}_{\tree}(x,y,q)}\right)^{\lrangle{-1}}_x,
\end{align}
and the compositional inverse is with respect to the variable $x$. Equivalently, 
\begin{align}\label{eq:mom-forest-GF-2}
	[x^n]\mathcal{G}^{\symbolMom}_{\forest}(x,y,q) = \frac{1}{n+1} [x^n] 
	\left(1+\mathcal{G}^{\symbolMom}_{\tree}(x,y,q) \right)^{n+1}.
\end{align}

It was checked by the authors of \cite{Moerman:2021cjg} that this rank generating function reproduces all results for $[x^ny^k]\mathcal{G}^{\symbolMom}_{\forest}(x,y,q)$ listed in Tables 1 and 2 of \cite{Ferro:2020lgp}. The results of \cite{Ferro:2020lgp} include as high as $(n,k)=(12,2)$. Table 1 of \cite{Moerman:2021cjg} records $[x^ny^k]\mathcal{G}^{\symbolMom}_{\forest}(x,y,q)$ for all $4\le n\le12$ and for all $2\le k\le \lfloor\frac{n}{2}\rfloor$. 

It follows from \cref{eq:mom-trees-GF,eq:mom-trees-C,eq:mom-forest-GF-1} that $\mathcal{G}^{\symbolMom}_{\tree}(x,y,q)$ and $\mathcal{G}^{\symbolMom}_{\forest}(x,y,q)$ are algebraic rank generating functions of degree $5$ and $6$, respectively, satisfying \eqref{eq:mom-GF-tree-rel} and \eqref{eq:mom-GF-forest-rel}, respectively. 

As a corollary of \cref{eq:mom-trees-GF,eq:mom-trees-C,eq:mom-forest-GF-1,eq:mom-forest-GF-2}, the number of $0$-dimensional boundary strata of $\tnnMom{k}{n}$ is $\binom{n}{k}$, i.e.\ $[x^n y^k q^0] \mathcal{G}^{\symbolMom}_{\forest}(x,y,q) = \binom{n}{k}$.

Based on the evidence given in \cite{Ferro:2020lgp}, it was conjectured in \cite{Moerman:2021cjg} that 
\begin{align}
	\tnnMom{k}{n} = \bigsqcup_{F\in\Forest_{k,n}^{\symbolG}}\Phi(\tpPos{F})
\end{align}
is a regular CW decomposition of $\tnnMom{k}{n}$, where the disjoint union is over contracted Grassmannian forests of type $(k,n)$. With this assumption, the Euler characteristic of $\tnnMom{k}{n}$, the coefficient of $x^ny^k$ in $\mathcal{G}^{\symbolMom}_{\forest}(x,y,-1)$, is one, as was calculated in \cite{Moerman:2021cjg}. This provides partial evidence that the Momentum Amplituhedron is homeomorphic to a closed ball.

\section{Summary}
The Momentum Amplituhedron connects two remarkable descriptions of tree-level amplitudes in SYM. On the one hand, it is the image of the non-negative Grassmannian via the Momentum Amplituhdedron map. On the other hand, it is the image of a positive geometry in the appropriate CHY moduli space via the four-dimensional twistor-string map. As we have seen in this chapter, the Grassmannian definition was integral in determining the boundary stratification of the Momentum Amplituhedron. Knowing the positroid stratification, we found the (induced) boundary stratification using the algorithm from \cref{chp:pos}. The data shows that contracted Grassmannian forests are the correct combinatorial labels for enumerating boundaries. The covering relations for contracted Grassmannian forests are precisely the allowed factorisation channels of tree-level amplitudes in SYM, thereby strengthening our results. Knowing the correct combinatorial labels, we determined an algebraic rank generating function for the boundaries of the Momentum Amplituhedron. In addition, we were able to use this generating function to prove some mathematical statements concerning the Momentum Amplituhdron. In the next chapter (\cref{chp:omom}), we will encounter a similar story for the orthogonal Momentum Amplituhedron.
	\chapter{The Orthogonal Momentum Amplituhedron}
\label{chp:omom}

\lettrine{T}{he orthogonal Momentum Amplituhedron} is the analogue of the Momentum Amplituhedron relevant to three-dimensional physics; it is (conjecturally) a positive geometry describing particle scattering in ABJM theory at the tree level \cite{Huang:2021jlh,He:2021llb}. In this chapter, we study the orthogonal Momentum Amplituhedron, repeating much of the analysis of \cref{chp:mom}. Again, our principal focus is on physical singularities encoded via the positive geometry’s boundary stratification. We will review the results of \cite{Lukowski:2021fkf} and explore this boundary stratification from two perspectives. On the one hand, it is an induced subposet of the orthitroid stratification. Recall that the orthitroid stratification is the non-negative orthogonal Grassmannian’s boundary stratification. Specifically, orthogonal Grassmannian forests (acyclic orthogonal Grassmannian graphs) provide combinatorial labels for orthogonal Momentum Amplituhedron boundaries. However, these faces also relate to those of the worldsheet associahedron via a simple diagrammatic map. We present a detailed analysis of both descriptions. Moreover, with a proper combinatorial description of orthogonal Momentum Amplituhedron boundaries, we derive an algebraic rank generating function using the results of \cref{chp:enum} and deduce the Euler characteristic. 

\section{ABJM Theory}

The theory of \emph{Aharony--Bergman--Jafferis--Maldacena (ABJM)} \cite{Aharony:2008ug,Hosomichi:2008jb} is a three-dimensional Chern-Simons matter theory with $\mathcal{N}=6$ supersymmetry. It is holographically dual to M-theory on $AdS_4\times S^7$ and it serves as a toy model in condensed matter physics. 

There are analogues in ABJM theory for many of the remarkable properties of $\mathcal{N}=4$ SYM. The counterpart to $\SU{4|4}$ dual-superconformal symmetry (and its Yangian extension) of tree-level amplitudes in $\mathcal{N}=4$ SYM \cite{Drummond:2008vq,Drummond:2009fd} is $\textrm{OSp}(6|4)$ dual-superconformal symmetry \cite{Bargheer:2010hn,Huang:2010qy,Gang:2010gy}. There is a twistor-string-inspired formula for tree-level amplitudes in ABJM theory \cite{Huang:2012vt} which resembles the well-known Rioban--Spradlin--Volovich--Witten (RSVW) formula \cite{Witten:2003nn,Roiban:2004vt,Roiban:2004yf}. Moreover, the leading singularities of amplitudes in ABJM theory can be expressed as contour integrals over the orthogonal Grassmannian \cite{Huang:2013owa,Huang:2014xza}. As before, the contour integrals represent a sum of residues. In this case, each residue selects an orthitroid cell in the non-negative orthogonal Grassmannian labelled by an OG graph.

The spectrum of ABJM theory consists of four complex scalars $X_I$, four complex fermions $\psi^I$, and their complex conjugates, represented by $\bar{X}^I$ and $\bar{\psi}_I$, transforming in the bi-fundamental representation of the colour group $\U{N}\times\U{N}$. The labels $I=1,\ldots,4$ refer to the $\SU{4}$ R-symmetry group. These fields can be collected into two superfields, defined as
\begin{align}
	\Phi^{\mathcal{N}=6}&=X_4 + \eta_A\psi^A - \frac{1}{2!}\eta_A\eta_B\varepsilon^{ABC}X_C - \frac{1}{3!}\eta_A\eta_B\eta_C\varepsilon^{ABC}\psi^4\,,\\
	\bar{\Psi}^{\mathcal{N}=6}&=\bar{\psi}_4 + \eta_A\bar{X}^A - \frac{1}{2!}\eta_A\eta_B\varepsilon^{ABC}\bar{\psi}_C - \frac{1}{3!}\eta_A\eta_B\eta_C\varepsilon^{ABC}\bar{X}^4\,,
\end{align}
where the indices $A,B,C$ run over $1,2,3$.

In three-dimension, the little group $\integer_2$ acts on the $D=3$ on-shell superspace variables $(\lambda_i^a|\eta_{i,A})$ as follows:
\begin{align}\label{eq:omom-LG}
	\lambda_i^a\mapsto-\lambda_i^a\,,\qquad\eta_{i,A}\mapsto-\eta_{i,A}\,.
\end{align}
Under a little group transformation $\Phi^{\mathcal{N}=6}$ is unchanged while $\bar{\Psi}^{\mathcal{N}=6}$ changes sign. Applying \eqref{eq:omom-LG} to the $i$\textsuperscript{th} particle produces the following transformation law for tree-level amplitudes:
\begin{align}
	A_n^{\tree} \mapsto
	\begin{cases}
		+A_n^{\tree}\,,&\text{if the $i$\textsuperscript{th} particle is a scalar}\\
		-A_n^{\tree}\,,&\text{if the $i$\textsuperscript{th} particle is a fermion}
	\end{cases}.
\end{align}
Only amplitudes with even multiplicity are non-vanishing. Tree-level amplitudes can be colour decomposed into partial amplitudes, for which there are only two classes:
\begin{align}
A_{2k}^{\tree}[\bar{\Psi}^{\mathcal{N}=6}_1,\Phi^{\mathcal{N}=6}_2,\bar{\Psi}^{\mathcal{N}=6}_3,\ldots,\Phi^{\mathcal{N}=6}_{2k}]\,,\qquad A_{2k}^{\tree}[\Phi^{\mathcal{N}=6}_1,\bar{\Psi}^{\mathcal{N}=6}_2,\Phi^{\mathcal{N}=6}_3,\ldots,\bar{\Psi}^{\mathcal{N}=6}_{2k}]\,.
\end{align}

The Grassmann degree of partial amplitudes is $3k$. In order to uplift partial amplitudes from rational functions to rational forms \cite{He:2018okq}, it is convenient to work with an $\mathcal{N}=4$ description of ABJM theory \cite{Huang:2021jlh,He:2021llb}, obtained form the familiar $\mathcal{N}=6$ formalism via SUSY reduction. In this new setup, there are four superfields given by
\begin{align}
	\Phi^{\mathcal{N}=4}&\coloneqq \Phi^{\mathcal{N}=6}|_{\eta_3\to0} = X_4+\eta_a\psi^a-\eta_1\eta_2X^3\,,\\
	\bar{\Phi}^{\mathcal{N}=4}&\coloneqq \int d\eta^3\bar{\Psi}^{\mathcal{N}=6} = \bar{X}^3+\eta_a\varepsilon^{ab}\bar{\psi}_b-\eta_1\eta_2\bar{X}^4\,,\\
	\Psi^{\mathcal{N}=4}&\coloneqq \int d\eta^3\Phi^{\mathcal{N}=6} = \psi^3+\eta_a\varepsilon^{ab}X_b-\eta_1\eta_2\psi^4\,,\\
	\bar{\Psi}^{\mathcal{N}=4}&\coloneqq \bar{\Psi}^{\mathcal{N}=6}|_{\eta_3\to0} = \bar{\psi}_4+\eta_a\bar{X}^a-\eta_1\eta_2\bar{\psi}_3\,,
\end{align}
where $\eta^{12}=-1$. The $\mathcal{N}=4$ versions of amplitudes are obtained by the same SUSY reduction: $\eta^3\to0$ for $k$ particles and an integration over $\eta^3$ is performed for the other $k$ particles. In this chapter, we focus on the partial amplitude
\begin{align}
	A_{2k}^{\tree}[\bar{1},2,\bar{3},\ldots,\bar{2k}]\coloneqq A_{2k}^{\tree}[\bar{\Phi}_1^{\mathcal{N}=4},\Phi_2^{\mathcal{N}=4},\bar{\Phi}_3^{\mathcal{N}=4},\ldots,\Phi_{2k}^{\mathcal{N}=4}]\,,
\end{align}
whose Grassmannian description is given by
\begin{align}
&A_{2k}^{\tree}[\bar{1},2,\bar{3},\ldots,\bar{2k}]\\
\nonumber &=\oint_{\gamma}\frac{d^{k\times2k}C}{\vol(\GL{k}{\complex})}\frac{(1,3,5,\ldots,2k-1)}{\prod_{i=1}^{k}(i,\ldots,i+k)}\delta^{k\times k}(C\eta C^T)\delta^{k\times 2}(C\lambda^T)\delta^{k\times2}(C\eta^T)\,.
\end{align}
Replacing
\begin{align}
\eta_i^a\mapsto d\lambda_i^a\,,
\end{align}
and stripping off $d^3q$, where $q^{a,b}\coloneqq\sum_{i=1}^{2k}\lambda_i^a(d\lambda_i^b)$, one obtains a non-vanishing $(2k-3)$-form which is the canonical form of the orthogonal Momentum Amplituhedron. 

In this chapter, we explore the orthogonal momentum amplituderon \cite{Huang:2021jlh,He:2021llb} which captures $A_{2k}^{\tree}[\bar{1},2,\bar{3},\ldots,\bar{2k}]$. While there is no known loop-level analogue of the orthogonal Momentum Amplituhedron, the all-loop integrand for the ABJM four-point amplitude can be calculated from the all-loop Amplituhedron by projecting onto three-dimensional kinematics \cite{He:2022sas}.

\section{The Orthogonal Momentum Amplituhedron}
\label{sec:omom-def}

\begin{figure}
	\centering
	\includegraphics{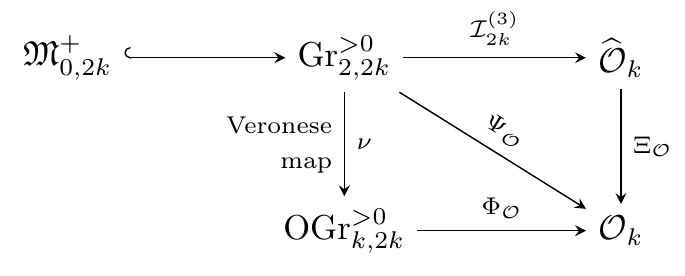}
	\caption{Web of connections between spaces associated with the orthogonal Momentum Amplituhedron.}
	\label{fig:omom-web-spaces}
\end{figure}

The definition of the orthogonal Momentum Amplituhedron parallels that of the Momentum Amplituhedron. It can be defined inside a Grassmannian or directly in the kinematic space of three-dimensional spinor-helicity variables \cite{Huang:2021jlh,He:2021llb}. In this section, we review these definitions. As was the case in \cref{chp:mom}, the former definition will facilitate an efficient classification of all boundaries of the orthogonal Momentum Amplituhedron \cite{Lukowski:2021fkf}. These faces correspond to all physical singularities of partial amplitudes in ABJM theory. Moreover, the orthogonal Momentum Amplituhedron is related to the worldsheet moduli space via the three-dimensional twistor-string map of \cite{He:2021llb}. \cref{fig:omom-web-spaces} depicts the multiple definitions of the orthogonal Momentum Amplituhedron. Throughout this section, we will let $k$ be an integer, greater than or equal to two.

\begin{figure}
	\centering
	\includegraphics{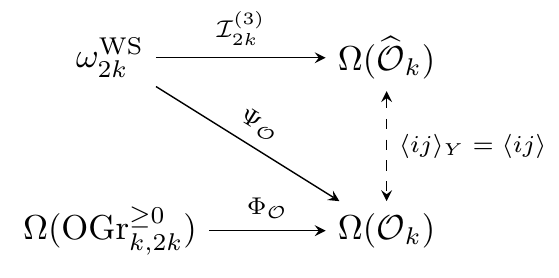}
	\caption{Web of connections between rational forms associated with the orthogonal Momentum Amplituhedron. Solid lines designate pushforwards.}
	\label{fig:omom-web-forms}
\end{figure}

\subsection{Grassmannian Definition}
\label{sec:omom-grass}

Let $\Lambda\in\tpMat{2k}{k+2}$ be a positive matrix, called the \emph{kinematic data}. Given an element $[C]\in\OG{k}{2k}$, let $Y \coloneqq C\Lambda$. The \emph{orthogonal Momentum Amplituhedron} $\tnnOMom{k}\coloneqq\OMomMap{\tnnOG{k}{2k}}$ is the image of $\tnnOG{k}{2k}$ through the \emph{orthogonal Momentum Amplituhedron map} \cite{Huang:2021jlh,He:2021llb}
\begin{align}
\symbolOMomMap: \OG{k}{2k}\to\G{k}{k+2}, [C]\mapsto[Y]\,.
\end{align}
Equivalently, the orthogonal Momentum Amplituhedron  $\tnnOMom{k}$ is the closure of $\tpOMom{k}\coloneqq\OMomMap{\tpOG{k}{2k}}$ in $\OMomMap{\OGReal{k}{2k}}$ where $\tpOMom{k}$ is the interior of $\tnnOMom{k}$. By comparison, we see that the orthogonal Momentum Amplituhedron can be viewed as half a Momentum Amplituhedron, projected onto three-dimensional kinematics \cite{Huang:2021jlh}. Since $\tnnOMom{2}$ is diffeomorphic to $\tnnOG{2}{4}$, their boundary stratifications are combinatorially equivalent. 

Natural coordinates for the orthogonal Momentum Amplituhedron correspond to analogues of three-dimensional spinor-helicity brackets. Let $[Y]$ be a point in $\G{k}{k+2}$. For $i,j\in[n]$, define the \emph{twistor coordinate}
\begin{align}
\lrangle{ij}_Y\coloneqq\det(Y_1,\ldots,Y_k,\Lambda_{i},\Lambda_{j})\,,
\end{align}
as the determinant of the $(k+2)\times(k+2)$ matrix with rows $Y_1,\ldots,Y_k,\Lambda_{i},\Lambda_{j}$. For each $I\subset[2k]$, define the \emph{Mandelstam variable} $S_I(Y)$ as the following quadratic polynomial in twistor coordinates:
\begin{align}
	S_I(Y)\coloneqq\sum_{\{i,j\}\in \binom{I}{2}}(-1)^{i+j+1}\lrangle{ij}_Y^2\,.
\end{align}
The following additional restriction applies to the kinematic data $\Lambda$:
\begin{align}\label{eq:omom-kinematic-data}
	S_I(Y)>0~&\text{for all $[Y]\in\tpOMom{k}$}\,,\\
	\nonumber &\text{for all CC subsets $I\subset[2k]$ with $2\le|I|\le 2k-2$}\,.
\end{align}
The restriction ensures that the codimension-one boundaries of the $\tnnOMom{k}$ capture all factorisation channels of partial amplitudes in ABJM theory \cite{Huang:2021jlh,He:2021llb}. One option for satisfying \eqref{eq:omom-kinematic-data}, as confirmed in \cite{Huang:2021jlh} for $k\le 5$, is given by  
\begin{align}
	\Lambda_{i,A} = x_i^{A-1}\,,
\end{align} 
where $i\in[2k]$, $A\in[k]$, and $x_1<\ldots<x_n$ are $n$ real points. Note that the rows of $\Lambda$ are vectors lying on moment curves. Since the combinatorial properties of $\tnnOMom{k}$ are conjecturally independent of our choice of kinematic data, we omit explicit dependence on $\Lambda$ throughout.

The twistor coordinates on $\G{k}{k+2}$ are projectively well-defined, embedding $\G{k}{k+2}$ into $\CP^{\binom{2k}{2}-1}$ where $\dim(\CP^{\binom{2k}{2}-1}) = (2k+1)(k-1)>2k = \dim(\G{k}{k+2})$. Moreover, given a point $V\in\OG{k}{2k}$, mapped to a point $\OMomMap{V}=[Y]\in\tnnOMom{k}$, it follows from the Cauchy-Binet formula that 
\begin{align}
	\lrangle{ij}_Y = \sum_{I\in\binom{[2k]}{k}}p_{I}(V)\,p_{I\cup\{i,j\}}(\Lambda)\,.
\end{align} 

For each point $[Y]\in\G{k}{k+2}$, let $Y^\perp$ be a matrix representing the orthogonal complement. The \emph{hypersurface of momentum conservation} is a codimension-$3$ algebraic subset of $\G{k}{k+2}$ defined as $P=\mathbbb{0}_{2\times 2}$ where
\begin{align}\label{eq:omom-P}
	P\coloneqq (Y^\perp\Lambda^T)\eta(Y^\perp\Lambda^T)^T\,,
\end{align}
and $\eta=\text{diag}(+1,-1,\ldots,+1,-1)$ is the non-degenerate symmetric bilinear form introduced in \cref{sec:ograss-tnn}. \cref{eq:omom-P} defines $3$ independent equations because $P$ is a symmetric $2\times2$ matrix. From 
\begin{align}
	0=Y^\perp Y^T = (Y^\perp\Lambda^T)C^T\,,
\end{align}
it follows that the row span of $Y^\perp\Lambda^T$ is contained in $[C^\perp]=[C\eta]$. Since $C$ is isotropic with respect to $\eta$, $C\eta$ is too. Hence, the orthogonal Momentum Amplituhedron lies on the hypersurface of momentum conservation \cite{Huang:2021jlh,He:2021llb}. 	

The orthogonal Momentum Amplituhedron $\tnnOMom{k}$ is (conjecturally) a $(2k-3)$-dimensional positive geometry whose canonical form encodes the partial amplitude $A^{\tree}_{2k}[\bar{1},2,\bar{3},\ldots,2k]$ of ABJM theory \cite{Huang:2021jlh,He:2021llb}. Moreover, the orthogonal Momentum Amplituhedron map $\symbolMomMap$ is (conjecturally) a morphism between $(\OG{k}{2k},\tnnOG{k}{2k})$ and $(\OMomMap{\OG{k}{2k}},\tnnOMom{k})$. This implies that the pushforward of $\Omega(\tnnOG{k}{2k})$ via $\symbolOMomMap$ equals $\Omega(\tnnOMom{k})$. To calculate this pushforward, suppose $\mathcal{C}=\{\tnnOrth{f}\}_{f\in F}$ is a $\symbolOMomMap$-induced tiling of $\tnnOMom{k}$ for some indexing set $F$ of matching affine permutations. For each $f\in F$, $\symbolOMomMap|_{\tpOrth{f}}:\tpOrth{f}\to\symbolOMomMap(\tpOrth{f})$ is a diffeomorphism and $\Omega(\tnnOrth{f})=\bigwedge_{i=1}^{2k-3}d\log(\tanh(t_i))$ as per \eqref{eq:ograss-orthitroid-Omega}. Then
\begin{align}\label{eq:omom-Omega}
	\Omega(\tnnOMom{k}) = 
	\sign(\mathcal{C})\sum_{f\in F}\symbolOMomMap_\ast\Omega(\tnnOrth{f})\,.
\end{align} 
where $\sign(\mathcal{C})=1$ (resp., $\sign(\mathcal{C})=-1$) if $\mathcal{C}$ is orientation-preserving (resp., orientation-reversing).

To extract the partial amplitude $A^{\tree}_{2k}[\bar{1},2,\bar{3},\ldots,2k]$ from $\Omega(\tnnOMom{k})$, we need the \emph{canonical function} $\underline{\Omega}(\tnnOMom{k})$. To determine the latter, we wedge $\Omega(\tnnOMom{k})$ with $d^3P\delta^3(P)$:
\begin{align}
	\Omega(\tnnOMom{k})\wedge d^3P\,\delta^3(P) = \bigwedge_{\alpha=1}^{k}\det(Y,d^2Y_{\alpha})\,\delta^3(P)\,\underline{\Omega}(\tnnOMom{k})\,.
\end{align}
Thereafter, we introduce $2k$ auxiliary Grassmann parameters $\phi^\alpha_a$, where $a\in[2]$ and  $\alpha\in[k]$, and we define
\begin{align}
	(\Lambda^T_\ast)_{A,i} = 
	\begin{pmatrix}
		\lambda_i^a \\
		\phi^\alpha_b\eta_i^b
	\end{pmatrix},
\end{align}
where $A=(a,\alpha)\in[k+2]$. Choosing the reference point
\begin{align}
	Y_\ast \coloneqq
	\begin{pmatrix}
		\mathbbb{0}_{k\times 2} & \mathbbb{1}_{k\times k}
	\end{pmatrix},
\end{align}
we extract the partial amplitude $A^{\tree}_{2k}[\bar{1},2,\bar{3},\ldots,2k]$ as follows
\begin{align}
	A^{\tree}_{2k}[\bar{1},2,\bar{3},\ldots,2k]=\int d^{2k}\phi\,\delta^3(P)\,\underline{\Omega}(\tnnOMom{k})\Big|_{(Y,\Lambda)\to(Y_\ast,\Lambda_\ast)}\,.
\end{align}

As with the Momentum Amplituhedron, we can equivalently define the orthogonal Momentum Amplituhedron as the intersection of two spaces: the hypersurface of momentum conservation and a winding space. To this end, let $\OMomV{k}$ be the set of points $[Y]$ in $\G{k}{k+2}$ which satisfy $P=\mathbbb{0}_{2\times2}$ and let $\OMomVReal{k}$ be its real part. Define the \emph{winding space} $\OMomW{k}$ as the set of points $[Y]$ in $\GReal{k}{k+2}$ satisfying 
\begin{subequations}
	\begin{align}
		&\lrangle{ii+1}_Y>0~\text{for}~i\in[n-1]~\text{and}~\lrangle{n\hat{1}}_Y>0\,,\\
		&\var(\lrangle{12}_Y,\lrangle{13}_Y,\ldots,\lrangle{1n}_Y)=k\,,\\
		&S_I(Y) > 0~\text{for  all CC subsets $I\subset[n]$ with $2\le|I|\le n-2$}\,,
	\end{align}
\end{subequations}
where $n=2k$ and $\Lambda_{\hat{1}}=(-1)^{k+1}\Lambda_1$. The symbol $\var$ counts the number of sign flips. Then $\OMomMap{\OG{k}{2k}}=\OMomV{k}$, $\tpOMom{k}=\OMomV{k}\cap\OMomW{k}$, and $\tnnOMom{k}$ is the closure of $\OMomV{k}\cap\OMomW{k}$ in $\OMomVReal{k}$. We have already proven $\OMomMap{\OG{k}{2k}}\subset\OMomV{k}$ and \cite{Huang:2021jlh,He:2021llb} establishes $\tpOMom{k}\subset\OMomV{k}\cap\OMomW{k}$. Nevertheless, the above equalities remain conjectures.

\subsection{Kinematic Definition}
\label{sec:omom-kinematic}

In analogy with Momentum Amplituhedron, the orthogonal Momentum Amplituhedron can be defined in the kinematic space $\OnShell{3}{n}$ of three-dimensional spinor-helicity variables:
\begin{align}
	\OnShell{3}{n}\coloneqq\left\{[\lambda]\in\frac{\Mat{2}{n}{\real}}{\SL{2}{\real}}:\sum_{i=1}^n\lambda_i^a\eta_{i,j}\lambda_j^b=0\right\},
\end{align}
where $\eta=\text{diag}(+1,-1,\ldots,+1,-1)$. As usual, $\SL{2}{\real}$ acts of $\Mat{2}{n}{\real}$ by left multiplication. The dimension of $\OnShell{3}{n}$ is given by
\begin{align}
	\dim(\OnShell{3}{n}) = \dim(\Mat{2}{n}{\real})-\dim(\SL{2}{\real}) - 3 = 2n -6\,.
\end{align}
Equivalently, $\OnShell{3}{n}$ is the zero set of a polynomial ideal in $\real[\lambda]\coloneqq\real[\lambda_1^1,\lambda_1^2,\ldots,\lambda_{n}^1,\lambda_{n}^2]$ generated by the \emph{Schouten identities}
\begin{align}
	\lrangle{ij}\lrangle{kl} + \lrangle{il}\lrangle{jk} + \lrangle{ik}\lrangle{lj} = 0\,,
\end{align}
and momentum conservation $\sum_{i=1}^{n}(-1)^{i+1}\lambda_i^a\lambda_i^{b}=0$ for $a,b\in[2]$. Consequently, $\OnShell{3}{n}$ is an affine algebraic subset of $\real^{2n}$. The complexification of $\OnShell{3}{n}$ is denoted by $\OnShellComplex{3}{n}$.

To motivate the kinematic space definition, recall that the Grassmannian $\G{k}{k+2}$ is naturally covered by (Zariski) open sets consisting of those points for which some Pl\"{u}cker coordinate is non-zero. Consider the open set $U\subset\G{k}{k+2}$ where
\begin{align}
	U\coloneqq\{V\in\G{k}{k+2}: p_{[k]+2}(V)\ne0\}\,.
\end{align}
Given a point $V$ in $U$, one can always choose a representative $Y$ of the form
\begin{align}\label{eq:omom-kin-chart}
	Y\coloneqq
	\begin{pmatrix}
		\mathbbb{1}_{k\times k} & -y^T
	\end{pmatrix},
\end{align}
where $V=[Y]$, for some $y\in\Mat{2}{k}{\complex}$. In fact, \eqref{eq:omom-kin-chart} furnishes a smooth affine chart for $U$. A canonical choice of representative $Y^\perp$ for the point $V^\perp$ in $U^\perp$ is given by
\begin{align}\label{eq:omom-kin-chart-perp}
	Y^\perp\coloneqq
	\begin{pmatrix}
		y & \mathbb{I}_{2\times 2}
	\end{pmatrix},
\end{align}
where $V^\perp=[Y^\perp]$. \cref{eq:omom-kin-chart-perp} defines a smooth affine chart on $U^\perp$. It is chosen such that the maximal minors of $Y^\perp$ are related to those of $Y$ as follows: 
\begin{align}\label{eq:omom-kin-minors}
	p_{[2k]\setminus I}(Y^\perp) = \epsilon_{[2k]\setminus I,I}\,p_I(Y)\,,
\end{align}
for all $I\in\binom{k+2}{2}$. With these charts as hand, define the map
\begin{align}
	\symbolOMomHatMap: U\to\Mat{2}{2k}{\complex}, V\mapsto\lambda\,.
\end{align}
where $\lambda\coloneqq Y^\perp\Lambda^T$. By construction, each point $V$ in $U\cap\OMomV{k}$ maps to a point $\lambda=\OMomHatMap{V}$ which satisfies $\lambda\eta\lambda^T = \mathbbb{0}_{2\times 2}$. Consequently, $\lambda$ represents a spinor-helicity variable in $\OnShellComplex{3}{2k}$ where $[\lambda]=\SL{2}{\complex}\lambda$. Moreover, twistor coordinates and spinor-helicity brackets coincide on $U$ \cite{Arkani-Hamed:2017vfh}:
\begin{align}\label{eq:omom-bracket-identification}
\lrangle{ij} = \lrangle{ij}_Y\,,
\end{align}
for every $i,j\in[2k]$ where $\lambda = \OMomHatMap{[Y]}$.

Altogether, these observations motivate the following definition \cite{Huang:2021jlh,He:2021llb}. Let $\OMomHatV{k}$ be the image of $\OMomV{k}\cap U$ under $\symbolOMomHatMap$. Define the \emph{winding space} $\OMomHatW{k}$ as the image of $\OMomW{k}$ under $\symbolOMomHatMap$; it consists of all points $\lambda$ in $\OMomHatMap{U(\real)}$ satisfying 
\begin{subequations}
	\begin{align}
		&\lrangle{ii+1}>0~\text{for}~i\in[n-1]~\text{and}~\lrangle{n\hat{1}}>0\,,\\
		&\var(\lrangle{12},\lrangle{13},\ldots,\lrangle{1n})=k\,,\\
		&s_I > 0~\text{for all CC subsets $I\subset[n]$ with $2\le|I|\le n-2$}\,,
	\end{align}
\end{subequations}
where $n=2k$ and $\lambda_{\hat{1}}=(-1)^{n-k+1}\lambda_1$. The symbol $\var$ counts the number of sign flips. Then the \emph{orthogonal Momentum Amplituhedron} $\tnnOMomHat{k}$ is the closure of $\tpOMomHat{k}\coloneqq\OMomHatV{k}\cap\OMomHatW{k}$ in $\OMomHatVReal{k}$, the real part of $\OMomHatV{k}$, where $\tpOMomHat{k}$ is the interior of $\tnnOMomHat{k}$. The canonical form $\Omega(\tnnOMomHat{k})$ is given by \eqref{eq:omom-Omega} with twistor coordinates replaced by spinor-helicity brackets as per \eqref{eq:omom-bracket-identification}:
\begin{align}
	\Omega(\tnnOMomHat{k})=\Omega(\tnnOMom{k})\Big|_{\lrangle{ij}_{Y}\to\lrangle{ij}}\,.
\end{align}
Finally, the partial amplitude $A^{\tree}_{2k}[\bar{1},2,\bar{3},\ldots,2k]$ can be extracted directly from $\Omega(\tnnOMomHat{k})$ via the replacement
\begin{align}
	A^{\tree}_{2k}[\bar{1},2,\bar{3},\ldots,2k]=\delta^3(p)\left(\Omega(\tnnOMomHat{k})\wedge d^3p\Big|_{d\lambda_i^a\to\eta_i^a}\right).
\end{align}

\subsection{\texorpdfstring{$D=3$}{Three-dimensional} Twistor-String Map}
\label{sec:omom-string}

The three-dimensional twistor-string map relates the orthogonal Momentum Amplituhedron $\tnnOMom{k}$ to the worldsheet moduli space $\tnnModuli{0}{2k}$ \cite{He:2021llb}. Towards defining this map, consider the Veronese map from $\tnzG{2}{n}$ to $\tnzG{k}{n}$ where $n=2k$. The orthogonality conditions $C_\nu(t,\sigma)\eta C_\nu^T(t,\sigma)=\mathbbb{0}_{k\times k}$ produces $n-1$ equations for $n$ unknowns $t=(t_1,\ldots,t_n)$:
\begin{align}
	\sum_{i=1}^{n}(-1)^{i+1}t_i^{n-2}\sigma_{i}^{\mu-1}=0\,,
\end{align}
where $\mu\in[n-1]$. The $\GL{1}{\complex}$ redundancy associated with the $t$-variables allows us to fix one of them, say $t_n=1$, in which case
\begin{align}\label{eq:omom-t}
	t_i^{2n-2}\coloneqq(-1)^i\frac{\prod_{j\ne n}\sigma_{n,j}}{\prod_{j\ne i}\sigma_{i,j}}\,,\qquad t_n=1\,.
\end{align}
Define $X(\sigma)\coloneqq X(t,\sigma)$ and $C_\nu(\sigma)\coloneqq C_\nu(t,\sigma)$ where the $t$-variables are fixed according to \eqref{eq:omom-t}. Then $\nu:\Moduli{0}{2k}\to\tnzOG{k}{2k},[X(\sigma)]\mapsto[C_\nu(\sigma)]$ is the \emph{Veronese map} from $\Moduli{0}{2k}$ to $\tnzOG{k}{2k}$ \cite{He:2021llb}. The  \emph{$D=3$ twistor-string map} is the rational map
\begin{align}
	\symbolOMomTwistorMap:\Moduli{0}{2k}\to\G{k}{k+2},[X(\sigma)]\mapsto[Y(\sigma)]\,,
\end{align}
where $Y(\sigma)\coloneqq C_\nu(\sigma)\Lambda$. The composition of $\symbolOMomTwistorMap$ and $\symbolOMomHatMap$ leads to the \emph{$D=3$ scattering equations} of \cite{Cachazo:2013iaa}
\begin{align}\label{eq:omom-SE}
	C_\nu(\sigma)\lambda^T = 0\,,
\end{align}
where $\lambda$ is a function of $y\in\Mat{2}{k}{\complex}$. Moment conservation $\lambda\eta\lambda^T=\mathbbb{0}_{2\times2}$ implies that \eqref{eq:omom-SE} contains only $2k-3$ independent equations. By gauge-fixing any three $\sigma$-variables (e.g.\ $(\sigma_1,\sigma_{n-1},\sigma_n)=(0,1,\infty)$), \eqref{eq:omom-SE} defines a zero-dimensional complete intersection ideal $\mathcal{I}_{2k}^{(4)}\subset\complex(y)[\sigma]$. Its zero-set contains $E_{2k-3}$ distinct points in $\Moduli{0}{n}$ \cite{Cachazo:2013iaa} where $E_{2k-3}$ is the \emph{Euler zag number} defined as
\begin{align}
	\tan(x)=\sum_{p=2}^{\infty}\frac{E_{2k-3}x^{2k-3}}{(2k-3)!} = x + \frac{2x^3}{3!} + \frac{16x^5}{5!} + \cdots\,.
\end{align}
Numerical evidence suggests that \smash{$\symbolOMomTwistorMap|_{\tpModuli{0}{2k}}:\tpModuli{0}{2k}\to\tpOMom{k}$} is a bijection \cite{He:2021llb}: for each point $\lambda$ in $\tpOMomHat{k}$, there are $E_{2k-3}$ (generally complex) solutions to the $D=3$ scattering equations of which only one is in $\tpModuli{0}{n}$. This naturally leads to the conjecture that the $D=3$ twistor-string map $\symbolOMomTwistorMap$ is a morphism between $(\Moduli{0}{n},\tnnModuli{0}{n})$ and $(\OMomMap{\OG{k}{2k}},\tnnOMom{k})$. Recall that the canonical form of the worldsheet moduli space is the Parke-Taylor form $\omega_n^\text{WS}=\Omega(\tnnModuli{0}{n})$. Consequently, one can express the canonical form $\Omega(\tnnOMom{k})$ as
\begin{align}\label{eq:omom-pushforward-twistor-string}
	\Omega(\tnnOMom{k}) = \symbolOMomTwistorMap_\ast\omega_n^\text{WS}\,,
\end{align}
and the canonical form $\Omega(\tnnOMomHat{k})$ as
\begin{align}\label{eq:omom-pushforward-scattering-equations}
	\Omega(\tnnOMomHat{k}) = (\mathcal{I}_n^{(3)})_\ast\omega_n^\text{WS}\,.
\end{align}

\section{Boundary Stratification and Rank Generating Function}
\label{sec:omom-boundaries}

As discussed in \cref{sec:mom-boundaries}, the boundaries of $\tnnMom{k}{n}$ are (conjecturally) enumerated by contracted Grassmannian forests of type $(k,n)$. Remarkably, an analogous claim exists for the orthogonal Momentum Amplituhedron: the boundaries of $\tnnOMom{k}$ are (conjecturally) enumerated by orthogonal Grassmannian forests or OG forests of type $\underline{k}$ \cite{Lukowski:2021fkf}. In this section, we summarise the results of \cite{Lukowski:2021fkf}, deriving a rank generating function for $\tnnOMom{k}$ and using it to show that the Euler characteristic of $\tnnOMom{k}$ is one.

Our combinatorial description of the orthogonal Momentum Amplituhedron is based on data for the $\symbolOMomMap$-induced boundary stratification of $\tnnOMom{k}$. Since the orthogonal Momentum Amplituhedron map $\symbolOMomMap$ is conjectured to be a morphism of positive geometries, we assume that the boundary stratification induced by $\symbolOMomMap$ is the boundary stratification of $\tnnOMom{k}$. Consequently, we will drop the phrase ``$\symbolOMomMap$-induced''. Moreover, we will label boundaries of $\tnnOMom{k}$ using reduced OG graphs of type $\underline{k}$ since these labels enumerate orthitroid cells in $\tnnOG{k}{2k}$.

\begin{figure}
	\centering
	\null
	\hfill
	\begin{subfigure}[b]{0.3\textwidth}
		\centering
		\includegraphics[scale=0.45]{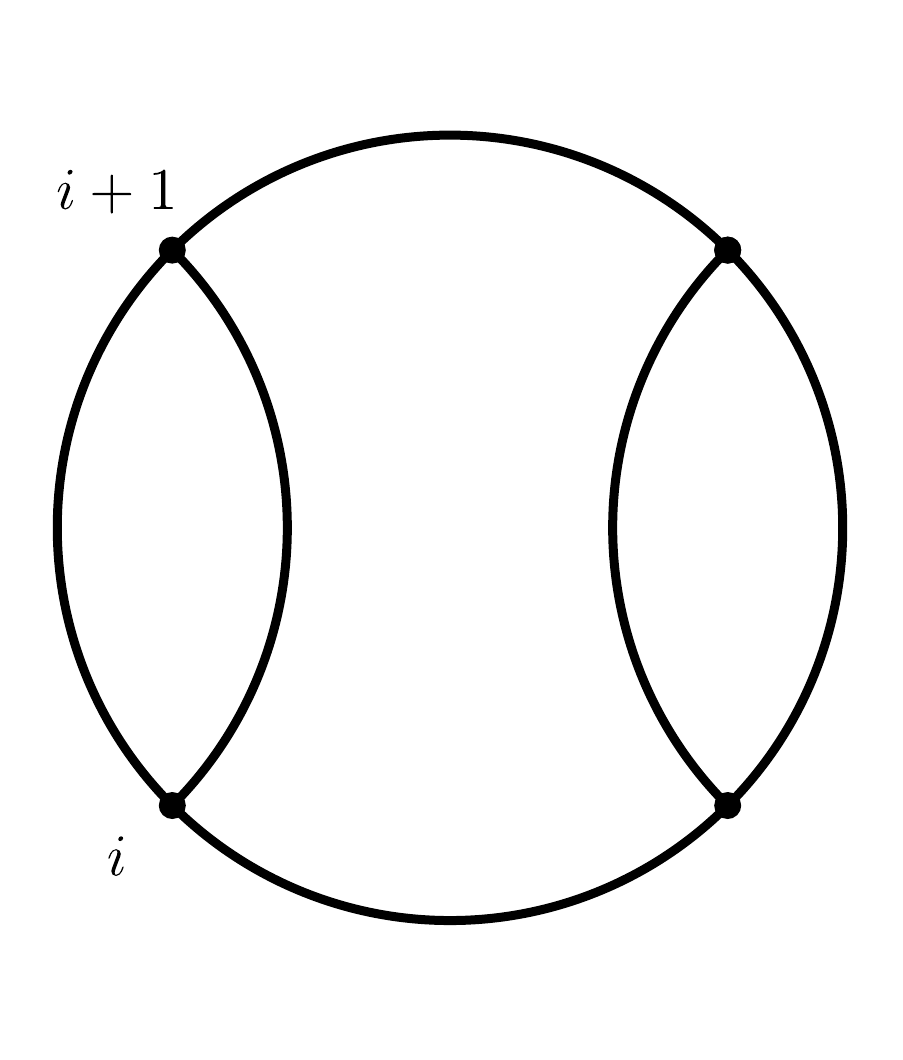}
		\caption{for $k=2$: $\lrangle{ii+1}_Y=0$}
		\label{fig:omom-facet-k=2}
	\end{subfigure}
	\hfill
	\begin{subfigure}[b]{0.3\textwidth}
		\centering
		\includegraphics[scale=0.45]{mom-facet-factorisation}
		\caption{for $k>2$: $S_I(Y)=0$}
		\label{fig:omom-facet-k>2}
	\end{subfigure}
	\hfill
	\null
	\caption{OG graph labels for facets of the orthogonal Momentum Amplituhedron. Labels inside vertices denote helicities. In \cref{fig:omom-facet-k>2}, $I = \{i,i+1,\ldots,j\}$ is a cyclically consecutive subset of $[2k]$, $|I|$ is odd with $2<|I|<2k-2$, and $k_1+k_2=k+1$.}
	\label{fig:omom-facets}
\end{figure}

For $k>2$, the facets (codimension-one boundaries) of the orthogonal Momentum Amplituhedron correspond to factorisation channels where odd-particle planar Mandelstams vanish \cite{Huang:2021jlh,He:2021llb}. An exception occurs for $k=2$: there are only two facets associated with vanishing two-particle planar Mandelstams. Using the first procedure presented in \cref{sec:pos-boundaries}, which takes as input any tiling (e.g.\ any BCFW tiling), one can generate the facets of $\tnnOMom{k}$. \cref{fig:omom-facets} displays the OG graph labels for the facets of $\tnnOMom{k}$. These OG graphs are precisely the on-shell diagrams for the corresponding amplitude singularities.

To generate higher codimension boundaries, we recursively applying the second algorithm appearing in \cref{sec:pos-boundaries}. The \Mathematica{} package \orthitroids{} includes an implementation thereof \cite{Lukowski:2021fkf}. The authors of \cite{Lukowski:2021fkf} used this package to analyse the boundary stratification of $\tnnOMom{k}$ for all $k\le 7$. From their observations, they concluded the following characterization: $\Phi(\tnnOrth{\Gamma})$ is a boundary of $\tnnOMom{k}$ if and only if $\Gamma$ is an OG forest of type $\underline{k}$. The boundary stratification of $\tnnOMom{k}$ is an induced subposet of the orthitroid stratification of $\tnnOG{k}{2k}$ and the corresponding covering relations are drawn in \cref{fig:ograss-covering}.

\begin{figure}
	\centering
	\centering
	\null
	\hfill
	\begin{subfigure}[t]{0.49\textwidth}
		\centering
		\includegraphics[width=\textwidth]{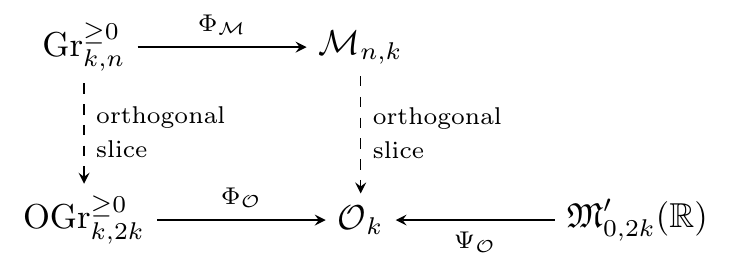}
		\caption{Boundaries}
	\end{subfigure}
	\hfill
	\begin{subfigure}[t]{0.49\textwidth}
		\centering
		\includegraphics[width=\textwidth]{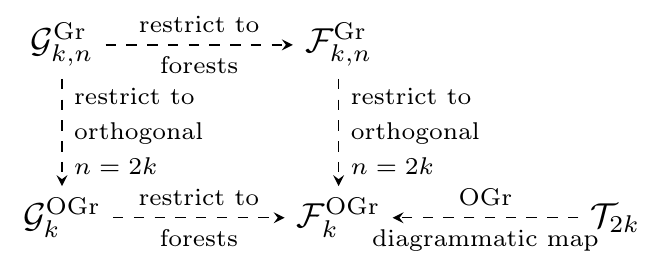}
		\caption{Labels}
	\end{subfigure}
	\hfill
	\null
	\caption{Web of connections between posets of boundaries and their corresponding labels.}
	\label{fig:omom-subposet}
\end{figure}

Combinatorially speaking, the boundary stratification of $\tnnOMom{k}$ is an induced subposet of the boundary stratification of $\tnnMom{k}{2k}$, by which we mean that $(\Forest^{\symbolOG}_k,\preceq_{\symbolOG})$ is an induced subposet of $(\Forest^{\symbolG}_{k,2k},\preceq_{\symbolG})$. This relationship is depicted in \cref{fig:omom-subposet}

The above characterization enables us to determine a rank generating function for the orthogonal Momentum Amplituhedron, along the lines of \cite{Moerman:2021cjg}. To this end, let us define the following statistic for OG forests. This statistic, motivated by empirical findings, captures the $\symbolOMomMap$-dimension of orthitroid cells. The \emph{orthogonal Momentum Amplituhedron dimension} or \emph{$\symbolOMom$-dimension} of an OG tree $T$ of type $\underline{k}$ is defined as
\begin{align}\label{eq:omom-dimension-T}
	\dim_{\symbolOMom}(T) \coloneqq\begin{cases}
		0\,,&\text{if}~k=1\\
		\sum_{v\in\Vertices_\text{int}(T)}(\deg(v)-3)\,,&\text{otherwise}
	\end{cases},
\end{align}
and the $\symbolOMom$-dimension of an OG forest $F$ is the sum of the $\symbolOMom$-dimensions of the OG trees in $F$:
\begin{align}\label{eq:omom-dimension-F}
	\dim_{\symbolOMom}(F)\coloneqq\sum_{T\in\Trees(F)}\dim_{\symbolOMom}(T)\,.
\end{align}
Crucially, it is conjectured that $\dim(\Phi(\tpOrth{F})) = \dim_{\symbolOMom}(F)$.

Since each OG forest is a Grassmannian forest, it has a $\symbolMom$-dimension. The relationship between the two dimensional formulae is as follows. Given an OG tree $T$ of type $\underline{k}$, its $\symbolMom$-dimension, given by \eqref{eq:mom-dimension-T}, equals
\begin{align}\label{eq:mom-dimension-T-omom}
	\dim_{\symbolMom}(T) = \begin{cases}
		1\,,& \text{if}~k = 1\\
		1 + \sum_{v\in\Vertices_\text{int}(T)}(2\deg(v)-5)\,,& \text{if}~k>1
	\end{cases}.
\end{align}
Comparing \eqref{eq:mom-dimension-T-omom} and \eqref{eq:omom-dimension-T}, we have that
\begin{align} \label{eq:mom-dimension-T-omom-comparison}
	\dim_{\symbolMom}(T) = 2\dim_{\symbolOMom}(T)+|\Vertices_\text{int}(T)|+1\,.
\end{align}
Moreover, the $\symbolMom$-dimension of an OG forest $F$ is related to its $\symbolOMom$-dimension via
\begin{align} \label{eq:mom-dimension-F-omom-comparison}
	\dim_{\symbolMom}(F) = 2\dim_{\symbolOMom}(F) + |\Trees(F)| + |\Vertices_\text{int}(F)|\,.
\end{align}

Knowing the $\symbolOMom$-dimension of all OG forests allows us to construct a rank generating function for enumerating the boundaries of the orthogonal Momentum Amplituhedron according to their dimensions, as done in \cite{Lukowski:2021fkf}. From \eqref{eq:omom-dimension-T}, it is clear that each internal vertex $v$ of an OG tree $T$ contributes $\deg(v)-3$ to $\dim_{\mathcal{O}}(T)$. Therefore, we define the statistic $f:\integer_{\ge 3}\to\rational(q)$ which maps $d$, the degree of an internal vertex, to $q^{d-3}$ when $d$ is even and $0$ when $d$ is odd. Suppose $h:\integer_{\ge 3}\to\rational(q)$ is defined in terms of $f$ as in \cref{thm:enum-tree}. Let $F(x)=\sum_{d\ge 3}f(d)x^d$ and $H(x)=x^2+\sum_{n\ge 3}h(n)x^n$. Then one can concretely compute $F(x)$ as
\begin{align}
	F(x)= q^{-3}\sum_{\ell=2}^{\infty}(xq)^{2\ell} = \frac{x^4q}{1-(xq)^2}\,.
\end{align}
From the proof presented in \cref{sec:enum-graphs} for \cref{thm:enum-tree} we have that
\begin{align*}
	h(n+2) = \frac{1}{n+1}\sum_{j=1}^{n}\binom{n+j}{j}[x^n]\left(\frac{F(x)}{x^2}\right)^j,
\end{align*}
where
\begin{align}
	\left(\frac{F(x)}{x^2}\right)^j = \sum_{\ell=0}^\infty\binom{j+\ell-1}{\ell}q^{2\ell+j}x^{2(\ell+j)}\,.
\end{align}
Clearly $[x^n]x^{2(\ell+j)}$ is non-zero only when $n$ is even. For $n=2i$, $[x^{2i}]x^{2(\ell+j)}$ equals one for $\ell = i-j$ provided $i\ge j$ since $\ell\ge0$. Combining this information, one can explicitly compute $H(x)$ as
\begin{align}
	H(x) &= x^2\left(1+\sum_{i=1}^{\infty}h(2i+2)x^{2i}\right).
\end{align}
Consequently, the number of OG trees of type $\underline{k}$ with $\symbolOMom$-dimension $r$ is given by $[x^{2k}q^r]\mathcal{G}^{\symbolOMom}_{\tree}(x,q)$ where
\begin{align}\label{eq:omom-trees-GF}
	\mathcal{G}^{\symbolOMom}_{\tree}(x,q)&=x^2\left(1+\sum_{i=1}^{\infty}\sum_{j=1}^{i}\frac{1}{i}\binom{2i+j}{2i+1}\binom{i}{j}q^{2i-j}x^{2i}\right),
\end{align}
is $H(x)$. Moreover, one can easily show that $\mathcal{G}^{\symbolOMom}_{\tree}(x,q)$ is an algebraic rank generating function of degree $3$ satisfying
\begin{align}
q (q+1) (\mathcal{G}^{\symbolOMom}_{\tree})^3-q^2 (\mathcal{G}^{\symbolOMom}_{\tree})^2 x^2-\mathcal{G}^{\symbolOMom}_{\tree} x^2+x^4 = 0\,.
\end{align}
From \cref{thm:enum-forest}, the number of OG forests of type $\underline{k}$ with $\symbolOMom$-dimension $r$ is given by $[x^{2k}q^r]\mathcal{G}^{\symbolOMom}_{\forest}(x,q)$ where
\begin{align}\label{eq:omom-forest-GF-1}
	x\, \mathcal{G}^{\symbolOMom}_{\forest}(x,q)=\left(\frac{x}{1+\mathcal{G}^{\symbolOMom}_{\tree}(x,q)}\right)^{\lrangle{-1}}_x,
\end{align}
and the compositional inverse is with respect to the variable $x$. Equivalently, by applying the Lagrange inversion formula \eqref{eq:enum-LIF} we have that
\begin{align}\label{eq:omom-forest-GF-2}
	[x^{2k}]\mathcal{G}^{\symbolOMom}_{\forest}(x,q) = \frac{1}{2k+1} [x^{2k}] 
	\left(1+\mathcal{G}^{\symbolOMom}_{\tree}(x,q) \right)^{2k+1}.
\end{align}
It follows from \eqref{eq:omom-forest-GF-1} that $\mathcal{G}^{\symbolOMom}_{\forest}(x,q)$ is an algebraic rank generating function of degree $4$ satisfying
\begin{align}
	\begin{split}
&(x^4-q^2 x^2)(\mathcal{G}^{\symbolOMom}_{\forest})^4 + ((2 q^2-1) x^2+q (q+1))(\mathcal{G}^{\symbolOMom}_{\forest})^3 \\*
&+ (x^2-q (q (x^2+3)+3))(\mathcal{G}^{\symbolOMom}_{\forest})^2+3 q (q+1)\mathcal{G}^{\symbolOMom}_{\forest}-q (q+1) = 0\,.
	\end{split}
\end{align}

As an immediate corollary, the number of vertices ($0$-dimensional boundaries) of $\tnnOMom{k}$ is $\frac{1}{k+1}\binom{2k}{k}$, i.e.\ $[x^{2k}q^0] \mathcal{G}^{\symbolOMom}_{\forest}(x,q) = \frac{1}{k+1}\binom{2k}{k}$. It was speculated in \cite{Lukowski:2021fkf} that 
\begin{align}
	\tnnOMom{k} = \bigsqcup_{F\in\Forest_{k}^{\symbolOG}}\Phi(\tpOrth{F})
\end{align}
is a regular CW decomposition, where the disjoint union is over OG forests of type $\underline{k}$. Moreover, by calculating the coefficient of $x^{2k}$ in $\mathcal{G}^{\symbolOMom}_{\forest}(x,-1)$, the authors of \cite{Lukowski:2021fkf} showed that the Euler characteristic of the orthogonal Momentum Amplituhedron is one. This bolsters the hypothesis that the orthogonal Momentum Amplituhedron is homeomorphic to a closed ball. 

\section{Diagrammatic Map}
\label{sec:omom-map}

The $D=3$ twistor-string map is a diffeomorphism of the interiors of positive geometries but not of their boundaries; the boundary stratification of the worldsheet associahedron $\tnnModuli{0}{2k}$ differs from that of the orthogonal Momentum Amplituhedron $\tnnOMom{k}$ for all $k>2$. It is, however, conjectured that the twistor-string map ``squashes'' the boundaries of $\tnnModuli{0}{2k}$ in a manner compatible with a simple pictorial rule \cite{Lukowski:2021fkf}. In this section, we describe the latter function. 

Recall that the faces of $\tnnModuli{0}{n}$ are in bijection with $\Tree_{n}$, the set of $n$-particle planar tree Feynman diagrams (i.e.\ series-reduced planar trees on $n$-leaves). Let $T$ be a tree in $\Tree_{n}$ with boundary vertices labelled by $1,\ldots,n$ in a clockwise fashion. Each internal edge $e\in\Edges_\text{int}(T)$ partitions the boundary vertices into two cyclically consecutive subsets, $A$ and $B$, each of size at least two. Consequently, this partitioning associates the Mandelstam variable $s_A=s_B$ to $e$.

\begin{figure}
	\centering
	\begin{align*}
	\vcenter{\hbox{\includegraphics[scale=0.3]{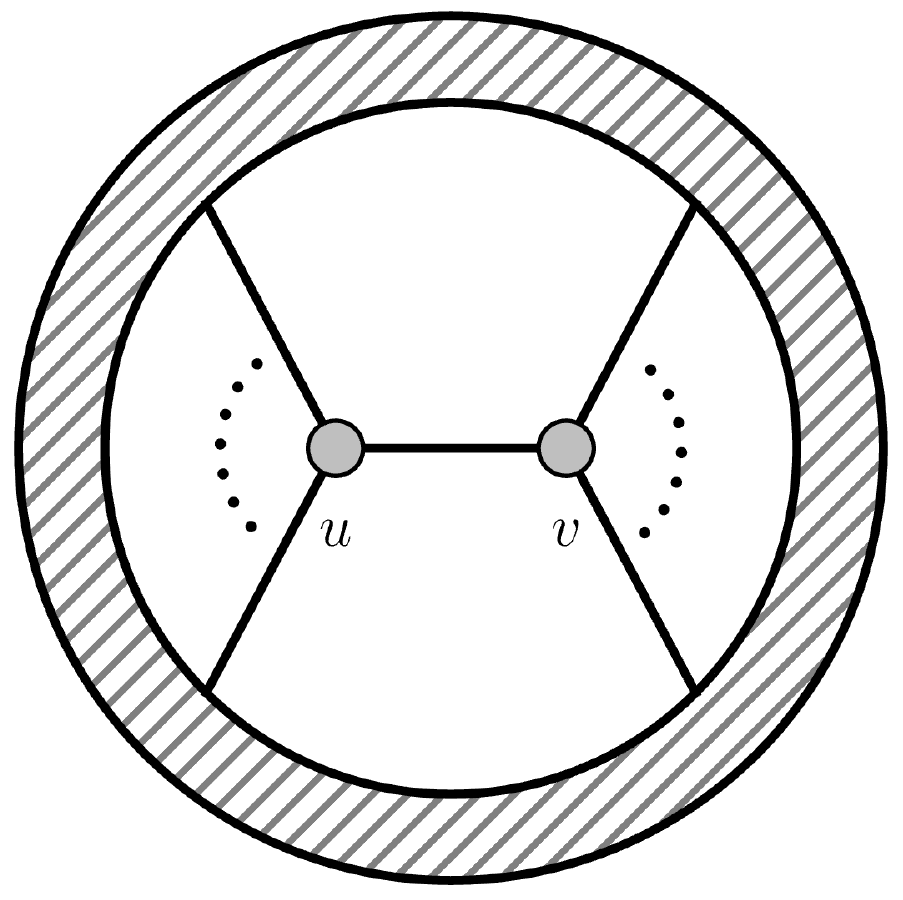}}}
	\precdot_{\symbolAbhy}	
	\vcenter{\hbox{\includegraphics[scale=0.3]{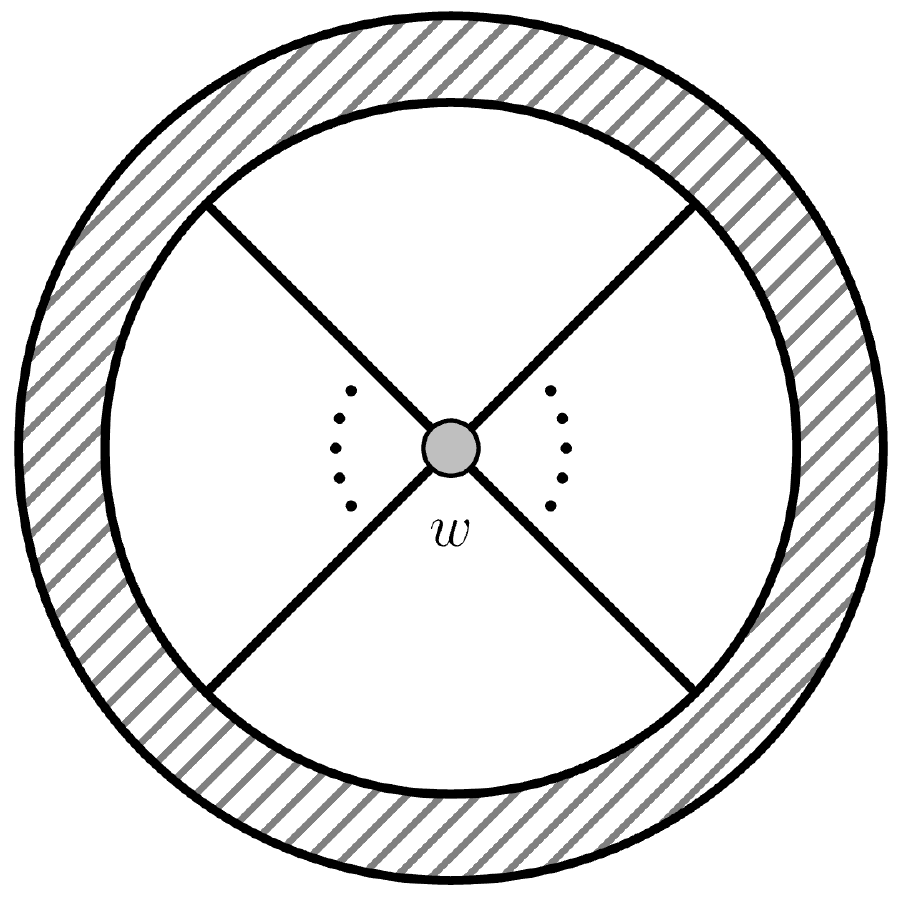}}}
	\end{align*}
	\caption{Covering relations for $\Tree_n$ with respect to $\preceq_{\symbolAbhy}$. Shaded regions represent unaffected graph components. Here $\deg(u)+\deg(v)=\deg(w)+2$ and $\deg(u),\deg(v)\ge3$.}
	\label{fig:omom-covering-WS}
\end{figure}

Set inclusion on the boundary stratification of $\tnnModuli{0}{n}$ translates as a partial order $\preceq_{\symbolAbhy}$ on $\Tree_{n}$. The covering relations, shown in \cref{fig:omom-covering-WS}, are defined as follows. Given $T_1,T_2\in\Tree_{n}$, we say that $T_2$ covers $T_1$, denoted by $T_1\precdot_{\symbolAbhy}T_2$, if $T_2$ can be obtained from $T_1$ by contracting two adjacent vertices $u,v\in\Vertices_\text{int}(T_1)$ into a single vertex $w\in\Vertices_\text{int}(T_2)$ where $\deg(u)+\deg(v)=\deg(w)+2$ and $\deg(u),\deg(v)\ge3$. Moreover, $(\Tree_{n},\preceq_{\symbolAbhy})$ is a graded poset since $\symbolAbhy$-dimension satisfies
\begin{align}
	T_1\precdot_{\symbolAbhy}T_2\implies\dim_{\symbolAbhy}(T_2)=\dim_{\symbolAbhy}(T_1)+1\,.
\end{align}

On the other hand, OG forests of type $\underline{k}$, the set of which is designated by $\Forest^{\symbolOG}_{k}$, enumerate the boundaries of $\tnnOMom{k}$. As a subset of $\mathcal{G}^{\symbolOG}_{k}$, $\Forest^{\symbolOG}_{k}$ inherits a partial order $\prec_{\symbolOG}$ from the orthitroid stratification of $\tnnOG{k}{2k}$. The inherited partial order is characterized by the covering relations depicted in \cref{fig:ograss-covering} and described as follows. Given $\Gamma_1,\Gamma_2\in\Forest^{\symbolOG}_{k}$, we write $\Gamma_1\precdot_{\symbolOG}\Gamma_2$, saying that $\Gamma_2$ covers $\Gamma_1$, if $\Gamma_2$ can be obtained from $\Gamma_1$ in one of two ways: either by replacing two edges belonging to a distinct trees in $\Gamma_1$ with a single vertex of degree four (ensuring that planarity is preserved), or by contracting two adjacent vertices $u',v'\in\Vertices_\text{int}(\Gamma_1)$ into a single vertex $w'\in\Vertices_\text{int}(\Gamma_2)$ where $\deg(u')+\deg(v')=\deg(w')+2$. Together with $\symbolOMom$-dimension, $(\Forest^{\symbolOG}_{k},\preceq_{\symbolOG})$ forms a graded poset because
\begin{align}
	\Gamma_1\precdot_{\symbolOG}\Gamma_2\implies\dim_{\symbolOMom}(\Gamma_2)=\dim_{\symbolOMom}(\Gamma_1)+1\,.
\end{align}

Given a series-reduced planar tree $T$ on $2k$-leaves, define $\symbolOG(T)$ by the following ordered sequence of operations:
\begin{axioms}{O}
	\item\label[step]{step:omom-OG-1} Remove all internal edges associated with even-particle Mandelstam variables. 
	\item\label[step]{step:omom-OG-2} Reduce all degree two vertices to single edges.
	\item\label[step]{step:omom-OG-3} Remove all connected subgraphs that are not incident on any boundary vertex.
\end{axioms}
By construction, $\symbolOG(T)$ is a series-reduced planar forest on $2k$-leaves. In addition, each of its internal vertices has even degree. The latter observation follows from the following argument. Given an internal vertex $v\in\Vertices_\text{int}(T)$, its incident edges can be labelled by Mandelstam variables $s_{A_1},\ldots,s_{A_d}$ where $d=\deg(v)$ and $A_1,\ldots,A_d$ forms a partition of $[2k]$. Since $|A_1|+\ldots+|A_d|$ is even, there must be an even number of blocks $A_i$ for which $|A_i|$ is odd. Consequently, by removing all incident edges associated with even-particle Mandelstam variables, \cref{step:omom-OG-1} maps $v$ to $v'$ which has even degree. If we declare the helicity of each internal vertex $v'\in\symbolOG(T)$ to be $h(v')=\deg(v')/2$, then $\symbolOG(T)$ is a OG forest of type $\underline{k}$.

Moreover, $\symbolOG:\Tree_{2k}\to\Forest^{\symbolOG}_k$ is surjective. Given a forest $\Gamma\in\Forest^{\symbolOG}_k$, one can construct a series-reduced planar tree on $2k$-leaves such that $\symbolOG(T)=\Gamma$. To this end, add a vertex $v_F$ to each face $F$ of $\Gamma$ (i.e.\ connected region separating elements of $\Trees(\Gamma)$). For each tree adjacent to $F$, add an edge from $v_F$ to an edge $e$ in the tree (by inserting a vertex somewhere along $e$) where $e$ is chosen such that the graph remains planar. After reducing all degree two vertices to single edges, the resulting graph is a tree $T\in\Tree_{2k}$ whose image under $\symbolOG$ is $\Gamma$.

Given trees $T_1,T_2\in\Tree_{2k}$ for which $T_1\precdot_{\symbolAbhy}T_2$, one can easily show that $\symbolOG(T_1)\preceq_{\symbolOG}\symbolOG(T_2)$. To this end, suppose $T_2$ is obtained from $T_1$ by contracting two adjacent vertices $u,v\in\Vertices_\text{int}(T_1)$ into a single vertex $w\in\Vertices_\text{int}(T_2)$. Let $e$ denote the internal edge between $u$ and $v$. We will restrict our attention to the subgraphs containing $u,v,w$ and their incident edges because the complements of these subgraphs are identical. Assume $u,v,w$ are mapped to $u',v',w'$ by \cref{step:omom-OG-1}. There are two outcomes to consider for $\symbolOG(T_1)$. If $e$ corresponds to an odd-particle Mandelstam variable, it persists as an internal edge of $\symbolOG(T_1)$ and $\symbolOG(T_2)$ can be obtained from $\symbolOG(T_1)$ by contracting the adjacent internal vertices $u',v'$ along $e$ to produce $w'$. Consequently, $\symbolOG(T_1)\precdot_{\symbolOG}\symbolOG(T_2)$. 

\begin{figure}
	\centering
	\begin{align*}
		\vcenter{\hbox{\includegraphics[scale=0.3]{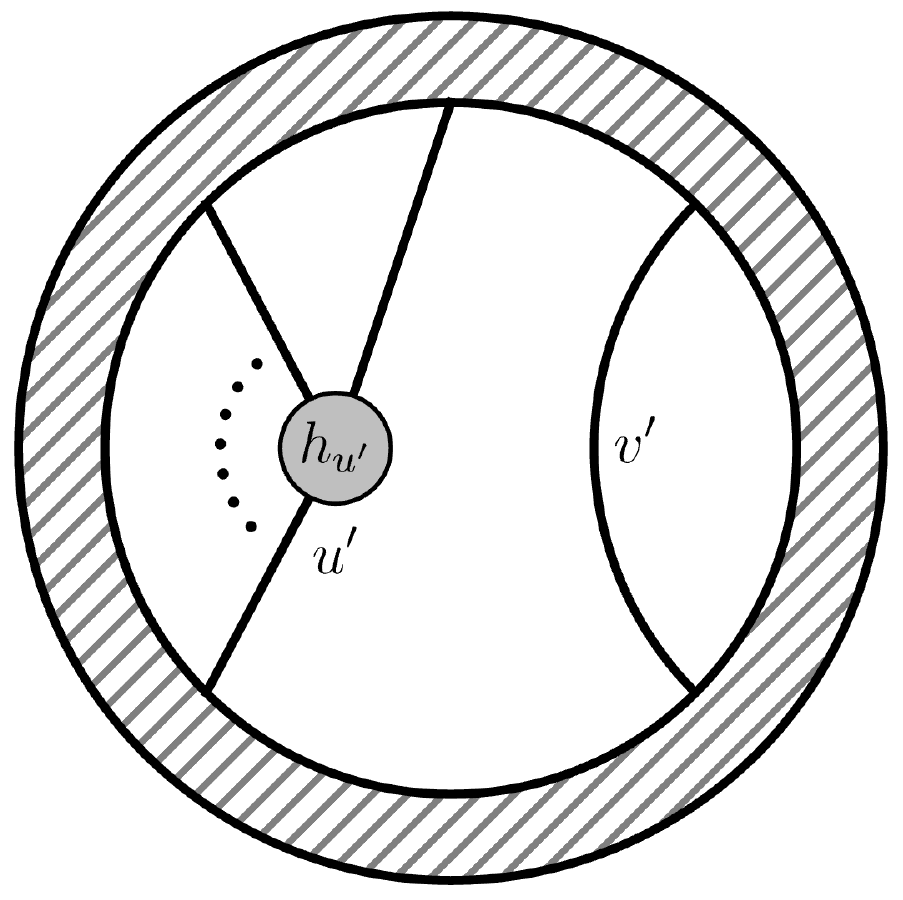}}}
		\precdot_{\symbolOG}	
		\vcenter{\hbox{\includegraphics[scale=0.3]{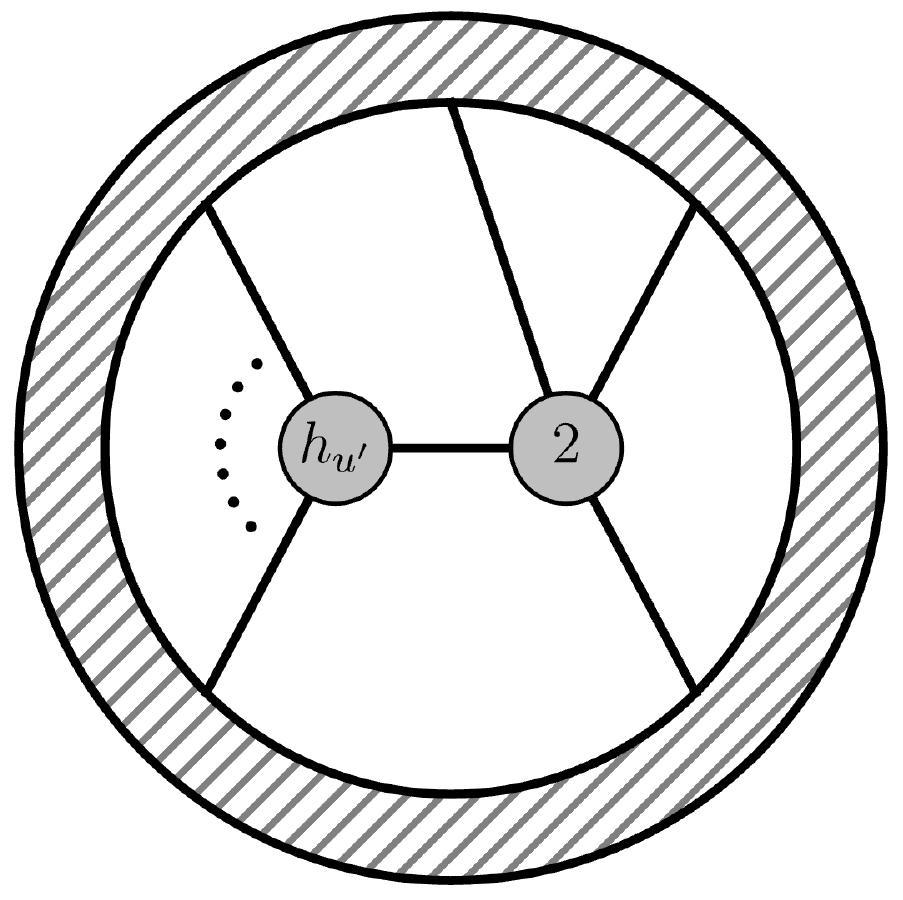}}}
		\precdot_{\symbolOG}	
		\vcenter{\hbox{\includegraphics[scale=0.3]{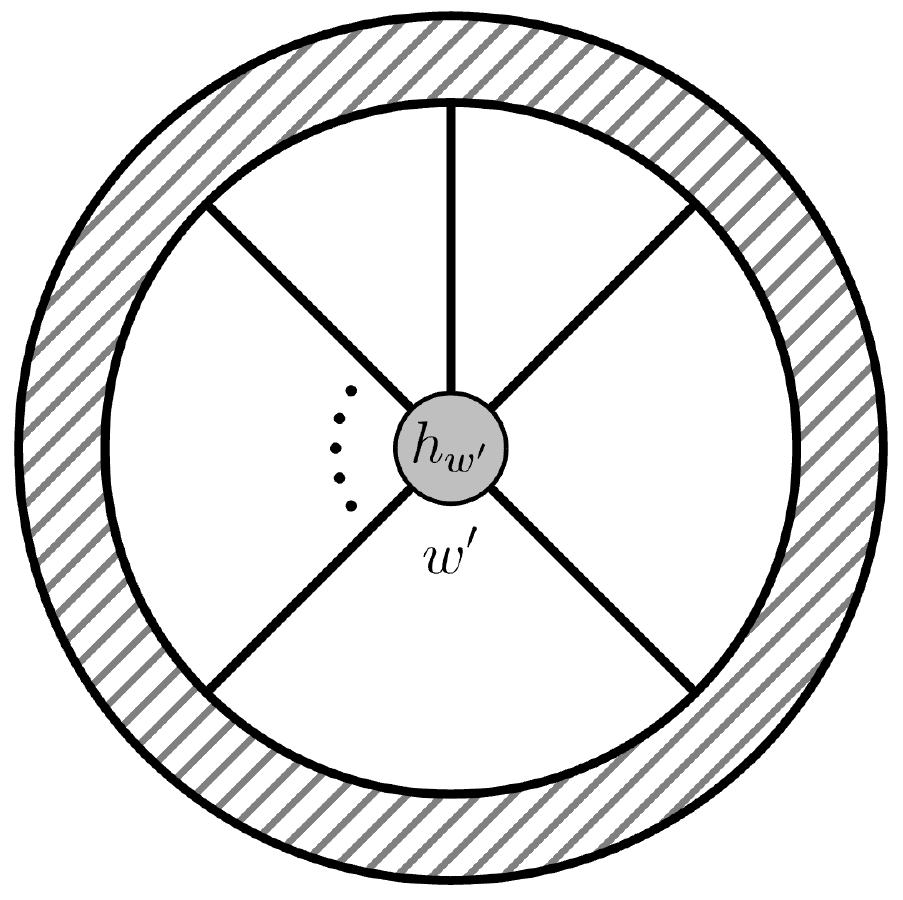}}}
	\end{align*}
	\caption{A sequence of covering relations establishing $\symbolOG(T_1)\prec_{\symbolOG}\symbolOG(T_2)$ when $\deg(u')\ge4$ and $\deg(v')=2$. Shaded regions represent unaffected graph components.}
	\label{fig:omom-covering-sequence-one}
\end{figure}

\begin{figure}
	\centering
	\begin{align*}
		\vcenter{\hbox{\includegraphics[scale=0.3]{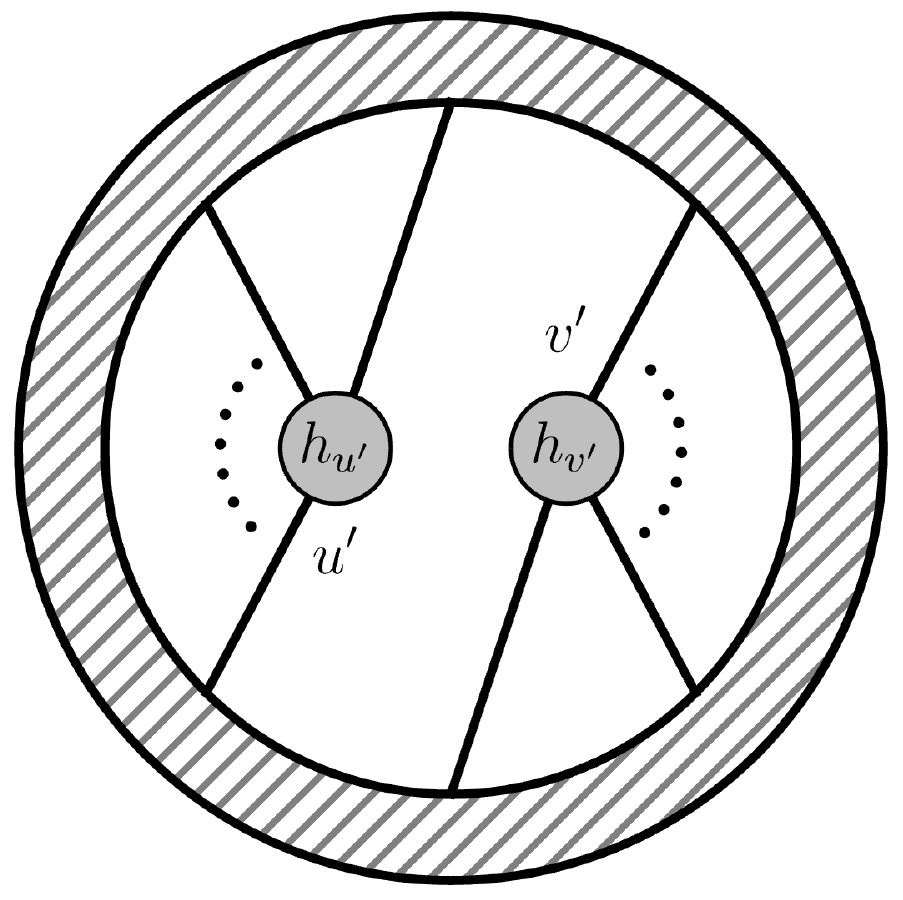}}}
		\precdot_{\symbolOG}	
		\vcenter{\hbox{\includegraphics[scale=0.3]{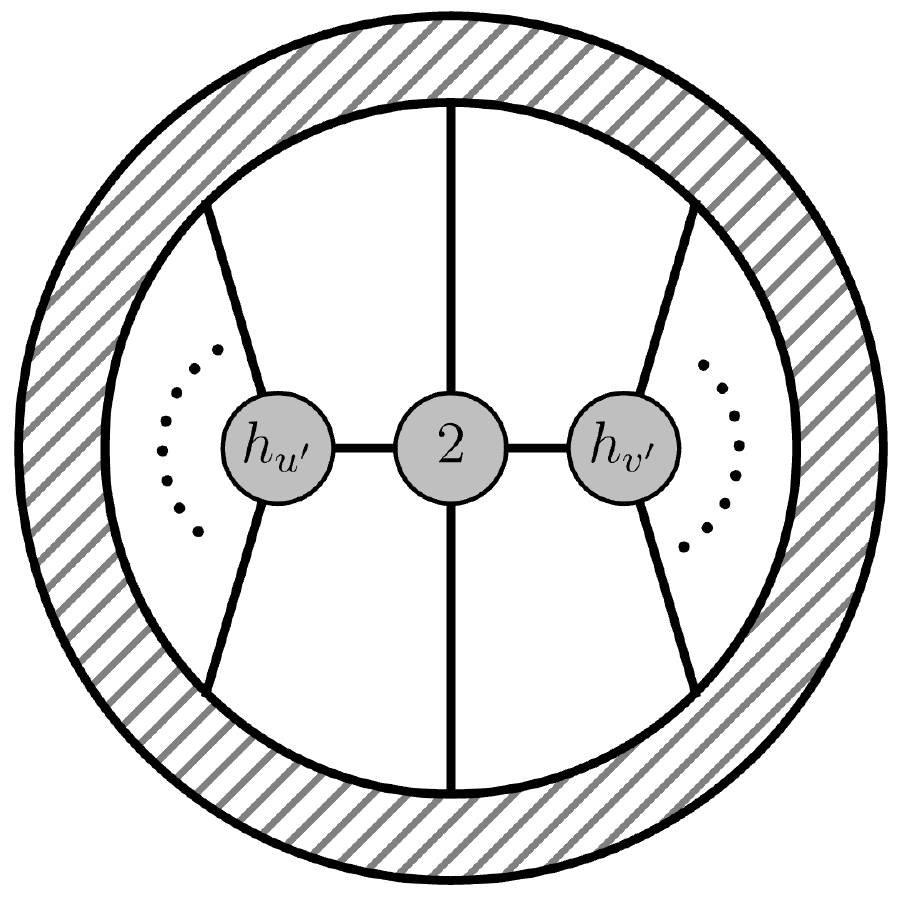}}}
		\precdot_{\symbolOG}	
		\vcenter{\hbox{\includegraphics[scale=0.3]{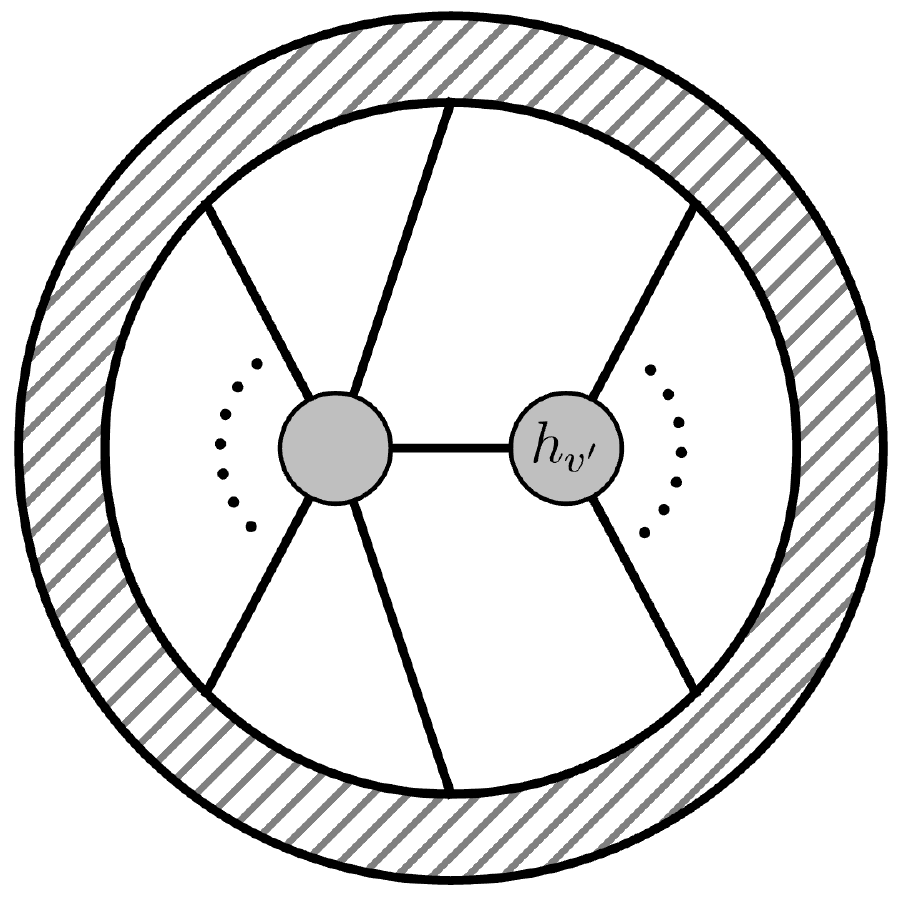}}\hbox{\includegraphics[scale=0.3]{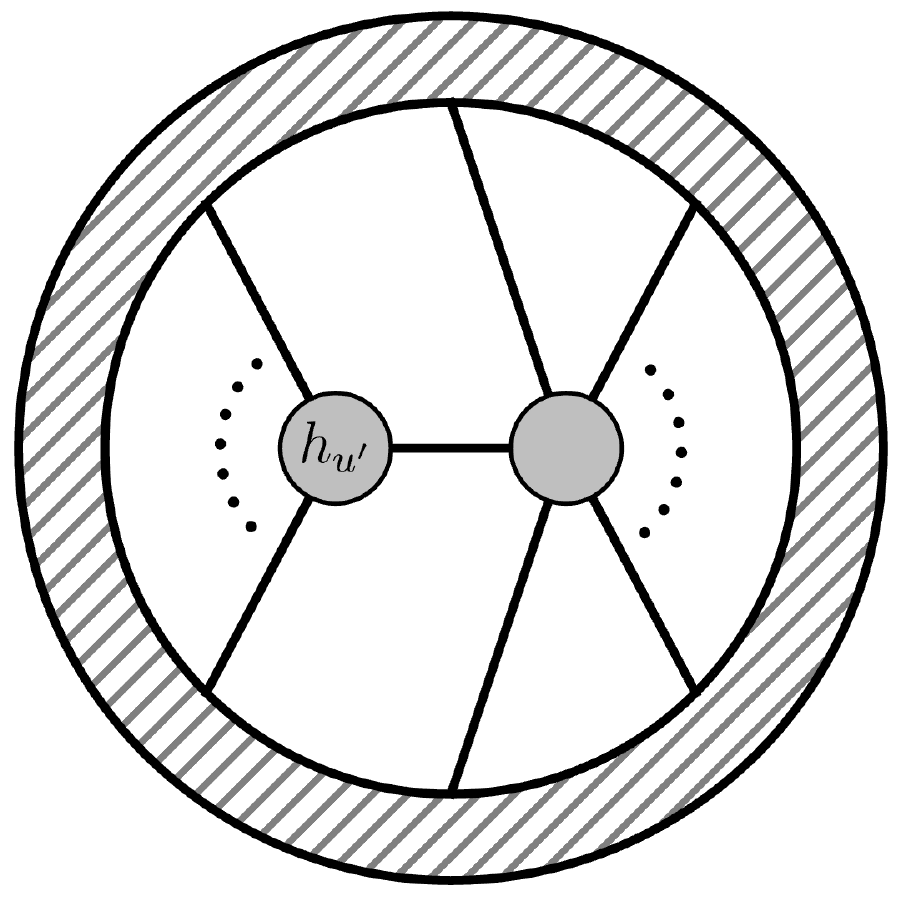}}}
		\precdot_{\symbolOG}	
		\vcenter{\hbox{\includegraphics[scale=0.3]{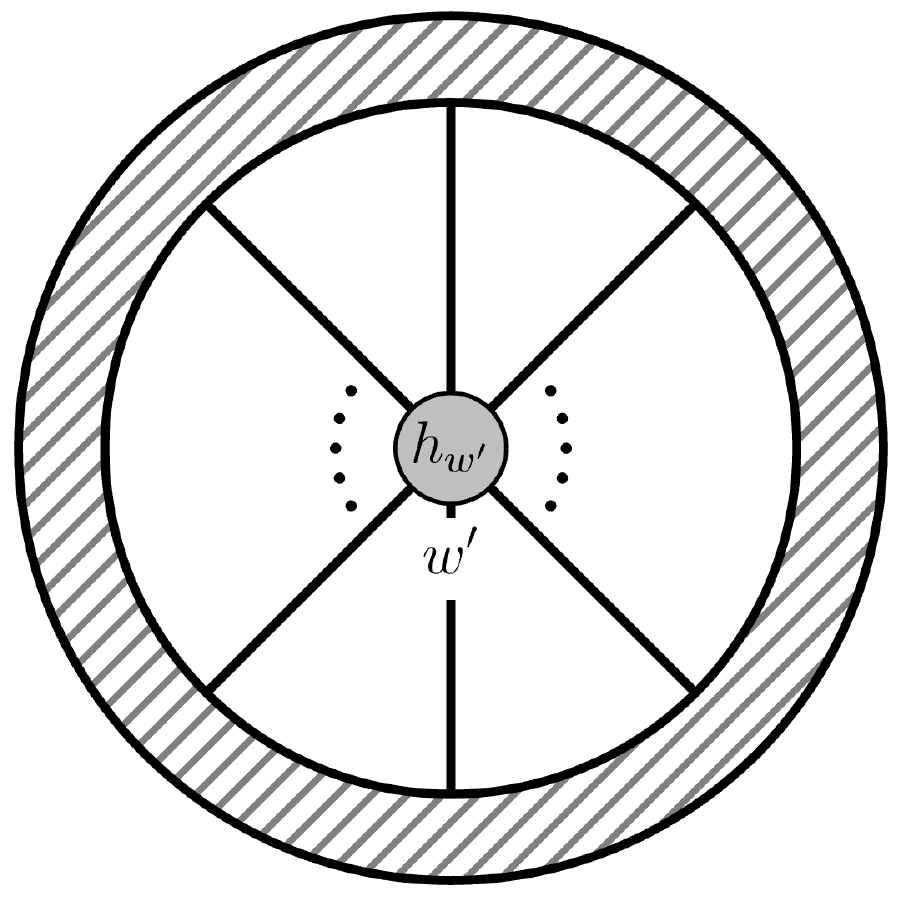}}}
	\end{align*}
	\caption{A sequence of covering relations establishing $\symbolOG(T_1)\prec_{\symbolOG}\symbolOG(T_2)$ when $\deg(u'),\deg(v')\ge 4$. Shaded regions represent unaffected graph components.}
	\label{fig:omom-covering-sequence-both}
\end{figure}

On the other hand, if $e$ corresponds to an even-particle Mandelstam variable, then $\symbolOG(T_2)$ can be obtained from $\symbolOG(T_1)$ by merging the internal vertices $u',v'$ of distinct trees in $\symbolOG(T_1)$ to produce $w'$ where $\deg(w')=\deg(u')+\deg(v')$. Let us inspect all possibilities in turn.
\begin{itemize}
	\item  If $\deg(w')=0$, i.e.\ $w'$ is removed by \cref{step:omom-OG-3}, then so are $u'$ and $v'$. Consequently, $\symbolOG(T_1)=\symbolOG(T_2)$.
	\item If $\deg(w')=4$, then $\deg(u'),\deg(v')=2$, \cref{step:omom-OG-2} replaces $u',v'$ with single edges and $\symbolOG(T_1)\precdot_{\symbolOG}\symbolOG(T_2)$ by definition. 
	\item If $\deg(u')\ge4$ and $\deg(v')=2$, then \cref{step:omom-OG-2} replaces $v'$ with a single edge and  $\symbolOG(T_1)\prec_{\symbolOG}\symbolOG(T_2)$ follows from a sequence of covering relations sketched in \cref{fig:omom-covering-sequence-one}. By symmetry, the same conclusion holds for $\deg(v')\ge4$ and $\deg(u')=2$.
	\item Finally, if $\deg(u'),\deg(v')\ge4$, then $\symbolOG(T_1)\prec_{\symbolOG}\symbolOG(T_2)$ follows from a sequence of covering relations shown in \cref{fig:omom-covering-sequence-both}.
\end{itemize}
Nevertheless, we always have that $\symbolOG(T_1)\preceq_{\symbolOG}\symbolOG(T_2)$. Consequently, $\symbolOG:\Tree_{2k}\to\Forest^{\symbolOG}_k$ is partial-preserving:
\begin{align}
T_1\preceq_{\symbolAbhy}T_2 \implies \symbolOG(T_1)\preceq_{\symbolOG}\symbolOG(T_2)\,.
\end{align}
While the reverse implication is not proven, \cite{Lukowski:2021fkf} postulates that the image of this diagrammatic map is the entire poset $(\Forest^{\symbolOG}_k,\preceq_{\symbolOG})$. In this sense, the boundary stratification of $\tnnOMom{k}$ can be regarded as an induced subposet of the boundary stratification of $\tnnModuli{0}{2k}$. This relationship is included in \cref{fig:omom-subposet}.

The above analysis bolsters the claim in \cite{Huang:2021jlh,He:2021llb} that vanishing even-particle Mandelstam variables correspond to higher-codimension faces of the orthogonal Momentum Amplituhedron \cite{Lukowski:2021fkf}. Assume $k>2$. (For $k=2$, $\tnnOMom{k}$ is one-dimensional and does not have faces beyond codimension-one.) Consider any facet of $\tnnModuli{0}{2k}$. It is labelled by a planar tree $T$ on $2k$-leaves with only two internal (adjacent) vertices $u,v\in\Vertices_\text{int}(T)$, connected via an internal edge $e$, where $\deg(u)+\deg(v)-2=2k$ and $\deg(u),\deg(v)\ge3$. Suppose $u,v$ are mapped to $u',v'$ by \cref{step:omom-OG-1}. If $e$ corresponds to an odd-particle Mandelstam variable, then $T$ and $\symbolOG(T)$ are identical as graphs and $\dim_{\symbolOMom}(\symbolOG(T)) = \dim_{\symbolAbhy}(T)=2k-4$, i.e.\ $\symbolOG(T)$ labels a facet of $\tnnOMom{k}$. Otherwise, $e$ corresponds to an even-particle Mandelstam variable. It is removed and $\symbolOG(T)$ contains two disconnected trees. Consequently, $\deg(u')=\deg(u)-1$ and $\deg(v')=\deg(v)-1$, which means that
\begin{align}
	\dim_{\symbolOMom}(\symbolOG(T))=
	\begin{cases}
		 \dim_{\symbolAbhy}(T)-1=2k-5\,,&\text{if $\deg(u')=2$ or $\deg(v')=2$}\\
		 \dim_{\symbolAbhy}(T)-2=2k-6\,,&\text{if $\deg(u'),\deg(v')\ge4$} 
	\end{cases},
\end{align}
i.e.\ $\symbolOG(T)$ represents a codimension-two or three face of $\tnnOMom{k}$. Our results corroborate the findings of \cite{Huang:2021jlh,He:2021llb} concerning physical singularities as boundaries of the orthogonal Momentum Amplituhedron.

\section{Summary}
The observations of this chapter mirror those of the previous chapter in multiple respects. Firstly, the Grassmannian-based definition of the orthogonal Momentum Amplituhedron facilitated a complete classification of tree-level physical singularities in ABJM in terms of orthogonal Grassmannian (OG) forests. To this end, we used the results of \cref{chp:pos}. The covering relations for OG forests, as displayed in \cref{fig:ograss-covering}, represent the allowed factorisation channels of the tree-level S-matrix. Secondly, using the results of \cref{chp:enum}, we derived an algebraic rank generating function for the boundaries of the orthogonal Momentum Amplituhedron. From this, we proved that its Euler characteristic is one, as is the case for the Momentum Amplituhedron. Thirdly, the orthogonal Momentum Amplituhedron is the image of the worldsheet associahedron via the three-dimensional scattering equations. The interiors of the two positive geometries are diffeomorphic. However, the scattering equations squash some of the worldsheet associahedron’s boundaries. Reference \cite{Lukowski:2021fkf} conjectures a simple diagrammatic map for the combinatorial labels of faces, compatible with the scattering equations. Again the scattering equations directly relate the CHY descriptions of one theory (in this case, ABJM) to the corresponding Grassmannian formulation.

	\chapter{All Roads Lead to ABHY}
\label{chp:meet}

\lettrine{T}{he scattering equations} relate the worldsheet associahedron to the Momentum Amplituhedron and the orthogonal Momentum Amplituhedron – this is the thesis of \cite{He:2021llb}, which we discussed in \cref{chp:mom,chp:omom}. In particular, the pushforward of the Parke-Taylor form (possibly augmented by a projectively well-defined rational form in little group parameters) via the scattering equations (for a specific spacetime dimension) gives the canonical form for each of the latter positive geometries; see \eqref{eq:omom-pushforward-scattering-equations} and \eqref{eq:mom-pushforward-scattering-equations}. Let us refer to the Momentum Amplituhedron and the orthogonal Momentum Amplituhedron as the four- and three-dimensional Momentum Amplituhedron, respectively. In this chapter, we consider going in the opposite direction. Starting with the canonical form of the $D$-dimensional Momentum Amplituhedron, we recover the planar scattering form, restricted to $D$ dimensions. Remember that the canonical form of the ABHY associahedron is a pullback of the planar scattering form. Our principal focus in this chapter is the results of \cite{Damgaard:2020eox}. We present a detailed comparison of the canonical forms of the Momentum Amplituhedron and the ABHY associahedron in two spaces. The first is the space of little group invariants, and the second is a subvariety of the space of Mandelstam invariants defined by Gram determinants conditions. The results of \cite{Damgaard:2020eox} suggest analogues for three and six dimensions where the planar scattering form and the ABHY associahedron are the common denominators. These findings originate from and highlight the common singularity structures of tree-level amplitudes in different theories and different spacetime dimensions. In this sense, all roads lead to the ABHY associahedron. As a by-product of our analysis, we consider the implications of restricting the planar scattering form to specific spacetime dimensions. Throughout this chapter, we denote the canonical form of the $D$-dimensional Momentum Amplituhedron by $\formOnShell[D]{n}$.

\section{Kinematic Spaces}
\label{sec:meet-spaces}

Let us begin by reviewing the kinematic spaces where canonical forms live and the intermediary settings where we can compare them.

\subsection{On-Shell Space}
\label{sec:meet-on-shell}

The success of modern on-shell methods hinges on having a ``good'' description of on-shell dynamics, i.e.\ kinematic variables that trivialise the on-shell condition and transform linearly under global symmetries of the theory. The spinor-helicity formalism, known in $D=3,4,6$ dimensions, is an example of a ``good'' description, parametrising massless scattering. In $(1,D-1)$ spacetime signature, the Lorentz group is $\Spin{1}{D-1}$ and the little group (for real momenta) is $\SO{D-2}$; see \cref{tbl:meet-spinor-helicity-formalism}. We denote the set of spinor-helicity variables for $n$ particles in $D$ dimensions by $\OnShell{D}{n}$. We refer to $\OnShell{D}{n}$ as \emph{on-shell space}; it is the bosonic part of on-shell superspace. 

\begin{table}[t]
	\centering
	\begin{tabular}{c|ccc}
		\toprule
		$D$& Lorentz Group & Little Group & Variables \\\midrule
		$3$& $\SL{2}{\real}$ & $\integer_2$ & $\lambda_i^a$ \\
		$4$& $\SL{2}{\complex}$ & $\SO{2}\cong\U{1}$ & $\lambda_i^{a},\tilde\lambda_i^{\dot{a}}$\\
		$6$& $\SL{2}{\quarternion}$ & $\SO{4}\cong\SU{2}\times\SU{2}$ & $\lambda_i^{Aa}$\\
		\bottomrule
	\end{tabular}
	\caption{Lorentz Group and Little Group for real momenta in $(1,D-1)$ spacetime signature where $D=3,4,6$.}
	\label{tbl:meet-spinor-helicity-formalism}
\end{table}

We have already encountered $\OnShell{D}{n}$ for $D=3,4$ in \cref{chp:omom,chp:mom}. In three dimensions, the Lorentz group is $\Spin{1}{2}=\SL{2}{\real}$, the little group is $\integer_2$, and the on-shell space is 
\begin{align}
	\OnShell{3}{n}\coloneqq\left\{[\lambda]\in\frac{\Mat{2}{n}{\real}}{\SL{2}{\real}}:\sum_{i,j=1}^n\lambda_i^a\eta_{i,j}\lambda_j^b=0\right\},
\end{align}
where $\eta=\text{diag}(+1,-1,\ldots,+1,-1)$ and $\SL{2}{\real}$ acts of $\Mat{2}{n}{\real}$ by left multiplication. This is the kinematic space of the orthogonal Momentum Amplituhedron $\tnnOMomHat{k}$ where $n=2k$. The dimension of $\OnShell{3}{n}$ is given by
\begin{align}\label{eq:meet-on-shell-3-dim}
	\dim(\OnShell{3}{n}) = \dim(\Mat{2}{n}{\real})-\dim(\SL{2}{\real}) - 3 = 2n -6\,.
\end{align}
Four-dimensional spinor-helicity variables are real and independent in $(2,2)$ spacetime signature. In this case, the Lorentz group is $\Spin{2}{2}=\SL{2}{\real}\times\SL{2}{\real}$, the little group is $\real_{\ne 0}$, and the on-shell space is 
\begin{align}
	\OnShell{4}{n}\coloneqq\left\{([\lambda],[\tilde\lambda])\in\frac{\Mat{2}{n}{\real}}{\SL{2}{\real}}\times\frac{\Mat{2}{n}{\real}}{\SL{2}{\real}}:\sum_{i=1}^n\lambda_i^a\tilde\lambda_i^{\dot{a}} = 0\right\},
\end{align}
with dimension
\begin{align}\label{eq:meet-on-shell-4-dim}
	\dim(\OnShell{4}{n}) = 2(\dim(\Mat{2}{n}{\real})-\dim(\SL{2}{\real})) - 4 = 4n -10\,.
\end{align}
The Momentum Amplituhedron $\tnnMomHat{k}{n}$ lives in $\OnShell{4}{n}$.

The spinor-helicity formalism in six dimensions was first developed by Cheung and O'Connell in \cite{Cheung:2009dc}. In $(1,5)$ spacetime signature, the Lorentz group is $\SL{2}{\quarternion}$ where $\quarternion$ denotes the quaternions \cite{Kugo:1982bn}. It is isomorphic to $\symbolSU^\ast(4)$ where $\ast$ indicates that the group is pseudo-real \cite{Elvang:2013cua}. The little group $\SU{2}\times\SU{2}$ acts on chiral and anti-chiral spinors. Six-dimensional on-shell space can be parametrised solely in terms of chiral spinors $\lambda_i^{Aa}$ (where $A\in[4]$ and $a\in[2]$) according to 
\begin{align}
	\OnShell{6}{n}\coloneqq\left\{[\lambda]\in\frac{\Mat{4\times2}{n}{\complex}}{\symbolSU^\ast(4)}:\sum_{i,j=1}^{n}\lambda_i^{Aa}\eta_{i,j}\lambda_j^{Bb}=0\right\}\,,
\end{align} 
where $\eta$ is the symplectic matrix
\begin{align}
	\eta=
	\begin{bmatrix}
		0 & \mathbbb{1}_k \\
		-\mathbbb{1}_k& 0
	\end{bmatrix}\,.
\end{align}
Since the momentum conservation constraint is anti-symmetric, the dimension of $\OnShell{6}{n}$ is given by
\begin{align}\label{eq:meet-on-shell-6-dim}
	\dim(\OnShell{6}{n})=\dim(\Mat{4\times2}{n}{\complex}) - \dim(\symbolSU^\ast(4)) - 15 = 8n-21\,.
\end{align}

There are known twistor-string/CHY-inspired formulae for scattering amplitudes in six-dimensional theories \cite{Heydeman:2017yww,Cachazo:2018hqa,Heydeman:2018dje,Schwarz:2019aat,Geyer:2018xgb} where the  relevant moduli space is
\begin{align}\label{eq:meet-moduli-6}
	\Moduli{0}{n}\stackbin[i=1]{n}{\times}\left\{W_i\in\Mat{2}{2}{\complex}:\det(W_i)=\prod_{\ell\in[n]\setminus\{i\}}\sigma_{\ell,i}\right\}/\SL{2}{\complex}\,.
\end{align}
Its dimension is $4n-6$. Presumably, this moduli spaces has a preferred positive part which constitutes a positive geometry with some canonical form which we denote by $\omega_n^{\text{WS}(6)}$. This reasoning lead the authors of \cite{Huang:2021jlh,He:2021llb} to hypothesise the existence of the ``symplectic Momentum Amplituhedron'', a positive geometry in $\OnShell{6}{n}$ whose canonical form $\formOnShell[6]{n}$ is the pushforward of $\omega_n^{\text{WS}(6)}$ via the $D=6$ scattering equations $\mathcal{I}_n^{(6)}$. Although the six-dimensional analogue of the Momentum Amplituhedron is as-yet-unknown, we include it in our analysis for completeness.

It is useful to note that \cref{eq:meet-on-shell-3-dim,eq:meet-on-shell-4-dim,eq:meet-on-shell-6-dim} are summarised by the following formula
\begin{align}
	\dim(\OnShell{D}{n}) = 2(D-2)n-\frac{D(D+1)}{2}\,.
\end{align}
We will use this formula later on.

\subsection{Little Group Invariant Space}
\label{sec:meet-little-group}

The space of little group invariants is obtained by modding out the on-shell space by the appropriate little group: 
\begin{align}\label{eq:meet-LG-D}
	\LittleGroup{D}{n}\coloneqq\OnShell{D}{n}/\SO{D-2}\,.
\end{align}
Its dimension is given by
\begin{align}\label{eq:meet-LG-dim-D}
	\dim(\LittleGroup{D}{n}) = (D-1)n-\frac{D(D+1)}{2}\,.
\end{align}
We will denote coordinates on $\LittleGroup{D}{n}$ by $\ell\in\real^{\dim(\LittleGroup{D}{n})}$. \cref{eq:meet-LG-dim-D} is recoded in \cref{tbl:meet-little-group-dimensions} for $D=3,4,6$ and for small values of $n$. While the dimensions of $\LittleGroup{D}{n}$ and $\Mandelstam{n}$ coincide for $D\le n\le D+1$, $\dim(\LittleGroup{D}{n})<\dim(\Mandelstam{n})$ for $n>D+1$. The origin of this inequality is explored in \cref{sec:meet-gram}.

In four dimensions, the appropriate little group is $\real_{\ne0}^n$. It acts on spinor-helicity variables via
\begin{align}\label{eq:meet-LG-4}
	\lambda_i\mapsto t_i\lambda_i\,,\qquad\tilde\lambda_i\mapsto t_i^{-1}\tilde\lambda_i\,,
\end{align}
where $t\coloneqq(t_1,\ldots,t_n)\in\real_{\ne0}^n$ is an $n$-tuple of little group parameters and $([\lambda],[\tilde{\lambda}])\in\OnShell{4}{n}$. Clearly $\dim(\LittleGroup{4}{n})=3n-10$. Let $([\lambda'],[\tilde{\lambda}'])$ be a point in $\LittleGroup{4}{n}$. There are many options for parametrising $\LittleGroup{4}{n}$. One approach begins by focusing first on $[\lambda']$. In particular, one can regard $[\lambda']\in\Mat{2}{n}{\real}/\SL{2}{\real}/\real_{\ne0}^n$ as a point in $\Moduli{0}{n}$ with $n-3$ degrees of freedom. In this case, it is natural to employ the parametrisation of Fock and Gonchorov \cite{fock2006moduli}, given by
\begin{align}
	\lambda'\coloneqq
	\begin{bmatrix}
		0 & 1 & 1 & 1 & \cdots & 1 \\
		-1 & 0 & 1 & 1 + \ell_1 & \cdots & 1 + \sum_{i=1}^{n-3}\prod_{j=1}^i\ell_j
	\end{bmatrix}.
\end{align}
Thereafter, $\tilde{\lambda}'$ requires $2n-7$ additional parameters: $\SL{2}{\real}$ invariance removes three degrees of freedom;  momentum conservation eliminates four degrees of freedom and introduces dependence on $\ell_1,\ldots,\ell_{n-3}$. This defines a chart for $\LittleGroup{4}{n}$ referred to as the \emph{extended Fock--Goncharov (FG) parametrisation} in \cite{Damgaard:2020eox}. Moreover, any chart for $\LittleGroup{4}{n}$ extends to a chart for $\OnShell{4}{n}$ via \eqref{eq:meet-LG-4}.

\begin{table}[t]
	\centering
	\begin{tabular}{c|*{8}{p{0.7cm}<{\centering}}}
		\toprule
		$n$&$3$&$4$&$5$&$6$&$7$&$8$&$9$&$10$\\\midrule
		$\dim(\Mandelstam{n})$&$0$&$2$&$5$&$9$&$14$&$20$&$27$&$35$\\\midrule
		$\dim(\LittleGroup{3}{n})$&$0$&$2$&$4$&$6$&$8$&$10$&$12$&$14$\\
		$\dim(\LittleGroup{4}{n})$&N/A&$2$&$5$&$8$&$11$&$14$&$17$&$20$\\
		$\dim(\LittleGroup{6}{n})$&N/A&N/A&N/A&$9$&$14$&$19$&$24$&$29$\\
		\bottomrule
	\end{tabular}
	\caption{Comparison of dimensions of $\LittleGroup{D}{n}$ and $\Mandelstam{n}$ for $D=3,4,6$.}
	\label{tbl:meet-little-group-dimensions}
\end{table}

\subsection{Gram Variety}
\label{sec:meet-gram}

The ABHY associahedron $\Abhy{n}$ lives in the \emph{Mandelstam space} $\Mandelstam{n}$, the space of (complex) Mandelstam variables for $n$-particle scattering. The latter is paramatrised by $\frac{n(n-3)}{2}$ planar Mandelstam variables, the set of which we denote by $\setMandelstam{n}$. In $D$ dimension, there can be at most $D+1$ independent momenta. So for $n>D+1$ there are additional restrictions on $\setMandelstam{n}$ coming from the following \emph{Gram matrix}
\begin{align}\label{eq:meet-gram-matrix}
	G_{i,j}\coloneqq s_{i,j} = X_{i,j+1}+X_{i+1,j}-X_{i,j}-X_{i+1,j+1}\,,
\end{align}
where $i,j\in[n]$. Specifically, the principal minors of $G$ of order $D+1$ must vanish:
\begin{align}\label{eq:meet-gram-minors}
	|G|_{I,I}=0\,,~\text{for all}~I\in\binom{[n]}{D+1}\,,
\end{align}
where $|G|_{I,I}$ denotes the minor of $G$ located in the row and column set $I$. When $n=D+1$, $\det(G)$ vanishes identically due to momentum conservation. \cref{eq:meet-gram-minors} defines the \emph{Gram ideal}
\begin{align}\label{eq:meet-gram-ideal}
	\idealRestrict{D}{n}\coloneqq\left\langle|G|_{I,I}:I\in\binom{[n]}{D+1}\right\rangle\subset\complex[\setMandelstam{n}]\,.
\end{align}
The letter ``r'', written in fraktur font, stands for ``restriction'' as the vanishing of the principal minors of $G$ of order $D+1$ restricts $\setMandelstam{n}$. The \emph{Gram variety} is the zero-set of this ideal
\begin{align}\label{eq:meet-gram-variety}
	\Gram{D}{n}\coloneqq\zero(\idealRestrict{D}{n})\subset\Mandelstam{n}\,.
\end{align}
Its dimension is precisely the dimension of the appropriate little group invariant space
\begin{align}
	\dim(\Gram{D}{n}) = \dim(\LittleGroup{D}{n})\,.
\end{align}
The difference between the dimensions of $\Mandelstam{n}$ and $\Gram{D}{n}$ is
\begin{align}
	\dim(\Mandelstam{n})-\dim(\Gram{D}{n}) = \binom{n-D}{2}\,.
\end{align}
Let $\setRestrict{D}{n}$ be a subset of $\setMandelstam{n}$ of size $\dim(\Gram{D}{n})$. Practically speaking, \eqref{eq:meet-gram-minors} allows us to parametrize $\Gram{D}{n}$ in terms of $\setRestrict{D}{n}$ by eliminating the remaining planar Mandelstam variables. This perspective motivates the definition of the following zero-dimensional ideal
\begin{align}\label{eq:meet-gram-ideal-prime}
	\idealRestrictPrime{D}{n}\coloneqq\left\langle|G|_{I,I}:I\in\binom{[n]}{D+1}\right\rangle\subset\complex(\setRestrict{D}{n})[\setMandelstam{n}\setminus\setRestrict{D}{n}]\,,
\end{align}
called the \emph{modified Gram ideal} where
\begin{align}
	|{\zero(\idealRestrictPrime{D}{n})}|= \binom{n-D}{2}\,.
\end{align}
The size of $\zero(\idealRestrictPrime{D}{n})$ for $D=3,4,6$ and for small values of $n$ is displayed in \cref{tbl:meet-gram-ideal-sizes}.

\begin{table}[t]
	\centering
	\begin{tabular}{c|*{8}{p{0.7cm}<{\centering}}}
		\toprule
		$n$&$3$&$4$&$5$&$6$&$7$&$8$&$9$&$10$\\
		\midrule
		$|\zero(\idealRestrictPrime{3}{n})|$&$\text{N/A}$&$0$&$1$&$3$&$6$&$10$&$15$&$21$\\
		$|\zero(\idealRestrictPrime{4}{n})|$&$\text{N/A}$&$\text{N/A}$&$0$&$1$&$3$&$6$&$10$&$15$\\
		$|\zero(\idealRestrictPrime{6}{n})|$&$\text{N/A}$&$\text{N/A}$&$\text{N/A}$&$\text{N/A}$&$0$&$1$&$3$&$6$\\
		\bottomrule
	\end{tabular}
	\caption{Comparison of sizes of $\zero(\idealRestrictPrime{D}{n})$ for $D=3,4,6$.}
	\label{tbl:meet-gram-ideal-sizes}
\end{table}

\section{Maps}
\label{sec:meet-maps}

\begin{figure}[t]
	\centering
	\includegraphics{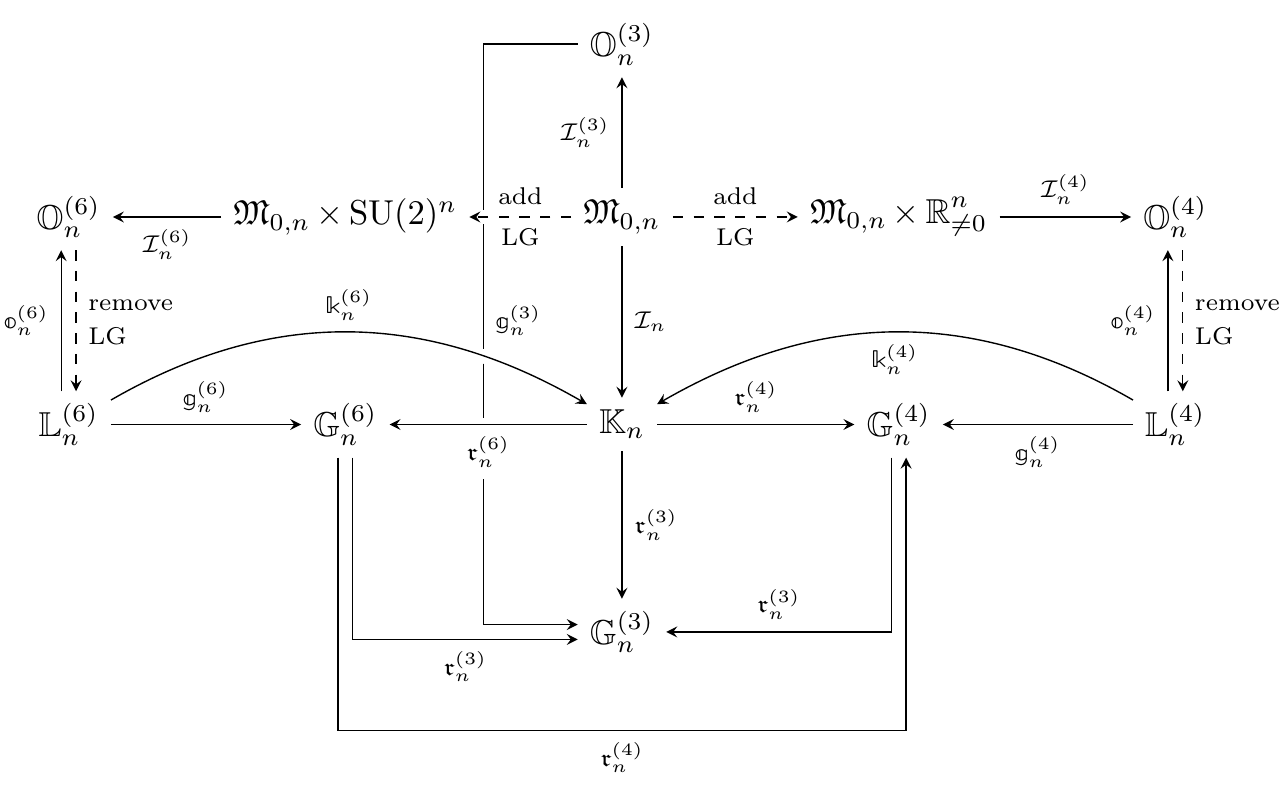}
	\caption{Web of kinematic spaces and maps relating them.}
	\label{fig:meet-maps}
\end{figure}

The kinematic spaces of \cref{sec:meet-spaces} relate to each other via particular maps. This web of relations is depicted in \cref{fig:meet-maps}. In this section, we define the maps between the kinematic spaces which are later used to perform pullback and pushforward operations. 

As previously discussed, any chart for the little group invariant space $\LittleGroup{D}{n}$ extends to a chart for the on-shell space $\OnShell{D}{n}$. In general, we will denote this extension by the rational map
\begin{align}
	\mapOnShell{D}{n}:\LittleGroup{D}{n}\times\SO{D-2}\to\OnShell{D}{n}\,.
\end{align} 
In four dimensions, the extended FG parametrisation furnishes a chart for $\LittleGroup{4}{n}$ with coordinates $\ell\coloneqq(\ell_1,\ldots,\ell_{3n-10})\in\real^{3n-10}$. Identifying $\LittleGroup{4}{n}$ with $\real^{3n-10}$, we have that \begin{align}
	\mapOnShell{4}{n}:\LittleGroup{4}{n}\times\real_{\ne0}^{n}\to\OnShell{4}{n}\,,(\ell,t)\mapsto([\lambda],[\tilde{\lambda}])\,,
\end{align} 
where $\lambda_i=t_i\lambda_i'(\ell)$ and $\tilde{\lambda}_i=t_i^{-1}\tilde{\lambda}_i'(\ell)$. The rational map $\mapOnShell{4}{n}$ is invertible for $t\in\real_{>0}^n$ \cite{Damgaard:2020eox}.

There are two obvious maps involving planar Mandelstam variables defined on $\LittleGroup{D}{n}$. One of them maps $\LittleGroup{D}{n}$ to the Mandelstam space $\Mandelstam{n}$:
\begin{align}
	\mapMandelstam{D}{n}:\LittleGroup{D}{n}\to\Mandelstam{n}\,,\ell\mapsto\setMandelstam{n}\,.
\end{align}
Here $\ell\in\real^{\dim(\LittleGroup{D}{n})}$ are coordinates on $\LittleGroup{D}{n}$. Another relates $\LittleGroup{D}{n}$ to the Gram variety $\Gram{D}{n}$:
\begin{align}
	\mapGram{D}{n}:\LittleGroup{D}{n}\to\Gram{D}{n}\,,\ell\mapsto\setRestrict{D}{n}\,.
\end{align}
 Both are rational maps defined via $X_{i,j}=\sum_{i\le i'<j'<j}\lrangle{i'j'}[i'j']$. For $\mapGram{D}{n}$, we restrict our attention to those $X_{i,j}$ contained in $\setRestrict{D}{n}$, a subset of $\setMandelstam{n}$ of size $\dim(\Gram{D}{n})$. The graph of $\mapGram{D}{n}$ defines the following zero-dimensional ideal
\begin{align}
	\idealGram{D}{n}\coloneqq\left\langle X_{i,j} - \sum_{i\le i'<j'<j}\lrangle{i'j'}[i'j']: X_{i,j}\in\setRestrict{D}{n}\right\rangle\subset\complex(\setRestrict{D}{n})[\ell]\,.
\end{align}
In the next section, we will use $\idealGram{D}{n}$ to perform pushforwards. In particular, $(\mapGram{D}{n})_\ast=(\idealGram{D}{n})_\ast$. The degrees of $\mapMandelstam{D}{n}$ and $\mapGram{D}{n}$ are non-trivial (i.e.\ larger than one) for $n\ge D+1$ \cite{Damgaard:2020eox}.

Lastly, the modified Gram ideal $\idealRestrictPrime{D}{n}\subset\complex(\setRestrict{D}{n})[\setMandelstam{n}\setminus\setRestrict{D}{n}]$, defined in \eqref{eq:meet-gram-ideal-prime}, relates $\Mandelstam{n}$ to $\Gram{D}{n}$. It allows us to pushforward rational forms defined on $\Mandelstam{n}$ such as $\omega_n^\text{ABHY}$.

\section{Rational Forms}
\label{sec:meet-forms}

The maps of \cref{sec:meet-maps} allow us to connect the canonical forms of different positive geometries. In particular, they allow us to define rational forms on $\LittleGroup{D}{n}$ and $\Gram{D}{n}$ related in natural ways. In this section, we will focus on four dimensions, reviewing the results of \cite{Damgaard:2020eox}. Some of these results were verified in \cite{Lukowski:2022fwz} using the methods discussed in \cref{chp:push}.

The Momentum Amplituhedron's canonical form $\formOnShell[4][k]{n}\coloneqq\Omega(\tnnMomHat{k}{n})$ is a rational form on $\OnShell{4}{n}$ of degree $2n-4$. It is the pushforward of $\omega_n^{\text{WS}(4)}\coloneqq\omega_n^\text{WS}\wedge\Omega(\projective(\real_{\ge0}^n))$ via the $D=4$ scattering equations (see \cref{eq:mom-pushforward-scattering-equations}). The pullback of $\formOnShell[4][k]{n}$ via $\mapOnShell{4}{n}$ defines a rational form in the variables $\ell=(\ell_1,\ldots,\ell_{3n-10})$ and $t=(t_1,\ldots,t_n)$. Empirical evidence \cite{Damgaard:2020eox} indicates that 
\begin{align}
	(\mapOnShell{4}{n})^\ast\formOnShell[4][k]{n}=\formLittleGroup[4][k]{n}\wedge\Omega(\projective(\real_{\ge0}^n))+\mathcal{O}(d^{n-2}t)\,,
\end{align}
where $\Omega(\projective(\real_{\ge0}^n))=\sum_{i=1}^n(-1)^{i+1}\bigwedge_{j\ne i}\frac{dt_j}{t_j}$ is the canonical form on $\projective(\real_{\ge0}^n)$ and $\mathcal{O}(d^{n-2}t)$ denotes rational forms of degree at most $n-2$ in $dt$. The terms inside $\mathcal{O}(d^{n-2}t)$ depend on the choice of chart for $\LittleGroup{4}{n}$ and are ignored. On the other hand, the \emph{reduced Momentum Amplituhedron form} $\formLittleGroup[4][k]{n}$ is parametrisation independent and hence well-defined on $\LittleGroup{4}{n}$. Moreover, it is invariant under the action of the little group. Its degree is $n-3$, because $\deg\formOnShell[4][k]{n}=2n-4$ and $\deg\Omega(\projective(\real_{\ge0}^n))=n-1$. The inverse-soft construction of \cite{He:2018okq} for $\formOnShell[4][k]{n}$ can be used to recursively derive $\formLittleGroup[4][k]{n}$ as demonstrated in \cite{Damgaard:2020eox}. An all-multiplicity expression for the reduced Momentum Amplituhedron form is known for MHV and $\overline{\text{MHV}}$ sectors \cite{Damgaard:2020eox}:
\begin{align}
	\formLittleGroup[4][2]{n}=\bigwedge_{i=2}^{n-2}d\log\frac{\lrangle{1,i}\lrangle{i+1,i+2}}{\lrangle{1,i+1}\lrangle{i,i+2}}\,,\quad\formLittleGroup[4][n-2]{n}=\bigwedge_{i=2}^{n-2}d\log\frac{[1,i][i+1,i+2]}{[1,i+1][i,i+2]}\,,
\end{align}
where the spinor brackets are evaluated on $\LittleGroup{4}{n}$.

\begin{figure}
	\centering
	\includegraphics{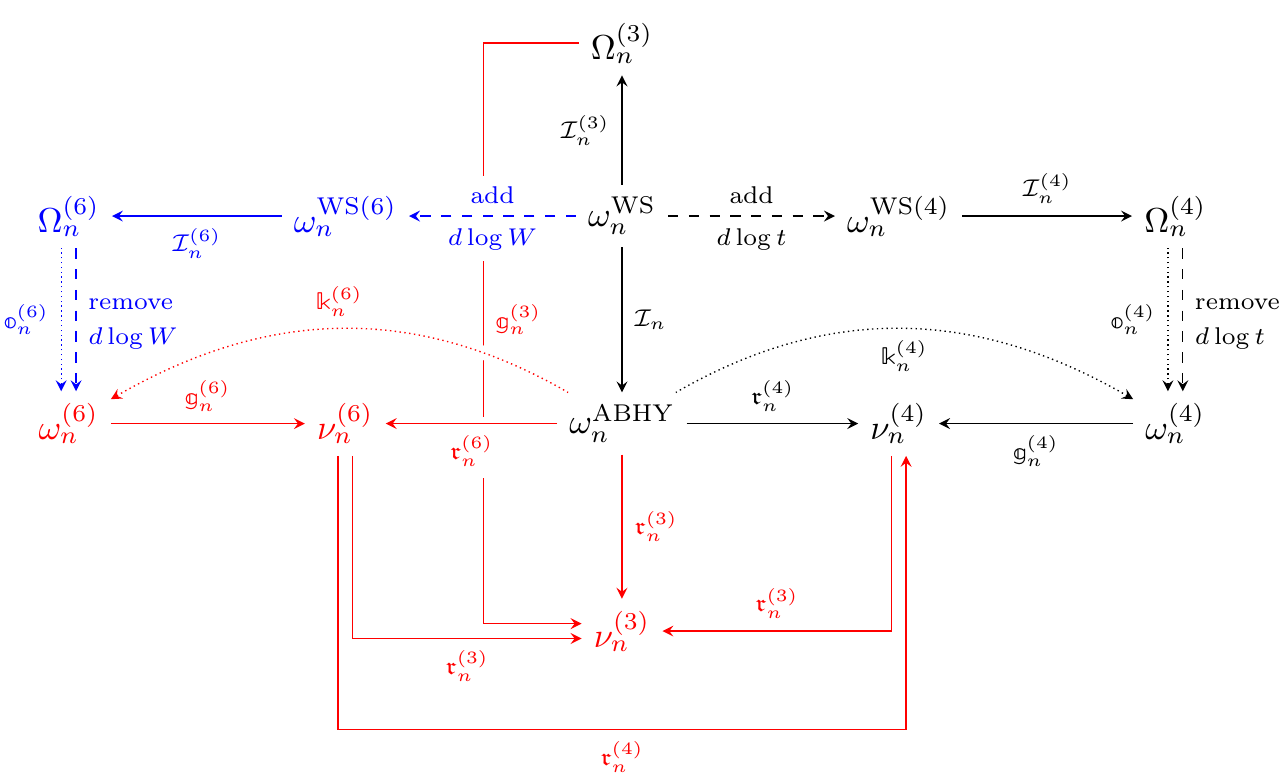}
	\caption{Web of rational forms and maps relating them. Solid lines designate pushforwards. Pullbacks are denoted by dotted lines. Black lines refer to relations already established in the literature, red lines are conjectures, and blue lines depend on the hypothetical symplectic Momentum Amplituhedron.}
	\label{fig:meet-forms}
\end{figure}

The pullback of the planar scattering form $\omega_n^\text{ABHY}$ via $\mapMandelstam{D}{n}$ produces  
\begin{align}
	\formLittleGroup[D]{n}\coloneqq(\mapMandelstam{D}{n})^{\ast}\omega_n^\text{ABHY}\,,
\end{align}
a well-defined $(n-3)$-form on $\LittleGroup{D}{n}$. Specialising to four-dimensions, $\formLittleGroup[4]{n}$ coincides with the reduced Momentum Amplituhedron form (summed over all helicity sectors) \cite{Damgaard:2020eox}:
\begin{align}\label{eq:meet-forms-little-group-4}
	\sum_{k=2}^{n-2}\omega_{n,k}^{(4)}=\omega_{n}^{(4)} \,.
\end{align}
\cref{eq:meet-forms-little-group-4} is the first of two foremost findings of \cite{Damgaard:2020eox}. It relates (the pullbacks of) the rational forms associated with tree-level scattering in $\mathcal{N} = 4$ SYM and BAS on $\LittleGroup{4}{n}$. The relationship stems from the fact that tree-level amplitudes in both theories have factorisation channels given by the vanishing of planar Mandelstam variables. \cref{fig:meet-forms} illustrates \eqref{eq:meet-forms-little-group-4} as a strand in a web of relations. 

The Gram variety provides an additional arena for comparing the Momentum Amplituhedron form and the planar scattering form. Pushing forward $\formLittleGroup[4][k]{n}$ via $\mapGram{4}{n}$ (equivalently $\idealGram{4}{n}$) produces the following $(n-3)$-form on $\Gram{4}{n}$: 
\begin{align}
	\formGram[4][k]{n}\coloneqq(\mapGram{4}{n})_\ast\formLittleGroup[4][k]{n}=(\idealGram{4}{n})_\ast\formLittleGroup[4][k]{n}\,.
\end{align}
Correspondingly, the pushforward of $\omega_n^\text{ABHY}$ via $\idealRestrictPrime{D}{n}$ yields 
\begin{align}
	\formGram[D]{n}\coloneqq(\idealRestrictPrime{D}{n})_\ast\omega_n^\text{ABHY}\,,
\end{align}
a rational form on $\Gram{D}{n}$ of the same degree. In four dimensions, $\formLittleGroup[4]{n}$ couples to $\sum_{k=2}^{n-2}\formGram[4][k]{n}$ through \cite{Damgaard:2020eox}
\begin{align}\label{eq:meet-forms-gram-4}
	\sum_{k=2}^{n-2}\formGram[4][k]{n} = \frac{|{\zero(\idealGram{4}{n})}|}{|{\zero(\idealRestrictPrime{4}{n})}|}\formGram[4]{n}
	\,.
\end{align}
\cref{eq:meet-forms-gram-4} serves as the second seminal discovery of \cite{Damgaard:2020eox}. It too is included in \cref{fig:meet-forms}. The ratio on the right-hand-side of \eqref{eq:meet-forms-gram-4} is given by
\begin{align}\label{eq:meet-forms-gram-ration-4}
\frac{|{\zero(\idealGram{4}{n})}|}{|{\zero(\idealRestrictPrime{4}{n})}|} = 2-\delta_{n,4}\,.
\end{align}
\cref{eq:meet-forms-gram-ration-4} was checked for $4\le n\le 7$ in \cite{Damgaard:2020eox}. It is a property of the pushforwards, not the rational forms. 

Starting at $n=6$, the leading singularities (i.e.\ iterated residues on simple poles that produces numbers) of $\formGram[4][k]{n}$ include, but are not limited to $\pm1$. Consequently, they cannot be canonical forms of positive geometries defined in $\Gram{4}{n}$. They may, however, correspond to weighted positive geometries, \`{a} la \cite{Dian:2022tpf}. 

It seems reasonable to expect that relationships analogous to \eqref{eq:meet-forms-little-group-4} and \eqref{eq:meet-forms-gram-4} exist for the orthogonal Momentum Amplituhedron and the ``symplectic Momentum Amplituhedron'' (if it exists). A web of these conjectured connections is contained in \cref{fig:meet-forms}, building on the graphic in the conclusions of \cite{He:2021llb}. All pushforwards in \cref{fig:meet-forms} can be calculated using the methods developed in \cref{chp:push}.

\section{Summary}
This chapter presented a network of interrelations between the various positive geometries studied in this dissertation. In particular, we considered relationships between the canonical forms of the $D$-dimensional Momentum Amplituhedron and the ABHY associahedron in two settings: the little group invariant space and the Gram variety. These relationships are confirmed in four dimensions (see \cite{Damgaard:2020eox}), and we expect them to hold in three and six dimensions. In the four-dimensional case, the restriction of the planar scattering form to the Gram variety no longer has uniform leading residues \cite{Damgaard:2020eox}, so it cannot be a canonical form. Nevertheless, it correctly captures the physics of tree-level scattering in four-dimensional BAS. This observation suggests a blueprint for describing scattering amplitudes in more realistic theories as some restriction of some positive geometry.
	\chapter{Conclusions and Outlook}
\label{chp:concl}

\section{Summary}

\lettrine{I}{n this dissertation}, we surveyed three intricately interrelated positive geometries: the Arkani-Hamed--Bai--He--Yan (ABHY) associahedron, the ($D=4$) Momentum Amplituhedron, and the orthogonal (or $D=3$) Momentum Amplituhedron. Each encodes tree-level scattering amplitudes in on-shell momentum space: the ABHY associahedron uses Mandelstam invariants; the three- and four-dimensional Momentum Amplituhedra employ spinor-helicity variables in three and four dimensions, respectively. Although relevant for different theories in different spacetime dimensions, the three positive geometries share multiple connections between their boundary posets and canonical forms. 

To synthesise and study the boundary stratifications of the three- and four-dimensional Momentum Amplituhedra, we formalised several ideas in \cref{chp:pos}. We introduced the notions of dissections and tilings induced by rational maps, generalising similar notions found in \cite{Lukowski:2020dpn}. We then modified the characterisation of morphisms to include rational maps between positive geometries of possibly differing dimensions. This modification facilitated the definition of morphism-induced boundaries (summarising definitions developed in \cite{Lukowski:2019kqi,Lukowski:2020bya,Ferro:2020lgp,Moerman:2021cjg,Lukowski:2021fkf}). Furthermore, we identified conjectures implicit in \cite{Lukowski:2019kqi,Lukowski:2020bya,Ferro:2020lgp, Moerman:2021cjg,Lukowski:2021fkf}, including the presumption that morphism-induced boundaries are actual boundaries.

In \cref{chp:mom}, we studied the Momentum Amplituhedron’s boundary stratification using the aforementioned conjecture. This boundary stratification is an induced subposet of the positroid stratification enumerated by contracted Grassmannian forests \cite{Ferro:2020lgp,Moerman:2021cjg}. Indeed, the inherited partial order mimics the singularity structure of partial amplitudes in $\mathcal{N}=4$ supersymmetric Yang-Mills (SYM) theory. \cref{fig:grass-covering} displays the covering relations. We found a similar characterisation for faces of orthogonal Momentum Amplituhedron in \cref{chp:omom}. In this case, the correct combinatorial labels are orthogonal Grassmannian forests \cite{Lukowski:2021fkf}. The covering relations, shown in \cref{fig:ograss-covering}, reflect the singularity structure of partial amplitudes in Aharony--Bergman--Jafferis--Maldacena (ABJM) theory. The boundary poset is an induced subposet of the orthitroid stratification \cite{Lukowski:2021fkf}. It is also an induced subposet of the Momentum Amplituhedron’s face poset, consisting of faces labelled by orthogonal Grassmannian forests \cite{Lukowski:2021fkf}. Thus, combinatorially speaking, the Momentum Amplituhedron contains the orthogonal Momentum Amplituhedron. \cref{fig:omom-subposet} summarises these connections.

Having precise diagrammatic descriptors for boundaries allowed us to engineer rank generating functions using results developed in \cite{Moerman:2021cjg} and summarised in \cref{chp:enum}. As a warm-up exercise, we used the series-reduced planar tree analogue of the Exponential formula to deduce the rank generating function for the ABHY associahedron in \cref{chp:abhy}. In \cref{chp:mom}, we presented the Momentum Amplituhedron's rank generating function \cite{Moerman:2021cjg}. Similarly, we deliberated over the rank generating function for the orthogonal Momentum Amplituhedron \cite{Lukowski:2021fkf} in \cref{chp:omom}. Both results use the series-reduced planar forest analogue of the Exponential formula. In each case, the rank generating function implies that the Euler characteristic is one.

The three positive geometries relate via the scattering equations to (appropriate extensions of) the worldsheet associahedron, a positive geometry in the genus zero moduli space \cite{Arkani-Hamed:2017mur,He:2021llb}; see \cref{fig:meet-maps}. The scattering equations concretely connect the CHY formalism in various dimensions to the framework of positive geometries. In each case, the scattering equations define a morphism between the momentum space positive geometry and the positive geometry in the space of worldsheet moduli; see \cref{chp:abhy,chp:mom,chp:omom}. Pushing forward (the appropriate extension of) the Parke-Taylor form via the scattering equations yields the canonical form of the momentum space positive geometry. 

In addition, the scattering equations in three dimensions provide an additional strand in the web of combinatorial connections between boundary posets. Faces of the worldsheet associahedron are in bijection with series-reduced planar trees. The same is true for the ABHY associahedron. The authors of \cite{Lukowski:2021fkf} conjecture that the three-dimensional scattering equations squash the boundaries of the worldsheet associahedron in a manner compatible with a simple partial-order-preserving pictorial rule; see \cref{chp:omom}. In this sense, the orthogonal Momentum Amplituhedron descends from the worldsheet associahedron. \cref{fig:omom-subposet} includes this combinatorial connection.

We explored three methods from \cite{Lukowski:2022fwz} in \cref{chp:push} for efficiently evaluating pushforwards. They utilize tools from computational algebraic geometry --- companion matrices and the duality of global residues --- to sidestep the problem of determining local inverses. Reference \cite{Lukowski:2022fwz} discusses the relative strengths and weaknesses of each approach. The three methods provide conceptual and practical resources for calculating pushforwards. They are suitable for constructing canonical forms in \cref{chp:abhy,chp:mom,chp:omom} as well as for performing pushforwards in \cref{chp:meet}.

In \cref{chp:meet}, we brought our discussion back to the ABHY associahedron, reviewing the results of \cite{Damgaard:2020eox} and their extensions in \cite{Lukowski:2022fwz}. To this end, we introduced $D$-dimensional analogues of the space of little group invariants and the Gram variety. The latter arises from so-called Gram determinant conditions on the space of Mandelstam invariants. We discussed the different maps which connect these auxiliary arenas, as summarised in \cref{fig:meet-maps}. As observed in \cite{Damgaard:2020eox}, the canonical forms of the ABHY associahedron and the Momentum Amplituhedron are precisely related in  these secondary settings. This phenomenon reflects the fact that factorisation channels of tree-level amplitudes in bi-adjoint scalar theory (BAS) and $\mathcal{N}=4$ SYM coincide. It is speculated in \cite{Lukowski:2022fwz} that similar comparisons hold for the orthogonal Momentum Amplituhedron. In effect, pulling back onto the space of little group invariants undoes the scattering equations, while pushing forward onto the Gram variety restricts the ABHY associahedron's canonical form to specific spacetime dimensions. \cref{fig:meet-forms} summarises the analysis of \cref{chp:meet}.

\section{Outlook}

The work presented in this dissertation provides multiple directions for future research. From a pure mathematics perspective, this dissertation makes several substantive claims which require proof. The characterisation of (orthogonal) Momentum Amplituhedron boundaries in terms of (orthogonal) Grassmannian forests rests on two conjectures: 
\begin{itemize}
	\item Morphism-induced boundaries coincide with actual boundaries;
	\item The algorithm developed in \cite{Lukowski:2019kqi} and discussed in \cref{chp:pos} determines all morphism-induced boundaries.
\end{itemize}
With these assumptions, a complicated topological problem (determining the face poset of a positive geometry) translates into a simple combinatorial one. These conjectures have wide-ranging applications. In particular, they are relevant for proving that the three-dimensional scattering equations ``squash'' the boundary poset of the worldsheet associahedron in a manner compatible with the diagrammatic map presented in \cite{Lukowski:2021fkf} (see \cref{chp:omom}).

Moreover, there are many unanswered questions concerning the boundary stratifications of the three- and four-dimensional Momentum Amplituhedra. For example, is the poset thin and shellable? If so, it is Eulerian and the face poset of a regular CW complex homeomorphic to a sphere \cite{bjorner1984posets}. \cref{fig:grass-covering,fig:ograss-covering} present the covering relations relevant for proving these properties.

On a practical point, all three methods for pushing forward rational forms, developed in \cite{Lukowski:2022fwz} and presented in \cref{chp:push}, use Gr\"{o}bner bases. This dependency poses a significant bottleneck for high-multiplicity examples or when working over fraction fields (as opposed to sampling over finite fields). Improving the efficiency of Gr\"{o}bner basis routines for fraction fields would significantly strengthen our results. Alternatively, it will prove invaluable to interface existing rational reconstruction algorithms (e.g.\ \texttt{FiniteFlow} \cite{Peraro:2019svx} and \texttt{FireFly} \cite{Klappert:2019emp}) with ultra-efficient algorithms for calculating Gr\"{o}bner bases over finite fields (e.g.\ Faug\`ere's $F4$/$F5$ algorithms \cite{faugere1999new,faugere2002new}). It is also worthwhile investigating methods for performing pushing forwards which do not require Gr\"{o}bner bases.

There are also many potential future physics projects related to this dissertation. Speculation on a $D=6$ Momentum Amplituhedron \cite{Huang:2021jlh,He:2021llb} warrants further investigation. Its existence is natural from the perspective of the $D=6$ scattering equations. (Recall that the $D=3,4$ scattering equations define maps onto the $D=3,4$ Momentum Amplituhedra \cite{He:2021llb}.) One idea for defining a $D=6$ Momentum Amplituhedron is to identify a suitable positive geometry in the moduli space given by \eqref{eq:meet-moduli-6} and to study its image under the $D=6$ scattering equations. To this end, the pushforward techniques discussed in this dissertation may prove useful. Alternatively, one can define the symplectic Momentum Amplituhedron as the image of the symplectic Grassmannian via some morphism. In this case, it is natural to expect that its boundaries are enumerated by suitable symplectic analogues of Grassmannian forests.

So far, we have shone a spotlight on positive geometries relevant for tree-level amplitudes. Recently, a loop-level version of the Momentum Amplituhedron was defined \cite{Ferro:2022abq}. It is the spinor-helicity analogue of the Amplituhedron, describing the integrands of amplitudes in planar $\mathcal{N}=4$ SYM. The loop Momentum Amplituhedron is not parity-symmetric with respect to helicity because its definition builds on that of the loop Amplituhedron. Nevertheless, it is desirable to analyse the loop Momentum Amplituhedron using the methodologies of this dissertation: to investigate and combinatorially characterise its boundary stratification; to consider analogues of its canonical form on the space of little group invariants and the Gram variety, obtained via pullback and pushforward operations. Furthermore, is it possible to define a parity-symmetric loop Momentum Amplituhedron using knowledge about factorisation channels and forward limits?

It is as-yet-unknown how to generalise the above constructions to non-planar loop amplitudes. The integrands of non-planar loop amplitudes in $\mathcal{N}=4$ SYM have logarithmic
singularities and uplift to logarithmic differential forms \cite{Arkani-Hamed:2014via}. The relevant combinatorial labels are non-planar on-shell diagrams. There are some results for MHV amplitudes \cite{Arkani-Hamed:2014bca} and six-particle NMHV amplitudes \cite{Bourjaily:2016mnp}. Reference \cite{Paranjape:2022ymg} initiates a systematic study of non-planar on-shell diagrams arising from non-adjacent BCFW shifts. Nevertheless, further investigation is required to understand the connection between generic non-planar on-shell diagrams and the Grassmannian. 

On-shell diagrams exist for other theories too, including gravity \cite{Heslop:2016plj,Herrmann:2016qea,Armstrong:2020ljm} where amplitudes are invariant under permutations of identical particles. Presently, there is no positive geometry describing gravity, despite promising progress towards a putative
Gravituhedron \cite{Trnka:2020dxl}. This thesis provides tools for finding such a positive geometry (or suitable generalisation thereof). In particular, one can use the methods of \cref{chp:push} to pushforward the CHY integrand for some gravitational theory  via the scattering equations to produce some rational form in on-shell momentum space. The resulting rational form will have non-trivial numerators and higher-order poles \cite{Herrmann:2016qea}. Reference \cite{Benincasa:2020uph} provides a possible framework for dealing with these difficulties. Alternatively, one can use point-splitting regularisation to obtain simple poles from higher-order poles. Nevertheless, we leave this important pushforward calculation for a later publication. 

Returning to the content of my thesis, the emergence of forest-like labels for boundaries of the Momentum Amplituhedron and the orthogonal Momentum Amplituhedron, positive geometries describing \emph{tree-level} scattering amplitudes, is to be expected. In fact, it bolsters our belief that they correctly capture the dynamics of the corresponding theories. Perhaps combinatorics provides an appropriate point of departure for defining (suitable generalisations of) positive geometries. Starting with an on-shell diagrammatic description for the singularity structure of some $S$-matrix (supplemented with covering relations), can one find a geometric realisation of the resulting poset in momentum space? 

These questions, while beyond the scope of this dissertation, represent but a snapshot of the research possibilities opened up through the paradigm of positive geometries.	
	
	\appendix
	
	\chapter{Results from Computational Algebraic Geometry}
\label{chp:misc}

In this appendix, we enclose a handful of results from computational algebraic geometry. These theorems are relevant for \cref{chp:push}.

\begin{theorem}[Shape Lemma \cite{sturmfels2002solving}]\label{thm:shape}
	Let $\mathcal{I}$ be a zero-dimensional radical ideal in $\complex[x_1,\ldots,x_m]$. Suppose that all $d$ complex roots of $\mathcal{I}$ have distinct $x_m$ coordinates. Then the unique reduced Gr\"{o}bner basis of $\mathcal{I}$ with respect to lex order ($x_1\succ\ldots\succ x_m$) has the shape
	\begin{align}
		\mathcal{G}_\text{lex}(\mathcal{I}) = \left\{x_1-q_1(x_m),\ldots,x_{m-1}-q_{m-1}(x_m),r(x_m)\right\},
	\end{align}
	where $r\in\complex[x_m]$ has degree $d$ and $q_i\in\complex[z_m]$ has degree strictly less than $d$ for each $i\in[m-1]$.
\end{theorem}

\begin{theorem}[Stickelberger's Theorem \cite{sturmfels2002solving}]\label{thm:stickelberger}
	Let $\mathcal{I}$ be a zero-dimensional ideal in $\complex[x_1,\ldots,x_m]$. Then the zero-set of $\mathcal{I}$ is the set of vectors of simultaneous eigenvalues of the companion matrices of $\mathcal{I}$.
\end{theorem}

The following corollary can be found in \cite{sturmfels2002solving}, for example.
\begin{corollary}\label{thm:simultaneous}
	Let $\mathcal{I}$ be a zero-dimensional ideal in $\complex[x_1,\ldots,x_m]$. The companion matrices of $\mathcal{I}$ are simultaneously diagonalizable if and only if $\mathcal{I}$ is a radical ideal.
\end{corollary}

\begin{theorem}[Elimination Theorem \cite{cox2013ideals}]\label{thm:elimination}
	Let $\mathcal{I}$ be an ideal in $\mathbb{C}[x_1,\ldots,x_m]$ and let $\mathcal{G}$ be a Gr\"{o}bner basis for $\mathcal{I}$ with respect to the lex order where $x_1\succ\ldots\succ x_m$. Then for every $0\le\ell\le m$, $\mathcal{G}\cap\mathbb{C}[x_{\ell+1},\ldots,x_m]$ is a Gr\"{o}bner basis for the $\ell$\textsuperscript{th} elimination ideal $\mathcal{I}\cap\mathbb{C}[x_{\ell+1},\ldots,x_m]$.
\end{theorem}

\begin{theorem}[Hilbert's Weak Nullstellensatz \cite{cox2013ideals}]\label{thm:nullstellensatz}
	Let $\mathcal{I}$ be an ideal in $\mathbb{C}[x_1,\ldots,x_m]$ whose zero-set is empty. Then $\mathcal{I}=\mathbb{C}[x_1,\ldots,x_m]$.
\end{theorem}

\begin{theorem}[{Global Duality Theorem \cite{cattani2005introduction}}]\label{thm:global-duality}
	Let $\mathcal{I}=\lrangle{f_1,\ldots,f_n}$ be a zero-dimensional ideal in $\complex[x_1,\ldots,x_m]$, let $\mathcal{Q}=\complex[x_1,\ldots,x_m]/\mathcal{I}$, let $h\in\complex[x_1,\ldots,x_m]$ and let $\Res(h)$ denote the global residue of $h$ with respect to the divisors $f_1,\ldots,f_m$. Then the symmetric inner-product $\lrangle{\bullet,\bullet}$ defined by
	\begin{align}
		\lrangle{\bullet,\bullet}:\mathcal{Q}\times \mathcal{Q}\to\complex,\; ([h_1],[h_2])\mapsto\lrangle{[h_1],[h_2]}\coloneqq\Res(h_1h_2)\,,
	\end{align}
	is non-degenerate.
\end{theorem}
	\chapter{Algebraic Relations}
\label{chp:alg-rel}

The rank generating functions for contracted Grassmannian trees and forests presented in \cref{chp:mom} satisfy the following algebraic relations \cite{Moerman:2021cjg}:
\begin{align}
	\label{eq:mom-GF-tree-rel}
	\begin{autobreak}
		0 
		= 
		q (q
		+1) (\mathcal{G}^{\symbolMom}_{\tree})^5 
		+q x (q^2 x y
		-5 q (y
		+1)
		-5 (y
		+1)) (\mathcal{G}^{\symbolMom}_{\tree})^4
		+ x^2 (q (-3 q^2 x y (y
		+1)
		+q (10
		-y ((x
		-10) y
		+x
		-19))
		+y (10 y
		+21)
		+10)
		+y)(\mathcal{G}^{\symbolMom}_{\tree})^3
		+ x^3 (q^3 x y (y (3 y
		+7)
		+3)
		+q^2 (y
		+1) (y (2 (x
		-5) y
		+2 x
		-17)
		-10)
		-q (y (y (-x
		+10 y
		+32)
		+32)
		+10)
		-3 y (y
		+1))(\mathcal{G}^{\symbolMom}_{\tree})^2
		- x^4 (q^3 x y (y
		+1) (y (y
		+4)
		+1)
		+q^2 (y
		+1) (y ((x
		-5) y^2
		+3 (x
		-4) y
		+x
		-12)
		-5)
		-q (y
		+1) (y (y (-x
		+5 y
		+16)
		+16)
		+5)
		-y (y (3 y
		+5)
		+3)) \mathcal{G}^{\symbolMom}_{\tree}
		+x^5 (q^3 x y^2 (y
		+1)^2
		-q^2 (y
		+1)^2 (y (y (-x
		+y
		+2)
		+2)
		+1)
		-q (y (y (y (-x
		+y (y
		+5)
		+10)
		+10)
		+5)
		+1)
		-y (y
		+1) (y^2
		+y
		+1))\,,
	\end{autobreak}\\
	\label{eq:mom-GF-forest-rel}
	\begin{autobreak}
		0
		=
		q (x
		-1) x^2 y (x y
		-1) (q (x y
		+x
		-1)
		+x) (q (x y
		+x
		-1)
		+x y)(\mathcal{G}^{\symbolMom}_{\forest})^6
		+ (q^3 x^2 y (x y
		+x
		-1) (x (x y (y
		+4)
		+x
		-5 (y
		+1))
		+4)
		+q^2 (x (x (-(x^3 (y
		+1) (y (y
		+2) (y^2
		+1)
		+1))
		+x^2 (y
		+1)^2 (y (5 y
		+3)
		+5)
		-2 x (y
		+1) (y (5 y
		+7)
		+5)
		+y (10 y
		+19)
		+10)
		-5 (y
		+1))
		+1)
		-q (x^2 (y^2
		+y
		+1)
		-2 x (y
		+1)
		+1) (x (x (x (y
		+1) (y (y
		+3)
		+1)
		-y (3 y
		+8)
		-3)
		+3 (y
		+1))
		-1)
		-x^2 y (x y
		+x
		-1) (x^2 (y^2
		+y
		+1)
		-2 x (y
		+1)
		+1))(\mathcal{G}^{\symbolMom}_{\forest})^5
		+ (q^3 x^2 y (x^2 (y (3 y
		+7)
		+3)
		-9 x (y
		+1)
		+6)
		-q^2 (x (x (5 x^2 (y
		+1)^2 (y^2
		+y
		+1)
		-x (y
		+1) (y (20 y
		+31)
		+20)
		+30 y^2
		+57 y
		+30)
		-20 (y
		+1))
		+5)
		-q (x (x (x^2 (y (y (y (5 y
		+21)
		+31)
		+21)
		+5)
		-4 x (y
		+1) (y (5 y
		+11)
		+5)
		+30 y^2
		+63 y
		+30)
		-20 (y
		+1))
		+5)
		-x^2 y (x^2 (y (3 y
		+5)
		+3)
		-6 x (y
		+1)
		+3))(\mathcal{G}^{\symbolMom}_{\forest})^4
		+(-10 q (q
		+1) x^3 y^3
		+(q
		+1) x^2 y^2 ((q (3 q
		-29)
		-3) x
		+30 q)
		+x y ((q
		+1) (q (3 q
		-29)
		-3) x^2
		+(q (q (57
		-4 q)
		+63)
		+3) x
		-30 q (q
		+1))
		-10 q (q
		+1) (x
		-1)^3)(\mathcal{G}^{\symbolMom}_{\forest})^3
		+ (q
		+1) (q^2 x^2 y
		-10 q (x y
		+x
		-1)^2
		-x^2 y)(\mathcal{G}^{\symbolMom}_{\forest})^2
		-5 q (q
		+1) (x y
		+x
		-1)\mathcal{G}^{\symbolMom}_{\forest}
		-q (q
		+1)\,.
	\end{autobreak}
\end{align}
	
	\printbibliography[heading=bibintoc]	
\end{document}